\documentclass[11pt,a4paper]{article}

\usepackage[left=2cm,right=2cm,top=2cm,bottom=2cm]{geometry}
\usepackage{amsmath} 
\usepackage[utf8]{inputenc}
\usepackage[T1]{fontenc}
\usepackage{amsmath}
\usepackage{amsthm}
\usepackage{amsfonts}
\usepackage{hyperref}
\usepackage{subcaption}
\usepackage{graphicx}
\usepackage{dsfont} 
\usepackage{booktabs}
\usepackage{makecell}
\usepackage{multirow}

\usepackage{amssymb}
\usepackage{cite}

\theoremstyle{definition}

\theoremstyle{remark}

\author{Sommer M\textsuperscript{(a)}, Fenske N\textsuperscript{(a)}, Heumann C\textsuperscript{(b)}, Scholz-Kreisel P\textsuperscript{(a)}, Heinzl F\textsuperscript{(a)}}
\date{} 
\title{Methods to derive uncertainty intervals for lifetime risks for lung cancer related to occupational radon exposure}
\begin{document}
\maketitle
Affiliations: \\

\textsuperscript{(a)}Federal Office for Radiation Protection \\
Ingolstädter Landstrasse 1 \\
85764 Oberschleissheim Neuherberg \\
Germany. \\

\textsuperscript{(b)}Department of Statistics \\
LMU Munich \\
Ludwigstrasse 33 \\
80539 Munich \\
Germany.

\newpage
%
\section*{Abstract}
\subsection*{Introduction}
Lifetime risks are a useful tool in quantifying health risks related to radiation exposure and play an important role in the radiation detriment and – in the case of radon – for radon dose conversion. This study considers the lifetime risk of dying from lung cancer related to occupational radon exposure. For this purpose, in addition to other risk measures, the lifetime excess absolute risk (LEAR), is mainly examined. Uncertainty intervals for such lifetime risk estimates are only hardly presented in the literature.

\subsection*{Objectives}
The objective of this article is to derive and discuss uncertainty intervals for lifetime risk estimates for lung cancer related to occupational radon exposure.
\subsection*{Methods}
Based on previous work, uncertainties of two main components of lifetime risk calculations are modelled: uncertainties of risk model parameter estimates describing the excess relative risk for lung cancer and of baseline mortality rates. Approximate normality assumption (ANA) methods derived from likelihood theory and Bayesian techniques are employed to quantify uncertainty in risk model parameters. The derived methods are applied to risk models from the German "Wismut" uranium miners cohort study (full Wismut cohort with follow-up 2018 and  1960+ sub-cohort with miners first hired in 1960 or later). Mortality rate uncertainty is assessed based on information from the WHO mortality database. All uncertainty assessment methods are realized with Monte Carlo simulations. Resulting uncertainty intervals for different lifetime risk measures are compared.

\subsection*{Results}
Uncertainty from risk model parameters imposes the largest uncertainty on lifetime risks but baseline lung cancer mortality rate uncertainty is also substantial. Using the ANA method accounting for uncertainty in risk model parameter estimates, the LEAR in \% for the 1960+ sub-cohort risk model was 6.70 with a 95\% uncertainty interval of [3.26; 12.28] for the exposure scenario of 2 Working level Months from age 18-64 years, compared to the full cohort risk model with a LEAR in \% of 3.43 and narrower 95\% uncertainty interval [2.06; 4.84]. ANA methods and Bayesian techniques with a non-informative prior yield similar results. There are only minor differences across different lifetime risk measures.

\subsection*{Conclusion}
Based on the present results, risk model parameter uncertainty accounts for a substantial and sufficient share of lifetime risk uncertainty for radon protection. ANA methods are the most practicable and should be employed in the majority of cases. The explicit choice of lifetime risk measures is negligible. The derived uncertainty intervals are comparable to the range of lifetime risk estimates from uranium miners studies in the literature. These findings should be accounted for when developing radiation protection policies which are based on lifetime risks. \\

Keywords: radon, lung cancer, epidemiology, uncertainty, risk analysis

\clearpage

\section{Introduction}
Lifetime risks describe the probability of developing or dying from a specific disease (here: lung cancer death related to radon exposure) in the course of one’s lifetime and play an important role in the epidemiological approach for radon dose conversion \cite{ICRP65_1993, ICRP103_2007, ICRP115_2010} and radiation detriment \cite{ICRP_det}. Valid lifetime risk estimates are crucial for effective radiation protection strategies. A lifetime risk estimate depends on several components, each imposing possible errors or uncertainties on the result. Lifetime lung cancer risk estimates related to radon exposure depend on (a) the exposure scenario, (b) baseline mortality rates for all causes of death and lung cancer, and (c) complex risk models describing the shape of the exposure-response relationship between radon exposure and lung cancer mortality. For occupational radon exposure, these risk models are derived from uranium miners cohorts. We focus on the latter two components (b) and (c) in the upcoming analysis, as own previous work \cite{Sommer_2024_Sensi} demonstrated their major influence on lifetime risk estimates. \\

Exposure to radon and radon progeny is a recognized leading cause of lung cancer. This association has been confirmed in uranium miners and residential radon studies  \cite{ BEIR_VI,UNSCEAR_2020_AnnexB, Kreuzer_2018}. Both uranium miners and residential radon studies have demonstrated a linear relationship between radon exposure and lung cancer risk, which is in case of uranium miners studies additionally influenced by factors like age, time since exposure, and exposure rate. The intricate nature of these models can make comparisons across different cohorts challenging. Lifetime risk estimates offer a valuable tool for comparison and interpretation of risk models and enable clearer public risk communication. However, current literature often lacks uncertainty intervals for lifetime risk estimates. Lifetime Excess Absolute Risk (LEAR) estimates derived from various international miners studies range from $2.5 \times 10^{-4}$ to $9.2 \times 10^{-4}$ per working level month (WLM) as reported by the large pooled uranium miners study PUMA \cite{Kreuzer2024_PUMALEAR}. The present study employs advanced statistical methods to quantify lifetime risk uncertainty with 95\% uncertainty intervals. \\

Uncertainty intervals together with point estimates provide a more complete and more nuanced picture, allowing for informed decision-making, meaningful comparisons, and transparent communication. Uncertainty assessment for lifetime risks is a complex endeavor because final lifetime risk estimates are composite quantities depending on multiple, independently derived results. \\

The literature addressing lifetime risk estimation uncertainty for lung cancer related to radon exposure based on uranium miners cohorts is limited and generally does not prioritize uncertainty quantification \cite{Tomasek_2020, EPA_2003}. Existing studies typically employ Monte Carlo simulation techniques, incorporating various distributional assumptions for calculation components. For example, \cite{Tomasek_2020} investigated the LEAR and assumed a multivariate normal distribution for risk model parameters, while in the EPA Report \cite{EPA_2003} additionally a complex distribution for average residential radon exposure is assumed. A comparison of uncertainties across different lifetime risk measures could not be found in the literature. Uncertainties in risk measure estimates similar to lifetime lung cancer risks due to radon exposure, like “attributable cases”, are better understood: In the BEIR VI report \cite{BEIR_VI}, Monte Carlo simulations are used to comprehensively quantify uncertainties for “attributable cases”, similar to methods for lifetime lung cancer risks in \cite{Tomasek_2020}. \\ 

Several software tools have been developed for calculating lifetime cancer risks and associated uncertainties. These tools, however, primarily rely on risk models derived from the Atomic Bomb Survivors Life-Span Study (LSS) \cite{Preston2007}, which involves acute external exposure to different radiation types, fundamentally different from the chronic internal exposure to radon progeny in occupational uranium miners studies. Notable among these tools are “CONFIDENCE” \cite{2020_Walsh}, “RadRAT” \cite{Berrington_de_Gonzalez2012-ts}, and “LARisk” \cite{Lee_2022}. “CONFIDENCE” is a European software designed for cancer risk assessment post-radiation exposure from nuclear accidents. “RadRAT”, a free tool based on the BEIR VII report \cite{BEIR_VII}, estimates lifetime cancer risks for the US population based on a user-specific exposure scenario. “LARisk” extends “RadRAT” by adding flexibility, such as modified baseline incidence rates, to tailor risk calculations for specific populations. While “CONFIDENCE” and “RadRAT” rely on Monte Carlo simulations and sampling from probability distributions for uncertainty assessment, “LARisk” employs parametric bootstrap methods. \\

Despite these advanced tools, there is a notable gap in the literature regarding the uncertainty quantification of lifetime risks for lung cancer specifically related to occupational radon exposure. Addressing this, our study places special emphasis on refining the understanding of these uncertainties, utilizing risk models derived from uranium miners cohorts. We specifically focus on the LEAR measure while also examining alternative excess lifetime risk measures: Risk of Exposure Induced Death (REID), Excess Lifetime Risk (ELR), \cite{Thomas1992,Kellerer_2001} and Radiation-Attributed Decrease of Survival (RADS) \cite{Ulanowski2019}. We elaborate on sources of uncertainty for lifetime risks, discuss existing techniques, and introduce advanced methods to quantify these uncertainties by calculating 95\% uncertainty intervals. \\

To derive such intervals, both frequentist methods derived from likelihood theory and Bayesian techniques, chosen for their flexibility, are employed to quantify uncertainty in risk model parameter estimates. These methods are applied to risk models derived from the German Wismut uranium miners cohort study as a practical example. Mortality rate uncertainty is assessed based on information from the WHO mortality database \cite{WHO_mortality_database}. All uncertainty assessment methods are realized through Monte Carlo simulations. \\

By systematically analyzing and advancing the methodologies for quantifying uncertainties in lifetime risk estimates, this study aims to contribute to a more stable basis for radon protection strategies and provide a more comprehensive assessment of lifetime risks, especially for lung cancer related to occupational radon exposure.

\section{Methods}

\subsection{Lifetime risk definition and computation}
For a lifetime risk measure of choice, we employ the Lifetime Excess Absolute Risk (LEAR). This measure is often used \cite{Vaeth1990,Kellerer_2001,tomasek_dose_2008,Tomasek_2020,Kreuzer_2023_lc} and is defined as 
\begin{equation}
\label{LEAR_eq}
    LEAR_E=LR_E-LR_0=\int_0^{\infty} \left( r_E(t)-r_0(t)\right)S(t ) \, dt = \int_0^{\infty} r_0(t) ERR(t; \Theta) S(t) \, dt.
\end{equation}
The lifetime risk of dying from a specific disease (here: lung cancer) under exposure (here: radon and radon progeny) is $LR_E=\int_0^{\infty}r_E(t) S(t)\,dt$ and $LR_0=\int_0^{\infty} r_0(t) S(t) \, dt$ is the corresponding baseline lifetime risk without exposure. $r_0(t)$ is the baseline lung cancer mortality rate and $r_E(t)$ is the lung cancer mortality rate under exposure at age $t$. $S(t)=\mathbb{P}\left( T \geq t\right)$ is the survival function and describes the probability to attain age $t$ with $ T \geq 0$ the unknown random retention time until death. The survival function is modeled as $S(t)=e^{-\int_0^t q_0(u) \, du}$ with baseline mortality rates $q_0(u)$ at age $u$ for overall death (all-cause mortality rates). $ERR(t; \Theta)$ specifies the excess relative risk at age $t$ with unknown parameter set $\Theta$. Overall, we assume the following risk projection model:
\begin{equation}
    r_E(t)=r_0(t)\left( 1+ERR(t; \Theta)\right).
\end{equation}
The exact structure and complexity of the $ERR(t):=ERR(t; \Theta)$ term depends, in addition to age $t$, on the chosen risk model with parameters $\Theta$ and its included effect-modifying variables. The $LEAR$ from equation (\ref{LEAR_eq}) is estimated and finally calculated with the approximation \begin{equation}
    LEAR \approx \sum_{t=0}^{t_{max}}r_0(t) ERR(t; \hat{\Theta}) \Tilde{S}(t).
\end{equation}
For lifetime risk calculations, parameter estimates $\hat{\Theta}$ are plugged into the corresponding risk model structure. $\Tilde{S}(t)=e^{-\sum_{u=0}^{t-1} q_0(u)}$ approximates the true survival function $S(t)$ and is based on the Nelson-Aalen estimator of the cumulative hazard rate \cite{Nelson_1972}. The maximum age $t_{max}$ was set to $t_{max}=94$ for comparability with previous studies on lifetime risks \cite{tomasek_dose_2008, Kreuzer_2023_lc}. \\

A lifetime risk estimate further depends on the chosen risk model shaping $ERR(t; \Theta)$, the mortality rates $r_0(t),q_0(t)$ for all ages $0 \leq t \leq t_{max}$ and the assumed radon exposure in WLM for all ages $0 \leq t \leq t_{max}$ (exposure scenario). In line with all considered risk models, the latency time $L$ in risk models, i.e. the minimal time between age at exposure and age at lung cancer risk amplification, was chosen as $L=5$ years. \\

Here, a generic lifetime risk estimate for a given risk model without uncertainty quantification is always calculated with an exposure scenario of 2 WLM from age 18-64 years (94 WLM total cumulative exposure over lifetime) to represent an occupational scenario, in line with the literature \cite{Tomasek_mines_2008,Tomasek_2020,Kreuzer_2023_lc} and fixed mortality rates $r_0^{(ICRP)}(t), q_0^{(ICRP)}(t)$ for all ages $0 \leq t \leq 94$ from the Euro-American-Asian mixed reference population as presented in ICRP publication 103 \cite{ICRP103_2007}. Such a lifetime risk estimate does not reflect uncertainties and is referred to as “reference estimate (ICRP 103)” or simply “ref. estimate” in the following. Further, for enhanced readability, lifetime excess risk estimates such as LEAR are presented as percentages (LEAR in \%, i.e. LEAR $\times 100$). 

\subsection{Data and risk models}
To estimate the risk models associated with the German uranium miners cohort, this analysis employs cohort data, including cumulative radon exposure, age, and other variables of individual miners. This cohort data provides the basis for modeling and assessing the developed parametric (continuous) and categorical risk models, particularly for the quantification of uncertainties. All considered risk models include unknown parameters (indicated by Greek letters) which are estimated using Maximum-Likelihood methods. Access to the German uranium miners cohort data allows for re-fitting the subsequent described risk models and thereby for an uncertainty assessment of the resulting lifetime risks. \\

The following parametric (continuous) risk models are considered:
\begin{align}
ERR(t;  \beta_1, \beta_2, \dots , \beta_6 ,\alpha, \varepsilon) &= \sum_{j=1}^6 er_j \beta_j W(t) \exp\left\{ \alpha \left(AME(t)-30)\right)+\varepsilon\left(TME(t)-20\right)\right\}, \label{model3_para_full} \\
     ERR(t;  \beta,\alpha, \varepsilon) &= \beta W(t) \exp\left\{ \alpha \left(AME(t)-30)\right)+\varepsilon\left(TME(t)-20\right)\right\}, \label{model2_para_sublinear}\\ 
    ERR(t; \beta ) &= \beta W(t),  \label{model1_para_linear} 
\end{align}
with cumulative radon exposure $W(t)$ at age $t$ in WLM and continuous effect-modifying variables age at median exposure $AME(t)$ at age $t$ in years, time since median exposure $TME(t)$ at age $t$ in years, and binary variables $er_j$ for $j=1, \dots , 6$ for six categories of exposure rate at age $t$ in units of working level (WL). Further, categorical BEIR VI exposure-age-concentration models (cf. \cite{BEIR_VI}) are considered:
\begin{equation}
\resizebox{.9\hsize}{!}{$
   ERR(t; \theta^{(1)} , \phi_{age},\gamma_{rate}, \beta )= \beta \left( \theta_{5-14}W_{5-14}+\theta_{15-24}W_{15-24}+\theta_{25-34}W_{25-34}+\theta_{35+}W_{35+}\right)\phi_{age}\gamma_{rate}$}, \label{model5_BEIR_VI_full}
\end{equation}
\begin{equation}
     ERR(t; \theta^{(2)} , \phi_{age},\gamma_{rate}, \beta )= \beta \left( \theta_{5-14}W_{5-14}+\theta_{15-24}W_{15-24}+\theta_{25+}W_{25+}\right)\phi_{age}\gamma_{rate}, \label{model4_BEIR_VI_sub}
\end{equation}
with  $\theta^{(1)}=\left( \theta_{5-14},\theta_{15-24},\theta_{25-34},\theta_{35+}\right)$, $\theta^{(2)}=\left( \theta_{5-14},\theta_{15-24},\theta_{25+}\right)$, and where $W_{5-14}$, $W_{15-24}$, $W_{25-34}$, $W_{25+}$, $W_{35+}$ is the cumulative radon exposure in WLM in windows 5-14, 15-24, 25-34, 25+ or 35+ years ago and $\phi_{age}$ and $\gamma_z$ are factors for attained age in years and exposure rate in WL, respectively. \\

All presented risk models are derived from the German Wismut cohort of uranium miners with follow-up 2018 and have been initially introduced in \cite{Kreuzer_2023_lc}. These models are used in this study as an application example of the subsequent introduced methods to quantify risk model parameter uncertainty. The models (\ref{model2_para_sublinear}) “Parametric 1960+ sub-cohort”, (\ref{model1_para_linear}) “Simple linear 1960+ sub-cohort”, and (\ref{model4_BEIR_VI_sub}) “BEIR VI 1960+ sub-cohort” are derived from the Wismut sub-cohort with miners hired in 1960 or later (1960+ sub-cohort) and the models “Parametric full cohort” (\ref{model3_para_full}) and “BEIR VI full cohort” (\ref{model5_BEIR_VI_full}) are derived from the corresponding full Wismut cohort. The derived parameter estimates are shown partly in the Supplement or can be inspected in the source paper \cite{Kreuzer_2023_lc}. The simple linear model (\ref{model1_para_linear}) is included in this analysis for comparability. All other models were chosen because they showed the best fits for the corresponding cohort. The chosen risk models (categorical / BEIR VI structure, parametric / continuous with effect modifying variables, and simple linear risk model) offer a diverse range of risk models. Here, the terms “categorical” and “parametric / continuous” refer to the categorical or continuous nature of the effect-modifying variables. \\

All considered risk model parameters are estimated with maximum likelihood (ML) methods based on internal Poisson regression applied to grouped cohort data. The corresponding Likelihood function is based on the assumption that the number of lung cancer deaths $C_i$ in cell $i$ with $i=1, \dots , n$ are Poisson-distributed via
\begin{equation}
    C_i \sim Poi \left(PY_i e^{\delta_1 + \sum_{k=2}^K \delta_k \mathds{1}_{\{k\}}(x_i)}\left(1+ERR_i\left(\Theta\right)\right)\right) \label{Poi_ass}
\end{equation}
with offset for person-years at risk $PY_i$, excess relative risk $ERR_i\left(\Theta\right)$ and unknown parameter vector $\Theta$. The specific shape of $ERR_i\left( \Theta \right)$ depends on the prescribed risk model structure. The baseline risk predictor $\eta_i = \delta_1 + \sum_{k=2}^K \delta_k \mathds{1}_{\{k\}}(x_i)$ describes the stratified baseline with $K$ levels, where $x_i$ is a categorical variable with $K$ levels. Each level represents a unique combination of conditions or classifications that are relevant variables in assessing the baseline risk. Here, the levels correspond to different groups categorized by age, calendar year, and duration of employment. Setting $\Delta=\left( \delta_1, \delta_2 , \dots , \delta_K\right)$, the likelihood model consists of the parameters $\Omega = \left(\Delta, \Theta\right)$. 

\subsection{Uncertainty assessment}
\subsubsection{Uncertainty intervals for lifetime risks}
Uncertainties for a parameter of interest are quantified by deriving 95\% uncertainty intervals, which is a range of values that is calculated to cover the true unknown parameter value with 95\% certainty. It consists of a lower and an upper bound, enclosed in parentheses. The precise interpretation of the uncertainty interval depends on the underlying statistical inference system (frequentist or Bayesian). Lifetime risks are composite quantities depending on multiple, independently derived results. Hence, we rely on sampling techniques to derive lifetime risk uncertainty intervals. \\

Here, uncertainties are determined by quantifying the variability in statistical estimates that results from drawing a sample from the entire population (sampling uncertainty). Risk model parameter uncertainties and mortality rate uncertainties are considered here. Depending on the investigated lifetime risk calculation component, previously fixed values for calculation components are replaced by random variables. The uncertainty quantification is carried out with Monte Carlo simulations. Here, we focus on the excess lifetime risk measure LEAR. N = 100,000 samples from the underlying assumed probability distribution are drawn independently and a LEAR is calculated for each sample resulting in N independent LEAR estimate samples. The two-sided 95\% uncertainty interval is the span of observed LEAR samples by disregarding the 2.5\% lowest and 2.5\% highest samples. All upcoming methods use this approach to derive uncertainty intervals unless explicitly stated otherwise, but differ in the calculation component analyzed and the assumed probability distribution. Corresponding probability density functions for the LEAR distribution are derived from the histogram of LEAR samples with a kernel density estimate. Note that for lifetime risks, we additionally present the relative span of the uncertainty interval in brackets, calculated as the interval span divided by the reference estimate (relative uncertainty span). This enables easier comparison across various estimates irrespective of their absolute values.

\subsubsection{Mortality rates}
\label{section:methods_mr_uncertainty}
Uncertainties in the mortality rates $r_0(t), q_0(t)$ are assessed by assuming gamma distributions for all ages $t$, 
\begin{align}
    r_0(t) &\sim G\left(a^{(r_0)}_t,b^{(r_0)}_t\right), \label{mr_r0_ass}\\
    q_0(t) &\sim G\left(a^{(q_0)}_t,b^{(q_0)}_t\right). \label{mr_q0_ass} 
\end{align}
with age-dependent shape parameters $a^{(r_0)}_t,a^{(q_0)}_t$ and rate parameter $b^{(r_0)}_t,b^{(q_0)}_t$ (SC Table \ref{tab:WHO_r0_q0_estimates_5EAA}). The parameter estimates $a^{(r_0)}_t,b^{(r_0)}_t,a^{(q_0)}_t,b^{(r_0)}_t$ are derived from data from the WHO Mortality Database \cite{WHO_mortality_database} with maximum-likelihood (ML) methods. Observations from all available countries from Europe, America, and Asia for females and males from the years 2001, 2006, 2011, 2016, and 2021 are used. The derivation of probability distributions for mortality rates in (\ref{mr_r0_ass}) and (\ref{mr_q0_ass}) from WHO data is independent of the ICRP Euro-American-Asian reference mortality rates $r_0^((ICRP))(t)$ and $q_0^((ICRP))(t)$. However, the geographical alignment of the chosen WHO data with the ICRP mortality rates ensures the appropriateness of the data for estimating mortality rate variability. An observation used for fitting models (\ref{mr_r0_ass}) and (\ref{mr_q0_ass}) is characterized as the mortality rate $\frac{d}{n}$ where $d$ are the number of lung cancer deaths for $r_0$ or all-cause deaths for $q_0$ and $n$ is the mid-year population size. Each observation is uniquely defined by a specific country (out of 153 countries), sex, and calendar year. Only observations with a positive number of individuals at risk $n$ and a positive number of cases $d$ were considered. To obtain uncertainty intervals for lifetime risks, independent samples from the above described gamma distributions (\ref{mr_r0_ass}) and (\ref{mr_q0_ass}) are drawn (Monte Carlo simulation), analogously to drawing samples in the ANA approach for risk model parameters.

\subsubsection{Risk models - ANA approach}
\label{section:methods_rm_uncertainty_ml}
The idea behind the Approximate normality assumption (ANA) approach is the following: Risk model parameter estimates $\hat{\Theta}_0$ are obtained by fitting the $ERR(t;\Theta)$ model to miners cohort data with ML methods assuming Poisson-distributed numbers of lung cancer cases (\ref{Poi_ass}). The ML method provides statistically efficient estimates, meaning they tend to be close to the true values on average given a sufficiently large sample size. The estimated parameters are subject to sampling uncertainty. By statistical theory, under mild regularity conditions met in our case \cite{Amemiya1985-wn}, the parameter estimator $\hat{\Theta}$ in a given risk model is asymptotically (for an infinitely large cohort size) normally distributed with expectation equal to the unknown true parameter value $\Theta$ and covariance matrix equal to the inverse Fisher information, see \cite[Equation (5.14)]{Held2021-wg} or \cite[Theorem 9.7]{In_all_Likelihood_Pawitan_2013}. The true parameter $\Theta$ is approximated with the cohort-specific maximum likelihood estimate (MLE) $\hat{\Theta}_0$. The inverse Fisher information can be approximated by an estimate $\hat{\Sigma}_0$ of the parameter covariance matrix. \\

The ANA approach follows the frequentist inference where the true risk model parameter $\Theta$ is treated as a fixed but unknown value. The parameter estimator $\hat{\Theta}$ is considered as a random variable subject to variability depending on the specific sample used in the estimation process. The estimate $\hat{\Theta}_0$ is a realization of $\hat{\Theta}$ by applying specific sample data. Following this framework, risk model parameter uncertainty is quantified by assuming a multivariate-normal distribution on the parameter estimator $\hat{\Theta}$,
\begin{equation}
\label{mle_dist}
    \hat{\Theta} \sim \mathcal{N}\left( \hat{\Theta}_0, \hat{\Sigma}_0\right)
\end{equation}
with cohort-specific ML estimate $\hat{\Theta}_0$ and covariance matrix estimate $\hat{\Sigma}_0$. By generating a large number of samples from this approximate distribution, and calculating the corresponding lifetime risks for each sample, the distribution of lifetime risk estimates reflects the inherited parameter sampling uncertainty. This approach is called the “Approximate normality assumption” (ANA) approach. In this study, we employ the \textit{Epicure} software \cite{EPICURE} to obtain parameter estimates $\hat{\Theta}_0$  and their associated covariance matrix estimate $\hat{\Sigma}_0$. Note that this method implicitly incorporates knowledge from the estimation of baseline parameters $\Delta$, as the covariance estimate $\hat{\Sigma}_0$ is adjusted accordingly \cite{EPICURE}. Importantly, after this theoretical introduction, the manuscript simplifies notation by referring to estimates using $\hat{\Theta}$ without further distinguishing the estimator notation. Note that the ANA approach requires access to the ML estimate $\hat{\Theta}_0$ and covariance matrix estimate $\hat{\Sigma}_0$., but no access to underlying cohort data since a re-estimation of parameters is not necessary. 

\subsubsection{Risk models - Bayesian approach}
In contrast to frequentist inference underlying the above ANA approach, in the Bayesian approach (Bayesian inference) the unknown parameter $\Theta$ is interpreted as a random variable itself. To apply Bayesian statistics to account for prior information about the risk model parameters we assume the generic Bayesian framework
\begin{equation}
    P(\Omega \vert X) = \frac{P\left(\Omega\right) L(X)}{\int_{\Omega} L\left(X \vert \Omega \right) P\left( \Omega \right) \, d\Omega}  \propto P\left(\Omega\right) L(X),
\end{equation}
where $P(\Omega \vert X)$ is the posterior probability density function of observing the parameter $\Omega$ given cohort data $X$. $P\left(\Omega\right)$ is the prior probability density for $\Omega$ and $L(X)=L(X\vert \Omega)$ is the likelihood function. Here $\Omega=\left( \Delta, \Theta\right)$, compare (\ref{Poi_ass}). Using concepts from \cite{Higueras_2018} allows to derive the marginal posterior distribution $P(\Theta \vert X)$ for risk model parameters $\Theta$ of interest analytically. This requires assuming independence between prior distributions for $\Delta$ and $\Theta$ and a non-informative prior for $\Delta$. The resulting marginal posterior given cohort data $X$ reads 
\begin{align}
    P\left(\Theta \vert X \right) = \frac{P\left( \Theta\right) \left[ \prod_{i=1}^n \left(1+ERR_i\left( \Theta\right) \right)^{C_i}\right] \left[\sum_{i \vert x_i=1}PY_i\left(1+ERR_i\left(\Theta\right) \right)\right]^{-S_1}}{M \prod_{k=2}^K \left[\sum_{i \vert x_i = k} PY_i \left(1+ERR_i\left( \Theta \right) \right) \right]^{S_k}}
\end{align}
with lung cancer cases in strata $k$, $S_k=\sum_{i \vert x_i=k}C_i$ for $k=1, \dots , K$ and normalizing constant $M$. Parameter estimats are derived as the values that maximize the posterior distribution (mode), denoted as $Mod(P\left(\Theta \vert X \right))=\hat{\Theta}$, analogously to the ANA approach. Note that the Bayesian approach requires a re-estimation of parameters and therefore access to the original cohort data in contrast to the ANA approach. \\

This approach is applied to the 1960+ sub-cohort models (\ref{model1_para_linear}) and (\ref{model2_para_sublinear}). Identical model structures were used in \cite{Tomasek_mines_2008}, which we employ as prior information. Due to increasing computational complexity, it was not feasible to apply this approach to full cohort risk models (\ref{model3_para_full}), (\ref{model5_BEIR_VI_full}) and the sub-cohort model (\ref{model4_BEIR_VI_sub}), which involve larger cohort data size and/or more parameters. Non-informative, uniform prior distributions for $\Theta$ are applied to obtain the true marginal likelihood. Otherwise, the prior information about $\beta$ in the simple linear risk model (\ref{model1_para_linear}) is modelled with a gamma distribution $\beta \sim G\left(a, \frac{a-1}{\hat{\beta}_{CZ+F}}\right)$ with $\hat{\beta}_{CZ+F}=0.016$ from \cite{Tomasek_mines_2008}. The prior information about $\beta$, $\alpha$ and $\varepsilon$ for model (\ref{model2_para_sublinear}) is modelled as $\beta \sim G\left(a, \frac{a-1}{\hat{\beta}^{\prime}_{CZ+F}}\right)$, and normal distributions $\alpha \sim \mathcal{N}\left(\hat{\alpha}_{CZ+F}, \sigma^2\right)$, $\varepsilon \sim \mathcal{N}\left(\hat{\varepsilon}_{CZ+F}, \sigma^2\right)$ with $\hat{\beta}^{\prime}_{CZ+F}=0.042$, $\hat{\alpha}_{CZ+F}=-0.06539$, $\hat{\varepsilon}_{CZ+F}=-0.07985$ from \cite{Tomasek_mines_2008}. Adjustments to the certainty of the prior information are achieved by modifying the gamma shape parameter $a$ for $\eta$ and the standard deviation $\sigma$ for $\alpha$ and $\varepsilon$. . By construction, the modes of the marginal prior distributions align with the corresponding parameter estimate from \cite{Tomasek_mines_2008}. All three components are assumed to be independent in the prior, i.e. $P\left(\Theta\right)=P(\beta)\cdot P(\alpha) \cdot P(\varepsilon)$. Importantly, estimates from \cite{Tomasek_mines_2008} are not assumed as true “prior” knowledge, but illustrate integrating diverse cohort information using Bayesian methods (see Discussion). \\

For the simple linear model with (\ref{model1_para_linear}), Rejection Sampling \cite{Held2021-wg} was applied to obtain N = 100,000 samples of the posterior distribution of $\beta$ with uniform proposal distribution $\mathcal{U}(0,0.04)$. For the more complex risk model (\ref{model2_para_sublinear}), Markov Chain Monte Carlo (MCMC) techniques via the Metropolis-Hastings algorithm were applied \cite{Robert2005-rc}. The approximate multivariate normal distribution (\ref{mle_dist})) was chosen as the proposal distribution. The log acceptance ratio for a proposal sample for $\Theta \sim P(\Theta \vert X)$ was calculated as the difference of the log marginal posterior evaluated at the proposal sample and the current sample. The initial proposal sample for $\Theta$ was chosen in proximity of the cohort-specific ML estimate $\hat{\Theta}$. N = 100,000 samples of the posterior distribution are used for uncertainty assessment after generating 110,000 samples and discarding the first 10,000 samples to account for a burn-in period. Overall, the resulting MCMC sample paths of risk model parameters indicate rapid convergence to stationarity, likely due to an effective proposal distribution. Here, the presented 95\% LEAR uncertainty intervals are derived by choosing the narrowest interval that covers 95\% of the derived LEAR samples, which we refer to as the Highest Posterior Density Interval (HPDI).

\subsubsection{Joint effect of risk model and mortality rate uncertainty}

The joint effect of risk model parameter estimate uncertainty and mortality rate uncertainty on LEAR estimates is assessed by simultaneously sampling from gamma distributions for the mortality rates $r_0(t)$ for all ages $t$ according to (\ref{mr_r0_ass}), and from multivariate normal distributions for the risk model parameter estimator $\hat{\Theta}$ according to the ANA approach (\ref{mle_dist}), independently. Note that $q_0(t)$ variability is here not accounted for as initial analysis showed a negligible impact (see Results). Analogous to the general Monte Carlo simulation approach, N = 100,000 LEAR samples are calculated from N = 100,000 sets of sampled values for $r_0(t)$ for all ages $t$ and $\hat{\Theta}$. All computations were carried out with the statistical software \textit{R} \cite{RCore}.

\subsection{Sensitivity analysis}
The Supplemental Section \ref{comp_lr_measures_SDC_A} details uncertainties for other lifetime risk measures besides LEAR. Likewise, an explorative uncertainty assessment derived from a simple interpretation of lifetime risks based on the Kaplan-Meier estimator for survival curves \cite{Kaplan58} is explored (Suppl. Section \ref{kaplanmeier_curves_SDC_B}). \\

Sensitivity analyses (Suppl. Sections \ref{Risk_model_parameter_uncertainty_SDC_C} to \ref{Sex_specific_uncertainty_SDC_E}) explore the impact of assuming different probability distributions on lifetime risk uncertainties. Additional insights on risk model parameter uncertainty is shown in Suppl. Section \ref{Risk_model_parameter_uncertainty_simple_linear_SDC_C1} and \ref{Risk_model_parameter_uncertainty_loglinear_bayes_SDC_C3}. Furthermore, the Bayesian approach for risk model parameter uncertainty quantification is applied to the simple linear risk model (\ref{model1_para_linear}) with log-normal priors for $\beta$ (Suppl. Section \ref{Risk_model_parameter_uncertainty_simple_linear_bayes_SDC_C2}). Analyses for the specific influence of all-cause mortality rate uncertainties are conducted (Suppl. Section \ref{Mortality_rate_uncertainty_preliminaries_SDC_D1}). An alternative Bayesian approach to assess mortality rate uncertainty employing the WHO data is found in the Suppl. Section \ref{Mortality_rate_uncertainty_poisson_lc_deaths_SDC_D2}. Log-normal distributed mortality rates (Suppl. Section \ref{Mortality_rate_uncertainty_lognormal_rates_SDC_D3}) and sex-specific uncertainties regarding mortality rate and risk model effects are investigated (Suppl. Section \ref{Sex_specific_uncertainty_SDC_E}). Exposure scenario uncertainty is briefly investigated in the Suppl. Section \ref{Radon_exposure_uncertainty_SDC_F}.

\section{Results}
\subsection{Effect of mortality rates}
Introductory analyses revealed that all-cause mortality rates $q_0(t)$ impose considerably less uncertainty on the LEAR than lung cancer rates $r_0(t)$ for all risk models (Suppl. Section D.1). The empirical distribution of LEAR samples for gamma-distributed $q_0(t)$ is considerably narrower compared to the empirical distribution for gamma-distributed $r_0(t)$, which is also reflected in the 95\% uncertainty intervals. The relative uncertainty span of 95\% uncertainty intervals is very similar across all considered risk models with roughly 0.10 for uncertainties in all-cause rates $q_0(t)$ and roughly 0.50 for uncertainties in lung cancer rates $r_0(t)$ and joint uncertainties $r_0(t)$, $q_0(t)$, (Suppl. Table \ref{tab:LEAR_MR_intervals_LN_vs_G}). Hence, all-cause mortality rate uncertainty can be reasonably neglected when assessing overall mortality rate uncertainty, as addressed in the subsequent analyses.

\subsection{Effect of risk model parameter uncertainty}
LEAR estimates derived from full cohort risk models are notably lower than estimates derived with 1960+ sub-cohort models \cite{Kreuzer2024_PUMALEAR}. This discrepancy is particularly pronounced for the Wismut cohort and primarily comes from the significantly lower estimates for Excess Relative Risk per 100 WLM (ERR/100 WLM) observed in the full cohort \cite{Kreuzer_2023_lc}. The variation between these ERR/100 WLM estimates is similarly reflected in the LEAR calculations.

\subsubsection{Approximate normality assumption (ANA) approach}
\label{section:ANA_results}

Table \ref{tab:COVA_LEAR_KIs} shows the resulting 95\% uncertainty intervals.  For risk models fit on the full cohort the resulting intervals are comparable although the model structures differ considerably: The relative uncertainty span is 0.81 and 1.03 for the parametric (\ref{model3_para_full}) and the BEIR VI risk model (\ref{model5_BEIR_VI_full}), respectively. For the 1960+ sub-cohort models, however, the results vary relatively widely. Especially risk model (\ref{model4_BEIR_VI_sub}) implies very wide LEAR uncertainty intervals with a relative uncertainty span of 6.27 and a notable portion of implausible negative LEAR samples. The model (\ref{model2_para_sublinear}) implies considerably less uncertainty compared to the model (\ref{model4_BEIR_VI_sub}) with a relative uncertainty span of 1.34, although both models are derived from the 1960+ sub-cohort. Visually (Figure \ref{fig:LEAR_ERR_MR_Histogramm}), the empirical distribution of LEAR samples resembles approximately a normal distribution for risk model parameter estimates derived from the full cohort (models (\ref{model3_para_full}) and (\ref{model5_BEIR_VI_full})). Conversely, the empirical distribution of LEAR samples with parameter estimates derived from the 1960+ sub-cohort inherit considerably heavier tails and a slight right skewness. The 1960+ sub-cohort is smaller and comparably young with less person-years at risk and less lung cancer deaths. This results in higher statistical uncertainty which is reflected in wider LEAR uncertainty intervals. \\

Note that, for the simple linear risk model  (\ref{model1_para_linear}) we obtain analytically, without sampling, a normal distribution  
$$LEAR \sim \mathcal{N}\left(0.0571,1.65 \times 10^{-4}\right)$$ with corresponding 95\% uncertainty interval $[3.18; 8.22]$ and relative uncertainty span of $0.88$ for LEAR in \%. By definition, this uncertainty intervals is  proportional to the 95\% confidence intervals $[0.75; 1.93]$ for $\hat{\beta}_0 \times 100 = 1.34$ from \cite{Kreuzer_2023_lc} by a factor of 4.27 here. The underlying theory is explained in Suppl. Section \ref{Risk_model_parameter_uncertainty_simple_linear_SDC_C1}.  

\subsubsection{Bayesian approach}
\begin{figure}[htbp]
    \centering
    \includegraphics[scale=0.0925]{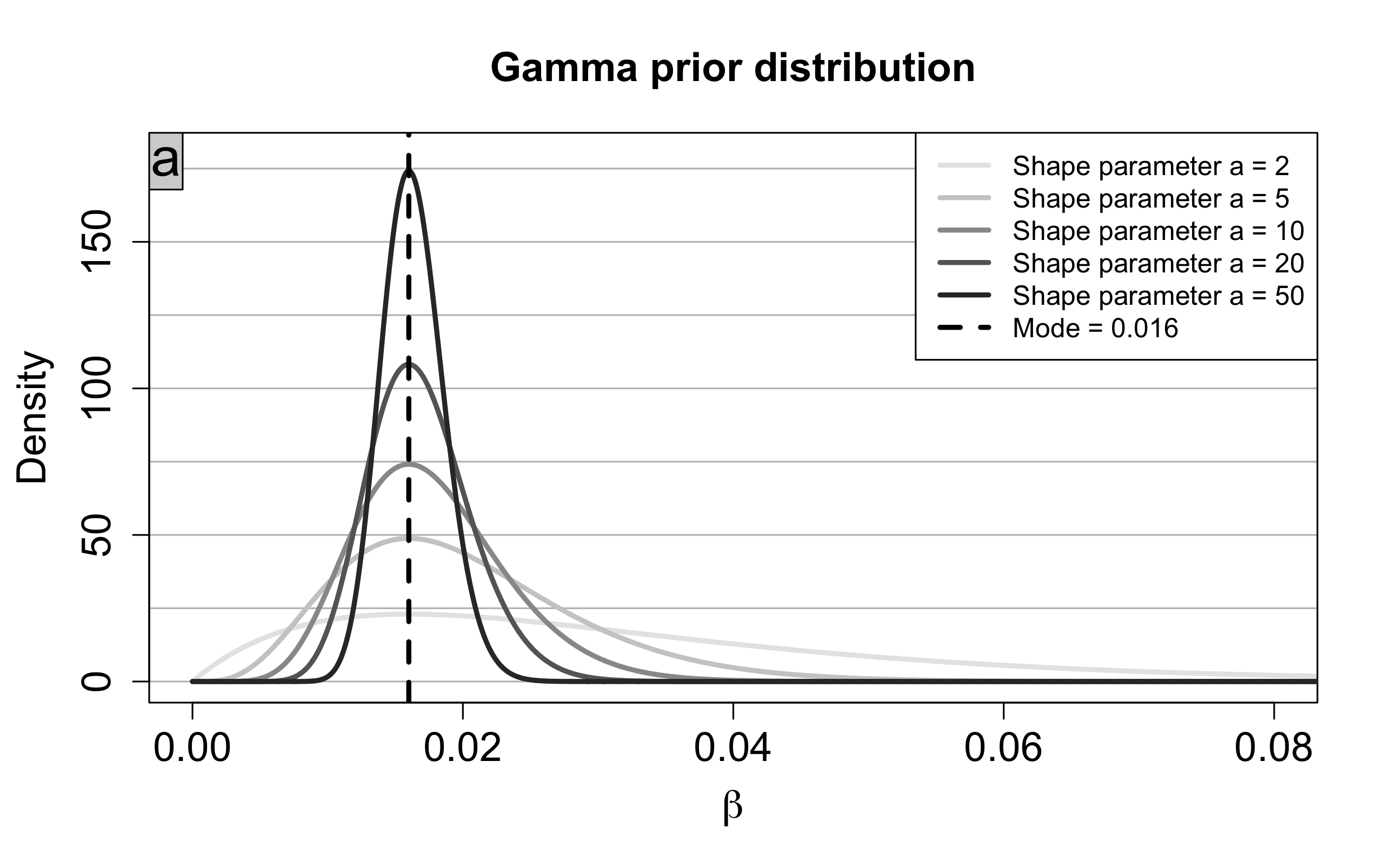}
    \includegraphics[scale=0.0925]{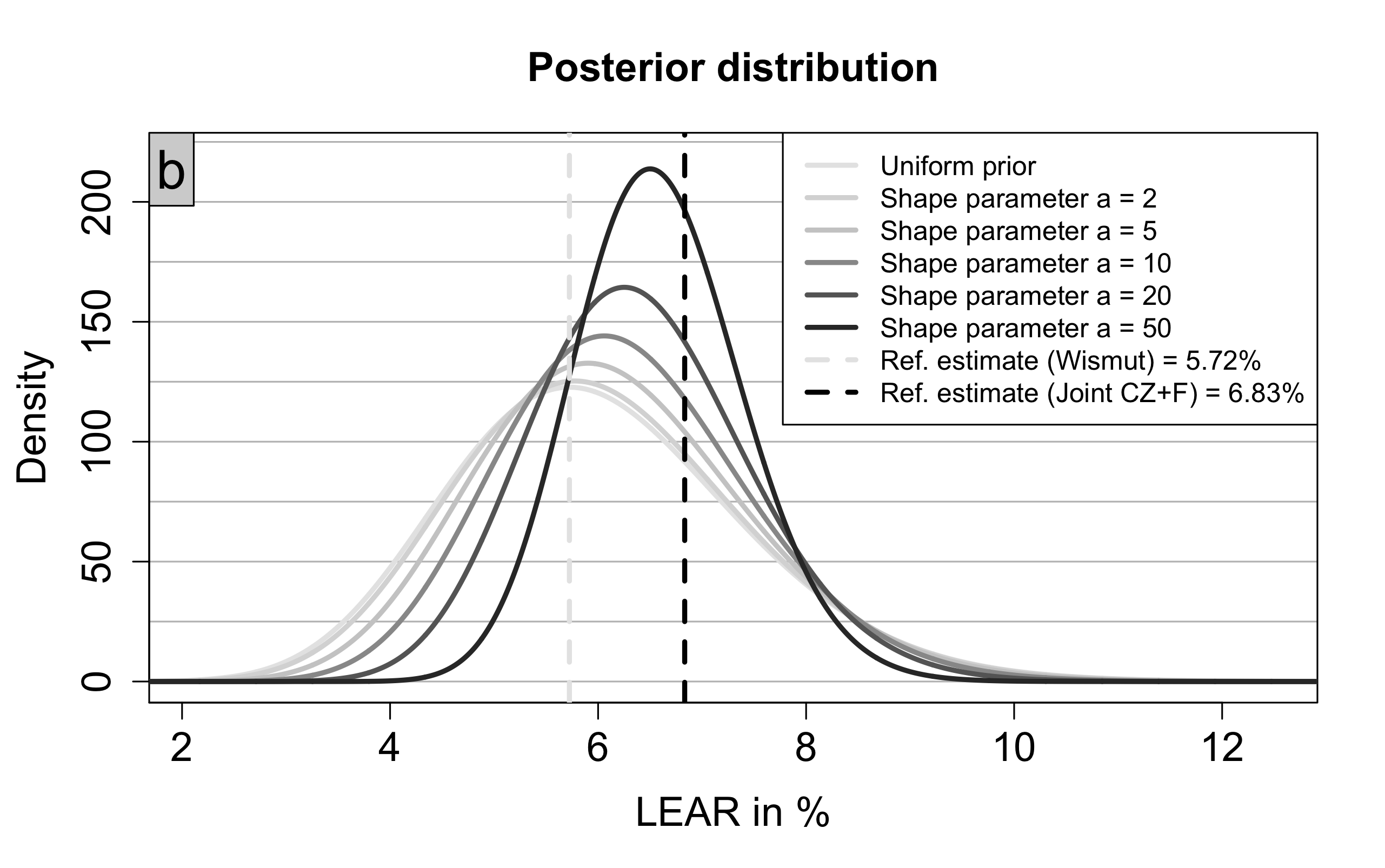}
    \caption{Gamma-distributed prior $P(\beta)$ centered at the parameter estimate from \cite{Tomasek_mines_2008} for different shape parameters $a$ and of corresponding posterior distributions LEAR for the simple linear 1960+ sub-cohort risk model (\ref{model1_para_linear}) in correspondence to Table \ref{tab:linERR_bayes_ki}. The vertical dashed lines represent the $\beta$ estimate from \cite{Tomasek_mines_2008} and the correspondingly derived reference LEAR estimate.}
    \label{fig:linERR_bayes_ki}
\end{figure}

Computational limitations restricted the applicability of this approach to models (\ref{model2_para_sublinear}) and (\ref{model1_para_linear}). For the simple linear model (\ref{model1_para_linear}), analogous to the ANA approach, LEAR uncertainty is directly proportional to $\beta$ uncertainty. The posterior mode of $P\left( \beta \vert X \right)$ with 95\% uncertainty intervals (HPDIs) for varying certainty in the prior information is shown in Table \ref{tab:linERR_bayes_ki} with corresponding plots in Figure \ref{fig:linERR_bayes_ki}. The resulting HPDI with a uniform prior are comparable to the conventional 95\% confidence interval for the estimate $\hat{\beta}$. However, the lower bound of the HPDI is notably higher than that of the classical confidence interval for $\hat{\beta}$. Further, the relative uncertainty span of the uncertainty interval decreases with increasing certainty in the prior information (increasing $a$) from 0.94 for $a=2$ up to 0.48 for $a=50$. \\

\begin{figure}[htbp]
    \centering
    \includegraphics[scale=0.0925]{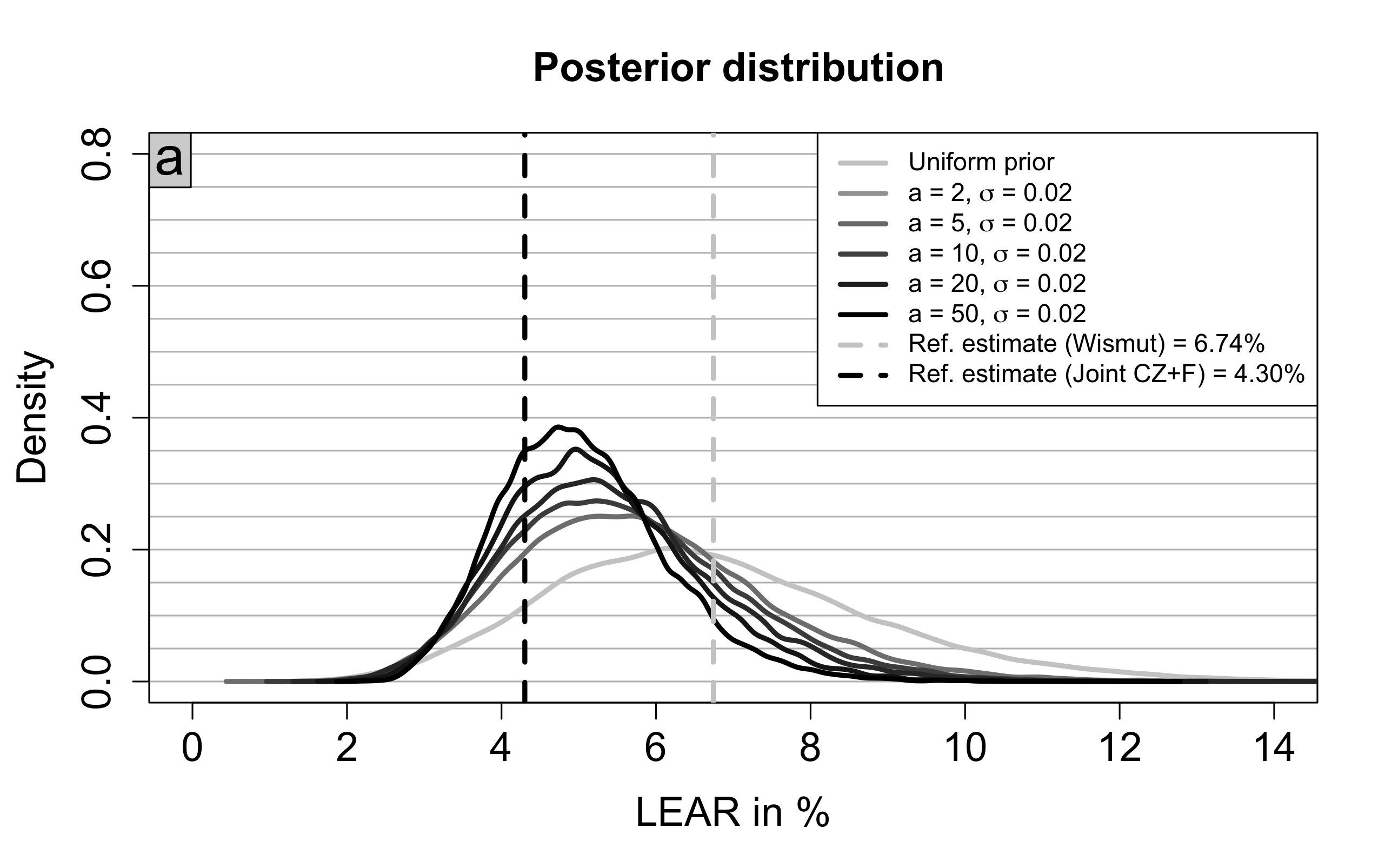}
    \includegraphics[scale=0.0925]{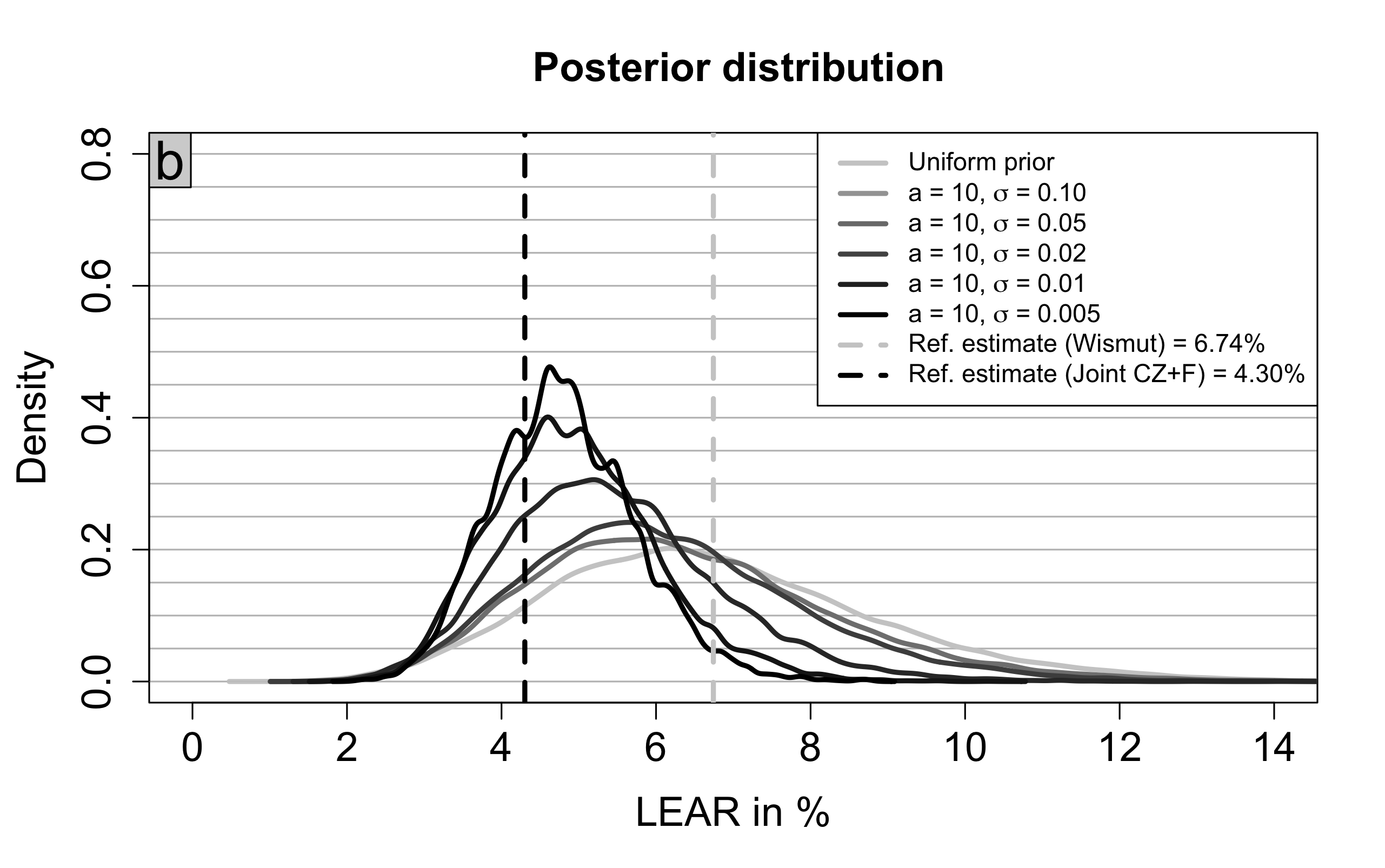}
    \caption{Results from Bayesian approach. Kernel density estimate of LEAR posterior distribution with uncertain risk model parameters $\Theta=\left(\beta, \alpha, \varepsilon\right)$ (model (\ref{model2_para_sublinear})) derived from sampling from the posterior distribution $P(\Theta \vert X)$ with prior $P(\Theta)=P(\beta)P(\alpha)P(\varepsilon)$  for gamma distributed $\beta$ and normally distributed $\alpha,\varepsilon$ centered at the corresponding parameter estimate from \cite{Tomasek_mines_2008} for different shape parameters $a$ and standard deviations $\sigma$ in correspondence to Table \ref{tab:linlogERR_bayes_ki}. The plots illustrate the effect of increasing prior certainty in parameter $\beta$ (Plot a) and in parameters $\alpha$, $\varepsilon$ (Plot b). Vertical dashed lines indicate the reference estimate (ICRP 103) with the risk model fit on the Wismut or the Joint CZ+F cohort, respectively.}
    \label{fig:linlogERR_bayes_full_LEAR}
\end{figure}

The resulting 95\% HPDIs for the more complex model (\ref{model2_para_sublinear}) are shown in Table \ref{tab:linlogERR_bayes_ki} for the special choice of a standard deviation $\sigma=0.02$ in the prior distributions for $\alpha$ and $\varepsilon$ (Suppl. Section \ref{Risk_model_parameter_uncertainty_loglinear_bayes_SDC_C3} for more choices for $\sigma$ and histograms of parameter samples). The corresponding distribution of LEAR samples is shown in Figure \ref{fig:linlogERR_bayes_full_LEAR} (Plot a). As expected, increasing the prior certainty (increasing shape parameter $a$) shrinks the uncertainty interval and moves the reference estimate (ICRP 103) of LEAR in \% from 6.74 (95\% HPDI [2.96; 11.09]) without prior influence towards the prior LEAR in \% estimate of 4.30 derived with the Joint Czech+French risk model. For example, $a=50$ results in a LEAR in \% of 4.99 and a 95\% HPDI of [3.14; 7.26]. A very similar effect is observed for decreasing the standard deviation $\sigma$ for a fixed shape parameter $a=10$ (Figure \ref{fig:linlogERR_bayes_full_LEAR} Plot b).

\subsection{Joint effect of mortality rate and risk model parameter uncertainty}

\begin{figure}[htbp]
    \centering
    \includegraphics[scale=0.0925]{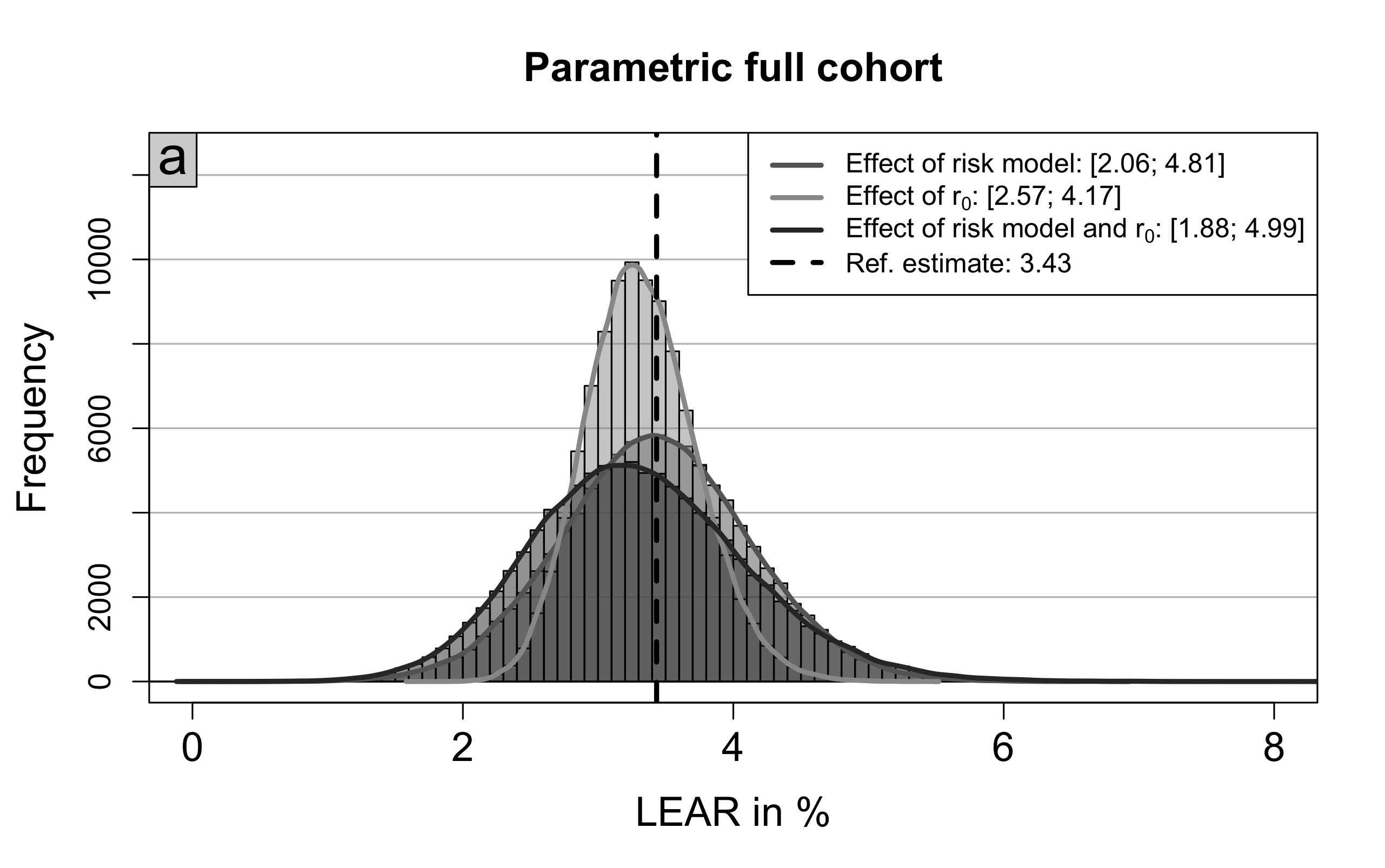}
    \includegraphics[scale=0.0925]{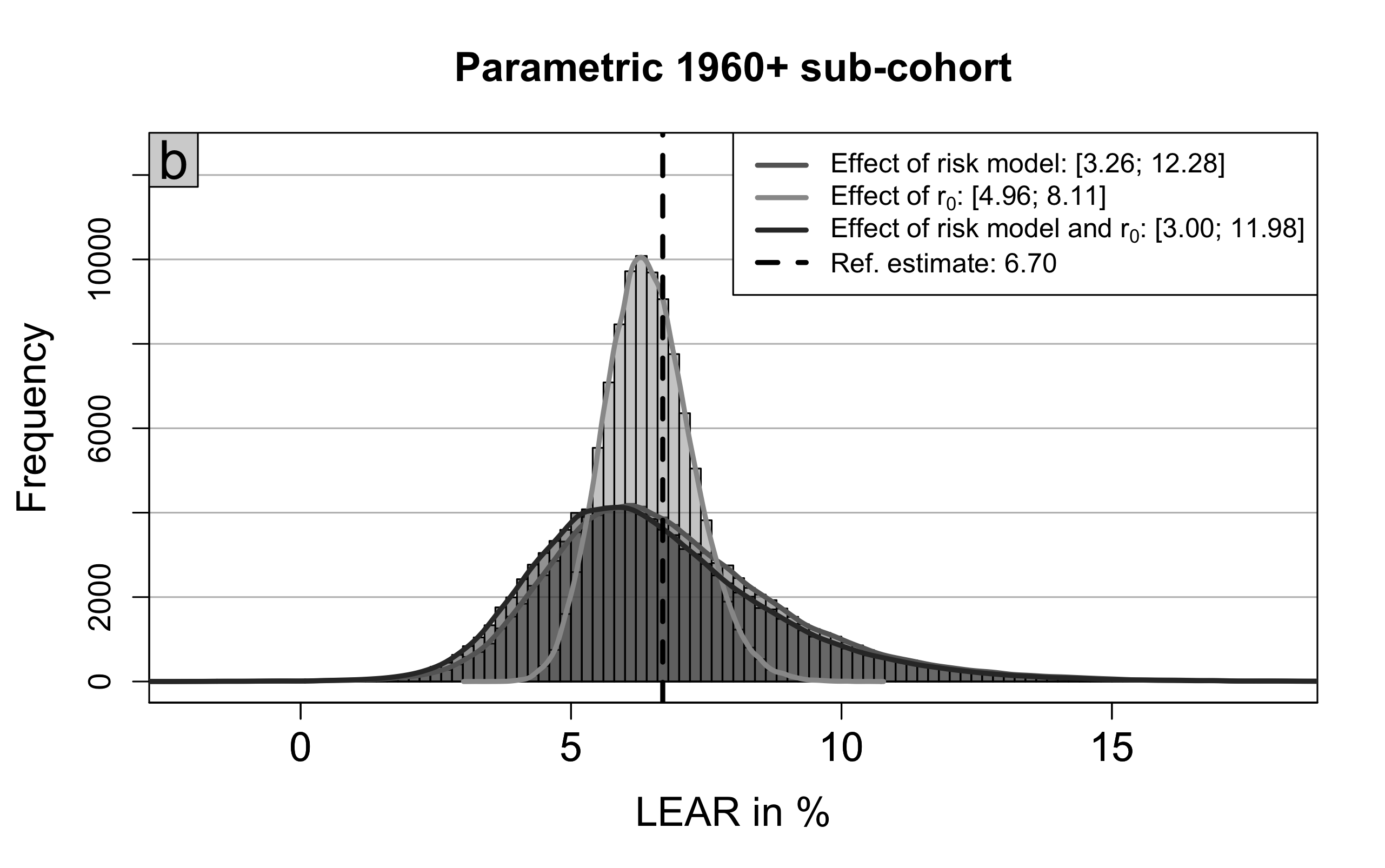}
    \includegraphics[scale=0.0925]{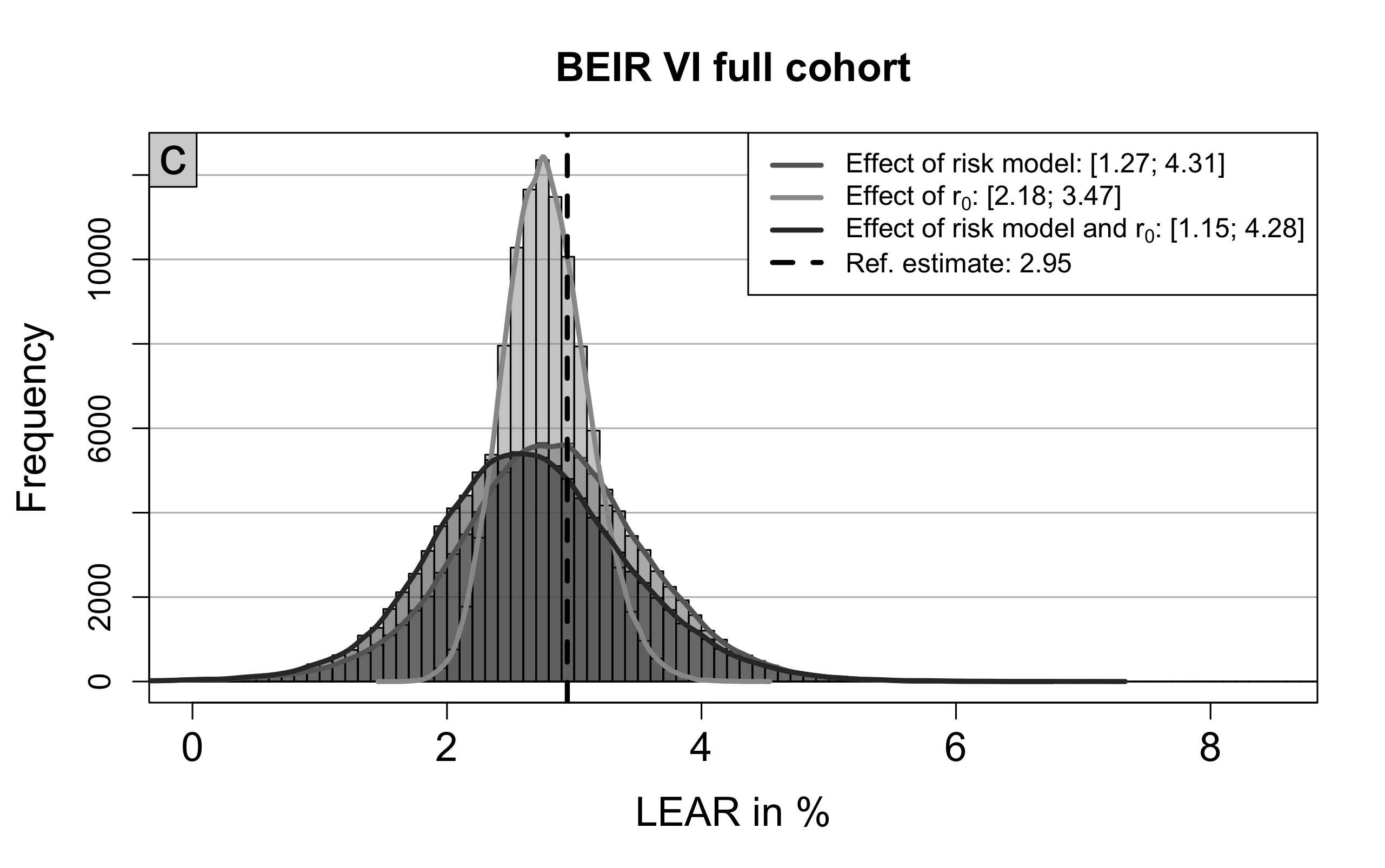}
    \includegraphics[scale=0.0925]{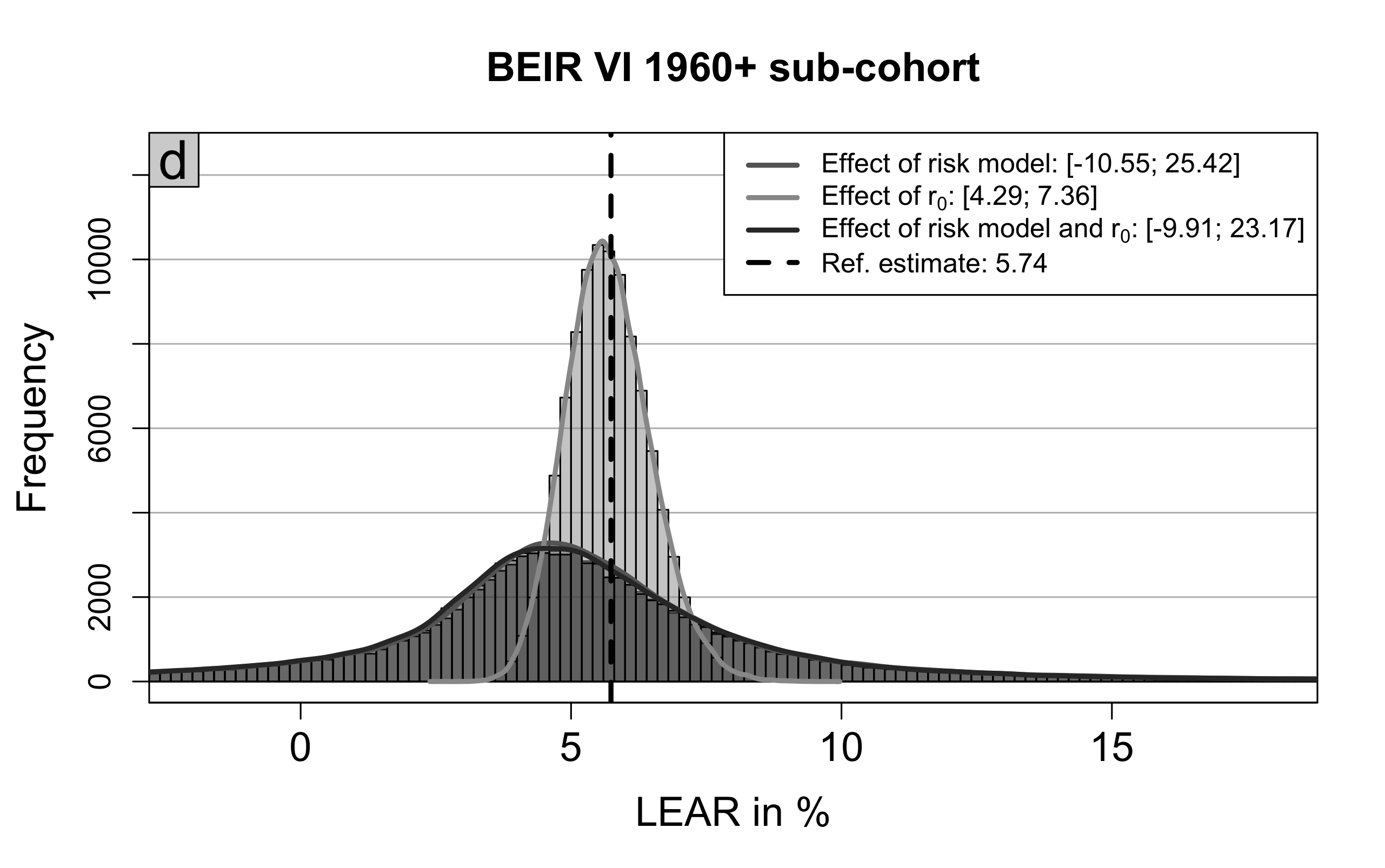}
    \includegraphics[scale=0.0925]{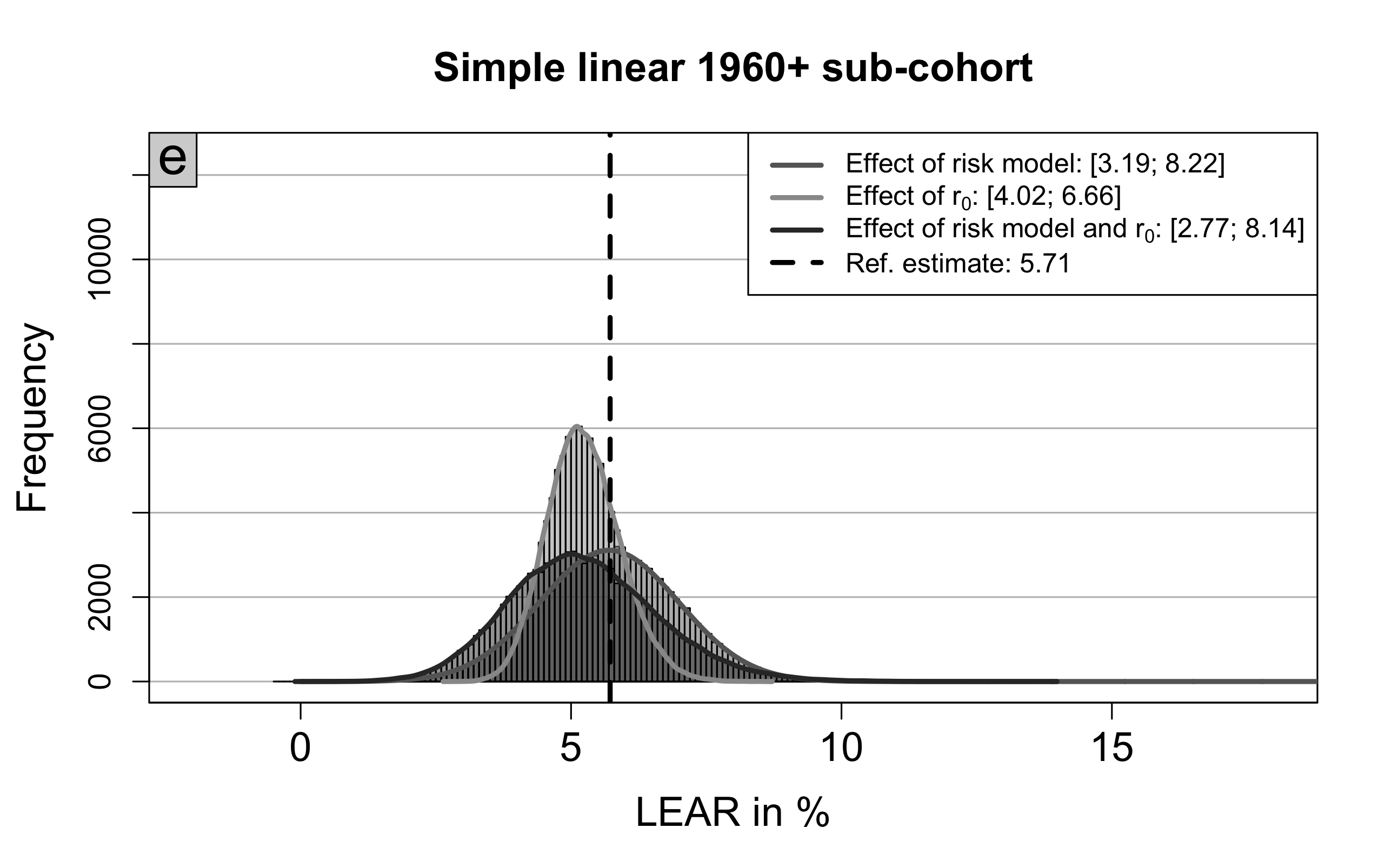}
    \caption{Results from ANA approach. Histograms of 100,000 resulting LEAR samples with kernel density (solid lines) for different risk models (Plot title a-e) and varying uncertain components in correspondence to Table \ref{tab:COVA_LEAR_KIs}. Risk model parameters (Effect of risk model) are assumed to follow a multivariate normal distribution (ANA approach). Lung cancer mortality rates $r_0(t)$ (Effect of $r_0$) are assumed to follow a gamma distribution with parameters as in Suppl. Table \ref{tab:WHO_r0_q0_estimates_5EAA}. The joint effect (Effect of risk model and $r_0$) results from independent sampling from both corresponding probability distributions. 95\% uncertainty interval are presented in the legend.}
    \label{fig:LEAR_ERR_MR_Histogramm}
\end{figure}

The empirical distribution of resulting LEAR samples, along with the corresponding 95\% uncertainty intervals, varies depending on the underlying risk model and its parameters, especially for the 1960+ sub-cohort models (Figure \ref{fig:LEAR_ERR_MR_Histogramm}, Table \ref{tab:COVA_LEAR_KIs}), last column). Risk model complexity and cohort size directly influence LEAR uncertainty intervals, as seen in the separate analyses. In contrast, the 95\% uncertainty intervals from uncertain $r_0(t)$ are overall narrower compared to those from uncertain risk model parameters and consistent across all risk models with a relative uncertainty span of roughly $0.5$. The joint effect of uncertain risk model parameters and uncertain lung cancer mortality rates is with a relative uncertainty span of roughly $1$ almost similar to the effect with only uncertain risk model parameters. Only the 95\% uncertainty interval [-9:91; 23.17] of the joint effect with relative uncertainty span of 5.76 for the BEIR VI 1960+ sub-cohort model (\ref{model4_BEIR_VI_sub}) remains implausible, although slightly narrower. Overall, accounting for lung cancer mortality rate uncertainty in addition to risk model parameter uncertainty has low impact on the overall LEAR uncertainty interval.

\section{Discussion}
This work provides a considerable methodological contribution in radiation protection research by successfully deriving uncertainty intervals for radon-induced lifetime lung cancer risk estimates. These intervals are grounded in a sound statistical framework, shifting the approach from solely assessing single lifetime risk estimates to quantifying the uncertainties around these estimates. For a comprehensive assessment of lifetime risks, it is advisable to consider both the point estimate and the uncertainty intervals. From this perspective, we investigate two key contributors to uncertainties: baseline mortality rates and risk models. We introduce advanced methods for quantifying these uncertainties, facilitating their application in radiation-related research questions. Our results confirm the BEIR VII report finding that risk model uncertainty is a major driver of overall lifetime risk uncertainty \cite{BEIR_VII}. Accounting for risk model parameter uncertainty in risk models from the Wismut cohort study as a practical example yields plausible lifetime risk uncertainty intervals that encompass the range of reported lifetime risk estimates from miners studies in the literature, as summarized in \cite[Table 4]{Kreuzer2024_PUMALEAR}. \\

This study specifically addresses quantifying uncertainties in lifetime lung cancer risks associated with protracted occupational radon exposure. Existing tools like “CONFIDENCE”, “RadRAT”, and “LARisk”, predominantly employ risk models derived from the Atomic Bomb Survivors Life-Span Study and focus on acute radiation exposure \cite{Preston2007}. The latter two build on methods from the BEIR VII report \cite{BEIR_VII} to quantify lifetime cancer risk uncertainties. Those methods consider sampling variability in risk model parameters similar to our ANA approach, uncertainties in the risk transfer between populations, and uncertainties in the dose and dose-rate effectiveness factor (DDREF) using the delta method \cite{Doob1935,VerHoef2012}. However, subjective variance inputs are needed for risk transfer and DDREF uncertainty assessment, which our study avoids. Results from the BEIR VII report differ from the results here due to different study populations and exposure metrics, but the span of uncertainty intervals are of comparable magnitude. \\

Moreover, in the context of occupational radon exposure, the present study adds to methods used in the BEIR VI report \cite{BEIR_VI}, the EPA report \cite{EPA_2003} and \cite{Tomasek_2020}, all of which use sampling techniques similar to the ANA approach here to address risk model parameter uncertainty. Although \cite{Tomasek_2020} does not provide direct lifetime risk uncertainty intervals, they can be derived from the results therein and align well with our findings. Overall, this adds credibility to both approaches. However, the methodology to derive uncertainty intervals has not been extensively introduced, discussed, and compared with other approaches so far, as was done in the present study. \\

\subsection{Sources of lifetime risk uncertainty}
Following \cite{Thomas1992}, three types of uncertainties arise in lifetime risk calculations: sampling uncertainty, uncertainty in choosing and deriving a suitable model structure (model uncertainty), and unspecified uncertainties like data errors and validity of assumptions, that cannot be formally specified with probability distributions. Quantifying uncertainties beyond sampling uncertainty is challenging, as acknowledged elsewhere \cite{BEIR_VI,Hoffmann_2021}. This analysis focuses on quantifiable sampling uncertainty. For a comprehensive overview of uncertainties, decisions, and potential errors associated with lifetime risk calculations see \cite[Table 3-13]{BEIR_VI}. \\

The risk model defining $ERR(t; \Theta)$ is a crucial component of LEAR calculations with inherent uncertainties, arising from factors such as disease classification, statistical power limitations, potential confounding, and particularly from exposure assessment – a major challenge in radiation research and risk model derivation \cite{UNSCEAR_2012_AnnexB}. Potential measurement errors, especially from early years of uranium mining are subject of ongoing research, e.g.  exploring their effects within the Wismut cohort \cite{Wismut_Unsicherheiten_Teil1_2018,Wismut_Unsicherheiten_Teil2_2022}. Assessing uncertainties in risk model derivation is beyond the scope of this study. \\ 

Different risk model structures (categorical or parametric) fitted to the same cohort data  result in comparable lifetime risk estimates \cite{Kreuzer2024_PUMALEAR}. The non-parametric reference mortality rates $r_0^{(ICRP)}(t),q_0^{(ICRP)}(t)$ and the corresponding survival function $S(t)$ avoid parameter uncertainty and leverage available population data. The discretization of lifetime risks from a theoretical integral to a calculable sum is mandated by data availability. In the scope of generic lifetime risk calculation, this all suggests low model uncertainty or limited model flexibility, allowing us to concentrate on quantifiable sampling uncertainty. \\

Preliminary sensitivity analyses indicate that many factors (e.g. the choice of the lifetime risk measure, latency time $L$ or the maximum age $t_{max}$) have only limited impact on lifetime risk variability \cite{Sommer_2024_Sensi}. Likewise, variability in the annual radon exposure is only briefly investigated (Suppl. Section \ref{Radon_exposure_uncertainty_SDC_F}) since the exposure scenario is fixed for most lifetime risk applications, especially for the important dose conversion considerations.  Thus, the present study focuses on the most influential components: sampling uncertainty in risk model parameters $\Theta$ and baseline mortality rates $r_0(t), q_0(t)$.

\subsection{Risk model parameter uncertainty}

\subsubsection{Approximate normality assumption (ANA) approach}
The ANA approach approximates the true underlying likelihood function for estimating risk model parameters using a normal distribution and follows the frequentist approach. The covariance matrix estimate $\hat{\Sigma}_0$ describes the amount of sampling uncertainty, which decreases as cohort size increases, resulting in narrower uncertainty intervals for lifetime risks. This explains why lifetime risk estimates with risk model parameters derived from the smaller, younger 1960+ sub-cohort have wider uncertainty intervals compared to those derived from the Wismut full cohort. This is particularly evident for models with BEIR VI structure \cite{BEIR_VI}, with dedicated parameters for miners data at higher age ranges. \\

The ANA method requires only parameter estimates and their covariance matrix to derive uncertainty intervals via Monte Carlo simulations. This makes the ANA approach practical and efficient, especially when complete access to cohort data is not available. While not entirely new, the idea of approximating the underlying likelihood function is rooted in earlier work \cite{1984_Rubin_bayes} later coined to the term "Approximate Bayesian Computation (ABC)" \cite{Beaumont2002-ld,Sisson2018_Handbook}. A similar method for uncertainty quantification of lifetime thyroid cancer risk related to radiation exposure was used in \cite{Xue2001-cy}.

\subsubsection{Bayesian approach}
In contrast to the frequentist method, Bayesian statistics incorporates prior knowledge or beliefs about model parameters through probability distributions, providing an alternative perspective on uncertainty. \\

Since the statistical software \textit{Epicure} \cite{EPICURE} does not support Bayesian methods, we implemented a solution in \textit{R} \cite{RCore}. Although \textit{R} can be computationally slower than \textit{Epicure}, which is specifically optimized for fitting $ERR(t;\Theta)$ risk model structures, individual \textit{R} solutions allow greater control over cohort data and model fitting. We calculate the marginal posterior distribution of risk model parameters, inspired by \cite{Higueras_2018}. Computing the full posterior is computationally expensive due to numerous baseline stratification parameters, but focusing on the marginal posterior simplifies this by reducing the parameter set to just the risk model parameters. Compared to \cite{Higueras_2018}, this technique is here extended to handle more complex risk models and protracted low exposure scenarios. Sampling from the true marginal posterior yields more nuanced uncertainty intervals that are not reliant on approximating asymptotic behavior as in the ANA approach. \\

A key strength of the Bayesian approach is the ability to integrate prior knowledge – e.g. the results from previous miners studies – through the selection of prior distributions. However, selecting these priors involves subjective judgment \cite{Goldstein2006-ln}. Decisions on prior distributions, in particular the degree of influence on the likelihood, have to be thoroughly considered, compare \cite{Greenland2006-yj} . Non-informative priors, like uniform distributions, have minimal influence on the posterior and lead to uncertainty intervals similar to the ANA approach, especially for large cohorts, while informative priors enable a more adaptive integration of diverse cohort data. Note that labeling estimates from the Joint Czech+French cohort as "prior knowledge" is a structured test of the methodology's adaptability and effectiveness of combining different cohort information rather than a strict integration of prior results. Users can tailor the approach to their needs, but selecting appropriate priors is a separate consideration beyond this discussion. \\

Although Bayesian methods offer flexibility and deeper insights, they require full access to cohort data and significant computational resources, especially for complex models with many parameters. While sampling methods like MCMC are efficient in this case due to high acceptance rates by choosing the approximate (asymptotic) normal distribution as a proposal density, the computational challenge lies in calculating the marginal posterior distribution itself.

\subsection{Mortality rate uncertainty}
Mortality rates also introduce notable uncertainty to lifetime risk estimates. Unlike for risk model parameter estimates, which are derived through a rigorous statistical framework (likelihood theory), the ICRP mixed Euro-American-Asian reference mortality rates $r_0^{(ICRP)}(t), q_0^{(ICRP)}(t)$ are presented as plain numbers sourced from a database. These mortality rates do not result from a statistical estimation process, requiring careful consideration when imposing probability distributions. We assessed this uncertainty by applying gamma distributions to baseline mortality rates $r_0(t)$ and $q_0(t)$ for all ages t incorporating observed variability in mortality rates across countries in Europe, America, and Asia from WHO data \cite{WHO_mortality_database}, which aligns geographically with ICRP reference rates. The gamma distribution was finally chosen as it fits the observed rates for numerous age groups well. This allowed us to quantify how mortality rate uncertainty influences lifetime risk estimates. In the described method, each rate derived from WHO data is assigned equal weight. Consequently, observations from smaller countries are given the same weight as those from larger countries. The focus is on estimating the variability of mortality rates themselves, irrespective of the population size. Alternatively, a population-weighted approach is applied in Suppl. Section \ref{Mortality_rate_uncertainty_poisson_lc_deaths_SDC_D2}. \\

Our analysis confirmed that uncertainties in all-cause mortality rates have negligible impact \cite{Sommer_2024_Sensi}, while lung cancer mortality rate uncertainties resulted in uncertainty intervals similar to those from full cohort risk model parameter uncertainty. \\

The derivation of parameter estimates in (\ref{mr_r0_ass}) and (\ref{mr_q0_ass}) from WHO data is independent from ICRP reference rates $r_0^{(ICRP)}(t)$ and $q_0^{(ICRP)}(t)$. Using a centered distribution with the expected value (mean) set equal to the ICRP rates had low effect on lifetime risk estimates and according uncertainties (Suppl. Section D.1.2), so we retained the un-centered distribution for Monte Carlo simulations to avoid constraining parameter estimation. \\

Here, lung cancer mortality rates introduce uncertainty comparable to that of full cohort risk model parameters for the Wismut cohort study. However, including both sources of uncertainty has little effect on the uncertainty intervals compared to just considering risk model parameters. Therefore, focusing solely on risk model parameter uncertainty is sufficient. Although unintuitive, this effect can be explained by acknowledging the product structure of $ERR(t;\Theta)$ and lung cancer rates $r_0($t) in the LEAR calculation. The variance of products of (independent) random variables is not necessarily larger than the variances of single factors \cite{Goodman_Var}.

\subsection{Interpretation of uncertainty intervals}
Uncertainty intervals capture the variability in lifetime risk estimates. For risk model effects, uncertainty intervals derived with ANA methods reflect sampling variation in estimated risk model parameters and may be referred to as (approximate) Wald-type confidence intervals. Bayesian credible intervals, such as Highest Posterior Density Intervals (HPDIs), represent the probability (e.g., 95\%) that the true value lies within the interval, incorporating prior beliefs in risk model parameters.\\

While both methods provide uncertainty intervals, their interpretations differ. ANA confidence intervals are less interpretable than Bayesian credible intervals. However, this theoretical distinction is practically less relevant for purposes in radiation protection research. In particular, credible intervals with non-informative priors are often similar to confidence intervals derived with the ANA approach.\\

Mortality rate uncertainty intervals are more difficult to interpret due to their dependence on external data and chosen distributions. They reflect sampling variability by accounting for observed mortality rate variation and can be considered as subjective confidence intervals, similar to \cite{BEIR_VII}. They provide a valuable quantitative sense on mortality rate variability. \\

The derived uncertainty intervals for lifetime risks like the LEAR reflect the expectable range of potential values that arise from the inherent variability in each calculation component. In particular, as lifetime risks are not directly estimated from data with sampling uncertainty, the intervals should not be interpreted as classical confidence intervals.
\subsection{Comparison with lifetime risk variation in the literature}
The uncertainty in risk model parameters significantly contributes to the overall uncertainty in lifetime risk estimates. Our analysis using the ANA approach applied to risk models from the Wismut cohort study reveals uncertainty intervals that align well with the literature: in the context of the PUMA study \cite{Kreuzer2024_PUMALEAR}, LEAR values from various studies were recalculated and summarized. Thereby, a range for the LEAR per WLM of $2.50 \times 10^{-4}$ to $9.22 \times 10^{-4}$ was reported across all published risk models of uranium miners studies that include time- and age-related effect modifiers. This range translates to an equivalent range for the LEAR in \% of 2.35 to 8.65. The 95\% uncertainty intervals derived in this study for the parametric 1960+ sub-cohort models, particularly [3.26; 12.28] for the best-fit model (\ref{model2_para_sublinear}) and [3.19; 8:22] for the simple linear model (\ref{model1_para_linear}), correspond well to the range of point estimates reported by the PUMA study group. To convert the reported LEAR per WLM values to the total LEAR, each value was multiplied by 94 (total cumulative exposure in WLM). The derived intervals for the full Wismut cohort exhibit a weaker alignment with this range: [2.06;4.84] for the parametric model (\ref{model3_para_full}) and [1.27;4.30] for the categorical model (\ref{model5_BEIR_VI_full}). However, the 1960+ sub-cohort Wismut models are preferred to full Wismut cohort models in order to estimate lung cancer risks at low protracted exposures due to high quality exposure assessment \cite{Kreuzer_2023_lc}. \\

Heterogeneity in radiation risk estimates between studies may explain differences in the LEAR and can likely be attributed to diverse factors such as structural differences in cumulative exposure range, duration of employment, and methods in mortality tracking and data analysis \cite{Kreuzer2024_PUMALEAR}. \\

While uncertainty intervals depend on the chosen confidence level (here 95\%), the close alignment of our derived intervals with literature values supports the reliability and appropriateness of our approach. This is especially remarkable as our results are solely derived from the Wismut miner cohort data as a practical example. The consistency across recognized miner studies confirms the reliability of the ANA methodology in assessing uncertainties in lifetime lung cancer risks from radon exposure. Further follow-up years for miners cohorts will refine our understanding of risk models and lifetime risks for radon-induced lung cancer and associated uncertainties.

\subsection{Strengths and weaknesses}
Our work benefits from the strengths of the German uranium miners cohort (Wismut cohort), the largest single cohort of uranium miners worldwide representing roughly half of all miners in the pooled PUMA cohort \cite{Rage2020-hj}. This large cohort (cf. \cite{Kreuzer_2009_cohort_profile}) allows us to achieve reliable estimates for risk model effects on lifetime risk uncertainties.  \\

Furthermore, the present study pioneers the application of the comprehensive WHO mortality database \cite{WHO_mortality_database} to assess mortality rate uncertainties in radon-induced lung cancer lifetime risks. These innovative approaches, along with the implementation of advanced Bayesian techniques, expand the methodological toolbox for uncertainty assessment in this field. To facilitate the application of the Bayesian technique, we developed a \textit{R} procedure for data grouping and model fitting, overcoming the limitations of existing software for this specific analysis. Transparency in reporting, with detailed descriptions of methods, statistical analysis, and results, facilitates replicability. \\

Both methods for assessing risk model uncertainties are reliable due to statistically grounded assumptions and show broad applicability. The ANA approach requires minimal assumptions, making it versatile. While the Bayesian approach with the analytical computed marginal Posterior distribution offers stronger rigor, it has specific data needs (Poisson-distributed numbers of lung cancer deaths). However, its concept of marginal posterior distributions likely extends beyond lifetime risk assessments. These techniques, given appropriate data, can be applied in radiation epidemiology to analyze various quantitative figures derived from likelihood functions, encompassing different exposures, health outcomes, and risk models. \\

However, limitations are acknowledged. The uncertainty intervals depend on the baseline mortality rates applied, the choice of risk model and the data used to estimate the risk models. The Bayesian approach, while statistically rigorous and contributing to the reliability of results, was here limited in applicability to models with few parameters due to computational constraints. However, generally, better computing power allows for the analysis of more complex models. This study prioritizes the development of methodology for uncertainty assessment. In particular, lifetime risk calculation inherits methodological limitations of risk model estimation, such as the incorporation of detailed smoking behaviour. Data for radon effects on females are sparse. Based on findings from residential radon studies in \cite{Darby_2005} we decided to assume the same $ERR(t;\Theta)$ for females and males. Finally, the analysis does not account for uncertainties arising from transferring risk estimates from miners cohorts to the general population (here: multiplicative risk transfer \cite{proZES_2020,UNSCEAR_2020_AnnexB}. Aligning mortality rates with the specific characteristics of the cohort data, such as using national mortality rates relevant to the cohort's origin, can help partially mitigate the risk transfer issue. The overall composite nature of a lifetime risk estimate depending on multiple independently conducted analyses, limits an all-encompassing uncertainty assessment.
\clearpage

\section{Conclusion}
Uncertainty quantification is crucial for a comprehensive understanding of lifetime risk estimates. This study demonstrates that uncertainty from risk model parameter estimates explains a substantial fraction of overall lifetime risk uncertainty. Two advanced methods to derive uncertainty intervals are developed and applied. The simple ANA approach proves to be a suitable and reliable uncertainty assessment technique for most cases. The more flexible Bayesian approach offers a more nuanced view of uncertainty. However, the approach is computationally more demanding and requires full access to grouped cohort data, limiting its wider applicability. From a practical perspective, additionally accounting for uncertainties in mortality rates is less critical. The explicit choice of lifetime risk measures is negligible for uncertainty assessment. The uncertainty intervals derived in this study correspond to the range of LEAR values from different miners studies in the literature, thus uncertainties derived by both methods are mutually confirmed. The introduced methods allow for a more complete comparison of lifetime risk estimates across uranium miners studies. These findings should be accounted for when developing radiation protection policies which are related to lifetime risks.

\clearpage
\subsection*{Conflicts of interest and sources of funding:}
The authors have no relevant financial or non-financial interests to disclose.
\clearpage

\section{Tables}

\begin{table}[htbp]
    \centering
\resizebox{\columnwidth}{!}{\begin{tabular}[h]{llllll}

\addlinespace
\hline
\addlinespace
\multicolumn{1}{l}{Risk model} & Equation &LEAR in \% & Effect of risk model& Effect of $r_0(t)$ & Joint effect  \\
\addlinespace
\hline
\addlinespace
Parametric full cohort & (\ref{model3_para_full})&$3.43$ & $[2.06;4.81]$ $(0.80)$& $[2.57;  4.17]$ $(0.47)$&$[1.88; 4.99]$ $(0.91)$\\
Parametric 1960+ sub-cohort & (\ref{model2_para_sublinear})&$6.70$ & $[3.26; 12.28]$ $(1.34)$& $[4.96; 8.11]$ $(0.47)$& $[3.00; 11.98]$ $(1.34)$\\
Simple linear 1960+ sub-cohort & (\ref{model1_para_linear})&$5.71$ & $[3.19; 8.22]$ $(0.88)$ & $[4.02; 6.66]$ $(0.46)$& $[2.77; 8.14]$ $(0.94)$\\
BEIR VI full cohort & (\ref{model5_BEIR_VI_full})&$2.95$ &$[1.27; 4.31]$ $(1.03)$& $[2.18; 3.47]$ $(0.44)$& $[1.15; 4.28]$ $(1.06)$\\
BEIR VI 1960+ sub-cohort & (\ref{model4_BEIR_VI_sub}) &$5.74$ &$[-10.55; 25.42]$ $(6.27)$& $[4.29; 7.36]$ $(0.53)$& $[-9.91; 23.17]$ $(5.76)$\\
\addlinespace
\hline
\addlinespace
\end{tabular}}
    \caption{Results from ANA approach. LEAR in \% estimates with corresponding 95\% uncertainty intervals (relative uncertainty span in brackets) for uncertain risk model parameters, lung cancer mortality rates $r_0(t)$ or both independently, in correspondence to Figure \ref{fig:LEAR_ERR_MR_Histogramm} Lung cancer rates $r_0(t)$ are sampled from a Gamma distribution according to Suppl. Table \ref{tab:WHO_r0_q0_estimates_5EAA}. Reference estimates for LEAR in \% are calculated with ICRP Euro-American-Asian mixed reference mortality rates \cite{ICRP103_2007} and an exposure scenario of 2 WLM from age 18-64 years.}
    \label{tab:COVA_LEAR_KIs}
\end{table}

\begin{table}[htbp]
\centering
 \begin{tabular}[h]{llc}
\addlinespace
\hline
\addlinespace
Prior information &$LEAR$ in \% & Relative uncertainty span \\
\addlinespace
\hline
\addlinespace
Uniform Prior & $6.74$ $[2.96; 11.09]$ & $1.20$\\
$a=2$ & $5.56$ $[2.76; 8.98]$ & $1.12$\\
$a=5$ & $5.41$ $[2.82; 8.45]$ & $1.04$\\
$a=10$ & $5.27$ $[2.89; 8.15]$ & $1.00$\\
$a=20$ & $5.13$ $[2.99; 7.58]$ & $0.89$\\
$a=50$ &$4.99$  $[3.14; 7.26]$ &$0.83$\\
\addlinespace
\hline
\addlinespace
\end{tabular}
    \caption{
    Results from Bayesian approach. LEAR in \% estimate with 95\% highest posterior density interval (HPDI) and relative uncertainty span for uncertain risk model parameters for risk model (\ref{model2_para_sublinear}) derived from sampling from the posterior distribution $P\left(\Theta \vert X \right)$. The prior distribution $P(\Theta)=P(\beta)P(\alpha)P(\varepsilon)$  for gamma-distributed $\beta$ for varying shape parameter $a$ and normally distributed $\alpha, \varepsilon$ with standard deviation $\sigma=0.02$ in correspondence to Figure \ref{fig:linlogERR_bayes_full_LEAR}. The LEAR in \% calculated with risk model parameters from the Joint Czech + French cohort \cite{Tomasek_mines_2008} is 4.30.}
    \label{tab:linlogERR_bayes_ki}
\end{table}

\begin{table}[htbp]
    \centering
  \begin{tabular}[h]{lllc}
\addlinespace
\hline
\addlinespace
Prior information & $\hat{\beta} \times 100$ &$LEAR$ in \% & Relative uncertainty span \\
\addlinespace
\hline
\addlinespace
Likelihood with Wald-type CI &	$1.34$ $[0.71;1.97]$ &	$5.72$ $[3.03;8.41]$ &$0.94$\\
Uniform Prior & $1.34$ $[0.79; 2.08]$ &$5.72$ $[3.37; 8.90]$&$0.96$\\ 
$a=2$ & $1.35$ $[0.81; 2.07]$ & $5.77$ $[3.44; 8.84]$ & $0.94$\\
$a=5$ & $1.38$ $[0.87; 2.06]$ & $5.90$ $[3.72; 8.82]$ &$0.86$\\
$a=10$ & $1.42$ $[0.94; 2.02]$ & $6.06$ $[4.00; 8.64]$ & $0.76$\\
$a=20$ & $1.46$ $[1.03; 1.99]$ & $6.25$ $[4.41; 8.48]$ & $0.65$\\
$a=50$ & $1.52$ $[1.18; 1.91]$ & $6.50$ $[5.05; 8.17]$ & $0.48$\\
\addlinespace
\hline
\addlinespace
\end{tabular}
    \caption{Parameter estimate $\hat{\beta}=Mod \left( P( \beta \vert X ) \right)$ of posterior distribution $P(\beta \vert X)$ with LEAR calculated with the risk model (\ref{model1_para_linear}) and corresponding 95\% highest posterior density interval (HPDI) and relative uncertainty spans for varying prior parameter settings. The gamma-distributed prior $P(\beta)$ is centered at the corresponding parameter estimate from \cite{Tomasek_mines_2008} for different values of prior gamma shape parameters $a$ in correspondence to Figure \ref{fig:LEAR_ERR_MR_Histogramm}. Wald-type confidence intervals (CI) are also shown for comparison and are calculated as $\hat{\beta} \pm 1.96 \times 0.003005$, where $0.003005$ represents the parameter standard error as stated in \cite{Kreuzer_2023_lc}.}
    \label{tab:linERR_bayes_ki}
\end{table} 
\clearpage

\section{Figures}

\clearpage

\bibliography{Literatur_Uns_arXive}
\bibliographystyle{unsrt}
\clearpage

\appendix

\renewcommand{\figurename}{Supplementary Figure}
\renewcommand{\tablename}{Supplementary Table}
\setcounter{table}{0}
\setcounter{figure}{0}
\section*{Supplementary information}
In this Supplement, additional information, alternative modeling approaches, and derivations of quantities from the main manuscript are presented regarding uncertainties in the lifetime risk for lung cancer related to occupational radon exposure. In particular, the basic methodology is identical to the main paper approach.  For a comprehensive overview of the conducted analyses, see Supplementary Table \ref{tab:Uncertainty_overview}. 

\begin{table}[htbp]
    \centering\resizebox{\columnwidth}{!}{ 
 \begin{tabular}[h]{llll}
\addlinespace
\hline
\addlinespace
Component& Methods & Sensitivities  & Section \\
\addlinespace
\hline
\hline
\addlinespace
\multirow{ 4}{*}{Mortality rates} & \multirow{ 3}{*}{Non-weighted regression} & Gamma distribution & Main paper, Suppl. \ref{Mortality_rate_uncertainty_preliminaries_SDC_D1}  \\
 & & Log-normal distribution & Suppl. \ref{Mortality_rate_uncertainty_lognormal_rates_SDC_D3}\\
  & & Sex-specific & Suppl. \ref{Sex_specific_uncertainty_mortality_rates_SDC_E1}\\
& Weighted Poisson regression (Bayesian) &  Varying prior certainty & Suppl. \ref{Mortality_rate_uncertainty_poisson_lc_deaths_SDC_D2}\\
\addlinespace
\hline
\addlinespace
\multirow{ 3}{*}{Risk model parameters} & ANA  & - & Main paper, Suppl. \ref{Risk_model_parameter_uncertainty_simple_linear_SDC_C1}\\
& \multirow{ 2}{*}{Bayesian}  &  Varying prior certainty  & Main paper, Suppl. \ref{Risk_model_parameter_uncertainty_simple_linear_bayes_SDC_C2}, Suppl. \ref{Risk_model_parameter_uncertainty_loglinear_bayes_SDC_C3}\\
& & Sex-specific & Suppl. \ref{Sex_specific_uncertainty_risk_model_SDC_E2}\\
\addlinespace
\hline
\addlinespace
\multirow{ 2}{*}{Exposure scenario} & \multirow{ 2}{*}{Log-normal annual exposure}  & \multirow{ 2}{*}{Exposure variability} & \multirow{ 2}{*}{Suppl. \ref{Radon_exposure_uncertainty_SDC_F}}\\
&  &   & \\
\addlinespace
\hline
\addlinespace
\multirow{ 3}{*}{Other lifetime risk measures} & ANA  & - &\multirow{ 2}{*}{Suppl. \ref{comp_lr_measures_SDC_A}}\\
& Bayesian  & Varying prior certainty &\\
& Kaplan-Meier  &  Windows of cumulative exposure & Suppl. \ref{kaplanmeier_curves_SDC_B}\\
\addlinespace
\hline
\addlinespace
\end{tabular}}
    \caption{Overview of lifetime risk uncertainty assessment by investigated component, methods, conducted sensitivity analyses, and where it is to find in the documents.}
    \label{tab:Uncertainty_overview}
\end{table}

\subsection*{Overview}
In Suppl. Section \ref{comp_lr_measures_SDC_A}, we derive uncertainty intervals for other lifetime risk measures ($ELR$, $REID$, and $RADS$) using ANA and Bayesian techniques. Across all measures, the relative uncertainty spans are comparable. Therefore, for practical purposes, relying on the $LEAR$ measure is sufficient. $RADS$ estimates are notably larger compared to the other measures. \\

We also present an exploratory Kaplan-Meier survival curve analysis \cite{Kaplan58}, stratified by cumulative exposure in categories (Suppl. Section \ref{kaplanmeier_curves_SDC_B}, Suppl. Fig. \ref{fig:KM_survival}). This approach relies on hardly any assumptions, using only miners cohort data, yields plausible uncertainty intervals and is useful for visual risk assessments. \\

Suppl. Section \ref{Risk_model_parameter_uncertainty_SDC_C}  shows additional insights for risk model parameter uncertainty. Particularly, the Bayesian approach for risk model parameter uncertainty for the simple risk model (\ref{simplelinearERR}) showed only slight differences between log-normal and gamma priors for $\beta$ (Suppl. Section \ref{Risk_model_parameter_uncertainty_simple_linear_bayes_SDC_C2}). \\

Suppl. Section \ref{Mortality_rate_uncertainty_SDC_D} further explores mortality rate uncertainties: An alternative Bayesian approach using Poisson-distributed numbers of lung cancer deaths for population-weighted mortality rate uncertainty (Suppl. Section \ref{Mortality_rate_uncertainty_poisson_lc_deaths_SDC_D2}) provided minimal new insights, as the extensive WHO data results in a highly peaked Likelihood function indicating low uncertainty in the parameter estimates and consequently overruled the prior’s influence. In Suppl. Section \ref{Mortality_rate_uncertainty_lognormal_rates_SDC_D3}, we compare log-normal and gamma distributions for mortality rates, showing similar uncertainty intervals but with slightly larger upper bounds for log-normal distributions. Both distributions fit the WHO data comparably well. This indicates that accounting for mortality rate uncertainty in lifetime risks involves a degree of subjectivity, influenced by the researcher’s methodological decisions and the need for careful interpretation of resulting uncertainty intervals. \\

For sex-specific lifetime risks with mortality rates derived from sex-specific WHO data (Suppl. Section \ref{Sex_specific_uncertainty_mortality_rates_SDC_E1}), estimates for males are roughly twice those of females due to overall higher male baseline all-cause and lung cancer mortality, though the relative uncertainty span remains similar. Lifetime risk estimates calculated with mortality rates from females must be interpreted cautiously, as the underlying excess relative risk term is based on male miners cohorts. Sex-specific $LEAR$ estimates using sex-specific ICRP reference rates (Suppl. Section \ref{Sex_specific_uncertainty_risk_model_SDC_E2}) accounting for risk model parameter uncertainty for the parametric 1960+ sub-cohort risk model (\ref{parametric_1960_rm}) with the Bayesian approach further confirm this pattern, with male $LEAR$ values being twice as high, while the relative uncertainty remains similar. \\

Finally, Monte Carlo simulations accounting for uncertainties in radon exposure scenarios (Suppl. Section \ref{Radon_exposure_uncertainty_SDC_F}) showed that lifetime risks become approximately normally distributed across all risk models and lifetime risk measures under low exposure uncertainty, with prescribed annual exposure variability strongly affecting results but maintaining consistent relative uncertainty intervals.

\section{Comparison and results for other lifetime risk measures}
\label{comp_lr_measures_SDC_A}
Besides the $LEAR$, further (excess) lifetime risk measures are used in the literature. Here, we compare the $LEAR$ to three additional lifetime risk measures: the  Risk of Exposure Induced Death ($REID$) (first introduced in \cite{UNSCEAR_1994} and employed in \cite{Little2008, Hunter2015}), the Excess Lifetime Risk ($ELR$) \cite{Vaeth1990} and the Radiation Attributable Decrease of Survival ($RADS$) \cite{Ulanowski2019}. All considered excess lifetime risk definitions emerge from the difference between a risk under exposure and a risk in the absence of exposure. Except for $RADS$, the lifetime risk is the sum of annual lung cancer risks weighted by a certain survival probability, see e.g. \cite{Thomas1992}. Note that $RADS$ is special as it is the only measure that describes no risk per se but rather a relative change in the survival function. In particular, it is the only measure calculated without all-cause mortality rates. However, we will see that all four quantities are connected. Before comparing uncertainty intervals, we introduce the additional lifetime risk measures shortly. 
\subsection{Definition}
The central difference of the additionally considered lifetime risk measures compared to $LEAR$ is the explicit accounting for radon exposure in the survival function $S(t )$. Survival under exposure shall be denoted by $S_E(t)$ and baseline survival by $S_0(t)$,  It holds, 
\begin{align}
    REID&=\int_{0}^{\infty} r_E(t) S_E(t ) \, dt - \int_{0}^{\infty} r_0(t) S_E(t ) \, dt = \int_{0}^{\infty} r_0(t)ERR(t) S_E(t ) \, dt, \label{REIDeq}\\
    ELR&=\int_{0}^{\infty} r_E(t) S_E(t ) \, dt - \int_{0}^{\infty} r_0(t) S_0(t ) \, dt, \label{ELReq} \\
    RADS&= \lim_{t \to \infty} \frac{S_0(t ) - S_E(t )}{S_0(t )}=1 - \lim_{t \to \infty} \frac{S_E(t )}{S_0(t)}. \label{RADSeq}
\end{align}

To calculate these additional lifetime risks, assumptions on the survival function affected by radon are necessary. Analogously to $S_0(t) =e^{-\int_0^tq_0(u)\, du} \approx \Tilde{S}_0(t) =e^{-\sum_{u=0}^{t-1}q_0(u)}$, we set $$S_E(t) =e^{-\int_0^tq_E(u) \, du} \approx \Tilde{S}_E(t)=e^{- \sum_{u=0}^{t-1} q_E(u)},$$  where $q_E(u)$ describes the all-cause mortality rate at age $u$ under exposure. Since there is currently no reliable evidence that radon can cause diseases other than lung cancer, we assume that radon exposure influences solely the lung cancer risk \cite{Henyoh2024-mp}. Hence $q_E(u)=q_0(u)+r_0(u)ERR(u)$ for all ages $u$ and 
\begin{align}
\label{S_Eformula}
S_E(t) = S_0(t)e^{-\int_0^t r_0(u)ERR(u)\, du} \approx \Tilde{S}_E(t) = \Tilde{S}_0(t)e^{-\sum_{u=0}^{t-1} r_0(u)ERR(u)}. 
\end{align}
Employing the same approach as for the $LEAR$, the final approximations for all considered lifetime risk measures read
\begin{align*}
    LEAR& \approx\sum_{t \geq 0} r_0(t) ERR(t) \Tilde{S}_0(t), \\
    REID& \approx \sum_{t \geq 0} r_0(t) ERR(t) \Tilde{S}_E(t)  , \\
    ELR&  \approx \sum_{ t \geq 0} r_0(t)(1+ERR(t))\Tilde{S}_E(t)  - r_0(t)\Tilde{S}_0(t), \\ 
    RADS& \approx  1- e^{-\sum_{t \geq 0}r_0(u)ERR(u)}.
\end{align*}

Without assuming a protective effect of radon exposure \cite{UNSCEAR_2000, ICRP103_2007}, it holds $r_E(t) \geq r_0(t)$ and $S_E(t ) \leq S_0(t )$ for all ages $t \geq 0$. Therewith one can easily deduce (proof in \cite{Sommer_2024_Sensi}), 
\begin{equation}
\label{central_inequality_3}
     ELR \leq REID \leq LEAR.
\end{equation}
The inequality (\ref{central_inequality_3}) is universal and holds for all choices and combinations of calculation components. For realistic excess absolute risks $r_0(t)ERR(t)$ (e.g. for moderate exposures observed in mines and reasonable lung cancer mortality rates) one can further deduce \begin{equation}
\label{central_inequality_4}
   ELR \leq REID \leq LEAR \leq RADS.
\end{equation}
The relationship (\ref{central_inequality_4}) is observed for the majority of reasonable exposure scenarios. However, technically $LEAR$ can exceed $RADS$, which is naturally bounded by one.

\subsection{Uncertainty assessment}
In the following, we employ the Bayesian approach and the approximate normality assumption (ANA) approach to assess risk model parameter effects in lifetime risk uncertainties across different previously introduced lifetime risk measures. We expand results from the main manuscript and give further insights. As in the main manuscript, a generic lifetime risk estimate without any uncertainty quantification is called "Reference estimate (ICRP 103)" or simply "ref. estimate". In addition to the 95\% uncertainty intervals we often present the span of uncertainty intervals relative to the reference estimate (relative uncertainty span). 
\label{section:lr_comp_ua_ba}

\subsubsection{Bayesian approach for risk model parameter effects} 
For this Bayesian risk model parameter uncertainty assessment, we consider the parametric 1960+ sub-cohort risk model from \cite{Kreuzer_2023_lc}, 
\begin{equation}
\label{parametric_1960_rm}
ERR(t;\beta, \alpha, \varepsilon)=\beta W(t)\exp\left\{ \alpha (AME(t)-30)+\varepsilon(TME(t)-20)\right\}
\end{equation} 
with parameter set $\Theta=\left(\beta, \alpha, \varepsilon \right)$ and estimates $\hat{\beta}=0.0466$, $\hat{\alpha}=-0.0301$, and $\hat{\varepsilon}=-0.0755$. 
The prior assumptions are analogous to those described in the Methods section of the main paper. Note that we do not investigate the simple linear risk model here, as it provides a poorer fit to the miner data. \\

\begin{figure}[htbp]
    \centering
    \includegraphics[scale=0.0925]{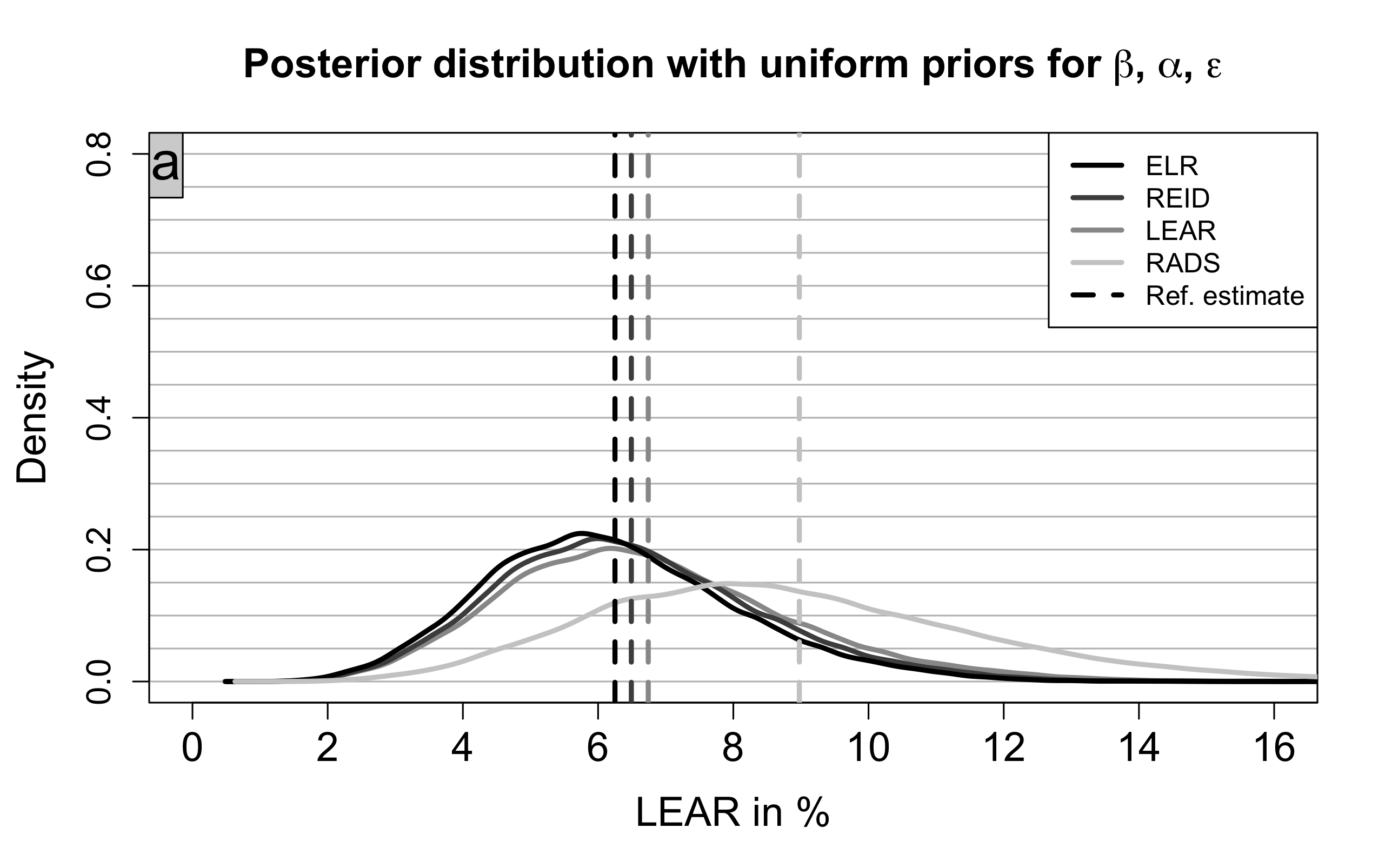}
    \includegraphics[scale=0.0925]{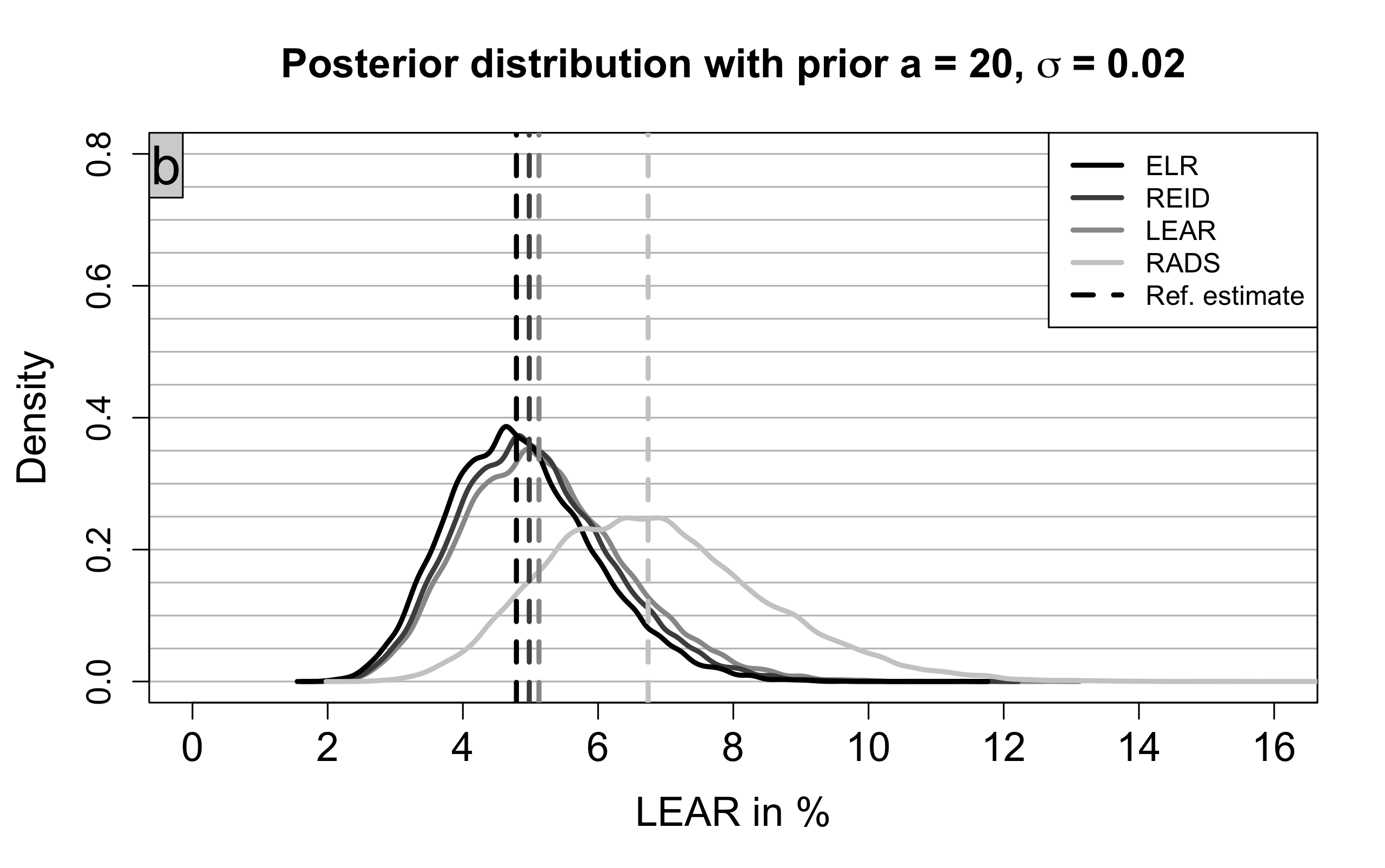}
    \includegraphics[scale=0.0925]{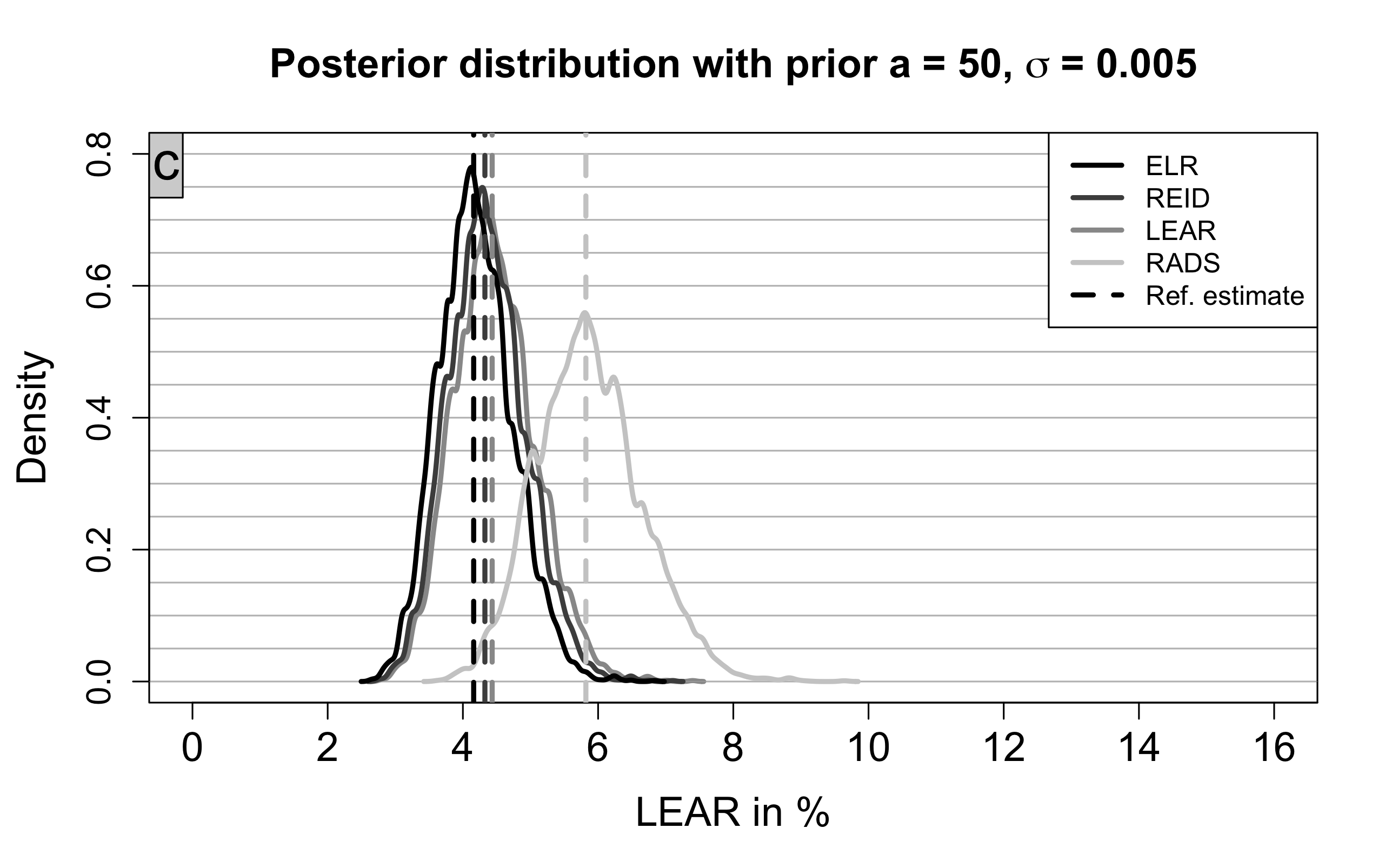}
    \caption{Distribution of excess lifetime risk measures $ELR$, $REID$, $LEAR$ and $RADS$ with risk model (\ref{parametric_1960_rm}) and parameter samples drawn from the posterior distribution with prior $P(\Theta)=P(\beta)P(\alpha)P(\varepsilon)$ for different combinations of gamma-distributed marginal prior for $\beta$ and normally distributed marginal priors for $\alpha$ and $\varepsilon$ for different shape parameters $a$ and standard deviations $\sigma$ in correspondence to Suppl. Table \ref{tab:linlogERR_bayes_LEAR_REID_ELR_RADS}. The reference estimate (ICRP 103) corresponds to the excess lifetime risk measures evaluated at the mode $\hat{\Theta}$.}
    \label{fig:linlogERR_bayes_LEAR_REID_ELR_RADS}
\end{figure}

Supplementary Figure \ref{fig:linlogERR_bayes_LEAR_REID_ELR_RADS} shows sample distributions across different lifetime risk measures obtained by drawing risk model parameter samples from the posterior distribution $P(\Theta \vert X) \propto P(\Theta) L(X \vert \Theta)$ for varying prior certainty. The reference estimate with corresponding 95\% highest density posterior intervals (HPDI) is shown in Suppl. Table \ref{tab:linlogERR_bayes_LEAR_REID_ELR_RADS}. The results are overall very similar and only $RADS$ estimates are considerably larger. However, the relative uncertainty spans are similar across all lifetime risk measures. 
As expected, the reference estimates shift towards the lifetime risk estimates calculated using point estimates of the risk model parameters derived from the Joint Czech-French cohort. This behavior reinforces the reliability of the Bayesian methods employed in this analysis. For comparison, the $ELR$, $REID$, $LEAR$, and $RADS$ in \% calculated with risk model parameters (\ref{parametric_1960_rm}) derived from this cohort are $4.04$, $4.20$, $4.30$, and $5.62$, respectively. \\

\begin{table}[htbp]
    \centering
    \centering\resizebox{\columnwidth}{!}{ 
 \begin{tabular}[h]{lllll}
\addlinespace
\hline
\addlinespace
Prior information & $ELR$ in \%  & $REID$ in \%  & $LEAR$ in \%  &$RADS$ in \%  \\
\addlinespace
\hline
\addlinespace
Uniform prior & $6.25$ $[2.80; 10.04]$ $(1.16)$ &$6.49$ $[2.89; 10.40]$  $(1.16)$& $6.74$ $[2.96; 11.09]$  $(1.21)$& $8.98$ $[3.57; 14.56]$  $(1.23)$\\
$a=20, \sigma =0.02$ & $4.79$ $[2.83; 7.00]$  $(0.87)$&$4.98$ $[2.94; 7.26]$  $(0.87)$& $5.13$ $[2.99; 7.58]$  $(0.89)$& $6.74$ $[3.98; 10.22]$  $(0.93)$\\
$a=50, \sigma = 0.005$ & $4.33$ $[3.16;5.27]$  $(0.49)$&$4.43$ $[3.32; 5.53]$  $(0.50)$& $4.43$ $[3.34; 5.66]$  $(0.52)$& $5.82$ $[4.40; 7.40]$  $(0.52)$\\
\addlinespace
\hline
\addlinespace
\end{tabular}}
    \caption{Excess lifetime risk measures evaluated at the mode $\hat{\Theta}$ of risk model parameters and their 95\% highest posterior density interval (HPDI) (relative uncertainty span in brackets) with prior $P(\Theta)=P(\beta)P(\alpha)P(\varepsilon)$ for different values of prior gamma shape parameters $a$ and standard deviations $\sigma$ in correspondence to Suppl. Figure \ref{fig:linlogERR_bayes_LEAR_REID_ELR_RADS}.}
    \label{tab:linlogERR_bayes_LEAR_REID_ELR_RADS}
\end{table} 

Note that lifetime risks evaluated at the mode $\hat{\Theta}$ of the risk model parameter distribution (reference estimate) do sometimes not align with the sample distribution mode in Suppl. Figure \ref{fig:linlogERR_bayes_LEAR_REID_ELR_RADS}. 
This discrepancy arises because a lifetime risk evaluated at a specific parameter estimate $\hat{\Theta}$ is essentially a non-linear parameter transformation of $\hat{\Theta}$. This affects the distribution characteristics. Here, the exponential nature of the employed risk model structure shifts the distribution mode slightly leftward depending on the magnitude of $\sigma$.\footnote{For example, the mode of $\exp\left\{Y\right\}$ for a normal distribution $Y \sim \mathcal{N}\left(\mu, \sigma ^2 \right)$ is $\exp\left\{\mu - \sigma^2\right\}$ instead of $\exp\left\{\mu\right\}$.} A comparable effect is observed in the upcoming analysis with the ANA approach.

\subsubsection{Approximate normality assumption (ANA) approach} 

\begin{figure}[htbp]
    \centering
    \includegraphics[scale=0.0925]{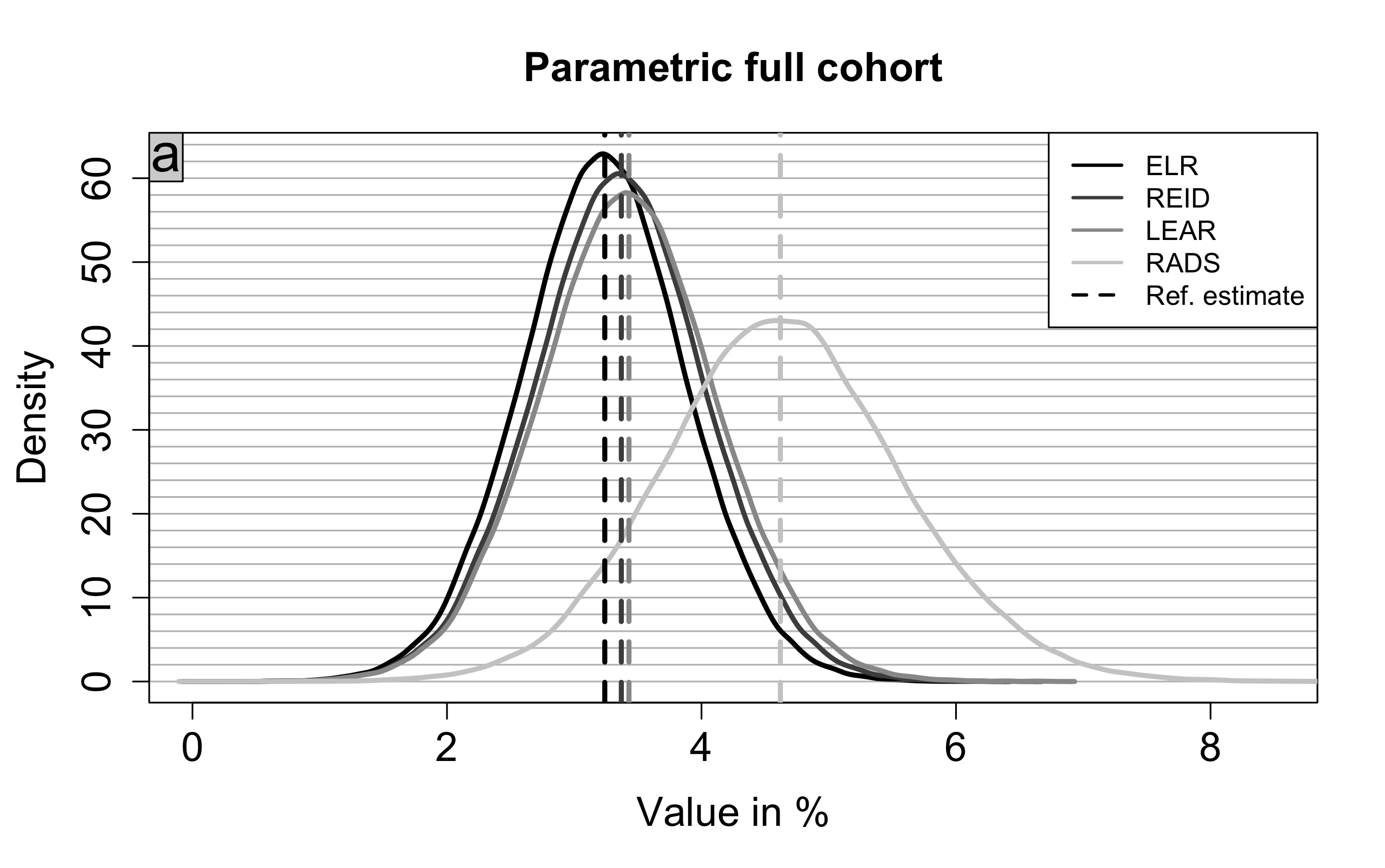}
    \includegraphics[scale=0.0925]{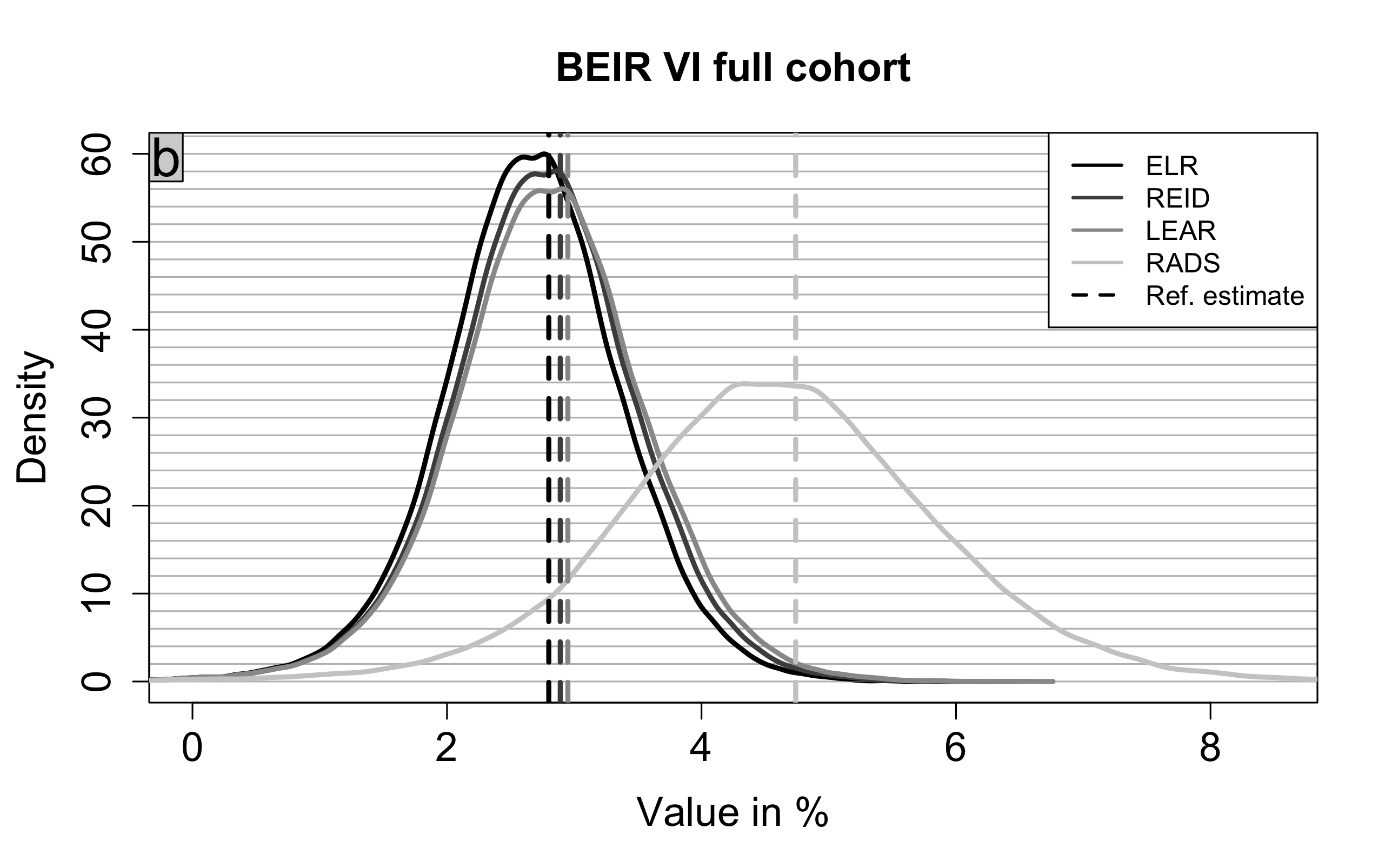}
    \includegraphics[scale=0.0925]{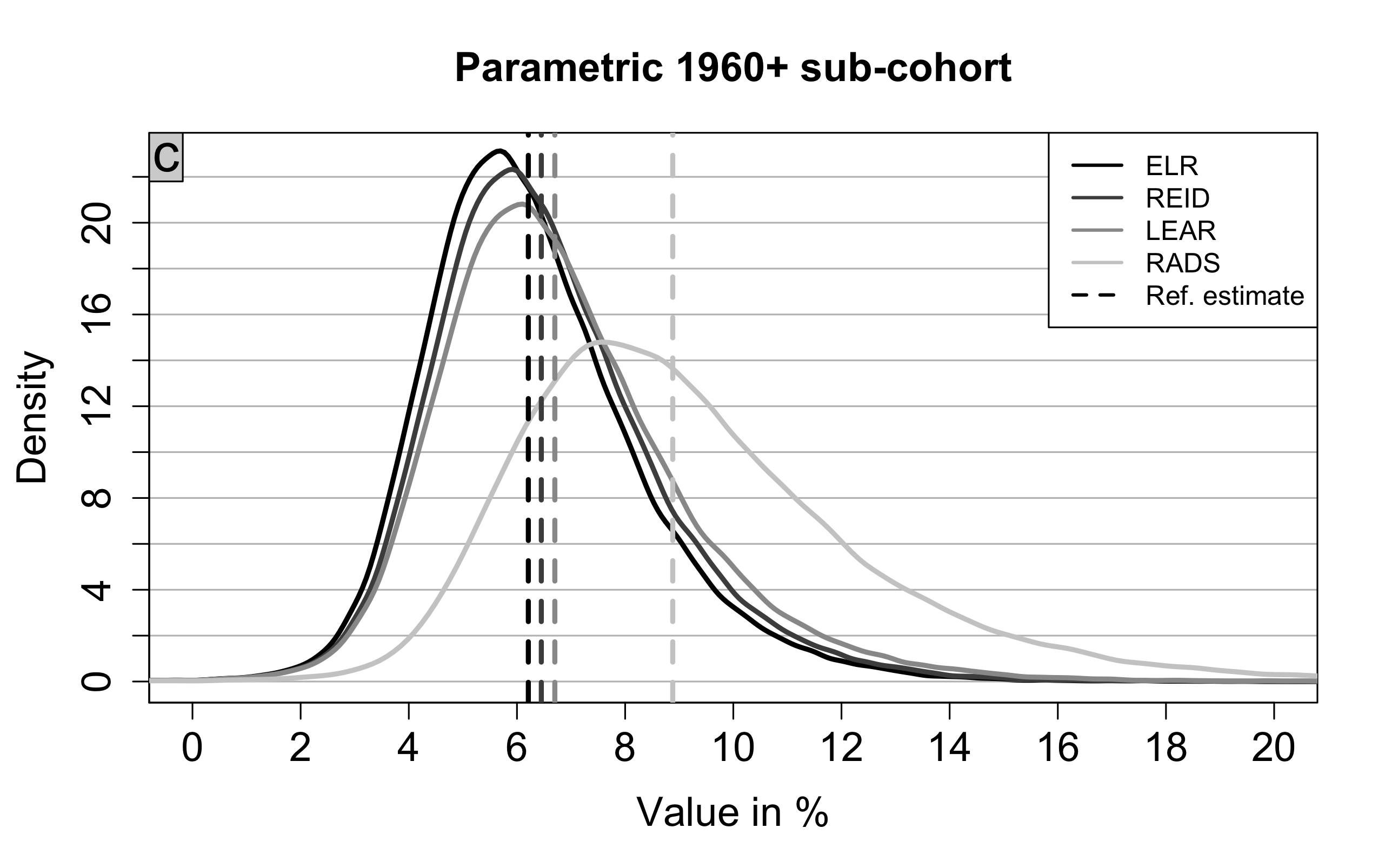}
    \includegraphics[scale=0.0925]{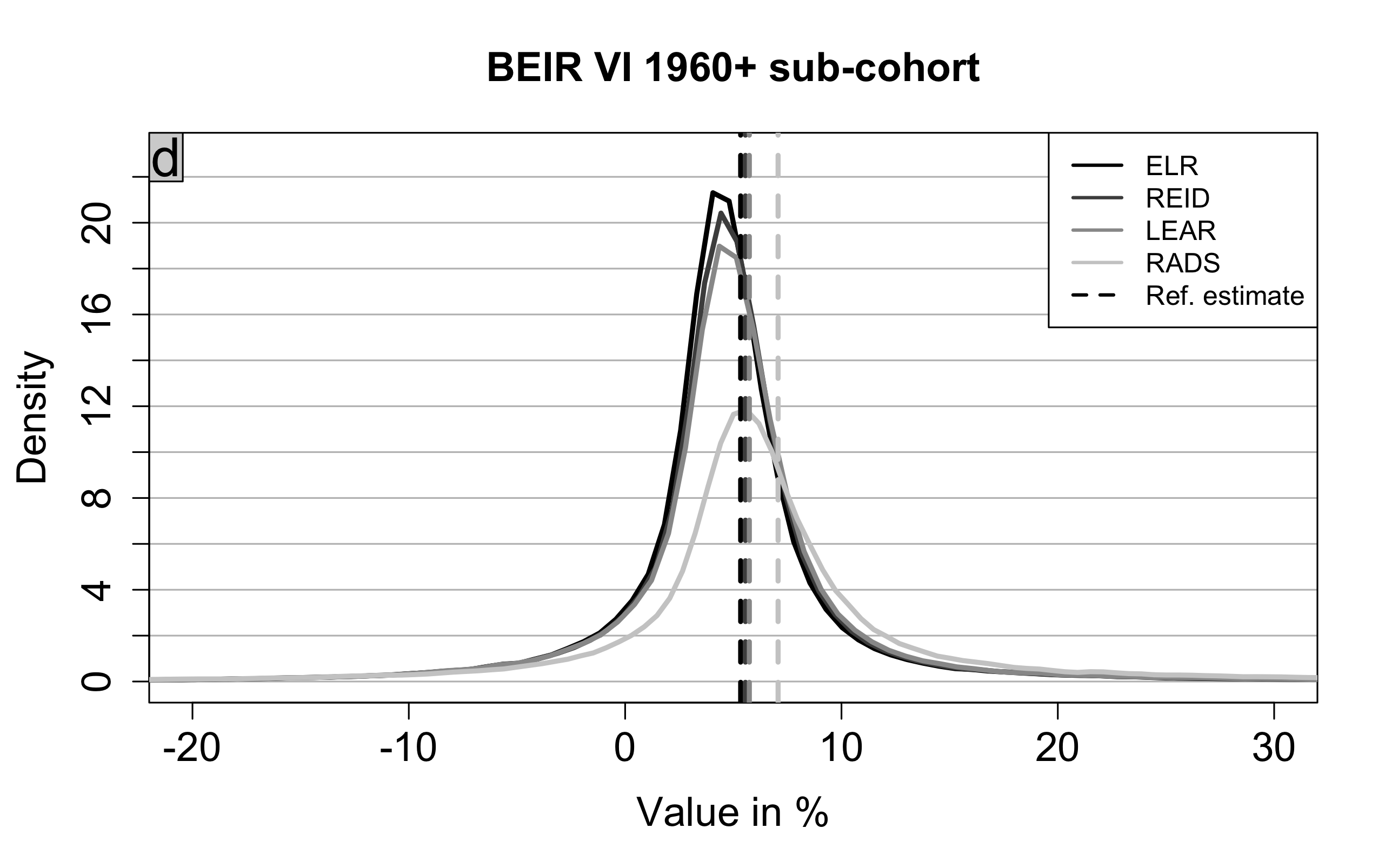}
    \includegraphics[scale=0.0925]{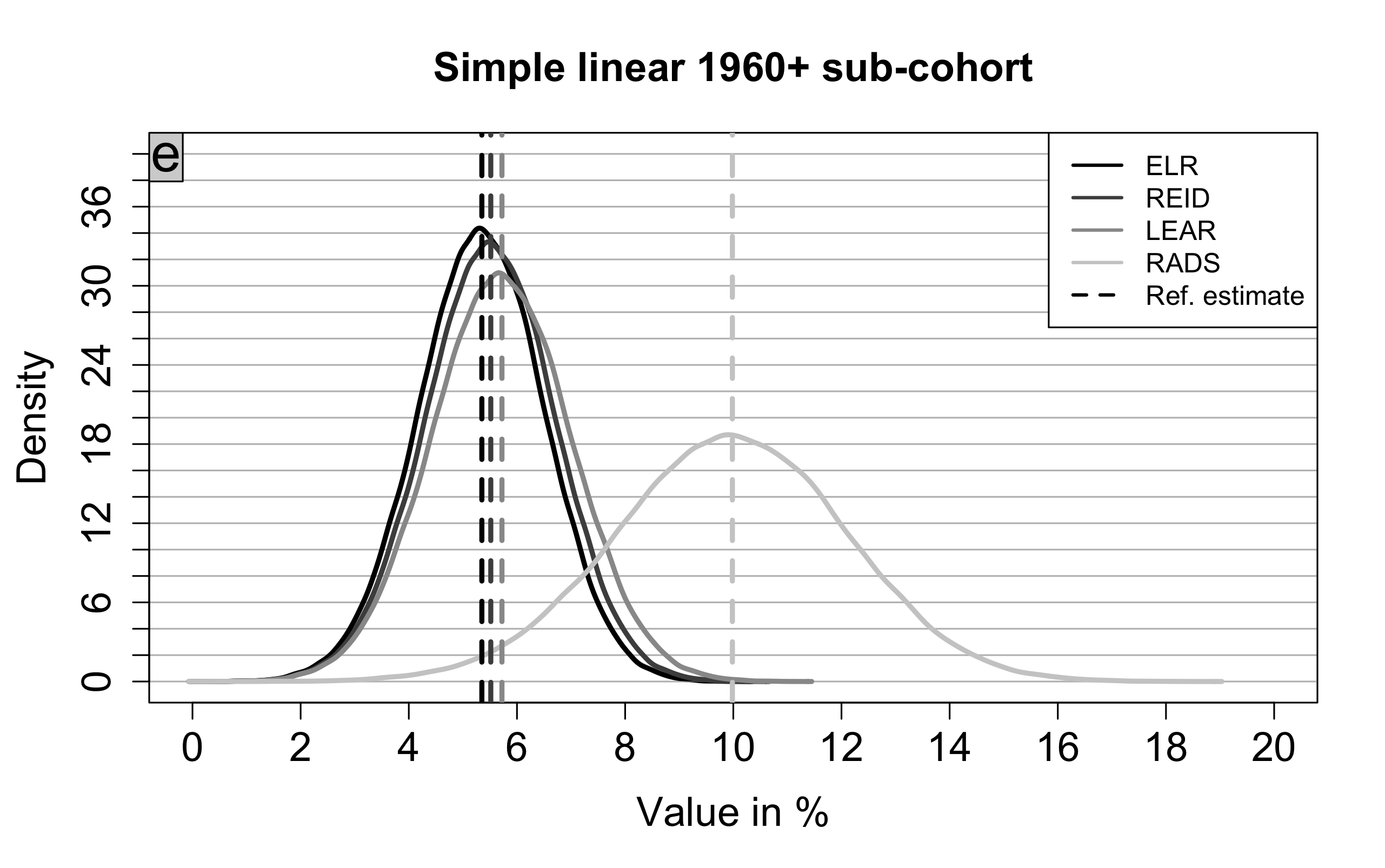}
    \caption{Density from the histogram of $100,000$ excess lifetime risk estimates calculated with risk model parameter estimates drawn from a multivariate normal distribution (ANA approach) for different lifetime risk measures and risk models in correspondence to Suppl. Table \ref{tab:Monte_Carlo_lr_vergleich}.}
    \label{fig:Monte_Carlo_lr_vergleich}

\end{figure}

Analyses with the ANA approach showed that the resulting differences in lifetime risk measures are comparably low for all considered risk models (Suppl. Figure \ref{fig:Monte_Carlo_lr_vergleich}, Suppl. Table \ref{tab:Monte_Carlo_lr_vergleich}). Again, only $RADS$ estimates are considerably larger, which is also reflected in the wider uncertainty intervals. However, the relative uncertainty span is similar across all considered lifetime risk measures. \\ 

The BEIR VI 1960+ sub-cohort risk model exhibits excessive parameter uncertainty. This is reflected in a high proportion of implausible (negative) $LEAR$ samples, likely due to the inherent uncertainty in parameters for greater ages fit on a young cohort.  This holds across all considered lifetime risk measures. However, the exponential nature of $RADS$ amplifies negative parameter samples, resulting in a considerably wider lower uncertainty bound compared to other lifetime risk measures. Overall, the excessive uncertainty in BEIR VI 1960+ sub-cohort risk model parameters makes corresponding uncertainty quantification currently impractical. With additional follow-up data, the uncertainty in risk model parameters, especially at higher ages,  will decrease, leading to more reliable estimates for the 1960+ sub-cohort. \\

\begin{table}[htbp]
    \centering\resizebox{\columnwidth}{!}{ 
 \begin{tabular}[h]{lllll}
\addlinespace
\hline
\addlinespace
Risk model& $ELR$ in \%  & $REID$ in \%  & $LEAR$ in \%  &$RADS$ in \%  \\
\addlinespace
\hline
\addlinespace
Parametric full cohort& $3.24$ $[1.96; 4.51]$ $(0.79)$&$3.37$ $[2.04; 4.69]$ $(0.79)$& $3.43$ $[2.06; 4.81]$ $(0.80)$& $4.62$ $[2.83; 6.55]$ $(0.81)$\\
BEIR VI full cohort & $2.80$ $[1.22; 4.06]$ $(1.01)$&$2.89$ $[1.26;4.20]$ $(1.02)$& $2.95$ $[1.27; 4.31]$ $(1.03)$& $4.74$ $[2.10; 7.11]$ $(1.06)$\\
Parametric 1960+ sub-cohort & $6.21$ $[3.08; 11.04]$ $(1.28)$&$6.45$ $[3.20; 11.44]$ $(1.28)$& $6.70$ $[3.26; 12.28]$ $(1.35)$& $8.88$ $[4.41; 17.10]$ $(1.43)$\\
BEIR VI 1960+ sub-cohort & $5.34$ $[-10.90; 20.77]$ $(5.93)$&$5.56$ $[-11.26; 21.31]$ $(5.86)$& $5.74$ $[-10.55; 25.42]$ $(6.27)$& $7.07$ $[-18.86;40.96]$ $(8.46)$\\
Simple linear 1960+ sub-cohort & $5.35$ $[3.03; 7.60]$ $(0.85)$&$5.52$ $[3.12; 7.84]$ $(0.86)$& $5.72$ $[3.19; 8.26]$ $(0.89)$& $9.98$ $[5.69; 14.09]$ $(0.84)$\\
\addlinespace
\hline
\addlinespace
\end{tabular}}
    \caption{Excess lifetime risks (reference estimates) with 95\% uncertainty interval (relative uncertainty span in brackets) from the $100,000$ risk model parameter estimates drawn from a multivariate normal distribution (ANA approach) for different lifetime risk measures and risk models in correspondence to Suppl. Figure \ref{fig:Monte_Carlo_lr_vergleich}.}
    \label{tab:Monte_Carlo_lr_vergleich}
\end{table}

In summary, the uncertainty intervals for the measures $ELR$, $REID$, and $LEAR$ are similar for both the ANA and the Bayesian approach. Notable differences between measures $ELR$, $REID$, and $LEAR$ are expected to occur only at higher exposures as they primarily differ in how they model the survival function, using either $S_0(t)$ or $S_E(t)$.  However, low exposure scenarios are particularly prevalent today and therefore play a major role in radiation protection purposes. Therewith, regarding uncertainties, no clear difference or benefit between the three lifetime risk measures is observed. $RADS$ being generally larger results in likewise shifted uncertainty intervals. The relative uncertainty span is similar across all lifetime risk measures. The BEIR VI 1960+ sub-cohort model results in peculiar uncertainty intervals and should, to date, be employed with care for uncertainty assessment. 
\clearpage
\section{Explorative uncertainty assessment: Kaplan-Meier lung cancer survival curves}
\label{kaplanmeier_curves_SDC_B}
We present a simple, explorative approach to deducing uncertainty intervals for lifetime risks. This approach requires minimal assumptions and is in particular independent of the typical (excess) lifetime risk structure incorporating $ERR$ risk models. Further, it only employs knowledge from miner cohorts and does not rely on external inputs like mortality rates or exposure scenarios. \\

\subsection{Introduction and definition}
Excess lifetime risks employed in the literature are the difference between a risk under exposure and a risk in the absence of exposure. In particular, for the $LEAR$ this relationship is expressed as $LEAR=LR_E-LR_0$ (see Methods section in the main manuscript for comparison). By interpreting $1-LR_E$ and $1-LR_0$ as (unknown) lifetime lung cancer survival probabilities, the excess lifetime risk can also be understood as a difference in survival probabilities, i.e.,
\begin{equation}
\label{LEARSidea}
   LEAR=LR_E-LR_0=\left(1-LR_0\right)- \left(1-LR_E \right)=: S_0 - S_E. 
\end{equation}

We present a simple approach to calculate estimates for lung cancer survival curves $S_0$ and $S_E$ for different radon exposures with uncertainty intervals. \\

Kaplan-Meier survival curves are such simple non-parametric estimates for survival probabilities in time-to-event analyses \cite{Kaplan58}. 
Using Kaplan-Meier survival curves to estimate the difference in lung cancer survival $S_0-S_E$ has little in common with the original definition of $LEAR$. While a $LEAR$ can be calculated for any explicit exposure scenario, the Kaplan-Meier survival curves rely on more rough windows of cumulative exposure. The resulting lifetime risk estimates are referred to as (naive) $LEAR$ estimates for simplicity. Here, Kaplan-Meier lung cancer survival functions are calculated for miners at Wismut with follow-up 1946-2018 stratified by certain exposure windows. The time axis is age $t$ in years whereas the event is lung cancer death. Due to the large number of miners in the Wismut cohort, we can calculate reliable Kaplan-Meier estimates $\hat{S}(t)$ for $S(t)$ for different radon exposures. Radon exposure is categorized into seven groups ("No exposure", 0-10, 10-50, 50-100, 100-500, 500-1000, and 1000+ WLM), and for each group, survival curves are calculated. For Kaplan-Meier survival estimates, it holds for the probability of not dying of lung cancer until age $t$,
\begin{align*}
    \hat{S}(t)= \prod_{t_k \leq t} \left( 1-\frac{d_k}{n_k}\right)
\end{align*}
with number of lung cancer deaths $d_k$ and individuals at risk $n_k$ at time point $t_k$. \\

Here, a confidence interval for $\hat{S}(t)$ is constructed with \textit{Greenwood's formula} \cite{greenwood1926report} via
\begin{equation}
  \widehat{Var}\left(\hat{S}(t)\right)=\hat{S}(t)^2 \sum_{t_k \leq t} \frac{d_k}{n_k(n_k-d_k)}   \label{GreenwoodFormula}
\end{equation} and the asymptotically normal distribution of $\hat{S}(t)$ yields the symmetric point-wise confidence interval at level $1-\alpha/2$, $$\hat{S}(t) \pm z_{1-\alpha/2} \sqrt{\widehat{Var}\left(\hat{S}(t)\right)}.$$
Note that Greenwood's formula provides point-wise confidence intervals at each age $t$, suitable for our analysis. However, researchers seeking simultaneous confidence intervals across the entire survival curve might consider using  Hall-Wellner confidence bands \cite{Hall_Wellner_1980} or equal probability bands \cite{Nair_1984_EP}. Detailed methodologies are available in \cite[Chapter 4.4]{Klein_Survival2005}, with a practical implementation by the "km.ci" R package \cite{km_ci}. \\

For each miner, we calculate the exact age at the end of follow-up (either 12/31/2018 or before, in case of death or loss-to-follow-up) in days in units of years: e.g. a person dying of lung cancer 10 days after its 85th birthday results in age at the end of follow-up of $85 + \frac{10}{365.25}=85.03$. Every observed lung cancer death marks a time point $t_k$. Hence $d_k=1$ in most cases. Rarely, multiple individuals died of lung cancer at the exact same age, resulting in $d_k>1$. \\
\subsection{Results}
\begin{figure}[htbp]
    \centering
    \includegraphics[scale=0.0925]{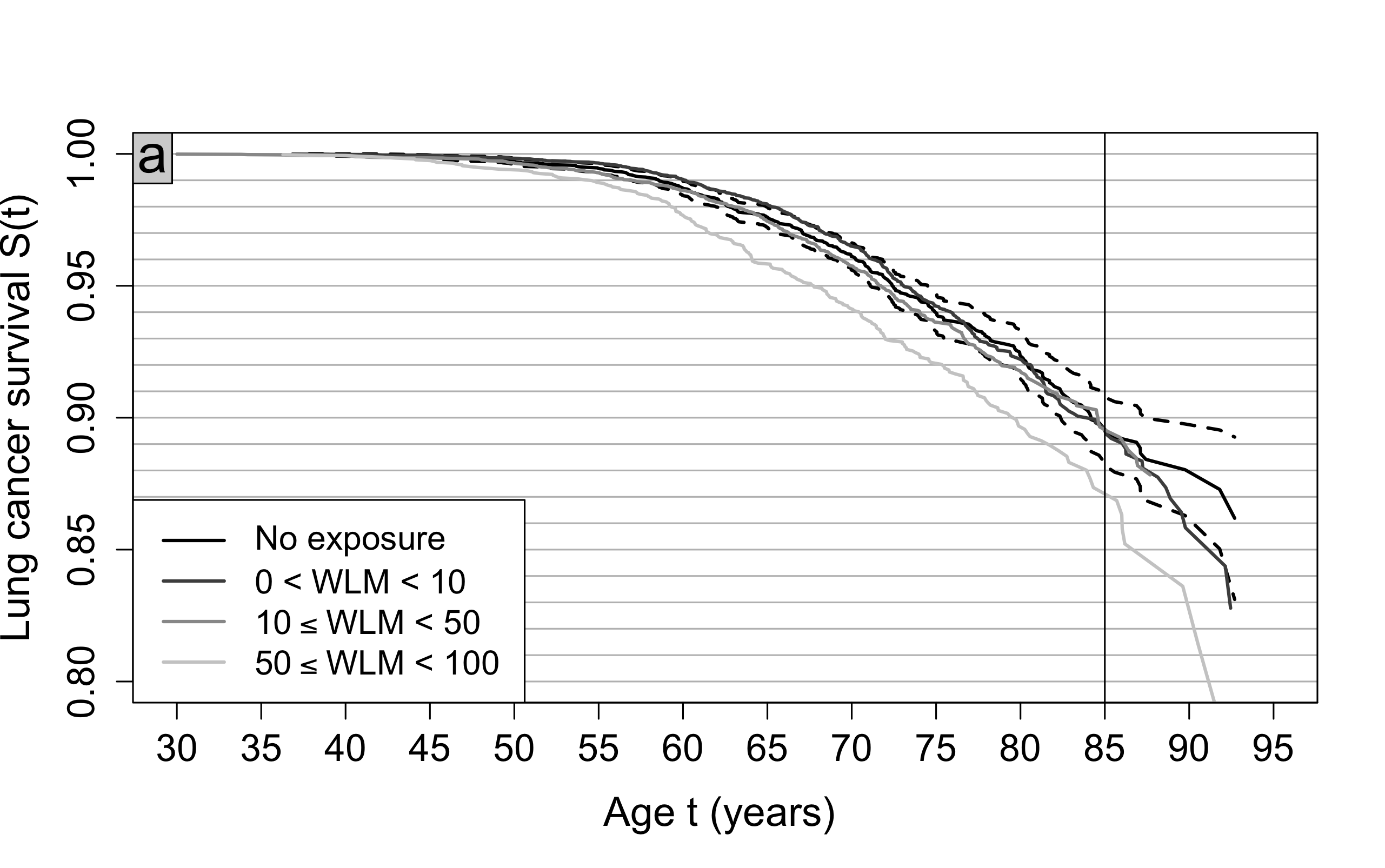}
    \includegraphics[scale=0.0925]{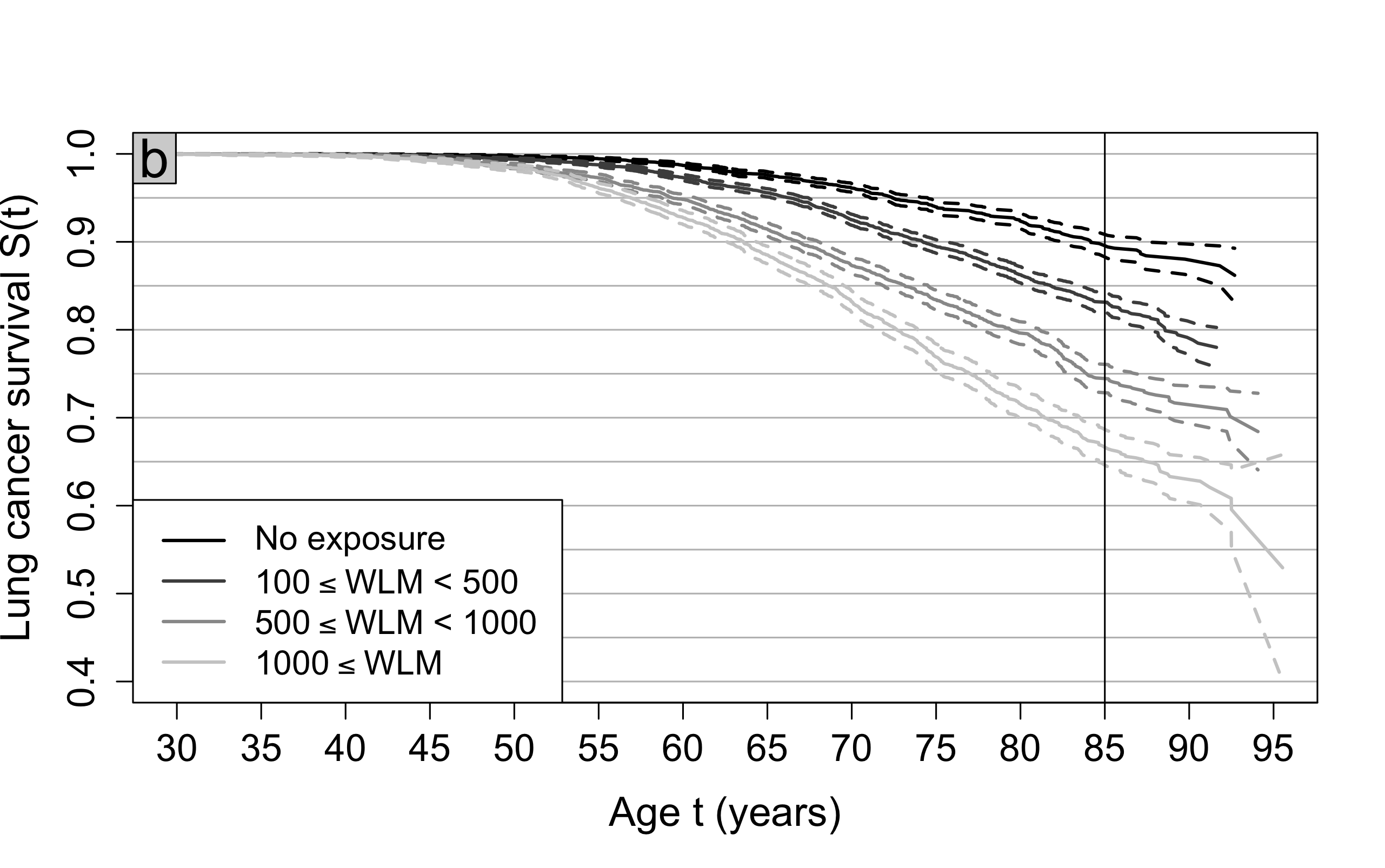}
        \caption{Kaplan-Meier lung cancer survival curve estimates for miners from the Wismut cohort for different windows of total cumulative exposure in WLM. The corresponding dashed lines represent the point-wise upper and lower bound of the $95\%$ uncertainty interval. A vertical line at age $85$ indicates the cut-point for lifetime risk calculations. The left-hand side plot shows lung cancer survival curve estimates for lower exposures, whereas the right-hand side plot shows lung cancer survival curve estimates for higher exposures.}
    \label{fig:KM_survival}
\end{figure}

Lung cancer survival curves were constructed for individuals with different levels of cumulative radon exposure (Suppl. Figure \ref{fig:KM_survival}). The three curves for no exposure, low exposure with at most 10 WLM, and slightly higher exposure with 10 to 50 WLM are close together and only the 50 to 100 WLM curve is notably below (left-hand side plot). Although the curve for 50-100 WLM is notably below the survival for unexposed individuals at practically all time points, the effect is not statistically significant as the point-wise $95\%$ uncertainty interval for the unexposed overlaps with the point-estimate for the 50-100 WLM curve. 
Likewise, there are age sections where the 0-10 WLM and the 10-50 WLM curves show a slightly higher survival probability than the "No exposure" curve. However, incorporating uncertainty intervals, this effect is not statistically significant. Note that the uncertainty bands for positive exposure groups in the low exposure plot are not shown for readability. The right-hand side plot shows survival curves for considerably higher exposures. There, the survival probabilities are all statistically significantly different at the $95\%$ confidence level. However, at the age of $95$, the uncertainties increase resulting in overlap for the category $500-1000$ WLM and $>1000$ WLM. \\

To calculate lifetime risks from these survival curves according to (\ref{LEARSidea}), setting a maximum age $t$ is necessary. Here, we refer to the calculated lifetime excess risk as $LEAR$ for simplicity. Here, the naive $LEAR$ estimate is the vertical distance between a survival curve for a certain exposure and the "No exposure" survival curve at $t=85$ (Suppl. Table \ref{tab:KM_Survival}), i.e.
 $$ LEAR = S_0(85)-S_E(85).$$  
 
The naive $LEAR$ confidence interval bounds are constructed by comparing the "No exposure" curve $S_0$ with the corresponding exposed survival curve $S_E$. The lower bound is the difference between the lower bound of the "No exposure" curve and the upper bound of the exposed survival curve. The upper bound is the difference between the upper bound of the "No exposure" curve and the lower bound of the exposed survival curve. $LEAR$ increases for higher exposure with large uncertainties at lower exposure. In particular, the $LEAR$ for exposures below $100$ WLM is not statistically significantly different from zero. For higher exposures, the confidence intervals get narrower relative to the reference estimate (relative uncertainty span). Note that in the German uranium miners cohort, the majority were smokers. This results in a presumably lower baseline survival probability compared to a general unexposed population. \\

    \begin{table}[htbp]
    \centering
  \begin{tabular}[h]{lllll}

\hline
\addlinespace
WLM interval & Cohort size&$\hat{S}_E(85)$ in $\%$ & $\widehat{LR}_E(85)$ in $\%$ & $LEAR$ in $\%$\\
\addlinespace
\hline
\addlinespace
No exposure & 8,213 (50.15\%)& $89.55$ $[88.28;90.84]$    &  $10.45$ $[9.16;11.72]$ & -\\
\addlinespace
$(0,10)$  & 16,544 (30.31\%)&$89.52$ $[88.28;90.78]$ &  $10.48$ $[9.22;11.72]$ & $0.03$  $[-2.50; 2.56]$ \\ 
\addlinespace
$[10,50)$    & 11,185 (39.16\%)&$89.47$ $[88.12;90.81]$ & $10.53$ $[9.19;11.88]$ & $0.09$   $[-2.54; 2.71]$\\
\addlinespace
$[50,100)$    & 3,314 (57.51\%)&$86.86$ $[84.67;89.05]$ & $13.14$ $[10.95;15.33]$ & $2.69$   $[-0.78; 6.17]$\\
\addlinespace
$[100,500)$    &8,979 (72.81\%)&$83.12$  $[81.97;84.27]$ & $16.88$ $[15.73;18.03]$ & $6.43$  $[4.01; 8.87]$\\
\addlinespace
$[500,1000)$    & 6,039 (80.54\%)&$74.42$ $[72.81;76.04]$ & $25.58$ $[23.94;27.19]$ & $15.13$ $[12.22;18.03]$\\
\addlinespace
$[1000,\infty)$ & 4,698 (85.89\%)& $66.58$ $[64.55;68.61]$ & $33.42$ $[31.39;35.45]$ & $22.97$ $[19.67;26.29]$\\
\addlinespace
\hline
\end{tabular}
     \caption{Values for Kaplan-Meier lung cancer survival estimates $\hat{S}_E(85)$, $\widehat{LR}_E(85)=1-\hat{S}_E(85)$ and $LEAR$ as $\widehat{LR}_E(85)$ minus the baseline $S_0(85)$ for different windows of total cumulative exposures in WLM with corresponding $95\%$ uncertainty interval in correspondence to Suppl. Figure \ref{fig:KM_survival}. For reference, the corresponding cohort size is shown, with the fraction of individuals deceased before age 85 in brackets. }
        \label{tab:KM_Survival}
\end{table}

This simple and naive approach to deriving uncertainty intervals for lifetime risks yields plausible results. Note that Log-Rank tests (compare \cite{Zhizhilashvili2024-uf}) confirmed that most survival curves in Suppl. Figure \ref{fig:KM_survival} are statistically significantly different from the "No exposure" curve at the 95\% confidence level, i.e. p-values below $0.05$. Only the "(0,10) WLM" and "[10,50) WLM" curves are not statistically significantly different from the "No exposure" curve with a p-value of $0.7$ and $0.4$, respectively. \\

Although the obtained intervals are rough, plotting the survival curves serves particularly well for an interpretable visual risk assessment to grasp the impact of radon exposure on lung cancer survival. Further, this approach can easily be applied to other endpoints than radon-induced lung cancer. 
\clearpage
\section{Risk model parameter uncertainty}
\label{Risk_model_parameter_uncertainty_SDC_C}
This supplementary analysis on risk model parameter uncertainty presents the ANA approach for the simple linear risk model and extends the Bayesian approach for the parametric 1960+ sub-cohort model by incorporating a wider range of prior information combinations. Further, the Bayesian approach is applied to the simple linear risk model using gamma-distributed and log-normally distributed priors for $\beta$. 

\subsection{ANA approach for the simple linear 1960+ sub-cohort risk model - analytical solution}
\label{Risk_model_parameter_uncertainty_simple_linear_SDC_C1}
For the simple linear risk model,
\begin{equation}
\label{simplelinearERR}
    ERR(t; \hat{\beta}) = \hat{\beta}W(t),
\end{equation}
which is derived from the Wismut 1960+ sub-cohort, the ANA approach employs:
\begin{equation}
\label{distip}
\hat{\beta}\sim \mathcal{N}\left( \hat{\beta}_0, \hat{\sigma}_0^2\right)
\end{equation}
with $\hat{\beta}_0=0.0134$ and  $\hat{\sigma}_0=0.003005$ as reported in \cite{Kreuzer_2023_lc}. Sampling from this distribution (\ref{distip}) is not necessary, as it holds 
\begin{equation}
\label{LEAR_linearERR_ana}
    LEAR \approx \sum_{t \geq 0} r_0(t) ERR(t; \hat{\beta}) \Tilde{S}(t) = \hat{\beta} \sum_{ t \geq 0} r_0(t) W(t) \Tilde{S}(t) =: \hat{\beta}\cdot  C,
\end{equation}
with $C=\sum_{t \geq 0} r_0(t) W(t) e^{-\sum_{u=0}^{t-1}q_0(u)}$.
Hence approximately, $LEAR \sim \mathcal{N} \left( \hat{\beta}_0 \cdot C,\hat{\sigma}^2_0 \cdot C^2 \right) $
with confidence bands at level $\alpha$, 
\begin{equation}
\label{ki_eq_simple_lin}
    \hat{\beta}_0 \cdot C \pm z_{1-\alpha/2} \cdot  \hat{\sigma}_0 \cdot  C,
\end{equation}
where $z_{1-\alpha/2}$ is the standard-normal quantile at level $1-\alpha/2$. It holds $C=4.27$ for ICRP reference mortality rates $r^{(ICRP)}_0(t),q^{(ICRP)}_0(t)$ for all ages $t$ and an exposure scenario of $2$ WLM from age 18-64 years. For this simple linear model, (\ref{ki_eq_simple_lin}) yields the estimate $5.71$ with 95\% uncertainty interval ($\alpha=0.05$), $[3.18;8.22]$ for LEAR in \%, as stated in the main manuscript. 
\subsection{Bayesian approach for the simple linear 1960+ sub-cohort risk model}
\label{Risk_model_parameter_uncertainty_simple_linear_bayes_SDC_C2}
Here, we investigate risk model parameter uncertainty and its influence on $LEAR$ estimates within the Bayesian framework for the simple linear risk model (\ref{simplelinearERR}). The general methodology and notation are analogous to the main manuscript. To assess the influence of prior choices on the inference of $\beta$, we explore choosing a log-normal distribution as a prior in contrast to the gamma distribution in the main manuscript. To test this framework, again the "prior" information is the $\beta$ estimate from \cite{Tomasek_mines_2008} with $\hat{\beta}_{CZ+F}=0.016$.  The corresponding "prior" $LEAR$ in \% estimate calculated with $\hat{\beta}_{CZ+F}$ is $6.83$. We assume for the prior distribution $P(\beta)$ of the risk model parameter $\beta$,
\begin{equation}
     \beta \sim \mathcal{LN}\left(\log\left(\hat{\beta}_{CZ+F}\right)+\sigma^2,\sigma^2\right)
\end{equation}
such that again $Mod(\beta)=e^{\log\left(\hat{\beta}_{CZ+F}\right)+\sigma^2-\sigma^2}=\hat{\beta}_{CZ+F}$ to match the mode with the prior information. The variance is controlled via $\sigma$. Compared to a gamma distribution, this distribution is characterized by heavier right tails.  \\

To be comparable with the assumed certainty with the gamma-distributed prior in the main manuscript, we assume an equal coefficient of variation between gamma- and log-normally distributed priors by prescribing $\sigma = \sqrt{\log \left(1+ \frac{1}{a} \right)}$. This allows for comparison with the gamma prior while maintaining similar variation (Suppl. Table \ref{tab:combined_linERR_bayes}). The values $\left(2,5,10,20,50\right)$ for $a$ translate to $\left(0.64,0.43,0.31,0.22,0.14\right)$ for $\sigma$. Log-normal priors lead to slightly narrower highest posterior density intervals (HPDIs) and a tendency toward larger values compared to gamma priors. This difference diminishes with increasing prior certainty (smaller $\sigma$).

\begin{table}[htbp]
    \centering
    \begin{tabular}{lll}
    \toprule
    Prior information & $\hat{\beta}=Mod(P(\beta \vert X)) \times 100$ & $LEAR$ in \% \\
    \midrule
    ML estimate with Wald-type CI &	$1.34$ $[0.71;1.97]$ &	$5.72$ $[3.03;8.41]$ $(0.94)$\\
    Uniform prior & $1.34$ $[0.79; 2.08]$ &$5.72$ $[3.37; 8.90]$ $(0.96)$\\
    \midrule
    \multicolumn{3}{c}{Log-normal-distributed prior} \\
    \midrule
    $\sigma=0.64$ & $1.37$ $[0.85; 2.07]$ &$5.85$ $[3.62; 8.81]$ $(0.89)$ \\ 
    $\sigma=0.43$ & $1.40$ $[0.90; 2.05]$ &$5.97$ $[3.85; 8.75]$ $(0.82)$\\
    $\sigma=0.31$ & $1.43$ $[0.97; 2.03]$ &$6.11$ $[4.16; 8.66]$ $(0.74)$\\
    $\sigma=0.22$ & $1.47$ $[1.06; 1.99]$ &$6.29$ $[4.54; 8.50]$ $(0.63)$\\
    $\sigma=0.14$ & $1.53$ $[1.20; 1.92]$ &$6.52$ $[5.13; 8.19]$ $(0.47)$\\
    \bottomrule
    \end{tabular}
    \caption{Mode $\hat{\beta}$ of posterior distribution $P(\beta \vert X)$ with $LEAR$ calculated with the model (\ref{simplelinearERR}) and their corresponding 95\% highest posterior density interval (HPDI) (relative uncertainty span in brackets) for varying prior parameter settings. The log-normal-distributed prior $P(\beta)$ is centered at the corresponding parameter estimate from \cite{Tomasek_mines_2008}  for different values of prior standard deviation parameters $\sigma$. Wald-type confidence intervals (CI) are also shown for comparison and are calculated as $\hat{\beta} \pm 1.96 \times 0.003005$, where $0.003005$ represents the parameter standard error as stated in \cite{Kreuzer_2023_lc}.}
    \label{tab:combined_linERR_bayes}
\end{table}

\subsection{Bayesian approach for the parametric 1960+ sub-cohort risk model - additional details}
\label{Risk_model_parameter_uncertainty_loglinear_bayes_SDC_C3}
\begin{figure}[htbp]
    \centering
    \includegraphics[scale=0.072]{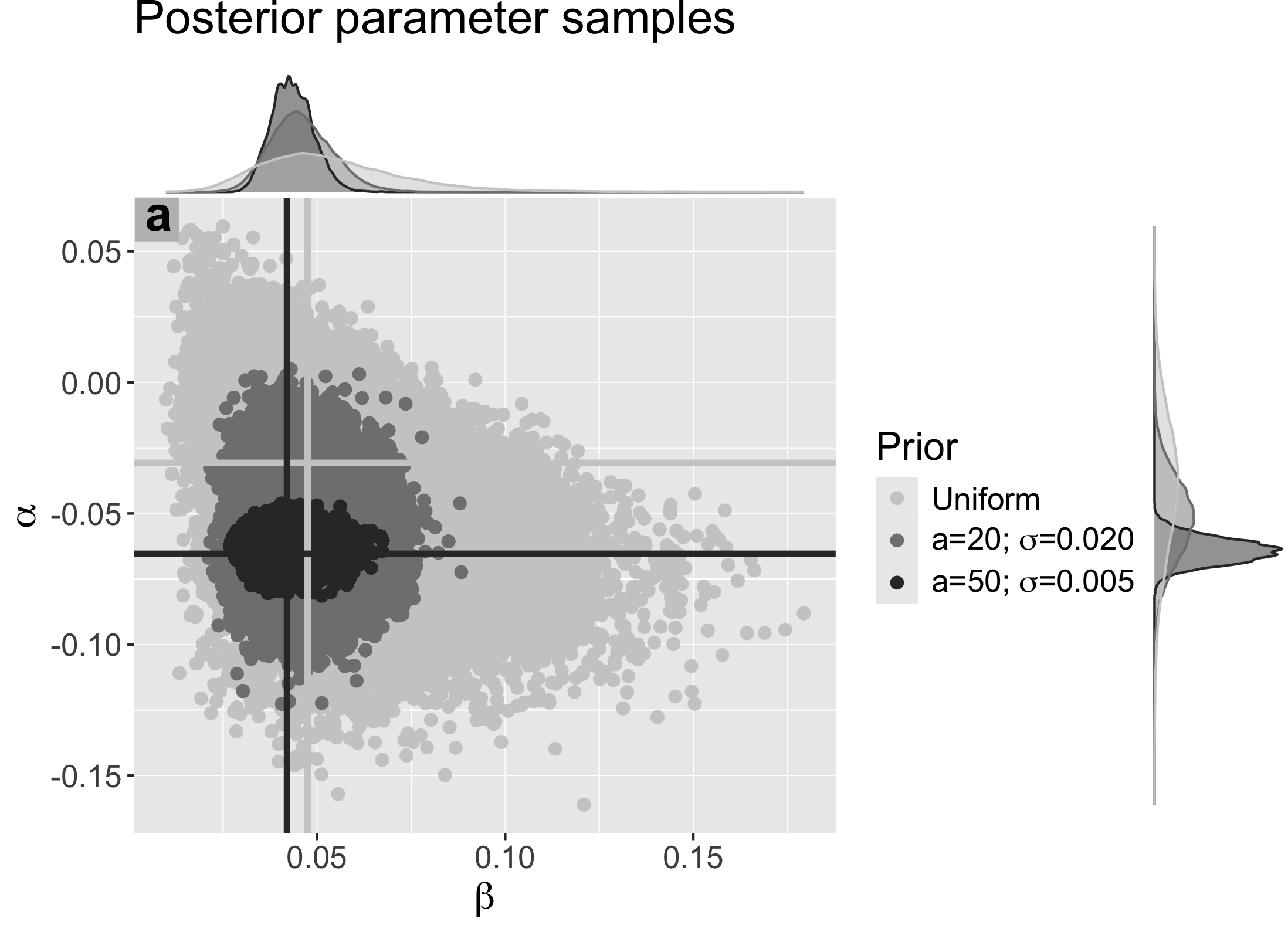}
    \includegraphics[scale=0.072]{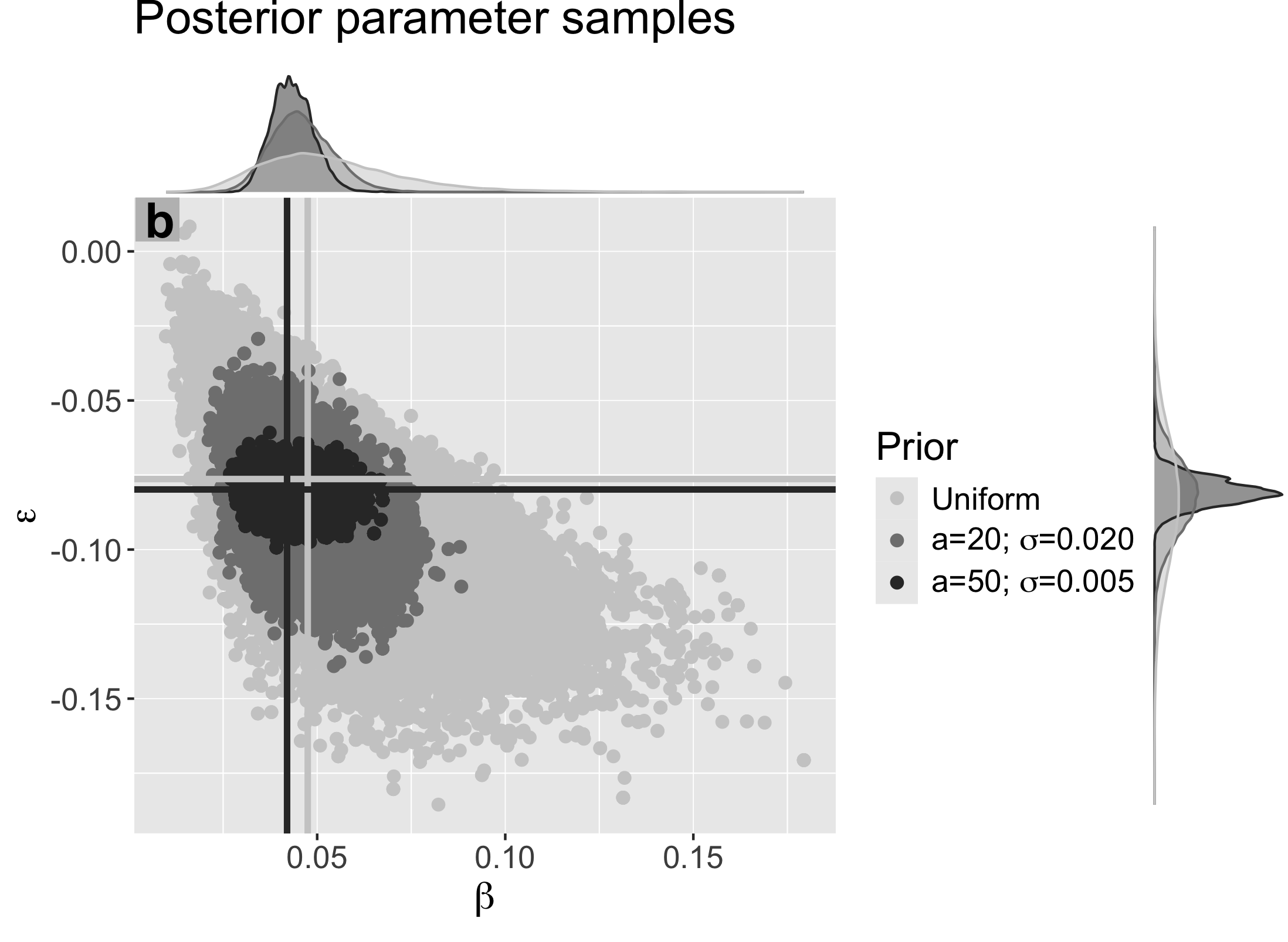}
    \includegraphics[scale=0.072]{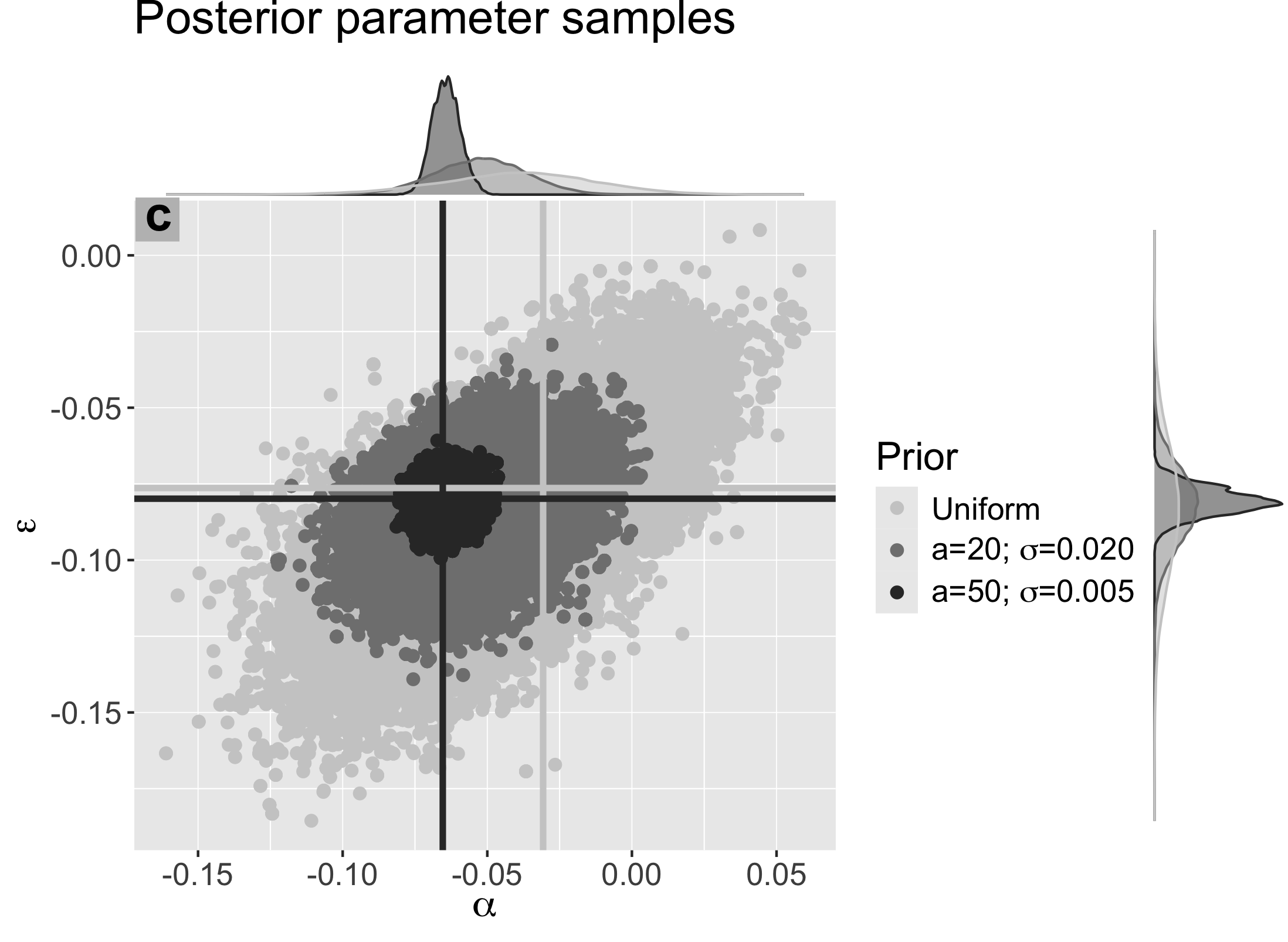}
    \includegraphics[scale=0.072]{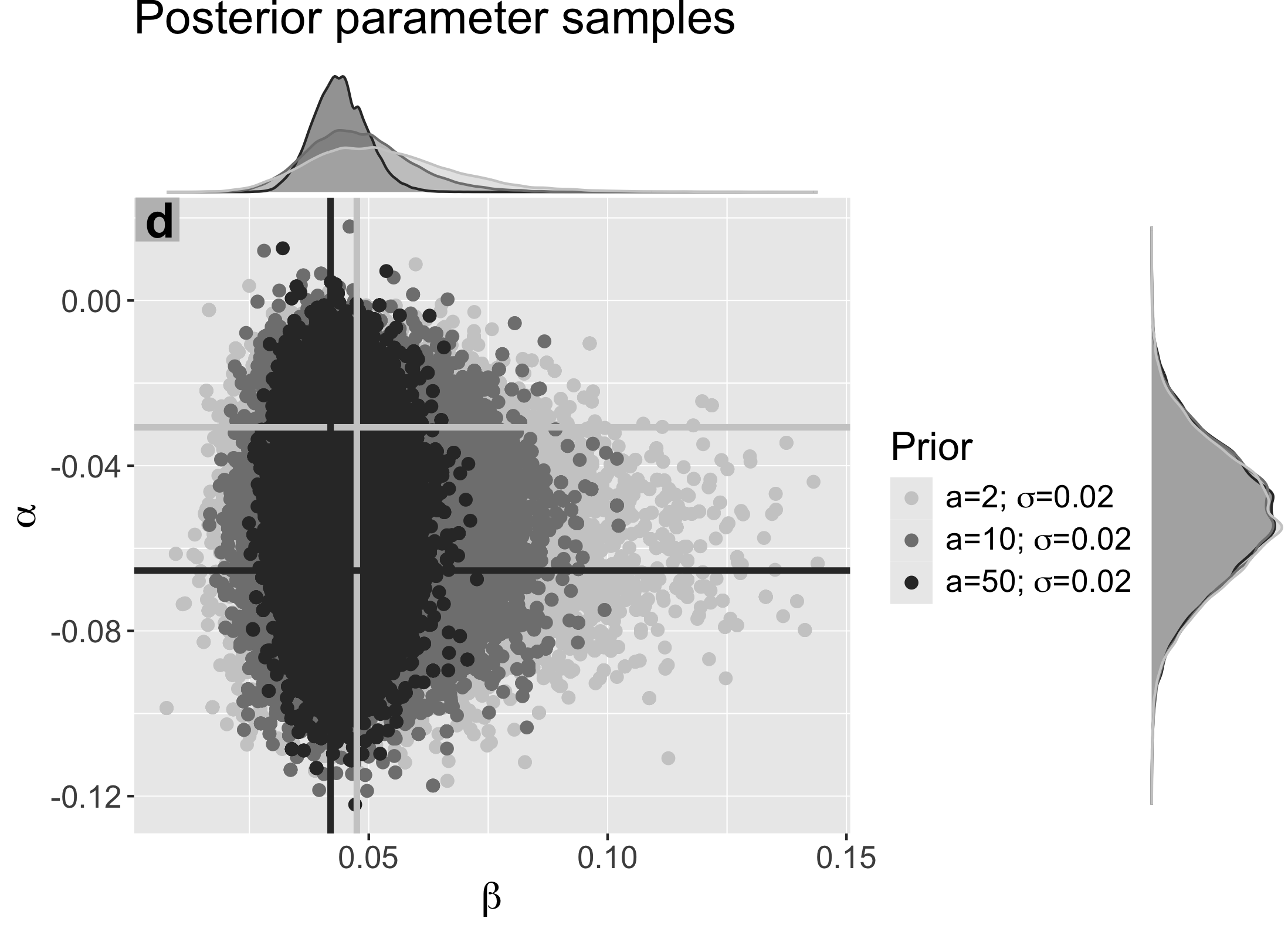}
    \includegraphics[scale=0.072]{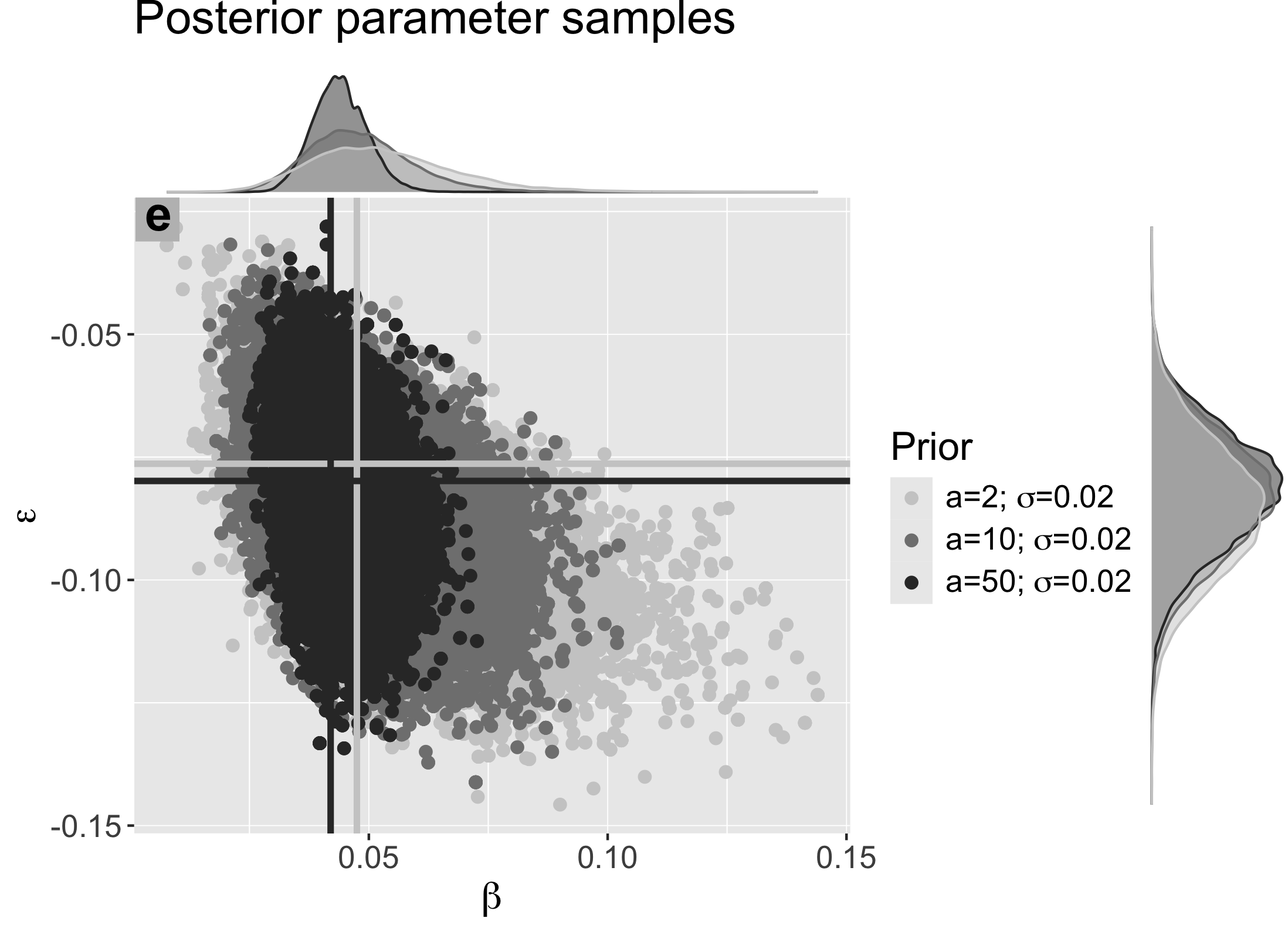}
    \includegraphics[scale=0.072]{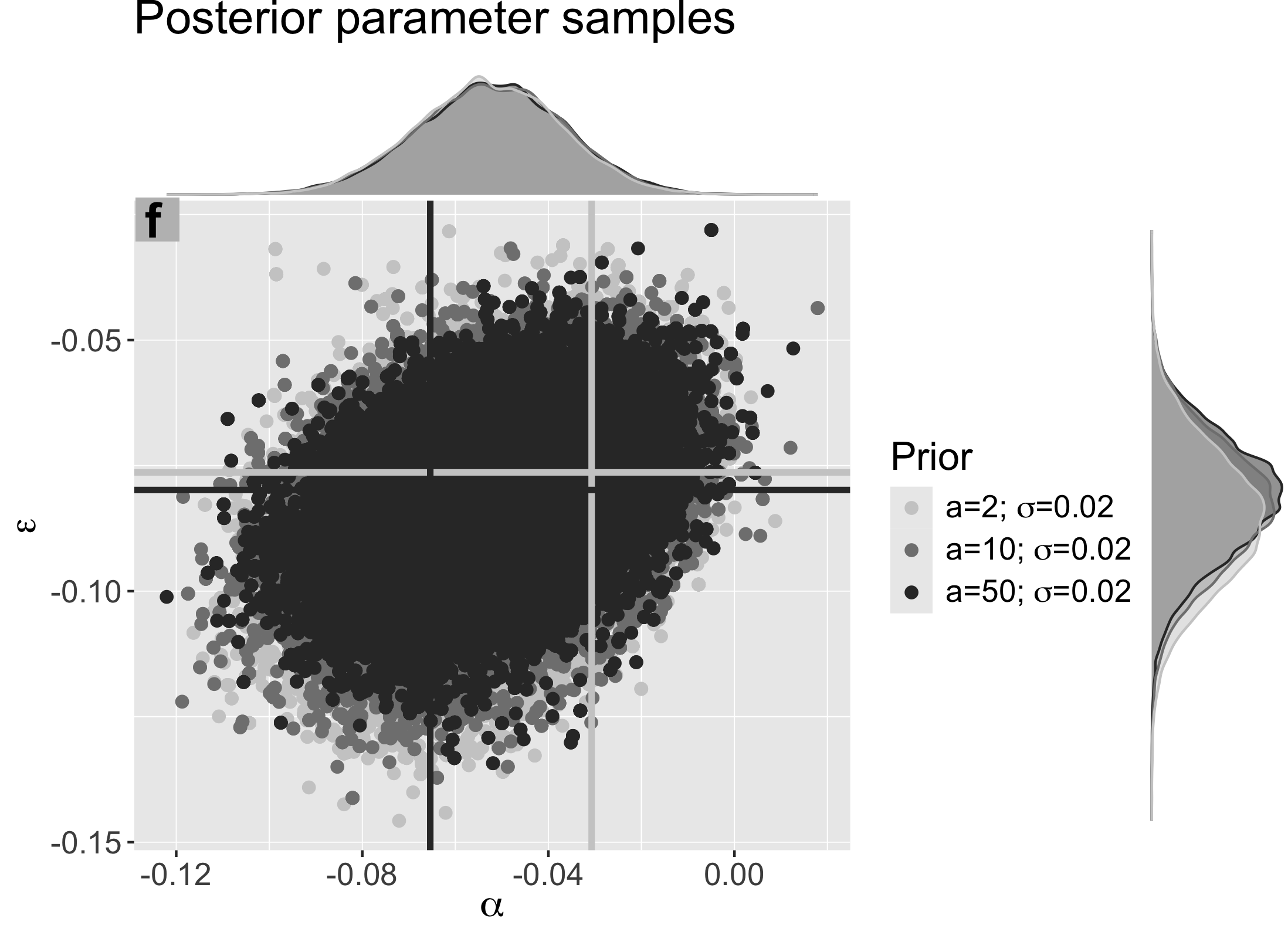}
    \includegraphics[scale=0.072]{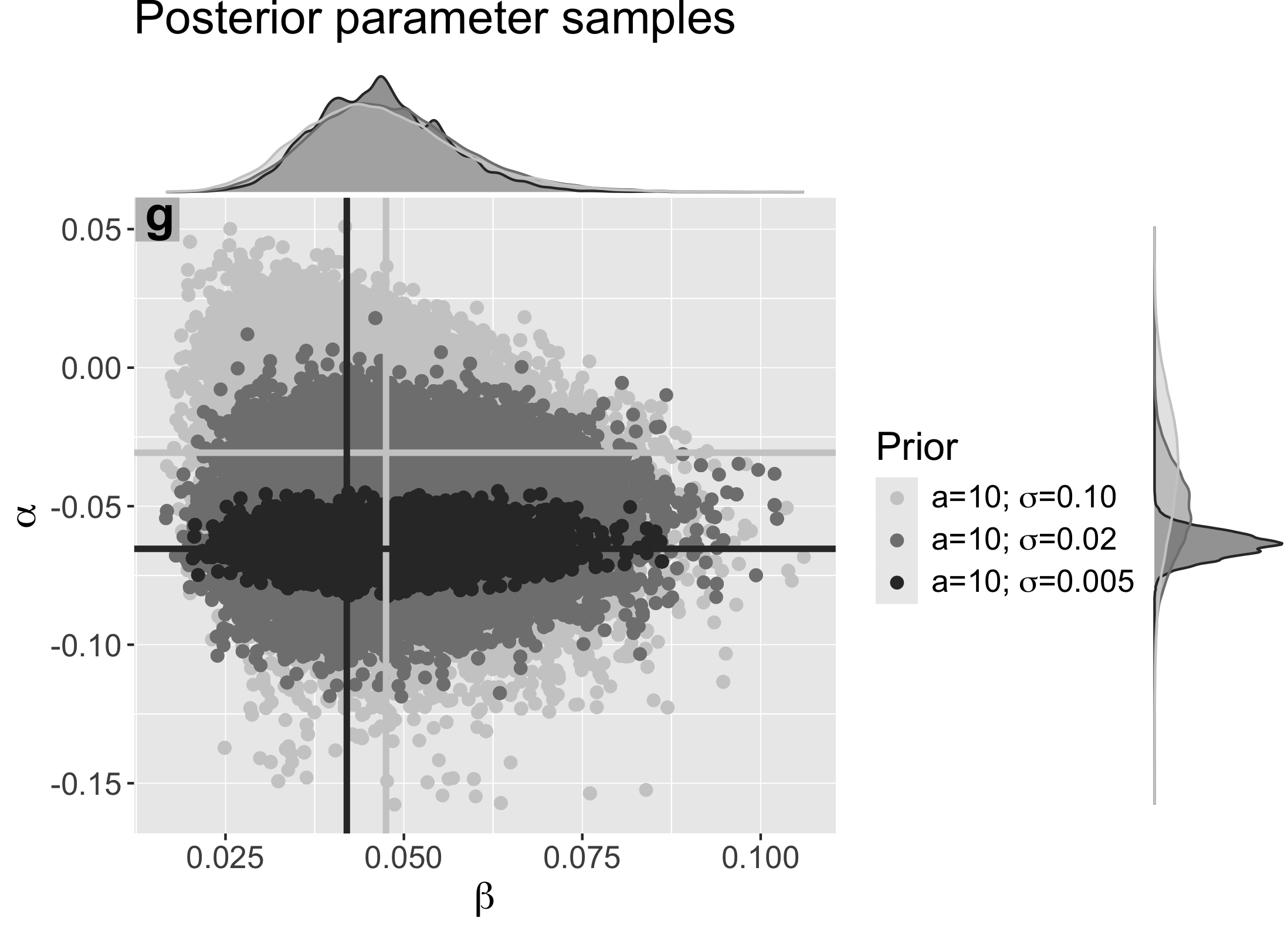}
    \includegraphics[scale=0.072]{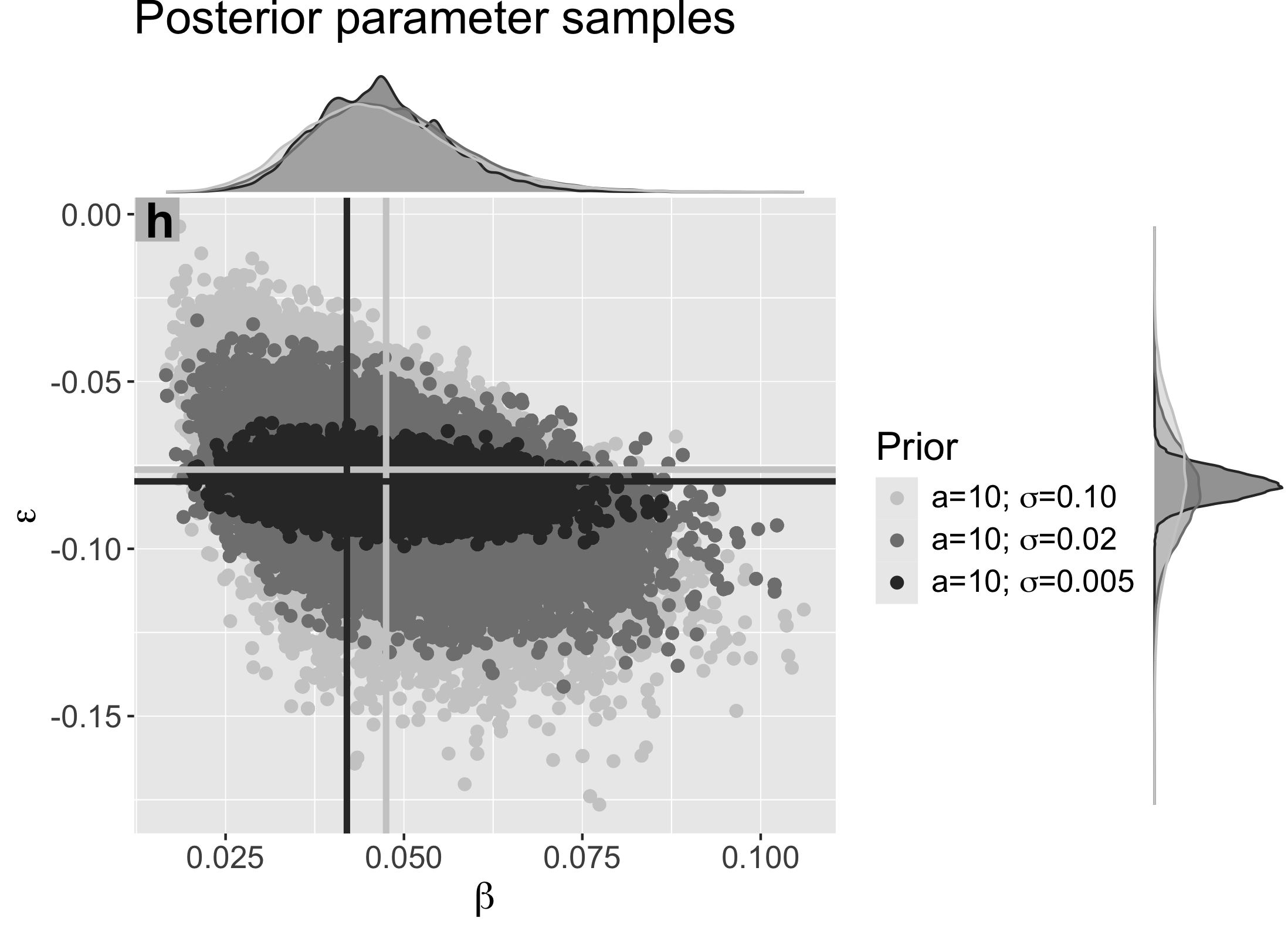}
    \includegraphics[scale=0.072]{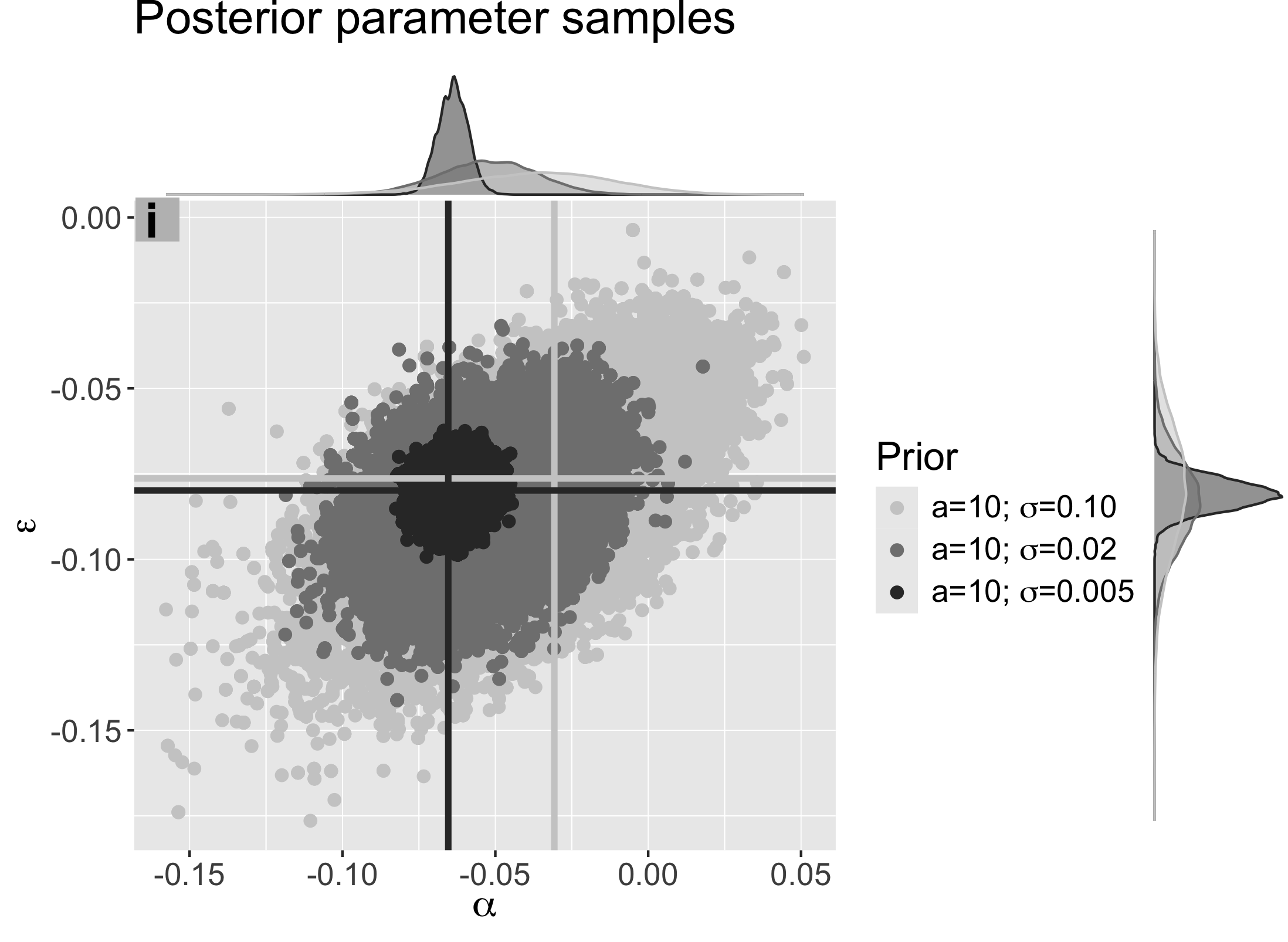}
    \caption{Scatterplots of $100,000$ parameter samples $(\beta,\alpha)$, $(\beta,\varepsilon)$ and $(\alpha, \varepsilon)$ from posterior distribution $P(\Theta \vert X)$ using Bayesian inference methods with marginal probability density on the sides in correspondence to Table \ref{tab:linlogERR_bayes_ki_sigma}. The model parameters $\Theta=\left(\beta, \alpha, \varepsilon\right)$ (model (\ref{model2_para_sublinear}) are estimated assuming prior distributions $P(\Theta)=P(\beta)P(\alpha)P(\varepsilon)$ for gamma-distributed $\beta$ and normally distributed $\alpha, \varepsilon$. The plots show the effect of increased certainty in all parameters $\beta, \alpha, \varepsilon$ (first row), only $\beta$ (second row), and $\alpha$, $\varepsilon$ (last row). The intersection of horizontal and vertical lines indicates the optimal parameter values in the prior distribution (black) and the likelihood function (light grey).}
    \label{fig:linlogERR_bayes_full}
\end{figure}

Here, supplemental results are shown for the Bayesian approach to quantify uncertainties for the parametric 1960+ sub-cohort risk model (\ref{model2_para_sublinear}) from \cite{Kreuzer_2023_lc},
\begin{equation*}
     ERR(t;  \beta, \alpha, \varepsilon) = \beta W(t) \exp\left\{ \alpha \left(AME(t)-30)\right)+\varepsilon\left(TME(t)-20\right)\right\}. 
\end{equation*}
The approach is completely analogous to the main manuscript. However, a more detailed view of results are shown with more choices for prior standard deviations $\sigma$. In particular, the derived risk model parameter estimates with 95\% HPDI are shown (Suppl. Table \ref{tab:linlogERR_bayes_ki_sigma} and Figure\ref{fig:linlogERR_bayes_full}).\\

Supplementary Figure \ref{fig:linlogERR_bayes_full} shows scatterplots for risk model parameters samples for varying prior parameter choices. The marginal probability densities are on the sides. To read Suppl. Figure \ref{fig:linlogERR_bayes_full}, the first row shows the scattering of parameters for increasing certainty in the priors for $\beta$, $\alpha$, $\varepsilon$. Varying gamma shape parameter $a$ affects the $\beta$ prior and varying $\sigma$ affects both the priors for $\alpha$ and $\varepsilon$. The second row shows the effect of increased certainty in $\beta$ without changing priors for $\alpha$ and $\varepsilon$. Analogously, the third row visualizes the effect of increased certainty in $\alpha$ and $\varepsilon$ without affecting $\beta$. Similar to the simple linear risk model (\ref{simplelinearERR}), we observe a concentration of samples for increased certainty in the prior information which likewise affects uncertainty intervals. Further, increasing certainty in one parameter barely affects the scattering for other parameters because the marginal prior distributions are mutually independent. 

\begin{table}[htbp]
    \centering
    \centering\resizebox{\columnwidth}{!}{ 
    \begin{tabular}[h]{lllll}
        \toprule
        Prior information & $\hat{\beta}\times 100$ & $\hat{\alpha} \times 100$ & $\hat{\varepsilon} \times 100$ &$LEAR$ in \%  \\
        \midrule
        ML estimate with Wald-type CI & $4.75$ $[1.60; 7.90]$ & $-3.07$ $[-8.16; 2.01]$ & $-7.63$ $[-11.80; -3.47]$ & $6.74$\\
        Uniform prior & $4.75$ $[2.13; 9.11]$ &$-3.07$ $[-9.36; 1.26]$ & $-7.63$ $[-12.89; -4.08]$ & $6.74$ $[2.96; 11.09]$ $(1.21)$\\
        \midrule
        \multicolumn{5}{c}{\textbf{Prior standard deviation $\sigma=0.1$}} \\
        \midrule
        $a=2$ & $4.75$ $[2.37; 8.55]$ &$-3.25$ $[-9.05; 0.86]$ & $-7.70$ $[-12.57; -4.42]$ & $6.59$ $[2.96; 10.57]$ $(1.15)$\\
        $a=5$ & $4.61$ $[2.40; 7.59]$ &$-3.17$ $[-8.83; 1.00]$ & $-7.57$ $[-12.05; -4.40]$ & $6.52$ $[2.91; 10.26]$ $(1.13)$\\
        $a=10$ & $4.49$ $[2.66; 6.88]$ &$-3.08$ $[-8.66; 1.20]$& $-7.45$ $[-11.59;-4.29]$  & $6.46$ $[2.83; 10.14]$ $(1.13)$\\
        $a=20$ & $4.38$ $[2.93; 6.18]$ &$-3.01$ $[-8.58; 0.99]$&$-7.35$ $[-11.41;-4.60]$  & $6.40$ $[3.04; 10.07]$  $(1.10)$\\
        $a=50$ & $4.29$ $[3.25; 5.45]$ &$-2.95$ $[-8.57; 1.14]$ & $-7.25$ $[-11.23; -4.65]$ & $6.35$  $[2.95; 9.91]$ $(1.10)$\\
        \midrule
        \multicolumn{5}{c}{\textbf{Prior standard deviation $\sigma=0.05$}} \\
        \midrule
        $a=2$ & $4.86$ $[2.41; 8.45]$ &$-3.78$ $[-8.89; 0.07]$ & $-7.96$ $[-12.16; -4.74]$ & $6.32$ $[2.93; 10.19]$ $(1.15)$\\
        $a=5$ & $4.70$ $[2.55; 7.72]$ &$-3.71$ $[-8.92; 0.01]$ & $-7.82$ $[-11.79; -4.74]$ & $6.22$ $[2.89; 9.89]$ $(1.13)$\\
        $a=10$ & $4.56$ $[2.79; 6.89]$ &$-3.64$ $[-8.67; 0.30]$& $-7.70$ $[-11.51;-4.73]$  & $6.12$ $[2.93; 9.66]$ $(1.10)$\\
        $a=20$ & $4.43$ $[2.92; 6.14]$ &$-3.58$ $[-8.60; 0.37]$&$-7.58$ $[-11.32;-4.86]$  & $6.04$ $[2.77; 9.23]$ $(1.07)$\\
        $a=50$ & $4.31$ $[3.28; 5.47]$ &$-3.52$ $[-8.50; 0.51]$ & $-7.48$ $[-10.96; -4.83]$ & $5.96$  $[2.88; 9.20]$ $(1.06)$\\
        \midrule
        \multicolumn{5}{c}{\textbf{Prior standard deviation $\sigma=0.02$}} \\
        \midrule
$a=2$ & $4.97$ $[2.55; 8.23]$ &$-5.17$ $[-8.37; -2.30]$ & $-8.36$ $[-11.26; -5.91]$ & $5.56$ $[2.76; 8.98]$ $(1.12)$\\
$a=5$ & $4.80$ $[2.72; 7.53]$ &$-5.16$ $[-8.41; -2.22]$ & $-8.27$ $[-11.11; -5.90]$ & $5.41$ $[2.82; 8.45]$ $(1.04)$\\
$a=10$ & $4.64$ $[2.81; 6.92]$ &$-5.15$ $[-8.31;-2.10]$& $-8.17$ $[-10.86;-5.81]$  & $5.27$ $[2.89; 8.15]$ $(1.00)$\\
$a=20$ & $4.49$ $[3.08; 6.21]$ &$-5.16$ $[-8.27;-2.05]$&$-8.10$ $[-10.67;-5.79]$  & $5.13$ $[2.99; 7.58]$ $(0.90)$\\
$a=50$ & $4.34$ $[3.30;5.50]$ &$-5.16$ $[-8.39;-2.10]$ & $-8.01$ $[-10.60; -5.82]$ & $4.99$  $[3.14; 7.26]$ $(0.83)$\\
        \midrule
        \multicolumn{5}{c}{\textbf{Prior standard deviation $\sigma=0.01$}} \\
        \midrule
$a=2$ & $4.76$ $[2.67; 7.68]$ &$-6.02$ $[-7.91; -4.23]$ & $-8.24$ $[-10.08; -6.69]$ & $5.00$ $[2.88; 8.13]$ $(1.05)$\\
$a=5$ & $4.65$ $[2.76; 7.12]$ &$-6.03$ $[-7.88; -4.22]$ & $-8.21$ $[-10.06; -6.71]$ & $4.90$ $[2.93; 7.53]$ $(0.94)$\\
$a=10$ & $4.55$ $[2.90; 6.58]$ &$-6.04$ $[-7.83; -4.21]$& $-8.17$ $[-9.85;-6.59]$  & $4.79$ $[2.96; 7.03]$ $(0.85)$\\
$a=20$ & $4.44$ $[3.05; 6.10]$ &$-6.05$ $[-7.88;-4.28]$&$-8.14$ $[-9.77;-6.58]$  & $4.68$ $[3.06; 6.63]$ $(0.76)$\\
$a=50$ & $4.32$ $[3.32;5.49]$ &$-6.06$ $[-7.83;-4.15]$ & $-8.11$ $[-9.78; -6.56]$ & $4.57$  $[3.34; 6.17]$ $(0.62)$\\
\midrule
\multicolumn{5}{c}{\textbf{Prior standard deviation $\sigma=0.005$}} \\
\midrule
        $a=2$ & $4.59$ $[2.79; 7.35]$ &$-6.40$ $[-7.36; -5.42]$ & $-8.07$ $[-9.03; -7.12]$ & $4.73$ $[2.74; 7.44]$ $(0.99)$\\
        $a=5$ & $4.52$ $[2.70; 6.76]$ &$-6.39$ $[-7.36; -5.46]$ & $-8.07$ $[-9.05; -7.19]$ & $4.66$ $[2.85; 7.06]$ $(0.90)$\\
        $a=10$ & $4.45$ $[3.01; 6.41]$ &$-6.39$ $[-7.34; -5.44]$& $-8.06$ $[-8.98;-7.17]$  & $4.59$ $[3.07; 6.57]$ $(0.76)$\\
        $a=20$ & $4.38$ $[3.11; 5.99]$ &$-6.40$ $[-7.35;-5.45]$&$-8.05$ $[-8.95;-7.11]$  & $4.52$ $[3.17; 6.27]$ $(0.69)$\\
        $a=50$ & $4.30$ $[3.30;5.39]$ &$-6.40$ $[-7.39;-5.50]$ & $-8.04$ $[-8.94; -7.12]$ & $4.43$  $[3.34; 5.66]$ $(0.52)$\\
        \bottomrule
    \end{tabular}}
    \caption{Components $\hat{\beta},\hat{\alpha},\hat{\varepsilon}$ of the mode vector $\hat{\Theta}=Mod(P\left(\Theta \vert X \right))$ of the posterior distribution  $P\left(\Theta \vert X \right)$ with $LEAR$ evaluated at mode $\hat{\Theta}$ and the corresponding 95\% highest posterior density interval (HPDI) (relative uncertainty span in brackets) for different values of prior gamma shape parameters $a$ and prior standard deviation $\sigma$. The $LEAR$ in \% with risk model parameter estimates derived from the Joint Czech+French cohort is $4.30$. The Wald-type confidence intervals (CI) are calculated as $\hat{\theta} \pm 1.96 \times \hat{\sigma}_{\theta}$, where $\hat{\sigma}_{\theta}$ represents the parameter standard error from \cite{Kreuzer_2023_lc} for $\theta=\beta, \alpha, \varepsilon$, respectively.}
    \label{tab:linlogERR_bayes_ki_sigma}
\end{table}

\clearpage

\section{Mortality rate uncertainty}
\label{Mortality_rate_uncertainty_SDC_D}
In this section, only mortality rates are assumed to be uncertain, analogously to the main manuscript (see uncertainty assessment in the Methods section). In particular, risk model parameters are fixed and do not influence uncertainty intervals.
\subsection{Preliminaries}
\label{Mortality_rate_uncertainty_preliminaries_SDC_D1}
This section investigates the effect of all-cause mortality rate uncertainty and the impact of centering mortality rate distributions derived from sex-averaged WHO data around the ICRP reference value. Both effects will be found to be negligible for upcoming analyses. 
\subsubsection{Influence of all-cause mortality rates}
Here, we compare the uncertainties in all-cause mortality rates $q_0(t)$ and lung cancer mortality rates $r_0(t)$ and their effect on the LEAR across various risk models. \\

As explained in the main manuscript, we assume a gamma distribution on both mortality rates for all ages $t$ as in (\ref{mr_r0_ass}) and (\ref{mr_q0_ass}),
\begin{align*}
    r_0(t) &\sim G\left(a^{(r_0)}_t,b^{(r_0)}_t\right), \\
    q_0(t) &\sim G\left(a^{(q_0)}_t,b^{(q_0)}_t\right), 
\end{align*}
with age-dependent shape parameters $a^{(r_0)}_t,a^{(q_0)}_t$ and rate parameter $b^{(r_0)}_t,b^{(q_0)}_t$. The main manuscript outlines the model fitting on WHO data with resulting parameter estimates shown in Suppl. Table \ref{tab:WHO_r0_q0_estimates_5EAA} and the corresponding derivation of LEAR uncertainty intervals. \\

\begin{figure}[htbp]
    \centering
    \includegraphics[scale=0.0925]{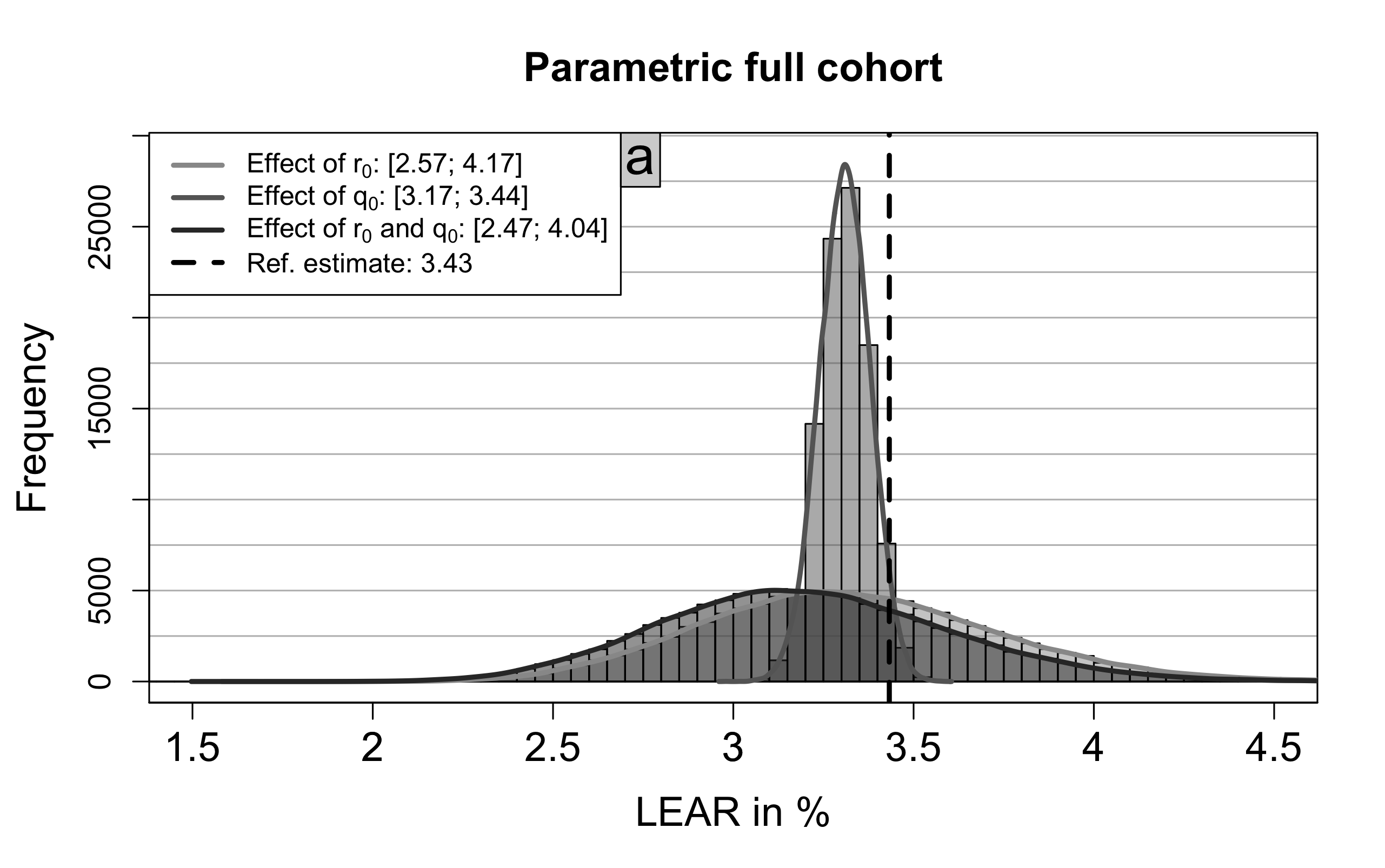}
    \includegraphics[scale=0.0925]{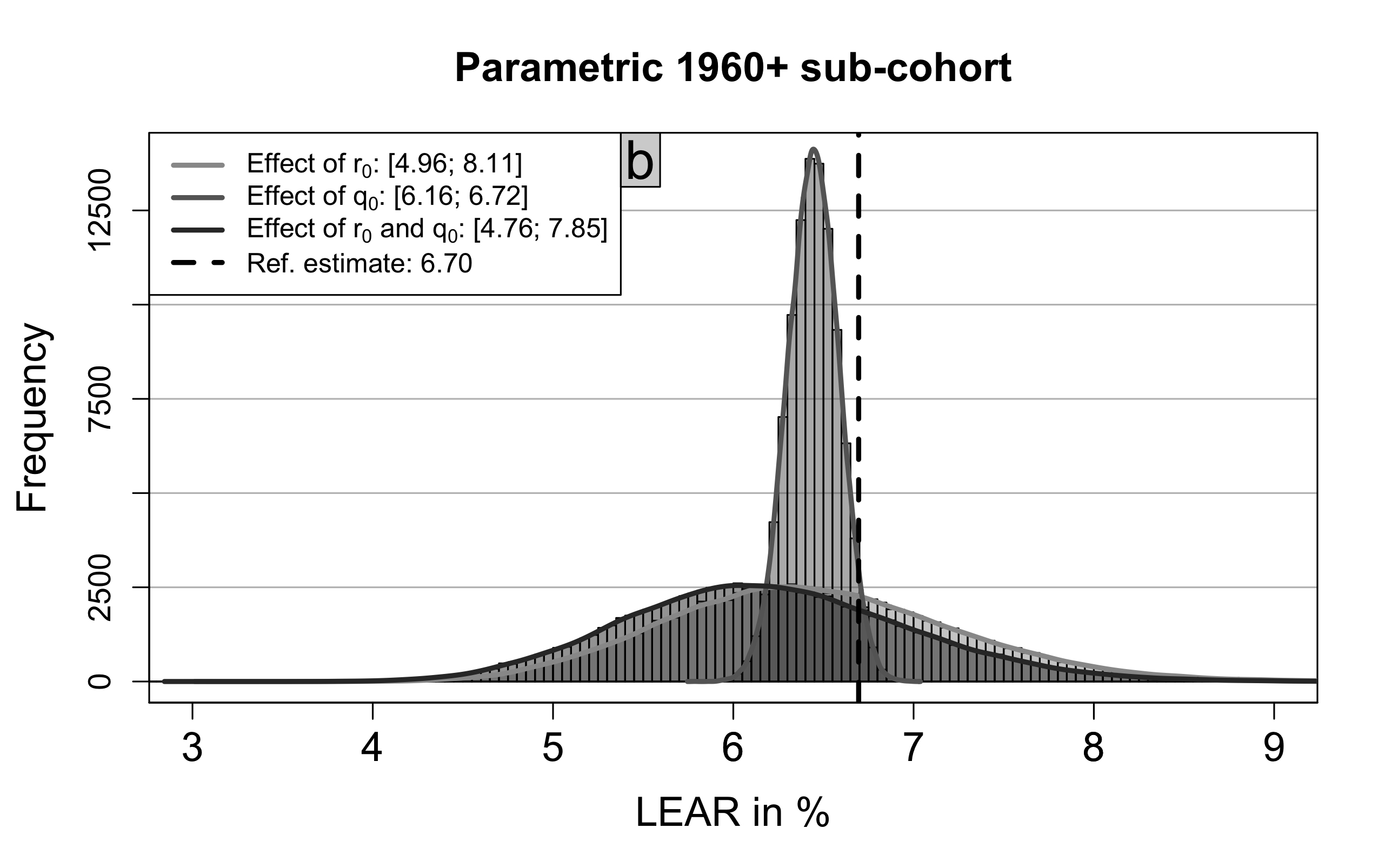}
    \includegraphics[scale=0.0925]{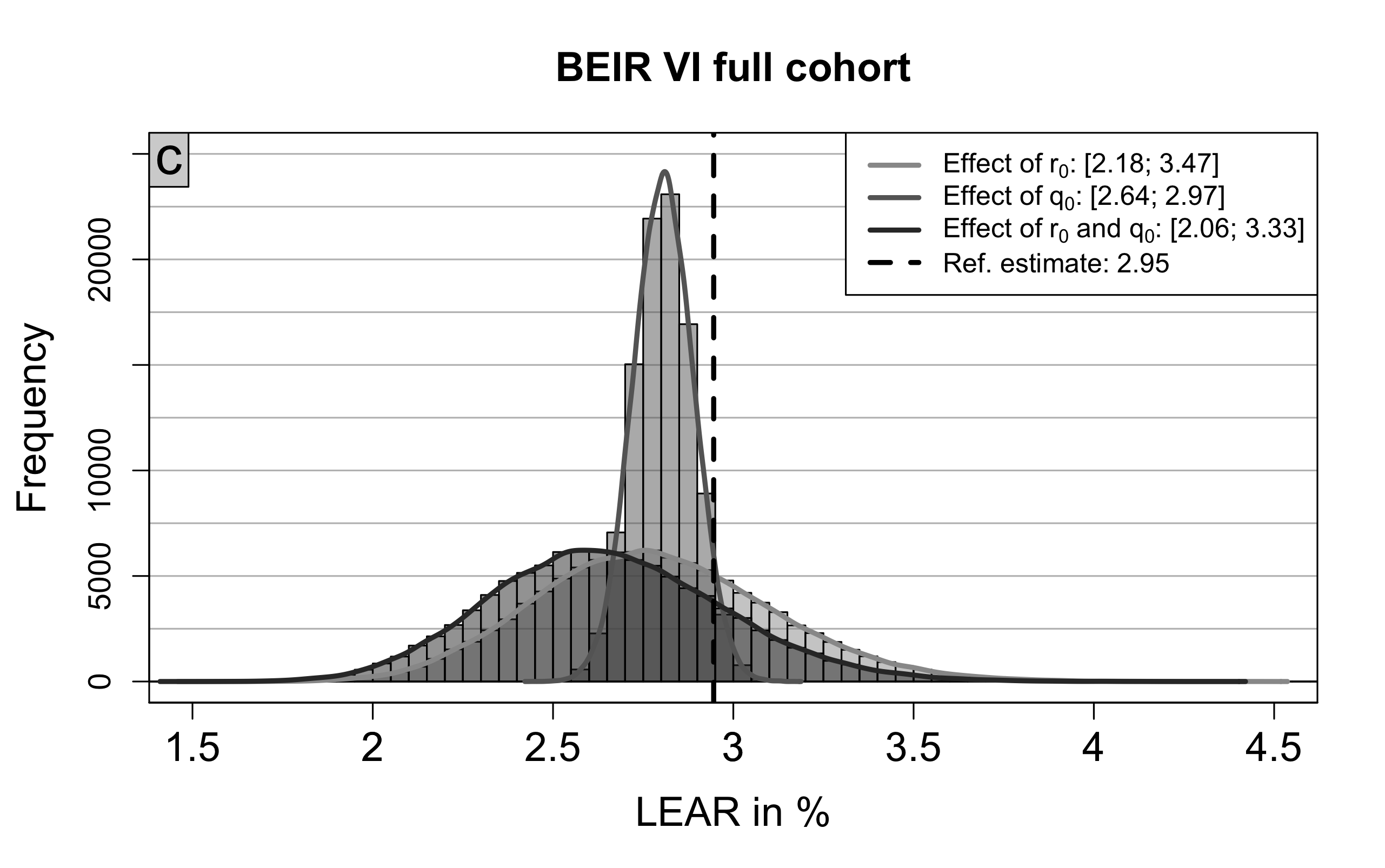}
    \includegraphics[scale=0.0925]{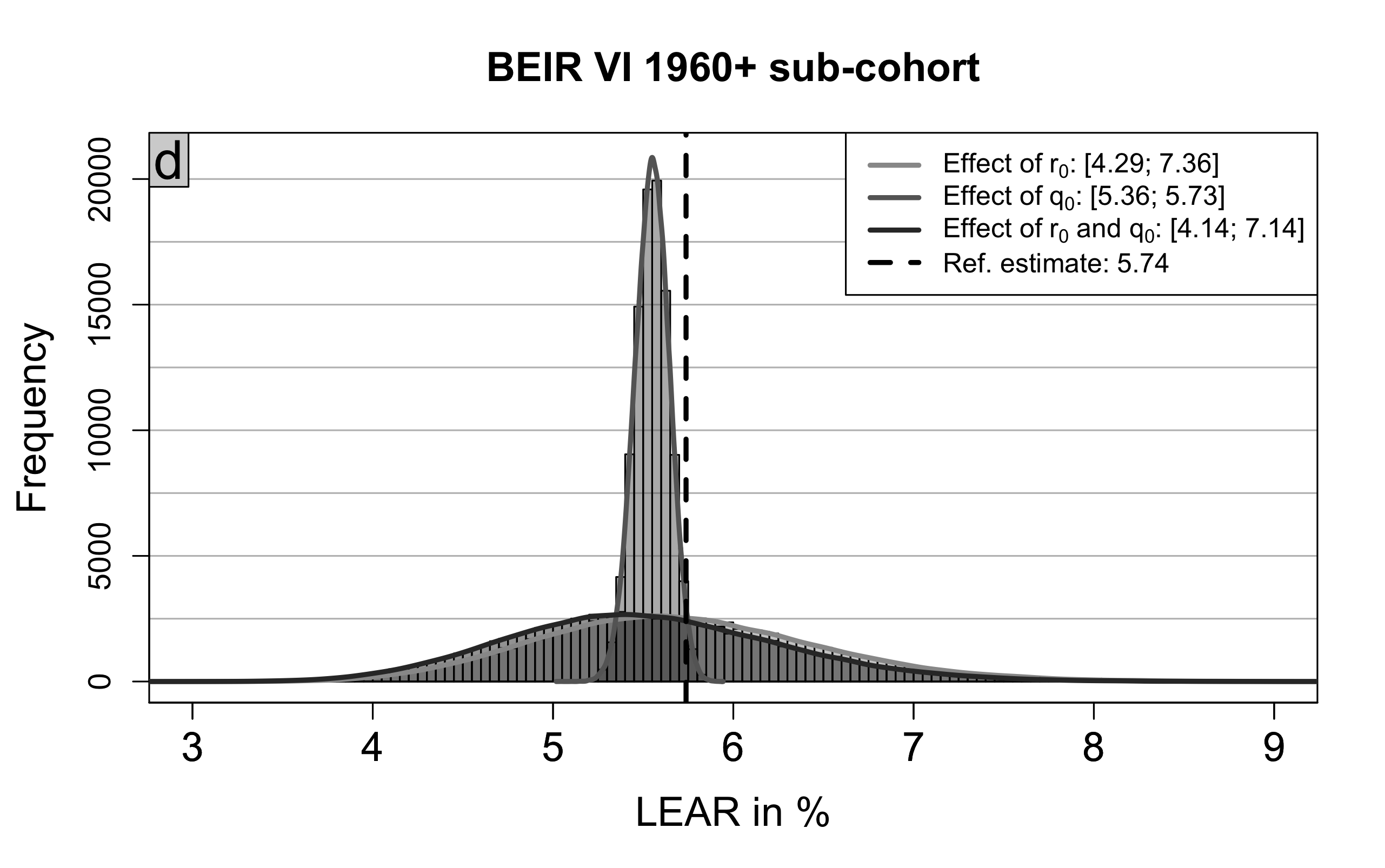}
    \includegraphics[scale=0.0925]{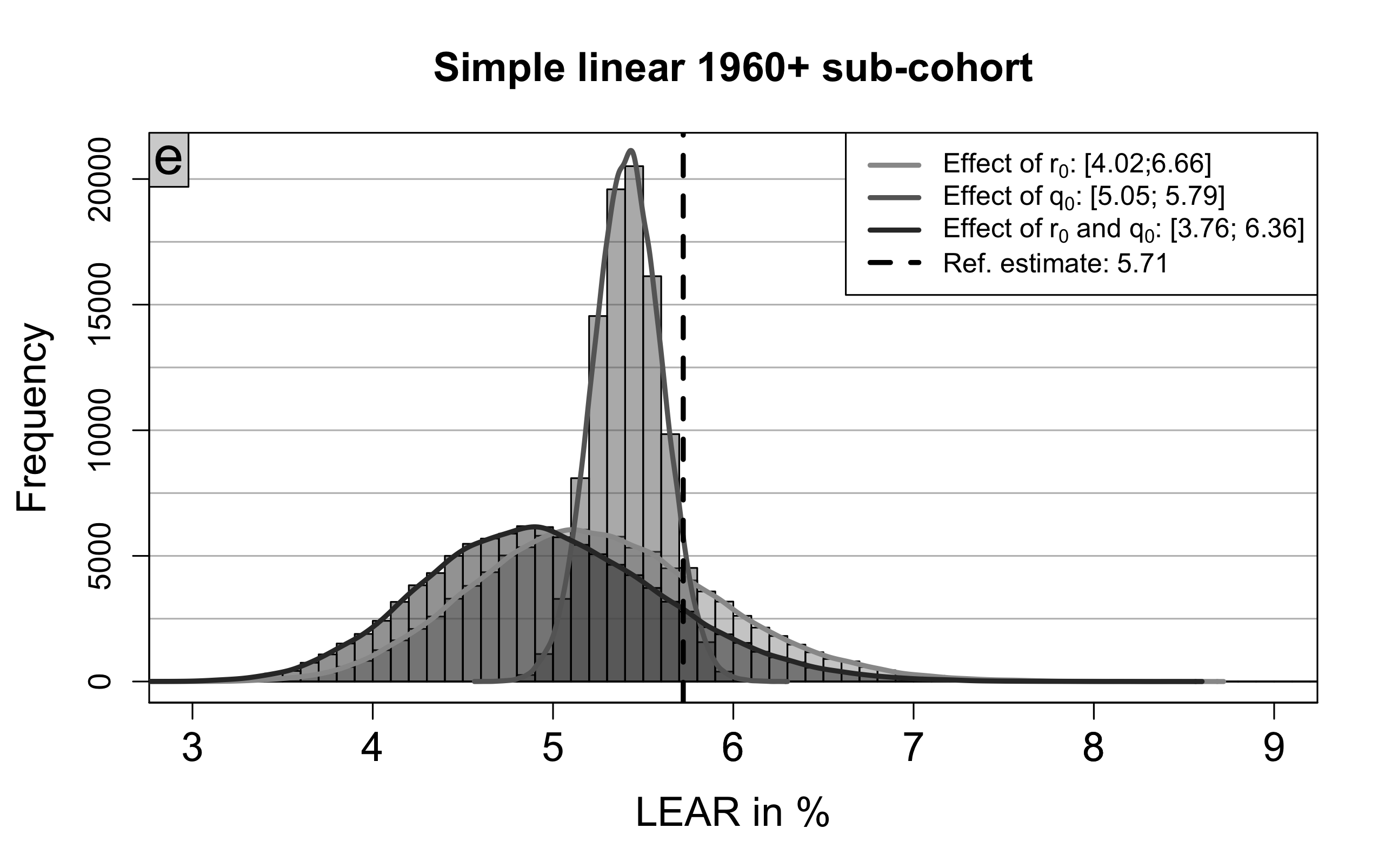}
    \caption{Histogram of $100,000$ resulting $LEAR$ estimates with kernel density (solid lines) for different risk models (plot title) and varying uncertainty in mortality rates (grayscale, different histograms per plot). Lung cancer mortality rates $r_0(t)$ (Effect of $r_0$) and all-cause mortality rates $q_0(t)$ (Effect of $q_0$) are assumed to follow gamma distributions with parameters as in Suppl. Table \ref{tab:WHO_r0_q0_estimates_5EAA}.  The joint effect (Effect of $r_0$ and $q_0$) results from independent sampling from both corresponding probability distributions. 95\% uncertainty intervals are presented in the legend.}
    \label{fig:LEAR_MR_Histogramm}
\end{figure}

The analysis reveals that all-cause mortality rates $q_0(t)$ impose considerably less uncertainty on the LEAR than lung cancer rates $r_0(t)$ for all risk models (Suppl. Figure \ref{fig:LEAR_MR_Histogramm}). The empirical distribution of sampled LEAR estimates for gamma-distributed $q_0(t)$ is considerably narrower compared to the empirical distribution for gamma-distributed $r_0(t)$, which is also reflected in the 95\% uncertainty intervals. The reference LEAR estimate is on the far right of the calculated uncertainty interval accounting for $q_0 (t)$ uncertainty and even outside the interval for the BEIR VI 1960+ sub-cohort risk model. This is possible because reference estimates (ICRP 103) are calculated with ICRP reference mortality rates independent of WHO data. The combined effect of $r_0(t)$ with $q_0(t)$ is very similar to the effect when only $r_0(t)$ is considered. The relative uncertainty span is very similar across all considered risk models with roughly 0.45 for $r_0(t)$ uncertainties and joint uncertainties $r_0(t)$, $q_0(t)$, and roughly 0.10 for only uncertain $q_0(t)$. \\

\begin{table}[htbp]
\centering
\centering\resizebox{\columnwidth}{!}{ 
\begin{tabular}[h]{r|rrrr|rrrr}
\hline
\addlinespace
 & \multicolumn{4}{c}{Lung cancer mortality }  & \multicolumn{4}{|c}{All-cause mortality } \\

\addlinespace
\hline
\addlinespace
\multicolumn{1}{c|}{Age $t$} & \multicolumn{1}{c}{$\hat{a}^{(r_0)}_t$} & \multicolumn{1}{c}{$\hat{b}^{(r_0)}_t$} & \multicolumn{1}{c}{$\frac{\hat{a}^{(r_0)}_t}{\hat{b}^{(r_0)}_t}\times 10^5$} & \multicolumn{1}{c|}{$r^{(ICRP)}_0(t) \times 10^5$ } & \multicolumn{1}{c}{$\hat{a}^{(q_0)}_t$} & \multicolumn{1}{c}{$\hat{b}^{(q_0)}_t$} & \multicolumn{1}{c}{$\frac{\hat{a}^{(q_0)}_t}{\hat{b}^{(q_0)}_t}\times 10^5$ } & \multicolumn{1}{c}{$q^{(ICRP)}_0(t) \times 10^5$ } \\
\addlinespace
\hline
\addlinespace
$20-24$ & $0.97$ & $244,677.63$ & $0.40$ &$0.14$ & $2.33$ & $3,952.30$ & $58.95$ &$51.48$\\
$25-29$ &$1.30$ & $219,930.33$& $0.59$ &$0.32$ & $2.26$ & $3,255.88$ & $69.41$&$58.12$\\
$30-34$ &$2.56$ & $248,034.10$&$1.03$&$0.98$ & $2.30$ & $2,568.26$ & $89.55$ &$76.26$\\
$35-39$ &$2.34$ & $100,097.58$& $2.34$&$2.64$ & $2.38$ & $1,867.51$ &$127.44$ &$104.91$\\
$40-44$ &$2.03$ & $30,061.06$& $6.75$&$6.99$ & $2.56$ & $1,278.99$ & $200.16$&$160.86$\\
$45-49$ &$1.79$ & $10,422.60$& $17.17$&$14.86$ & $2.80$ & $878.12$ & $318.86$ &$238.39$\\
$50-54$ &$1.83$ & $4,940.48$&$37.04$ &$29.98$ & $2.99$ & $592.00$ & $505.07$&$363.15$\\
$55-59$ &$1.66$ & $2,448.62$& $67.79$&$56.76$ & $3.29$ & $424.11$ & $775.74$&$589.27$\\
$60-64$ &$1.74$ & $1,595.66$& $109.05$&$107.21$ & $3.60$ & $302.02$ & $1,191.97$&$1,044.14$\\
$65-69$ &$1.69$ & $1,078.86$& $156.65$&$174.41$ & $4.13$ & $222.12$ & $1,859.36$&$1,717.99$\\
$70-74$ &$1.70$ & $830.72$& $204.64$&$240.85$ & $5.05$ & $170.36$ & $2,964.31$&$2,855.71$\\
$75-79$ &$1.69$ & $690.09$& $244.90$&$284.17$ & $5.88$ & $121.95$ & $4,821.65$&$4,618.33$\\
$80-84$ &$1.70$ & $627.56$& $270.89$ &$304.90$ & $9.65$ & $117.77$ & $8,193.94$&$7,807.24$\\
$85-89$ & $1.47$ & $594.61$& $247.22$&$274.92$ & $6.85$ & $48.92$ & $14,002.45$ &$11,369.96$\\
$90-94$ & $1.68$ & $714.17$ & $235.24$ &$233.50$ & $29.31$ & $127.69$ & $22,954.03$ &$20,897.22$\\
\addlinespace
\hline
\addlinespace
\end{tabular}}
\caption{Estimates $\hat{a}^{(r_0)}_t,\hat{a}^{(q_0)}_t$ and $\hat{b}^{(r_0)}_t,\hat{b}^{(q_0)}_t$ with corresponding fraction (mean of the distribution) for shape and rate parameter in the gamma distribution $\mathcal{G}\left(a^{(r_0)}_t,b^{(r_0)}_t\right)$ and $\mathcal{G}\left(a^{(q_0)}_t,b^{(q_0)}_t\right)$ based on observed lung cancer or all-cause mortality in the WHO data, respectively for different age groups with the corresponding ICRP mortality rates $r^{(ICRP)}_0(t)$, $q^{(ICRP)}_0(t)$.}
\label{tab:WHO_r0_q0_estimates_5EAA}
\end{table} 

\subsubsection{Centered gamma distributed mortality rates}
\label{Mortality_rate_uncertainty_preliminaries_centered_SDC_D12}
The derivation of gamma distribution parameters estimates in (\ref{mr_r0_ass}) and (\ref{mr_q0_ass}) from WHO data is independent of ICRP reference rates $r_0^{(ICRP)}(t)$ and $q_0^{(ICRP)}(t)$. In particular, the mean of the resulting gamma distributions does not match the ICRP reference rates. To address this, we investigated a centered gamma distribution with the expectation equal to the ICRP rate by setting $$\hat{b}_t^{(r_0)}=\frac{\hat{a}_t^{(r_0)}}{r_0^{(ICRP)}(t)}$$ and $$\hat{b}_t^{(q_0)}=\frac{\hat{a}_t^{(q_0)}}{q_0^{(ICRP)}(t)}$$ for all ages $t$.\footnote{The expectation of a generic gamma distribution $G\left(a_t,b_t\right)$ is $a_t/b_t$.} Preliminary analyses revealed that this adjustment yielded only minor differences and a negligible impact on lifetime risk estimates. For example, this can be inferred by comparing similar results for the values $r^{(ICRP)}_0(t)$, $\frac{\hat{a}^{(r_0)}_t}{\hat{b}^{(r_0)}_t}$ and $q^{(ICRP)}_0(t)$, $\frac{\hat{a}^{(q_0)}_t}{\hat{b}^{(q_0)}_t}$ for the important groups of older ages in Suppl. Table \ref{tab:WHO_r0_q0_estimates_5EAA}. Hence, we retained the un-centered gamma distributions for Monte Carlo simulations in the main paper to avoid constraining parameter estimation. The same applies to the log-normal distribution used in Section D.3. Note that this holds for sex-averaged mortality rates. For sex-specific mortality rates (Section E.1), larger discrepancies justify explicitly analyzing centered mortality rate uncertainties. Although centering may affect lifetime risk estimates, the relative uncertainty span is not notably affected.
\subsection{Poisson distributed lung cancer deaths}
\label{Mortality_rate_uncertainty_poisson_lc_deaths_SDC_D2}
We present an alternative, but similar approach,  to assess mortality rate uncertainty on lifetime risks connecting ICRP reference mortality rates (Euro-American-Asian mixed population) with WHO mortality data (Countries from Europe, America, and Asia from the calendar years 2001, 2006, 2011, 2016, and 2021). Ultimately, this approach aims at a Bayesian assessment of mortality rate uncertainty: \\

Non-parametric Poisson regression on WHO lung cancer mortality data yields age-specific rates $r_0(t)$ for all ages $t$ (Suppl. Table \ref{tab:WHOrates_Cy}). Only observations with a positive number of cases and a positive number of individuals at risk were included. Focusing on lung cancer mortality (heavily influencing $LEAR$s), this approach is analogous to less influential all-cause mortality rates $q_0(t)$. Each data point (country, sex, and calendar year) comprises deceased and alive individuals (mid-year population), interpreted as cases and person-years for Poisson regression. Correlations between years and countries are neglected, assuming independence between observations for simplicity.\footnote{This is no major issue because the upcoming analyses reveal that also uncorrelated WHO data from a specific country and year yield narrow $LEAR$ uncertainty intervals.}
\subsubsection{Likelihood}
We employ Poisson regression assuming the number of lung cancer deaths $D_i$ for a certain age group (compare e.g. Suppl. Table \ref{tab:WHO_r0_q0_estimates_5EAA}) follows a Poisson distribution with $D_i \sim Poi(\lambda_i) = Poi(N_i e^{\theta})$ for WHO population data $X=\left\{ N_i, d_i ~ \vert ~ i=1, \dots, M \right\}$.  Here, $\theta$ is the unknown parameter, $N_i$ are known person-years at risk, and $d_i$ are observed lung cancer cases for $M$ observations. Each observation $i$ is uniquely defined by a specific country, sex, and calendar year. \\

Note that $\theta$ does not depend on the observations $i$. So, regardless of the country and year under consideration, the expected number of deaths depends only on the number of people in that group. Therefore, country- and calendar year-specific characteristics are only accounted for through the random fluctuation of the Poisson distribution. This approach is applied for every age group separately. Here we suppress the age dependency for simplicity, but later we write for the relevant parameter $\theta_t=\theta$.  The probability of $d_i$ deaths with $N_i$ person-years at risk is:
\begin{equation}
\mathbb{P}\left( D_i=d_i \vert X, \theta \right) = \frac{\left( N_i e^{\theta}\right)^{d_i}}{d_i!}e^{-N_ie^{\theta}}.
\end{equation}
The likelihood function $L(X \vert \theta )$ is then
\begin{equation}
    L(X \vert \theta )= \prod_{i=1}^M \frac{\left( N_i e^{\theta}\right)^{d_i}}{d_i!} e^{- N_i e^{\theta}}
\end{equation}
with log-likelihood $\ell \left( X \vert \theta \right)$,
\begin{equation}
    \ell(X \vert \theta) = \sum_{i=1}^M d_i \left( \log \left(N_i\right) + \theta \right)- N_i e^{\theta}- \log \left(d_i!\right)  \propto \sum_{i=1}^M d_i \left( \log \left(N_i\right) + \theta \right)- N_i e^{\theta}.
\end{equation}

The simple structure yields the maximum likelihood estimate for $\theta$:
\begin{equation}
\hat{\theta} = \log \frac{\sum_{i=1}^M d_i}{\sum_{i=1}^M N_i}.
\end{equation}
This simplifies to $e^{\hat{\theta}} = \frac{d}{N}$ with $d = \sum_{i=1}^M d_i$ and $N = \sum_{i=1}^M N_i$ (estimates in Suppl. Table \ref{tab:WHOrates_Cy}).\\

Notably, $e^{\hat{\theta}}$ can be interpreted as a population-weighted average of single mortality rates $\frac{d_i}{N_i}$ with weights $w_i = \frac{N_i}{N}$ and $N=\sum_{i=1}^M N_i$ due to the Poisson likelihood. The weights $w_i$ correspond to the relative size of each population at risk $N_i$ compared to the total population $N$. Each $\frac{d_i}{N_i}$ is equivalent to the maximum likelihood estimate for the probability of a single binomial experiment with known population $N_i$ and observed successes $d_i$. Larger populations contribute more to the parameter estimation. This is a key difference to the other approach, utilizing a Gamma or Log-normal distribution, where observed mortality rates are all equally weighted for parameter estimation. In the above Poisson Regression, each observation is represented by the tupel $(N_i,d_i)$, whereas in the applied Gamma (or Log-normal) regression, it is defined by the ratio  $d_i/N_i$. In the latter approach, the specific information regarding the individual counts $d_i$,$N_i$ is not accounted for. \\

The most suitable approach will depend on the assumptions made about the variance of the observations. If larger populations are assumed to lead to more precise estimates, the Poisson approach may be preferable. Alternatively, if all observations are considered equal, gamma regression could be a more appropriate choice.

\begin{table}[htbp]
    \centering
    \small
    \begin{tabular}[h]{lllll}
        \toprule
        Age & Observations & $N \times 10^{-6}$ & $d$ & $\frac{d}{N} \times 10^6$  \\
        \midrule
        \multicolumn{5}{c}{\textbf{Europe, America and Asia; 2001, 2006, 2011, 2016 and 2021}} \\
        \midrule
    $20-24$ & $118$ & $195.89$ & $326$ & $1.66$ \\
    $25-29$ & $210$ & $265.12$ & $777$ & $2.93$ \\
    $30-34$ & $259$ & $285.62$ & $2,073$ & $7.26$ \\
    $35-39$ & $313$ & $300.78$ & $6,261$ & $20.82$ \\
    $40-44$ & $352$ & $302.67$ & $18,933$ & $62.55$ \\
    $45-49$ & $378$ & $293.17$ & $46,568$ & $158.84$ \\
    $50-54$ & $380$ & $274.87$ & $93,770$ & $341.14$ \\
    $55-59$ & $383$ & $245.26$ & $155,584$ & $634.37$ \\
    $60-64$ & $384$ & $213.69$ & $220,337$ & $1,031.10$ \\
    $65-69$ & $384$ & $184.51$ & $278,720$ & $1,510.60$ \\
    $70-74$ & $386$ & $154.22$ & $313,580$ & $2,033.39$ \\
    $75-79$ & $388$ & $126.34$ & $312,099$ & $2,470.33$ \\
    $80-84$ & $372$ & $86.59$ & $238,244$ & $2,751.36$ \\
    $85-89$ & $365$ & $61.16$ & $134,348$ & $2,196.53$ \\
    $90-94$ & $167$ & $8.99$ & $21,826$ & $2,428.93$ \\
        \midrule
        \multicolumn{5}{c}{\textbf{Europe, America and Asia; 2016}} \\
        \midrule
    $20-24$ & $36$ & $56.38$ & $125$ & $2.22$ \\
    $25-29$ & $60$ & $72.29$ & $261$ & $3.61$ \\
    $30-34$ & $73$ & $75.88$ & $545$ & $7.18$ \\
    $35-39$ & $88$ & $77.21$ & $1,274$ & $16.50$ \\
    $40-44$ & $100$ & $76.85$ & $3,141$ & $40.87$ \\
    $45-49$ & $107$ & $76.36$ & $7,391$ & $96.80$ \\
    $50-54$ & $108$ & $71.80$ & $16,868$ & $234.92$ \\
    $55-59$ & $109$ & $64.94$ & $32,291$ & $497.29$ \\
    $60-64$ & $109$ & $56.95$ & $50,737$ & $890.96$ \\
    $65-69$ & $109$ & $51.64$ & $68,089$ & $1,318.62$ \\
    $70-74$ & $110$ & $37.99$ & $67,823$ & $1,785.49$ \\
    $75-79$ & $111$ & $32.71$ & $68,777$ & $2,102.52$ \\
    $80-84$ & $103$ & $22.83$ & $57,269$ & $2,508.08$ \\
    $85-89$ & $104$ & $15.10$ & $37,619$ & $2,491.16$ \\
    $90-94$ & $62$ & $4.06$ & $10,072$ & $2,481.31$ \\
        \midrule
        \multicolumn{5}{c}{\textbf{Germany; 2016}} \\
        \midrule
        $20-24$ & $2$ & $4.58$ & $3$ & $0.65$\\
        $25-29$ & $2$ & $5.38$ & $8$ & $1.49$\\
        $30-34$ & $2$ & $5.19$ & $26$ & $5.01$\\
        $35-39$ & $2$ & $5.01$ & $60$ & $11.99$\\
        $40-44$ & $2$ & $4.91$ & $208$ & $42.40$\\
        $45-49$ & $2$ & $6.39$ & $755$ & $118.12$\\
        $50-54$ & $2$ & $6.97$ & $2,143$ & $307.48$\\
        $55-59$ & $2$ & $6.13$ & $4,037$ & $658.47$\\
        $60-64$ & $2$ & $5.24$ & $5,833$ & $1,112.81$\\
        $65-69$ & $2$ & $4.45$ & $6,832$ & $1,536.11$\\
        $70-74$ & $2$ & $3.81$ & $7,223$ & $1,894.77$\\
        $75-79$ & $2$ & $4.31$ & $8,535$ & $1,980.40$\\
        $80-84$ & $2$ & $2.61$ & $5,564$ & $2,132.05$\\
        $85-89$ & $2$ & $1.49$ & $3,415$ & $2,288.84$\\
        $90-94$ & $2$ & $0.60$ & $995$ & $1,659.25$\\
        \bottomrule
    \end{tabular}
    \caption{Sex-averaged lung cancer mortality rate estimates $e^{\hat{\theta}}=\frac{d}{N}$ for all age groups derived from WHO data with total observed deaths $d>0$ and individuals at risk $N>0$.}
    \label{tab:WHOrates_Cy}
\end{table}
\subsubsection{Bayesian posterior}
\begin{figure}[htbp]
    \centering
    \includegraphics[scale=0.0925]{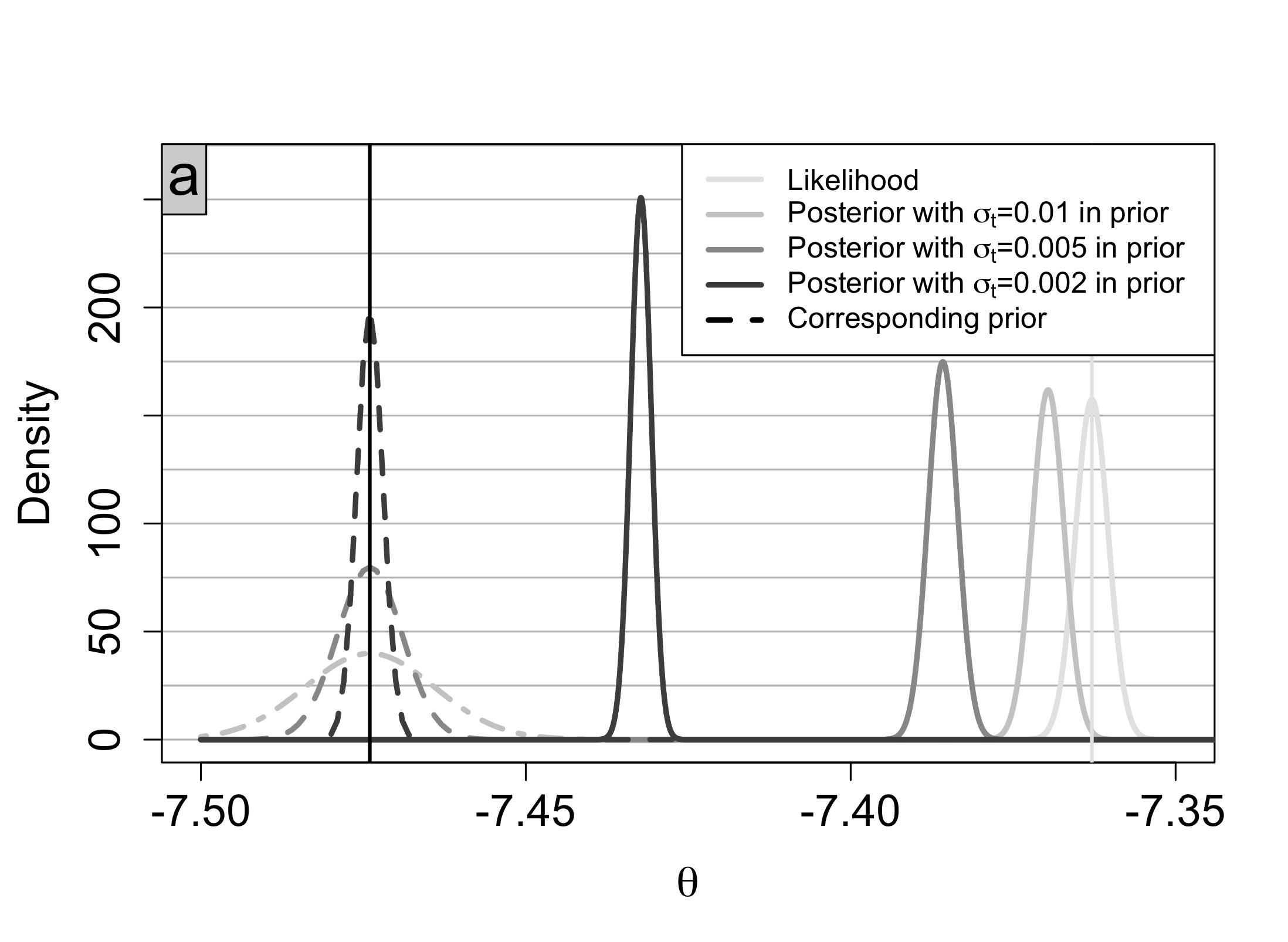}
    \includegraphics[scale=0.0925]{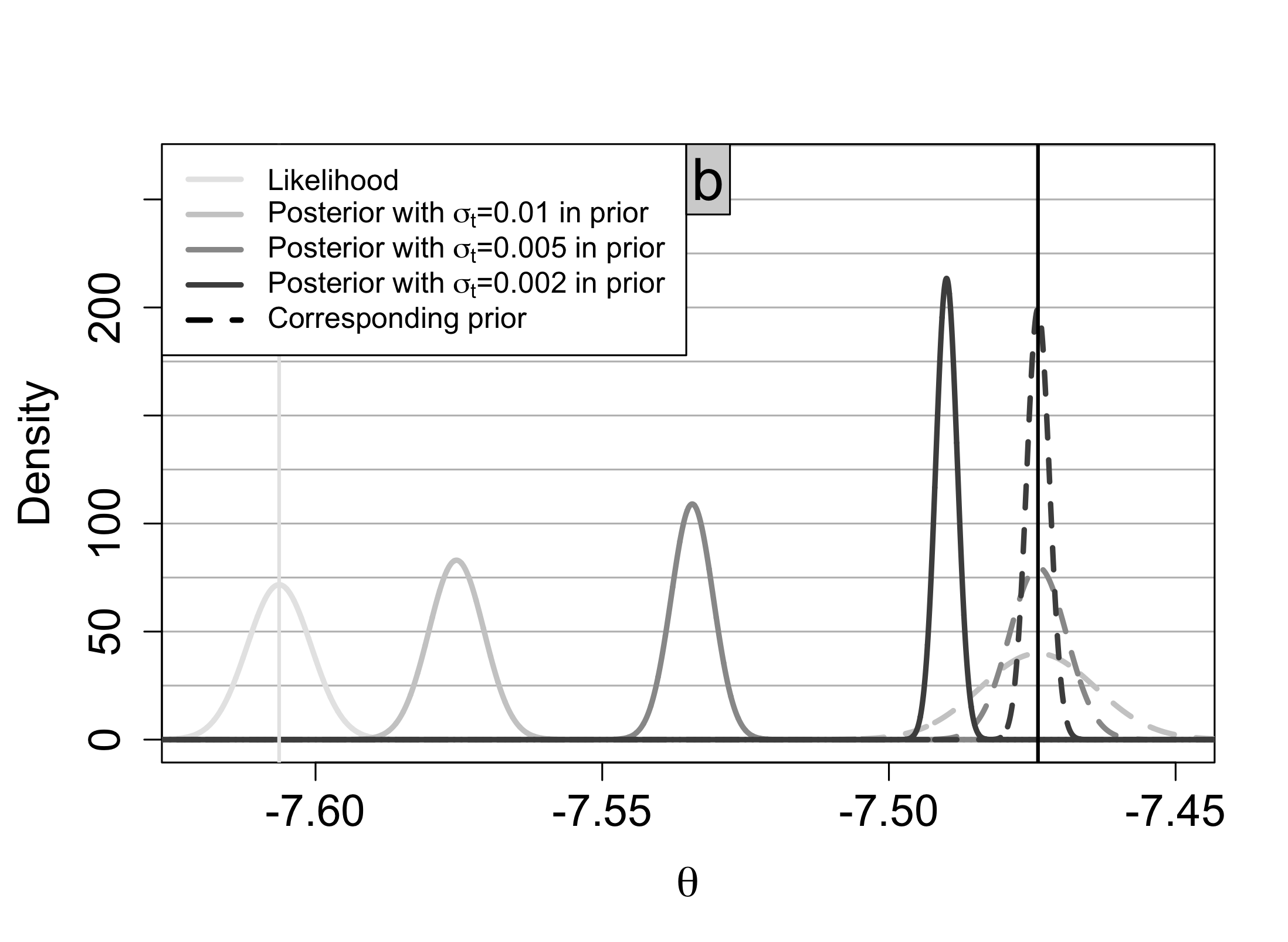}
    \includegraphics[scale=0.0925]{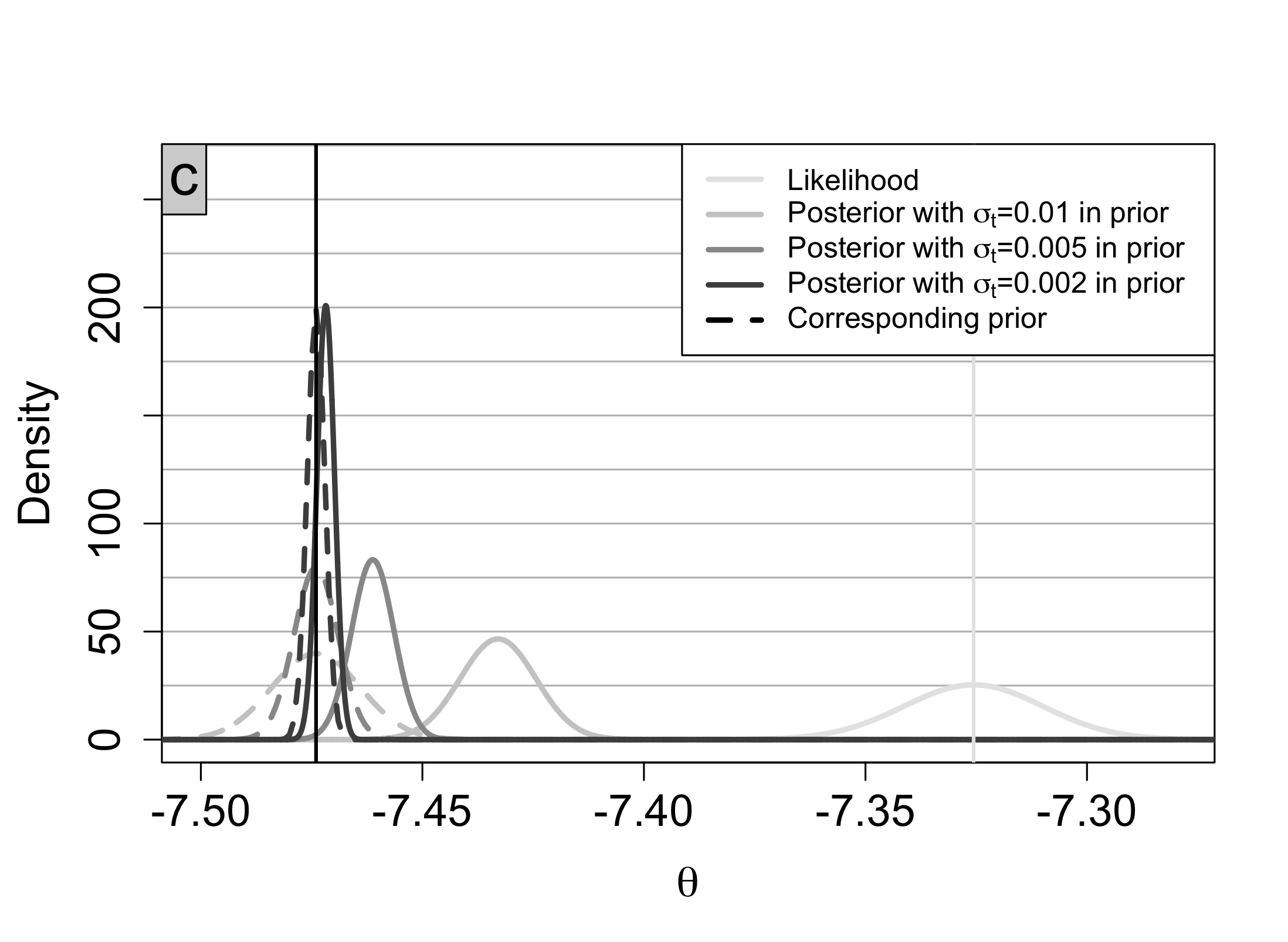}
    \caption{Distribution of $\theta_t$ for $55 \leq t \leq 59$ for varying prior standard deviation $\sigma_t$. The first plot shows results using WHO data from all countries in Europe, America, and Asia with available data from 2001, 2006, 2011, 2016, and 2021. The second plot is reduced to data from 2016 and the last plot only employs German rates from 2016. The vertical solid black (light gray) line represents the prior (likelihood) mode.}
    \label{fig:Bayes_MR_WHO}
\end{figure}

After setting up the Likelihood structure, we add the prior information for the Bayesian inference:
A normal prior $\mathcal{N}(\mu_t, \sigma_t^2)$ is assumed for $\theta=\theta_t$ with $\mu_t = \log(r^{(ICRP)}_0(t))$ incorporating ICRP reference lung cancer rates. Varying $\sigma_t^2$ controls the prior influence (visualized in Suppl. Figure~\ref{fig:Bayes_MR_WHO} for ages 55-59 and different subsets of the WHO data). The likelihood dominates and differences between WHO rates and ICRP rates are minor, resulting in near-normal posteriors with minimal uncertainty (confirmed by Kolmogorov-Smirnov tests). \\

\begin{figure}[htbp]
    \centering
    \includegraphics[scale=0.0925]{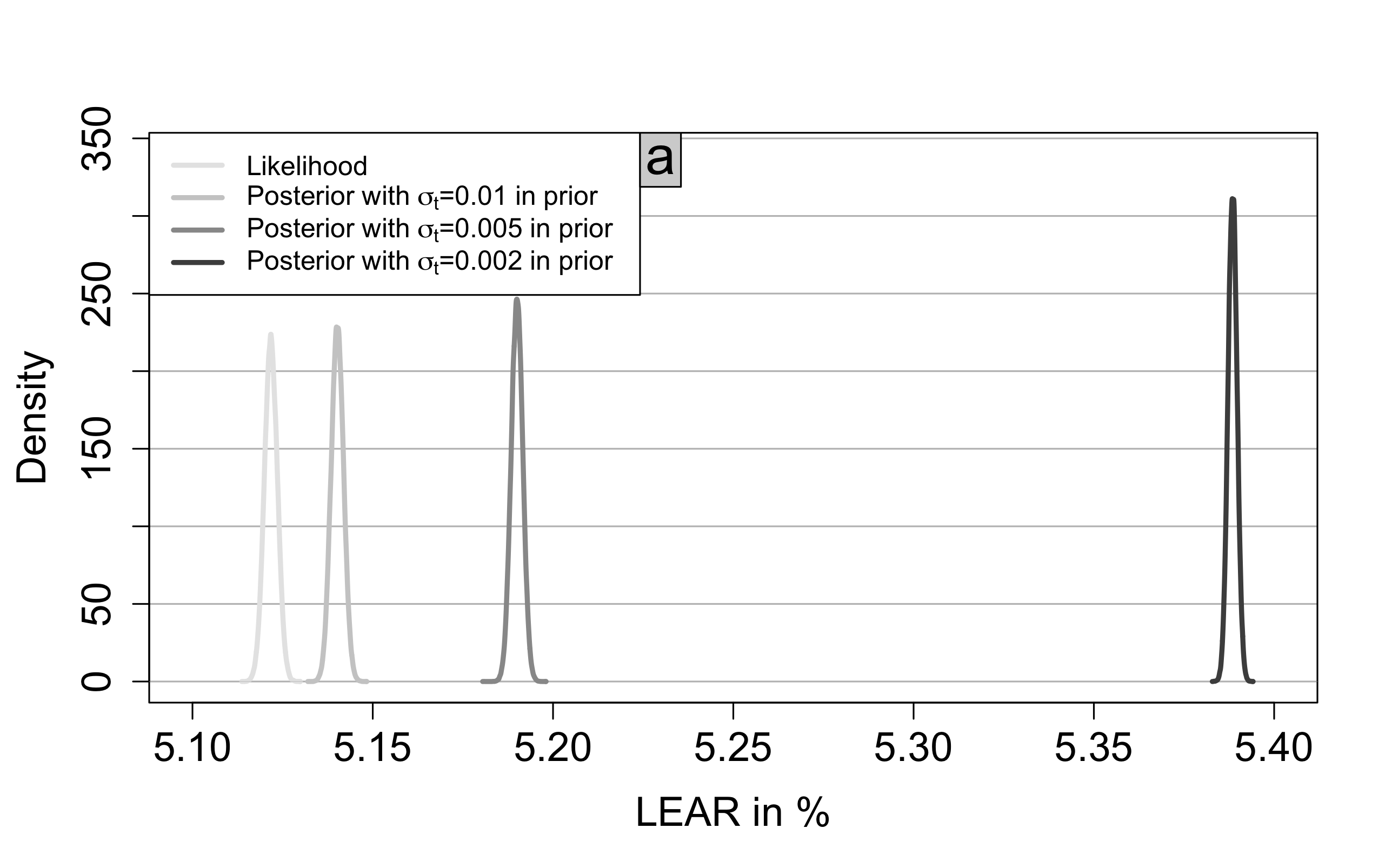}
    \includegraphics[scale=0.0925]{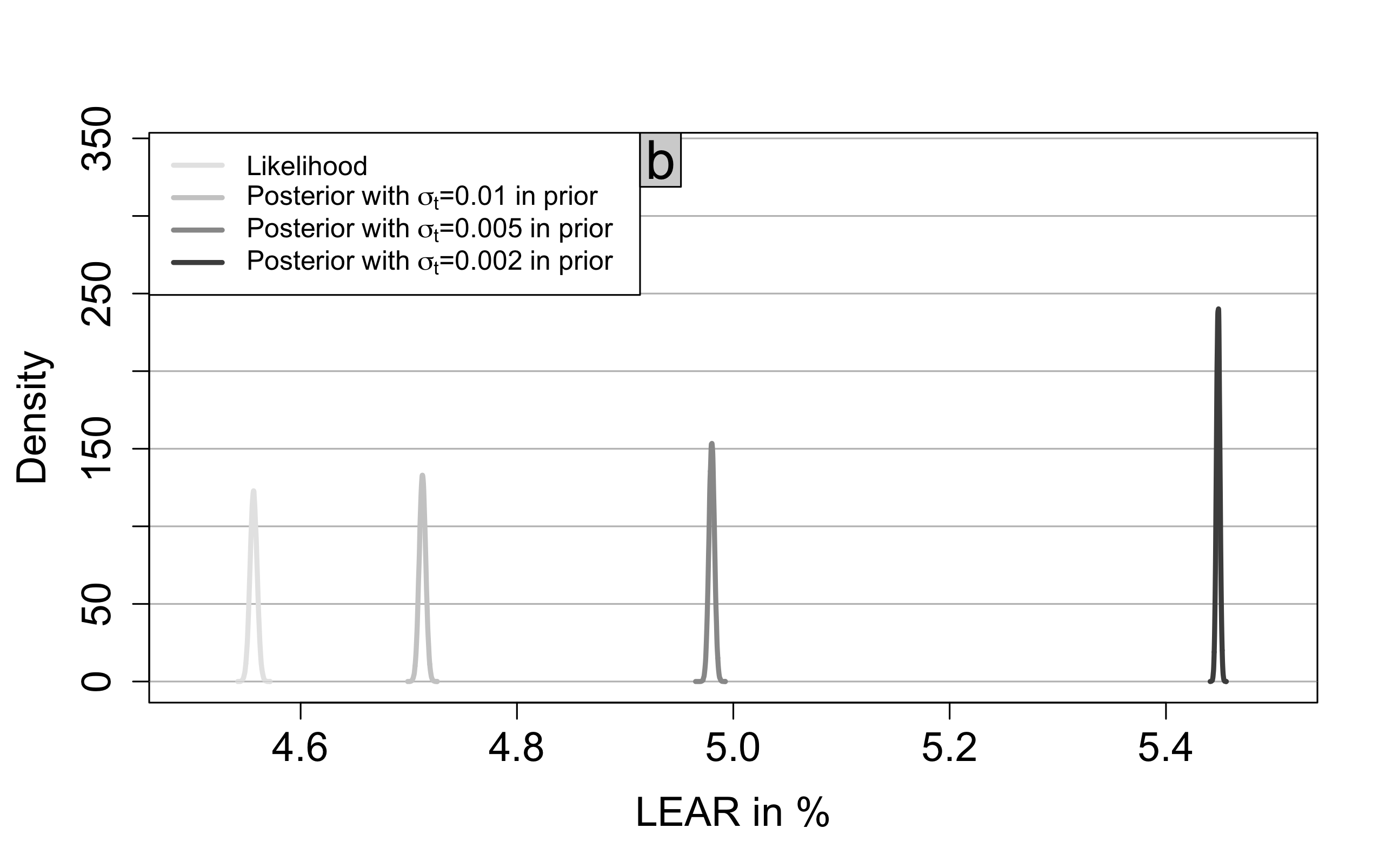}
    \includegraphics[scale=0.0925]{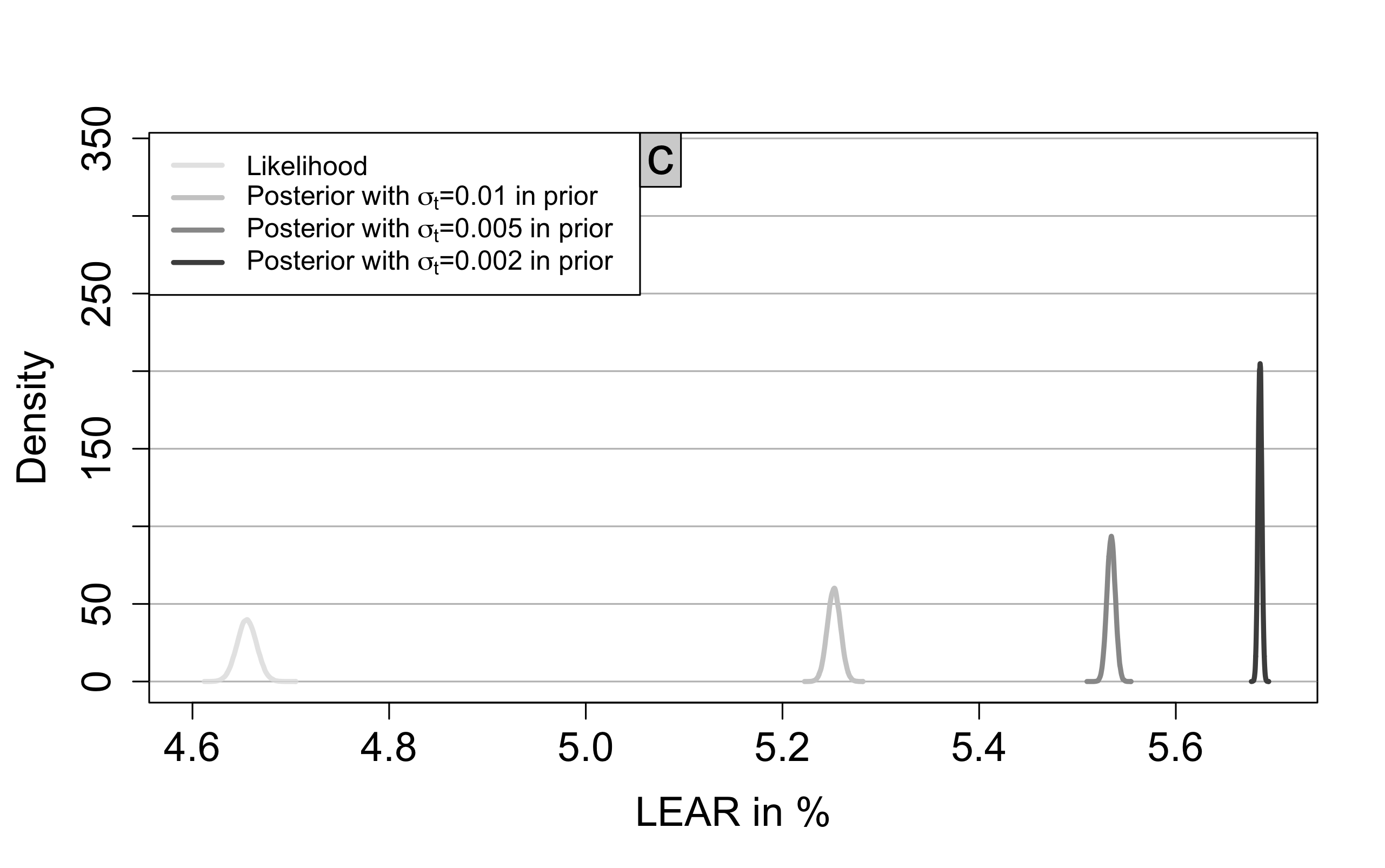}
    \caption{Posterior distribution of $LEAR$ in \% for uncertain lung cancer mortality rates and varying prior standard deviation $\sigma_t$ (Bayesian framework). The first plot shows results using WHO data from all countries in Europe, America, and Asia with available data from 2001, 2006, 2011, 2016, and 2021. The second plot is reduced to data from 2016 and the last plot only employs German rates from 2016. }
    \label{fig:Bayes_MR_WHO_LEARs}
\end{figure}

For the calculation of corresponding $LEAR$s, the simple linear risk model $ERR(t; \hat{\beta}) = \hat{\beta} W(t)$ with $\hat{\beta} = 0.0134$ is used. While decreasing the WHO data influence increases the impact of the prior distribution, the effect on $LEAR$ uncertainty remains negligible (Suppl. Table~\ref{tab:Bayes_MR_LEARs}) with relative uncertainty spans of order $10^{-3}$. Increasing prior certainty narrows uncertainty intervals and shifts the mode towards the $LEAR$ calculated with ICRP rates (Suppl. Figure~\ref{fig:Bayes_MR_WHO_LEARs}). The dominance of WHO data raises questions about the overall benefit of this approach.\\

In conclusion, the integration of a population-weighted Poisson approach with the comprehensive WHO dataset results in mortality rate estimates with little uncertainty. Likewise, employing a Bayesian approach with arbitrary prior encoding beliefs from other sources (i.e. ICRP reference rates) offers minimal new insights. The extensive dataset provided by the WHO considerably diminishes the influence of prior assumptions on the analytical outcomes. Only when using unrealistically low prior variances or substantially reducing the WHO data does any impact of the prior become noticeable. However, the scientific value of such an approach for understanding mortality rate uncertainties is limited. Instead, it is evident that the combination of the WHO database and the population-weighted Poisson method yields low uncertainties, and the WHO rates align well with ICRP reference rates.

\begin{table}[htbp]
    \centering
    \begin{tabular}{cccc}
\hline
\addlinespace
Prior & \multicolumn{3}{c}{$LEAR$ in \% (95\% HPDI)}  \\
\addlinespace
\hline
\addlinespace
$\sigma_t$ & Full & Full 2016 & Germany 2016 \\
\addlinespace
\hline
\addlinespace
Uniform prior & $5.1218$ $[5.1182;5.1254]$   & $4.5565$ $[4.5500;4.5629]$ & $4.6553$ $[4.6351; 4.6743]$  \\
$0.050$ & $5.1226$ $[5.1190;5.1261]$   & $4.5647$ $[4.5581;4.5710]$ & $4.7274$ $[4.7083; 4.7463]$  \\
$0.010$ & $5.1402$ $[5.1369;5.1437]$   & $4.7126$ $[4.7068; 4.7186]$ & $5.2523$ $[5.2392; 5.2654]$  \\
$0.005$ & $5.1900$ $[5.1869;5.1932]$   & $4.9800$ $[4.9748; 4.9849]$ & $5.5342$ $[5.5258; 5.5425]$  \\
$0.002$ & $5.3885$ $[5.3860;5.3909]$    & $5.4482$ $[5.4450;5.4514]$ & $5.6856$ $[5.6818; 5.6893]$  \\
\addlinespace
\hline
    \end{tabular}
    \caption{$LEAR$ evaluated at the mode of the posterior distribution with 95\% highest posterior density interval (HPDI) (relative uncertainty spans of order $10^{-3}$) for the simple linear risk model $ERR(t;\beta)=\beta W(t)$ with $\beta=0.0134$. The prior is modelled as $\log r_0(t) \sim \mathcal{N}\left(\mu_t,\sigma_t^2 \right)$ and $\mu_t=\log\left(r^{(ICRP)}_0(t)\right)$. The $LEAR$ in \% with ICRP reference rates is $5.7222$. "Full" corresponds to WHO data from countries in Europe, America, and Asia from the calendar years 2001, 2006, 2011, 2016, and 2021.}
    \label{tab:Bayes_MR_LEARs}
\end{table}


\subsection{Log-normal distributed mortality rates}
\label{Mortality_rate_uncertainty_lognormal_rates_SDC_D3}
Here, we compare utilizing a log-normal distribution for fitting sex-averaged mortality rates from WHO data, instead of a gamma distribution as in the main manuscript. Specifically, we model the mortality rates for all ages $t$ as follows:
 \begin{align}
     r_0(t) &\sim \mathcal{LN}\left(\mu^{(r_0)}_t,\left(\sigma^{(r_0)}_t\right)^2\right), \label{mr_r0_ln_ass}\\
     q_0(t) &\sim \mathcal{LN}\left(\mu^{(q_0)}_t,\left(\sigma^{(q_0)}_t\right)^2\right). \label{mr_q0_ln_ass} 
 \end{align}
 with age-dependent log-mean parameter $\mu^{(r_0)}_t,\mu^{(q_0)}_t$ and log-standard deviation  parameter $\sigma^{(r_0)}_t,\sigma^{(q_0)}_t$. The parameters are fit on data from the WHO Mortality Database \cite{WHO_mortality_database} with maximum-likelihood (ML) methods, just as for the gamma distribution in the main manuscript. \\

\begin{figure}[htbp]
    \centering
    \includegraphics[scale=0.0925]{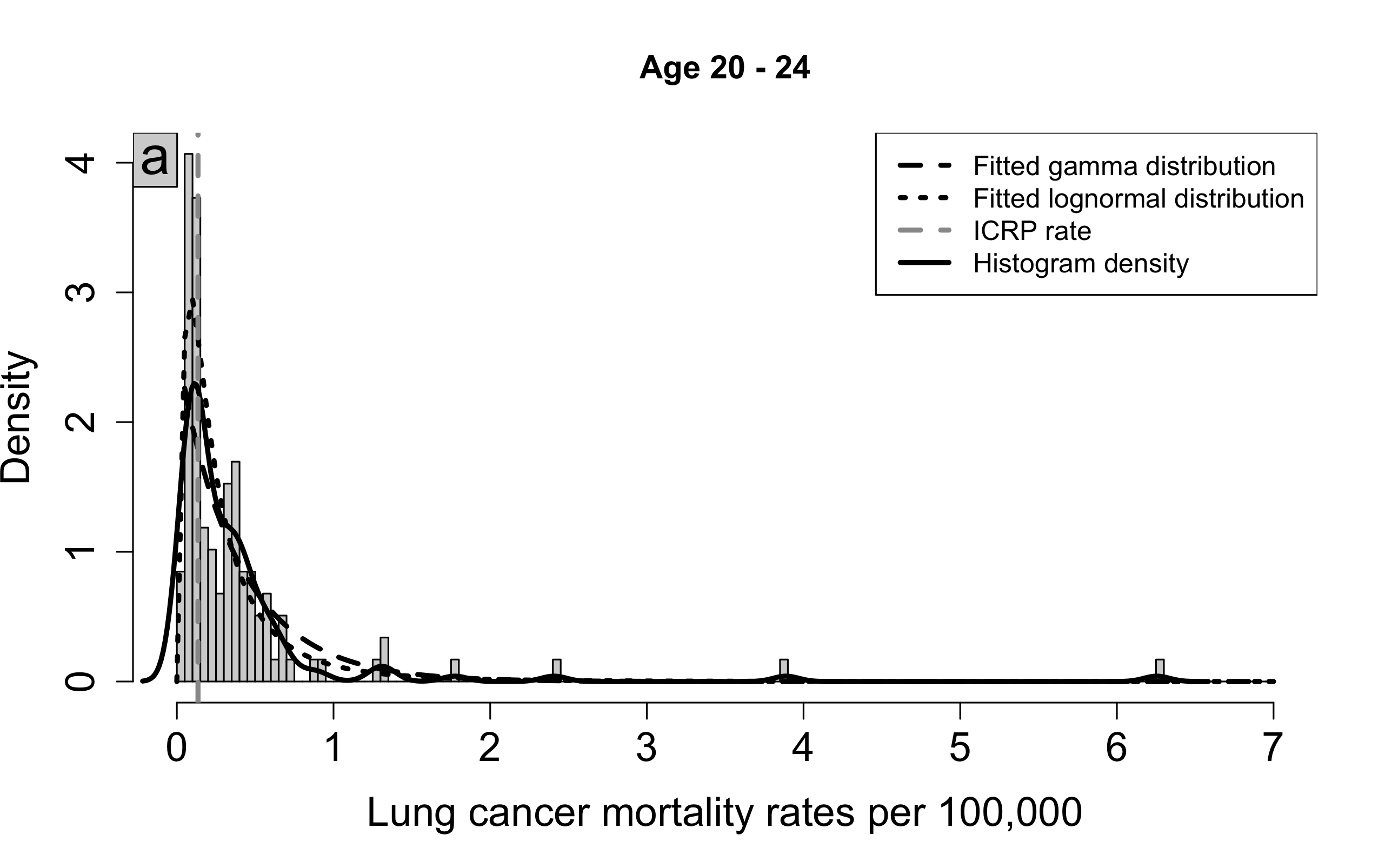}
    \includegraphics[scale=0.0925]{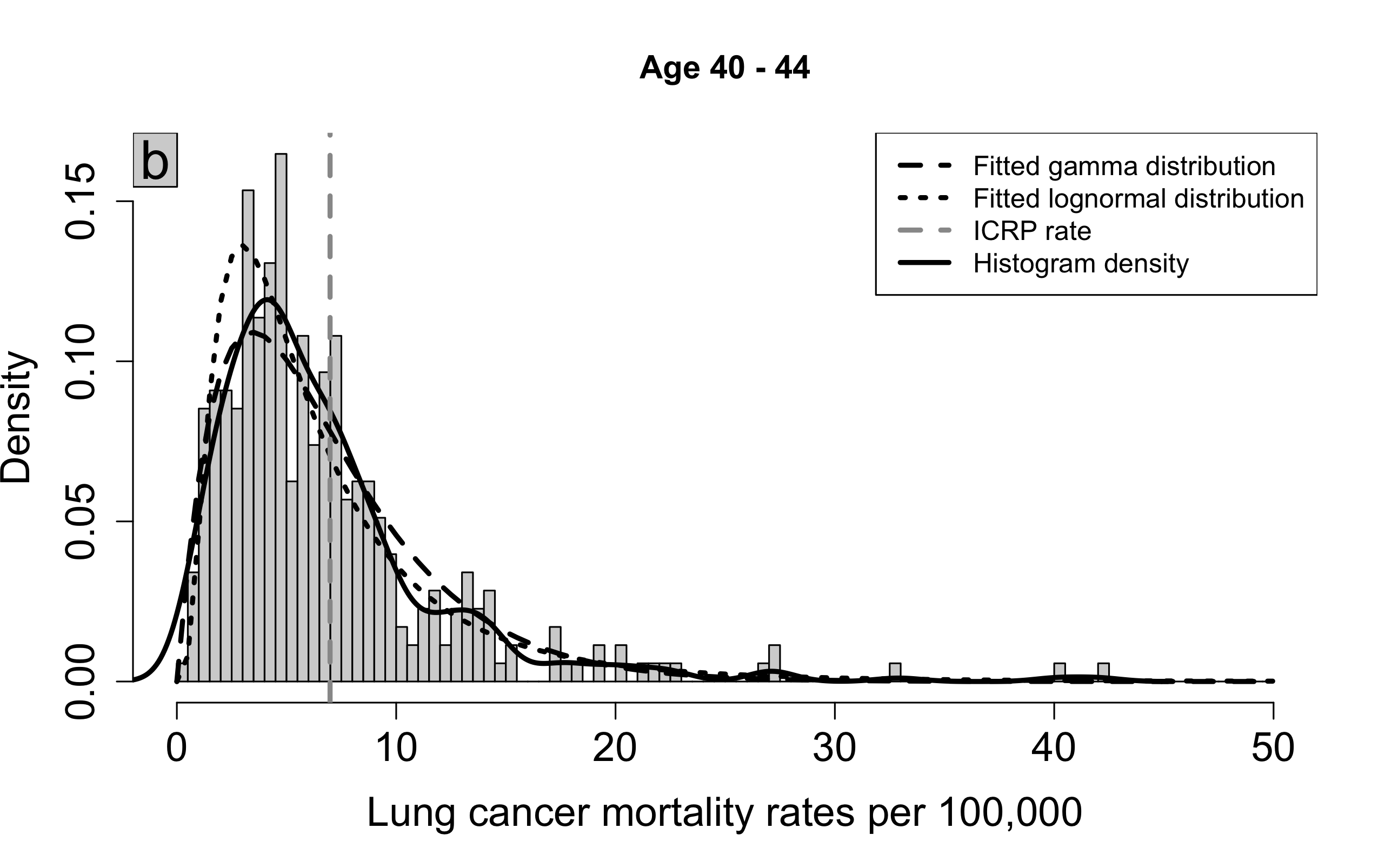}
    \includegraphics[scale=0.0925]{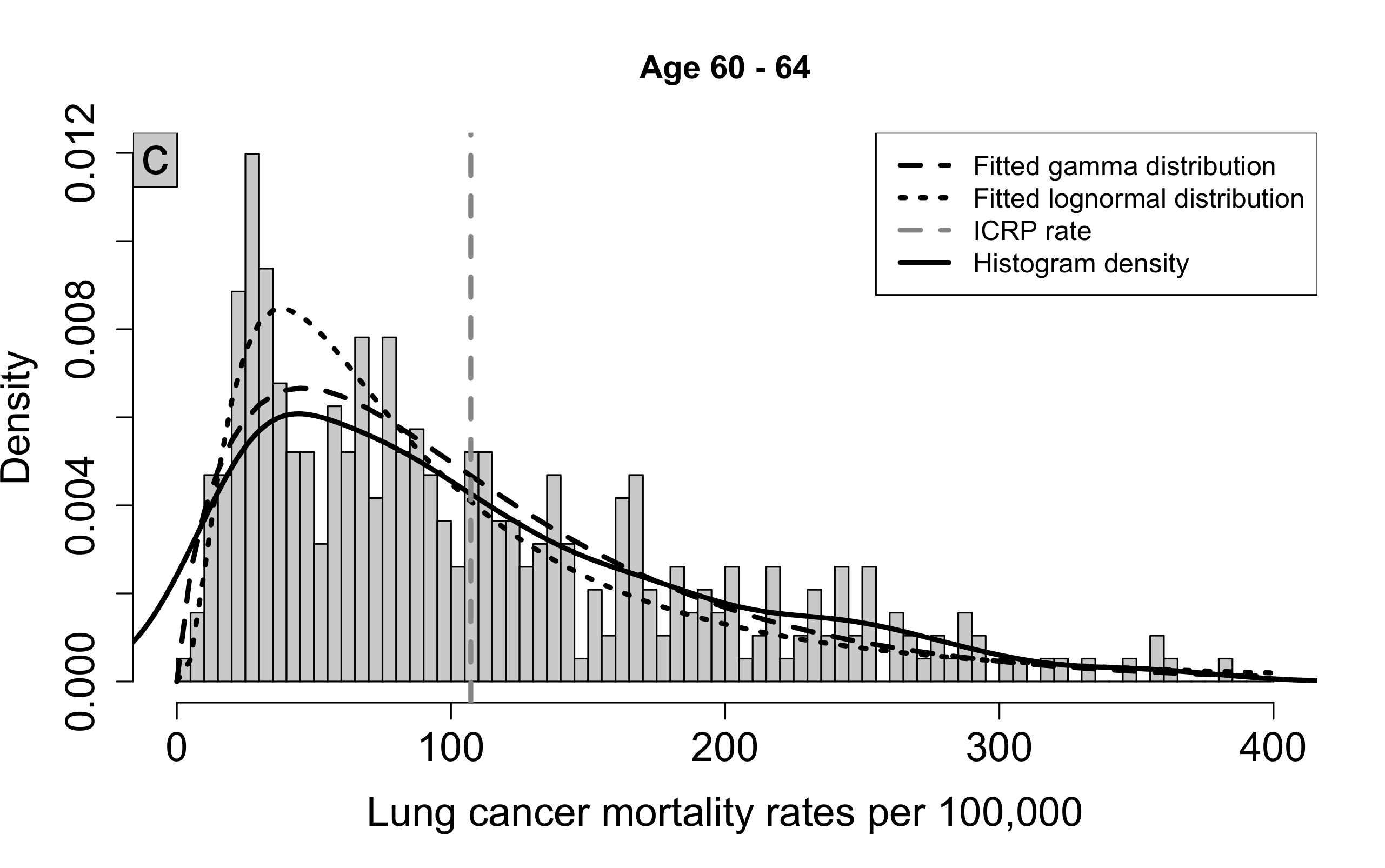}
    \includegraphics[scale=0.0925]{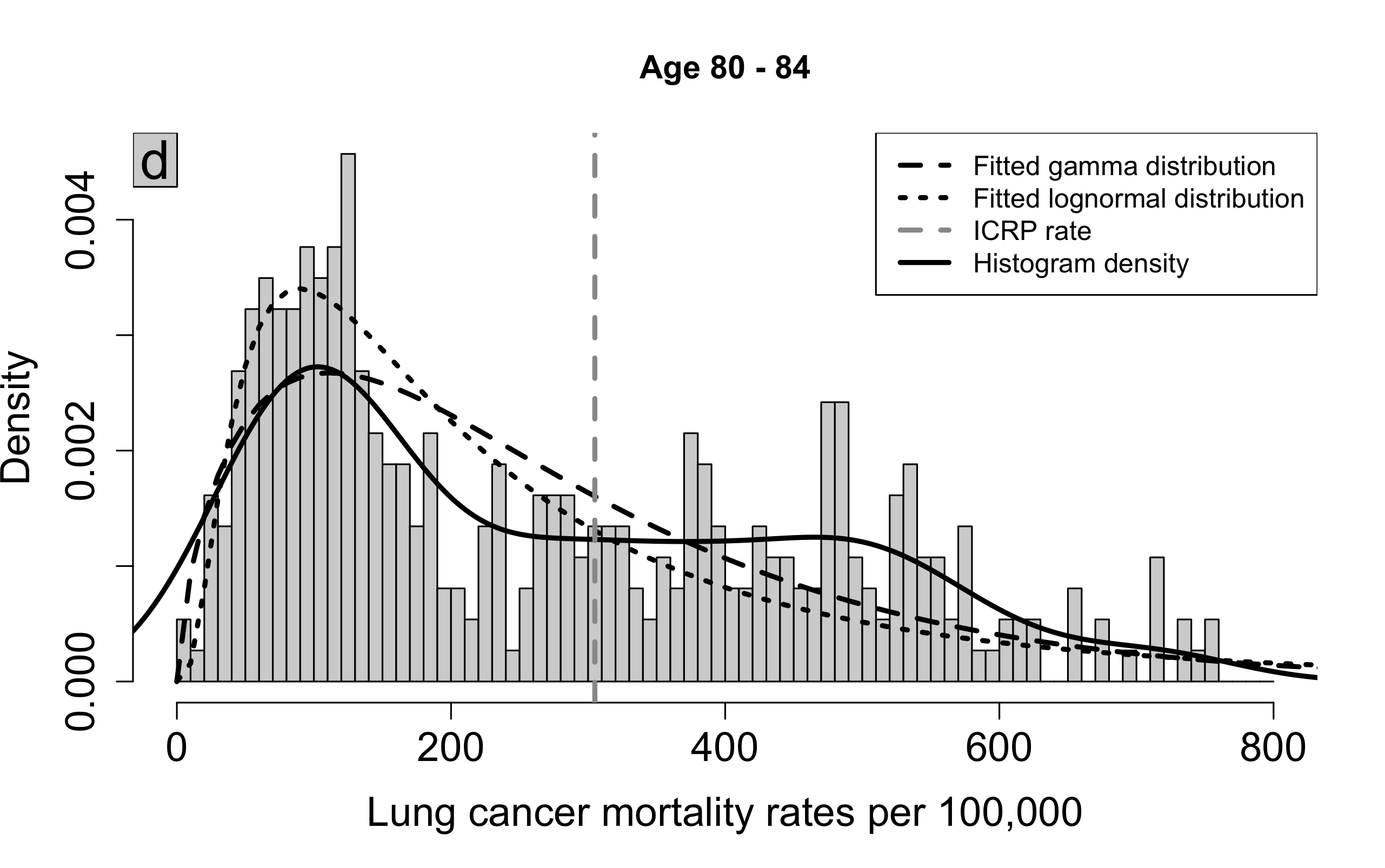}
    \caption{Histogram of lung cancer mortality rates for the ages $20-24,40-44, 60-64$ and $80-84$ derived from WHO data. For comparison, the reference lung cancer mortality rates from the ICRP Euro-American-Asian mixed population are shown with a vertical dashed line. The dashed (dotted) curve shows the density for a gamma (log-normal) distribution fitted to the histogram data.}
    \label{fig:WHO_lungrates_5EAA_ln_vs_g}
\end{figure}

\begin{figure}[htbp]
    \centering
    \includegraphics[scale=0.0925]{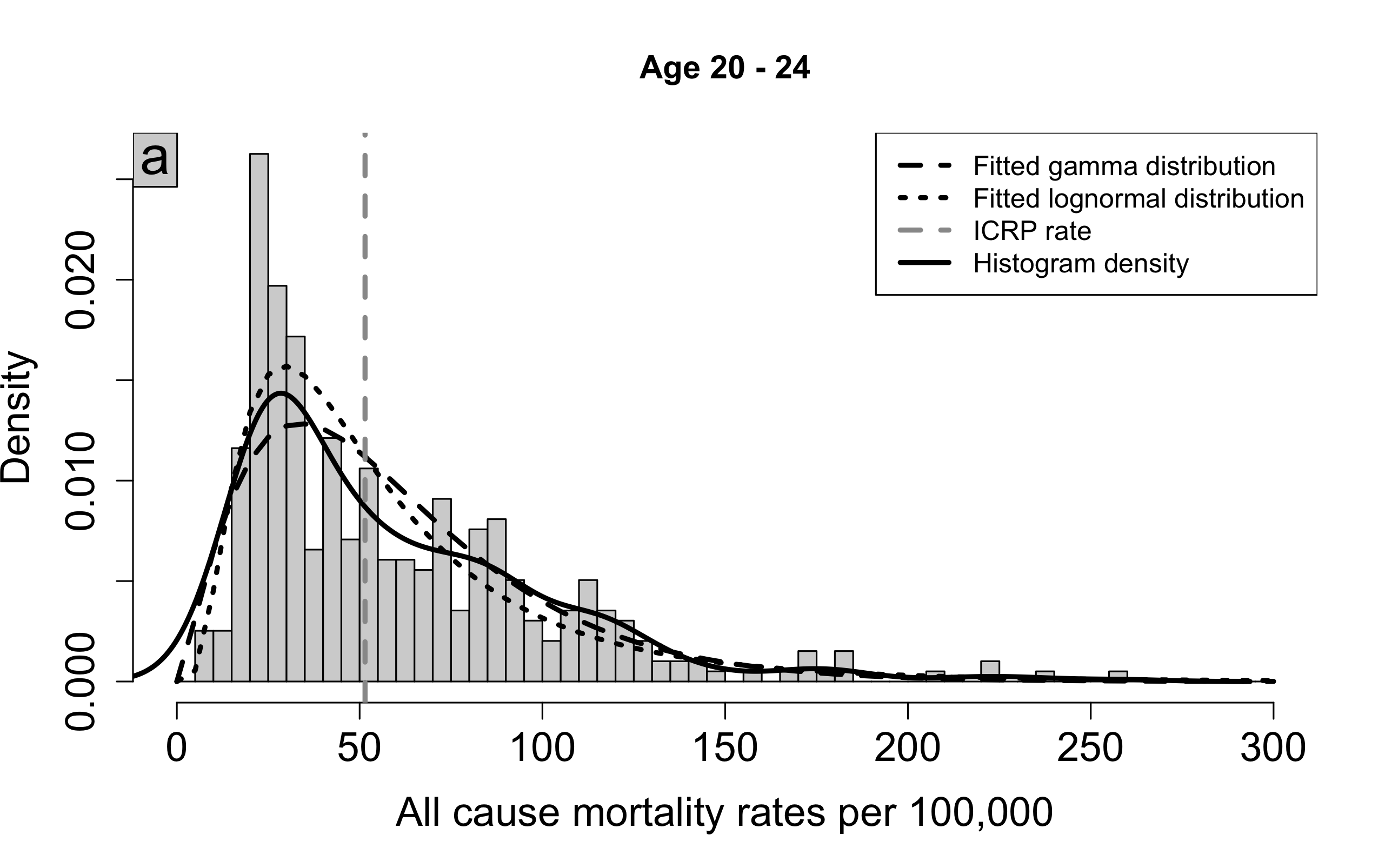}
    \includegraphics[scale=0.0925]{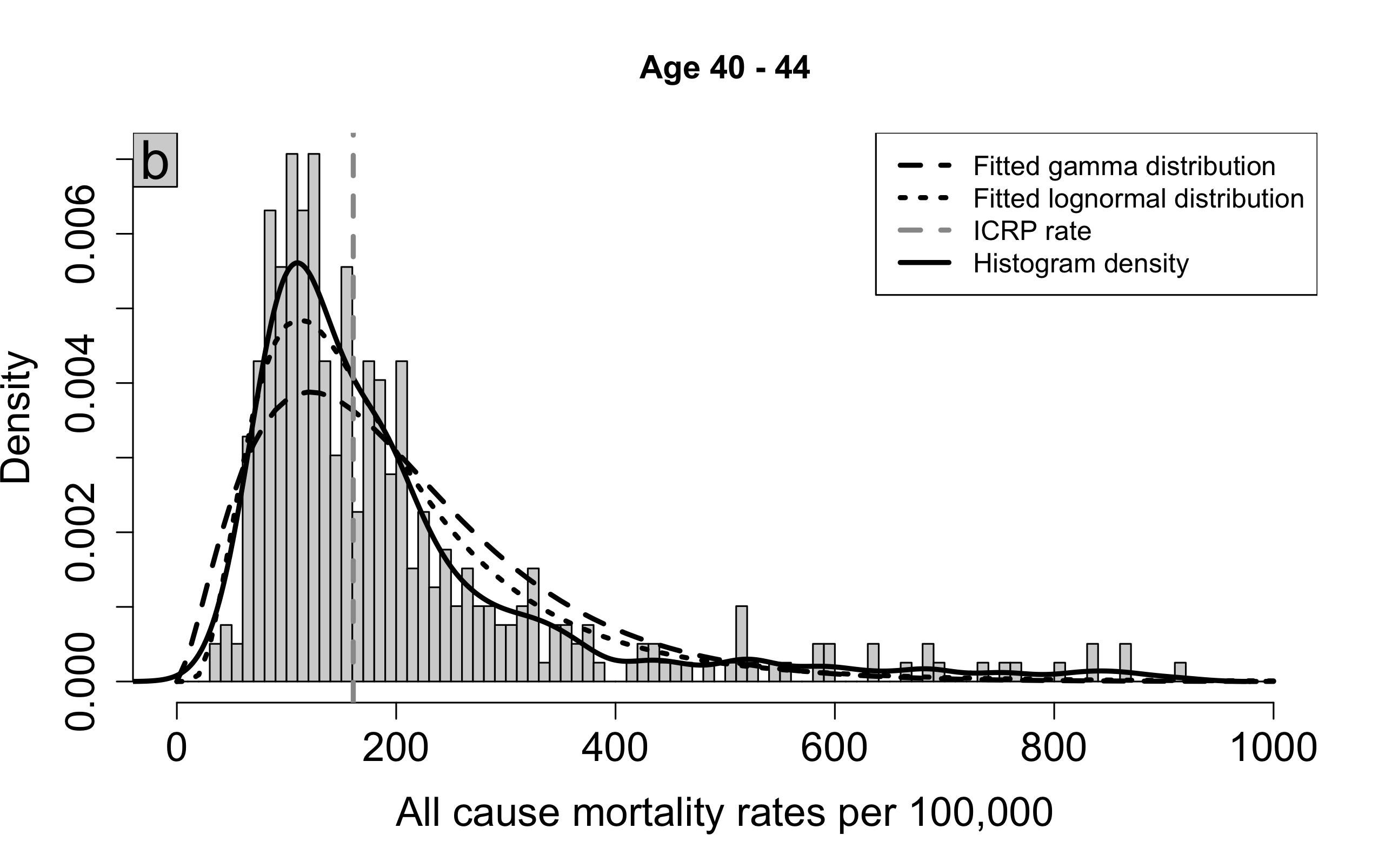}
    \includegraphics[scale=0.0925]{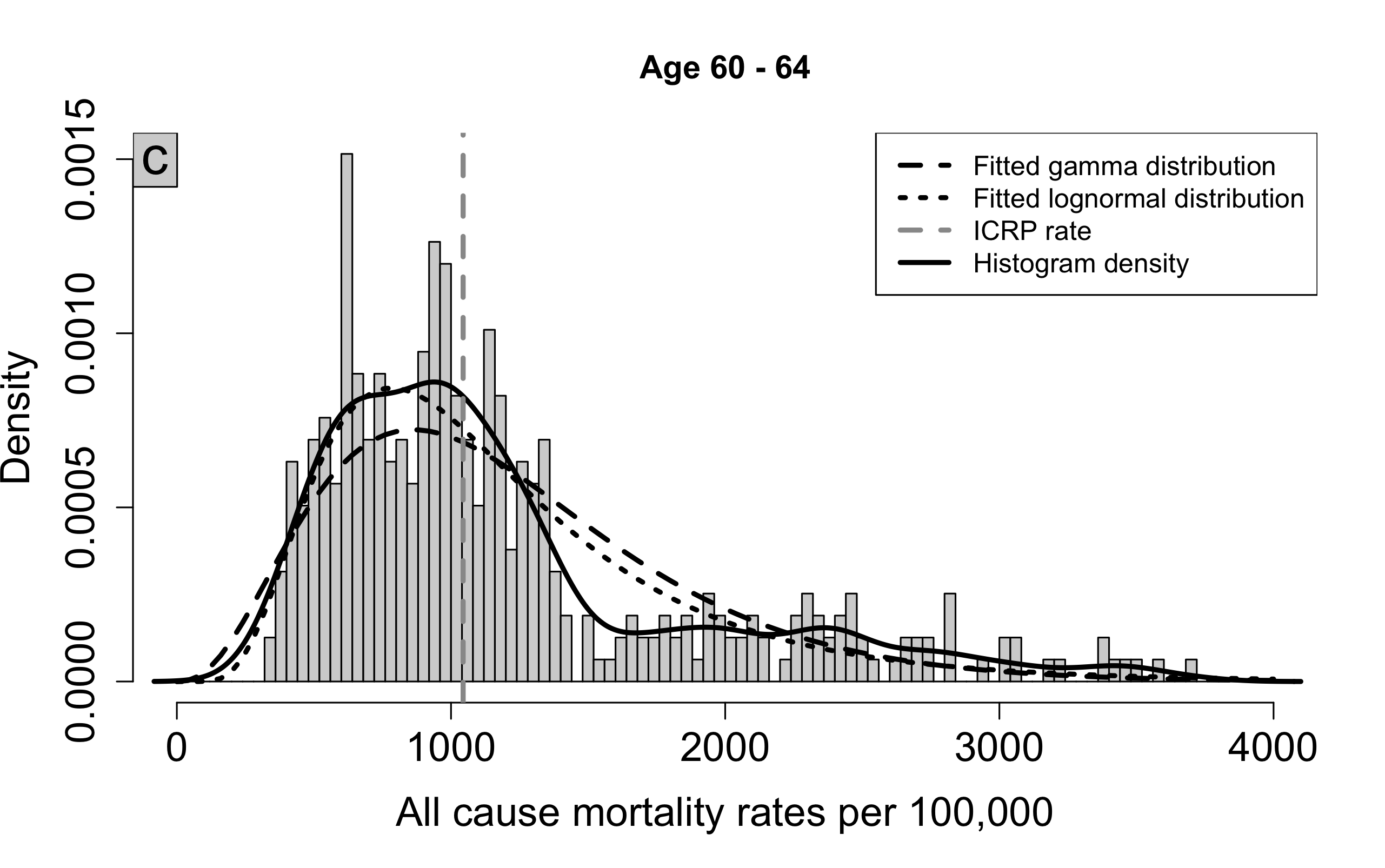}
    \includegraphics[scale=0.0925]{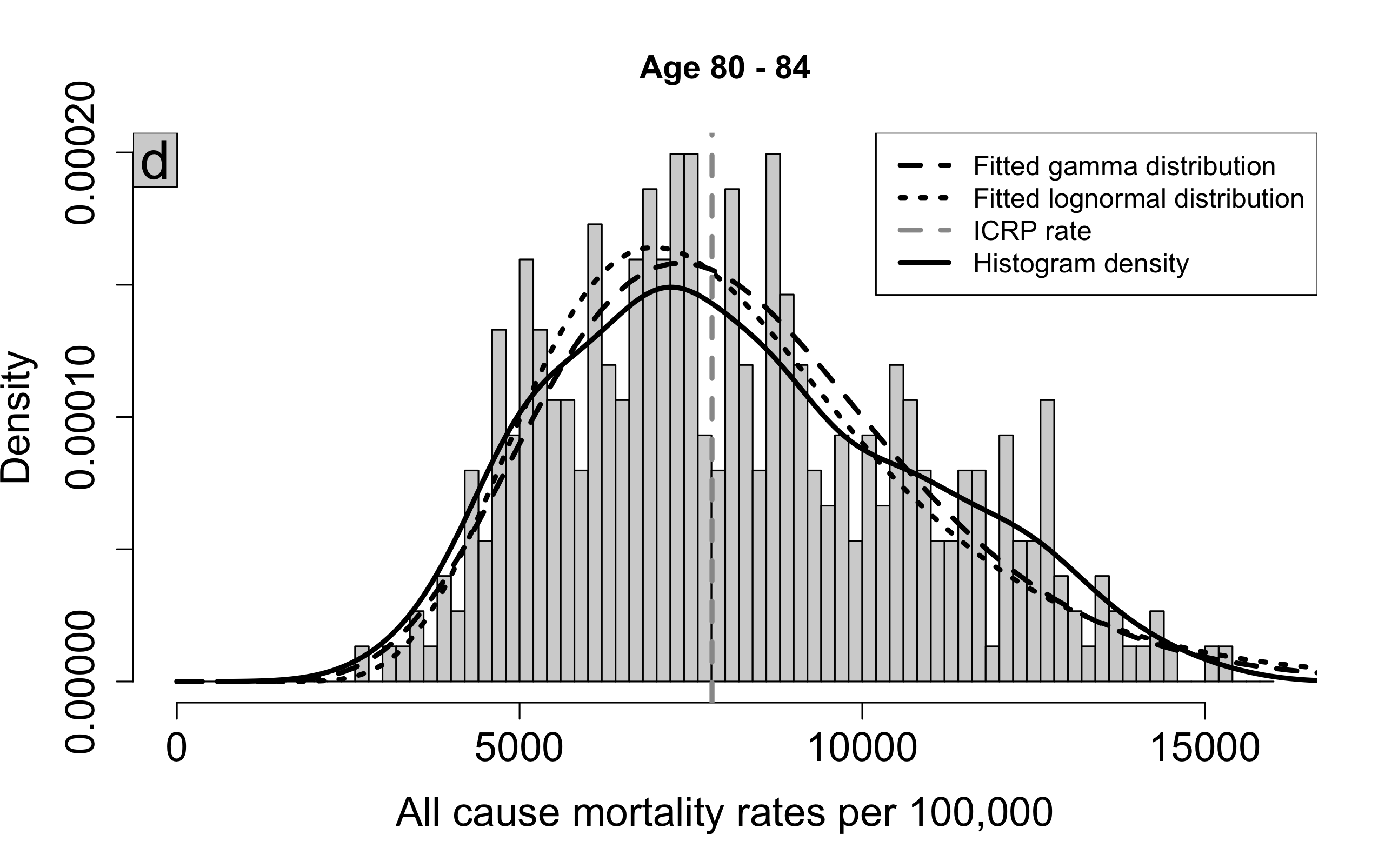}
    \caption{Histogram of all-cause mortality rates for the ages $20-24,40-44, 60-64$ and $80-84$ derived from WHO data. For comparison, the reference all-cause mortality rates from the ICRP Euro-American-Asian mixed population are shown with a vertical dashed line. The dashed (dotted)  curve shows the density for a gamma (log-normal) distribution fitted to the histogram data.}
    \label{fig:WHO_allcauserates_5EAA_ln_vs_g}
\end{figure}

\begin{figure}[htbp]
    \centering
    \includegraphics[scale=0.0925]{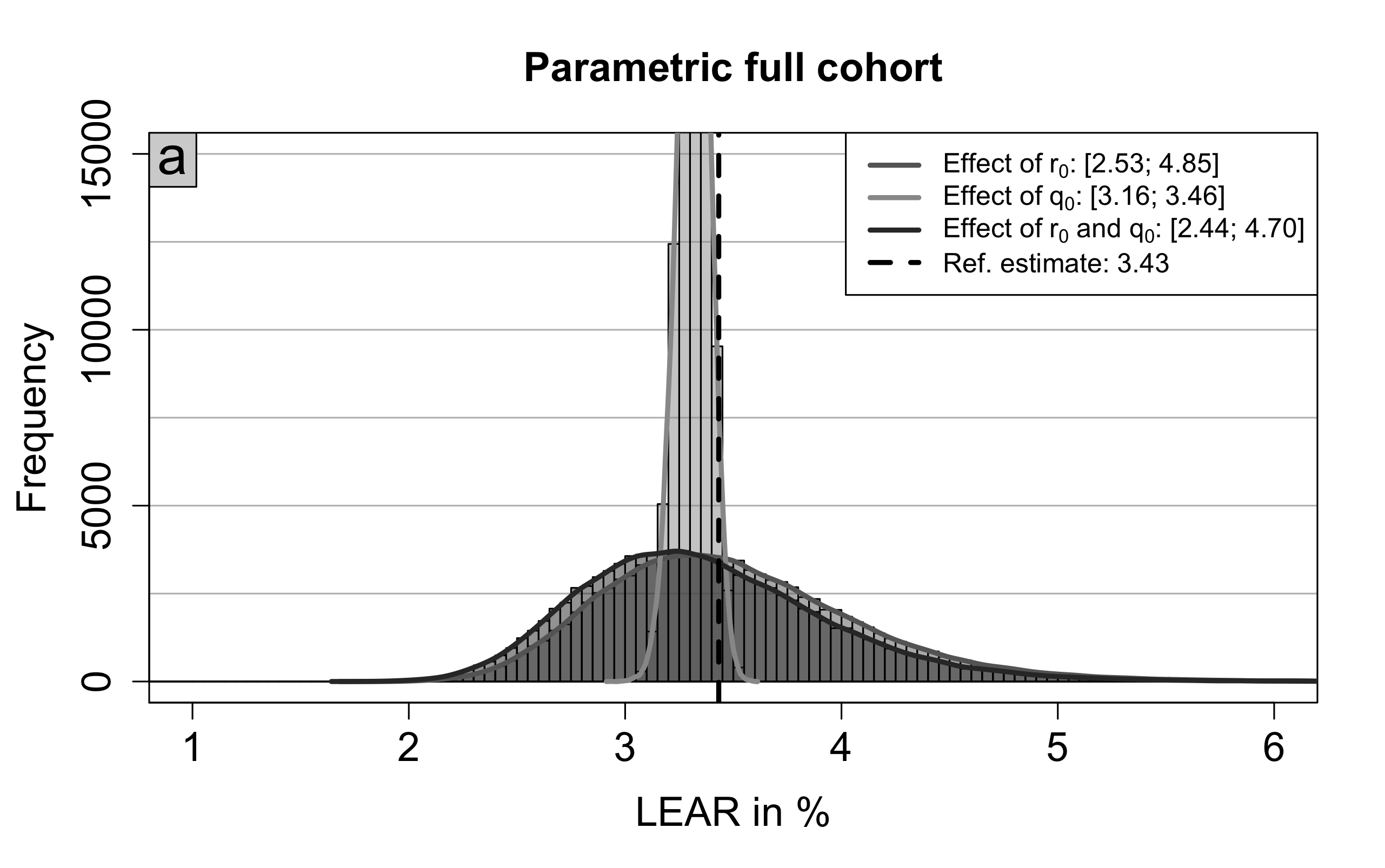}
    \includegraphics[scale=0.0925]{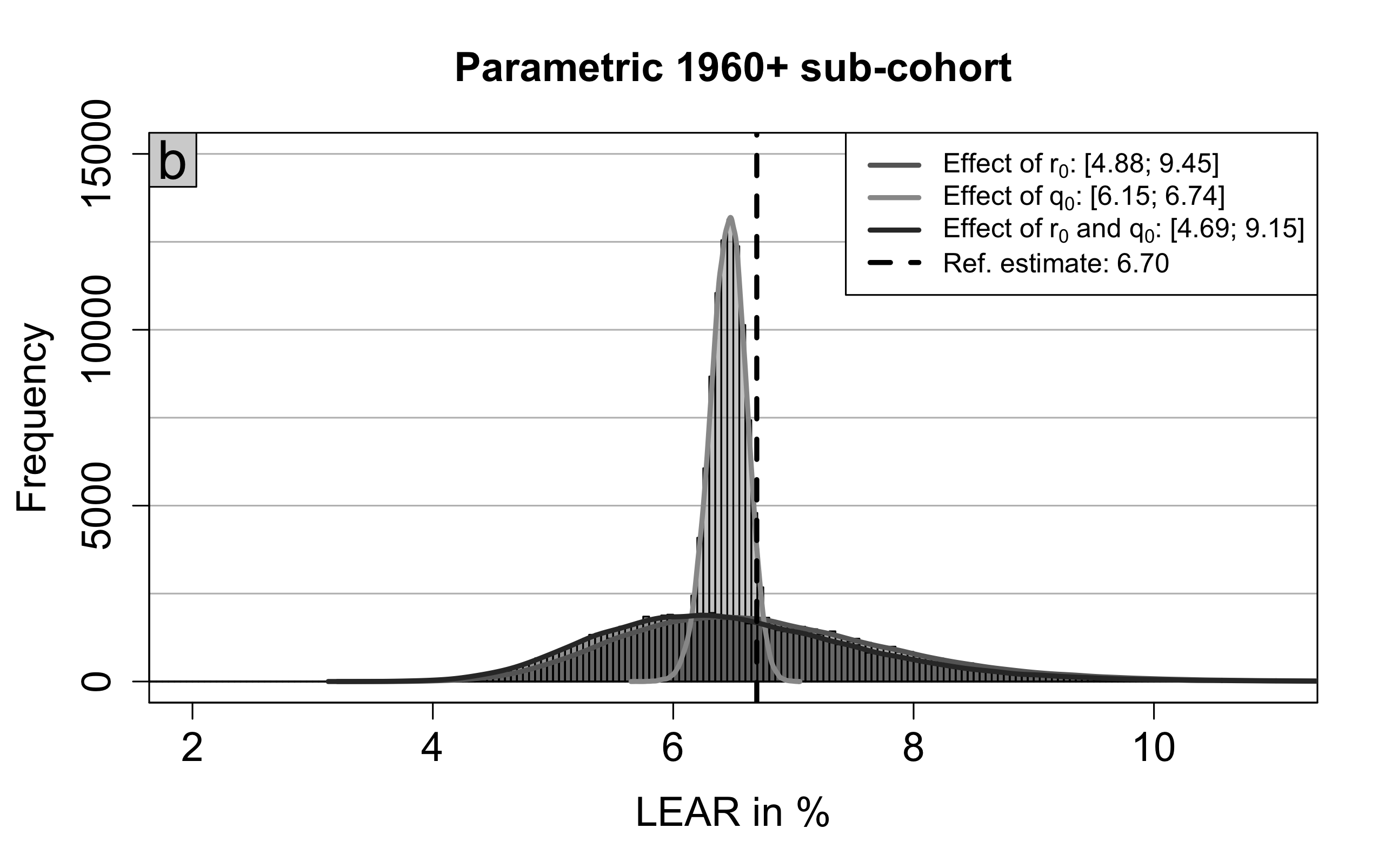}
    \includegraphics[scale=0.0925]{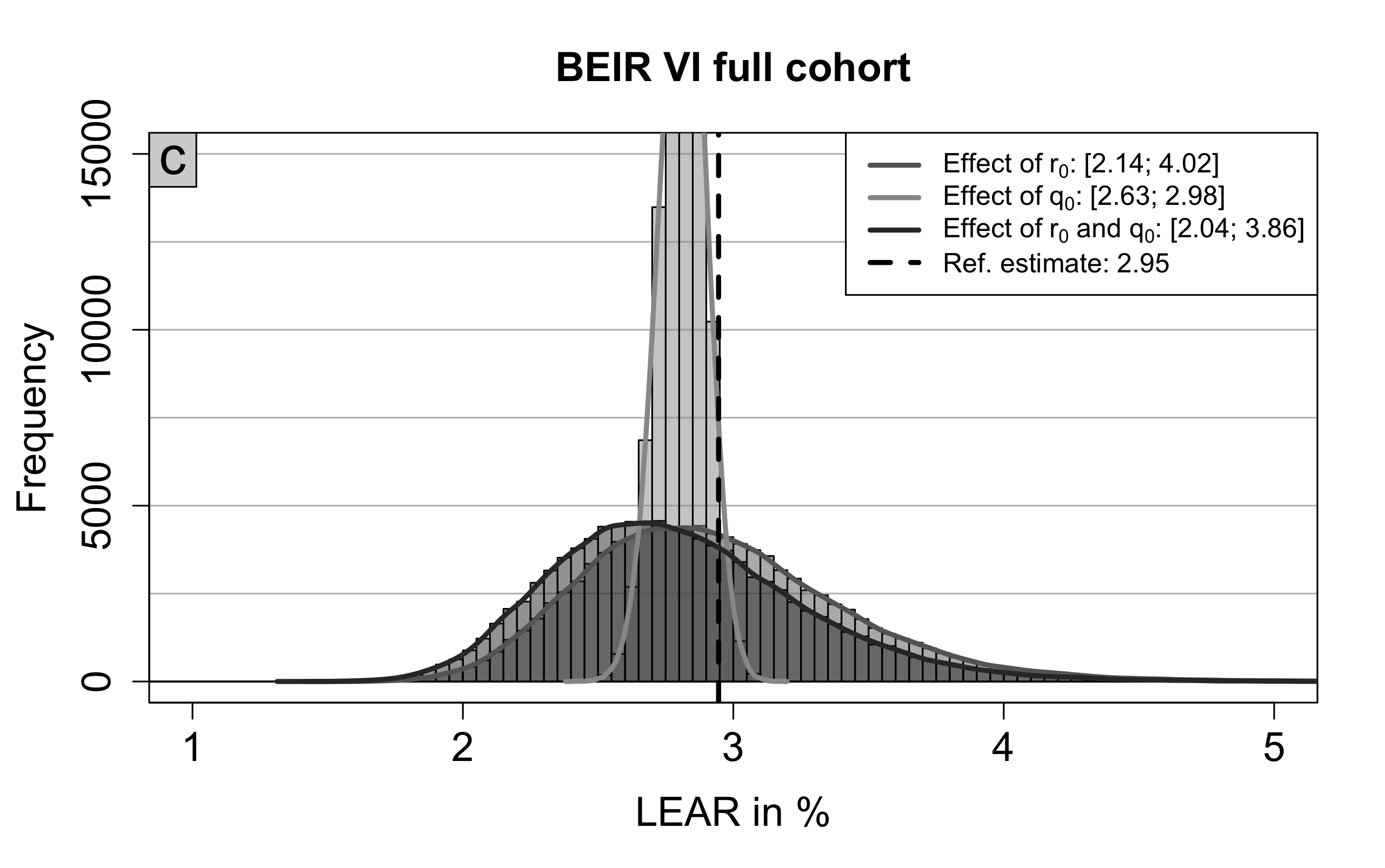}
    \includegraphics[scale=0.0925]{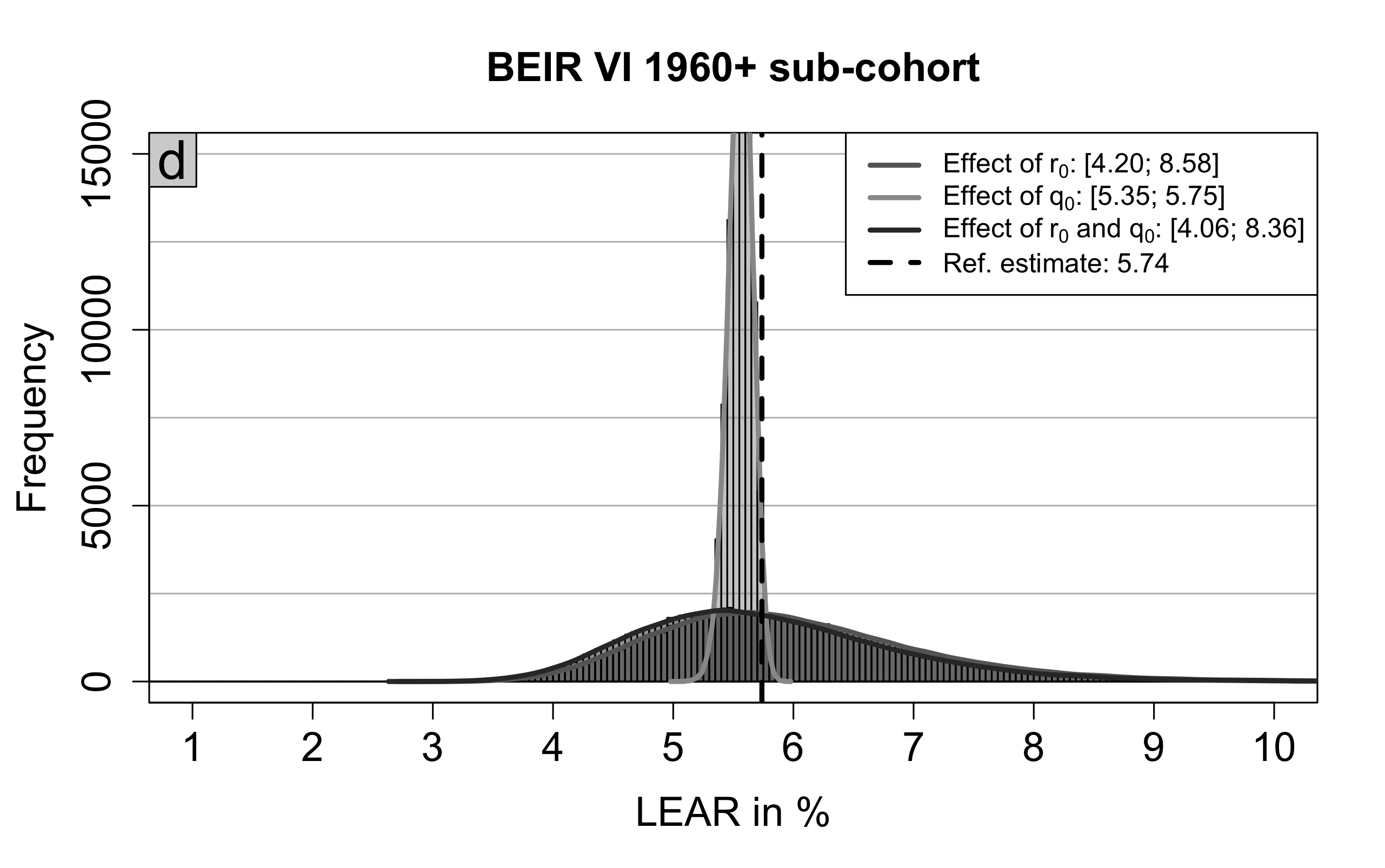}
    \includegraphics[scale=0.0925]{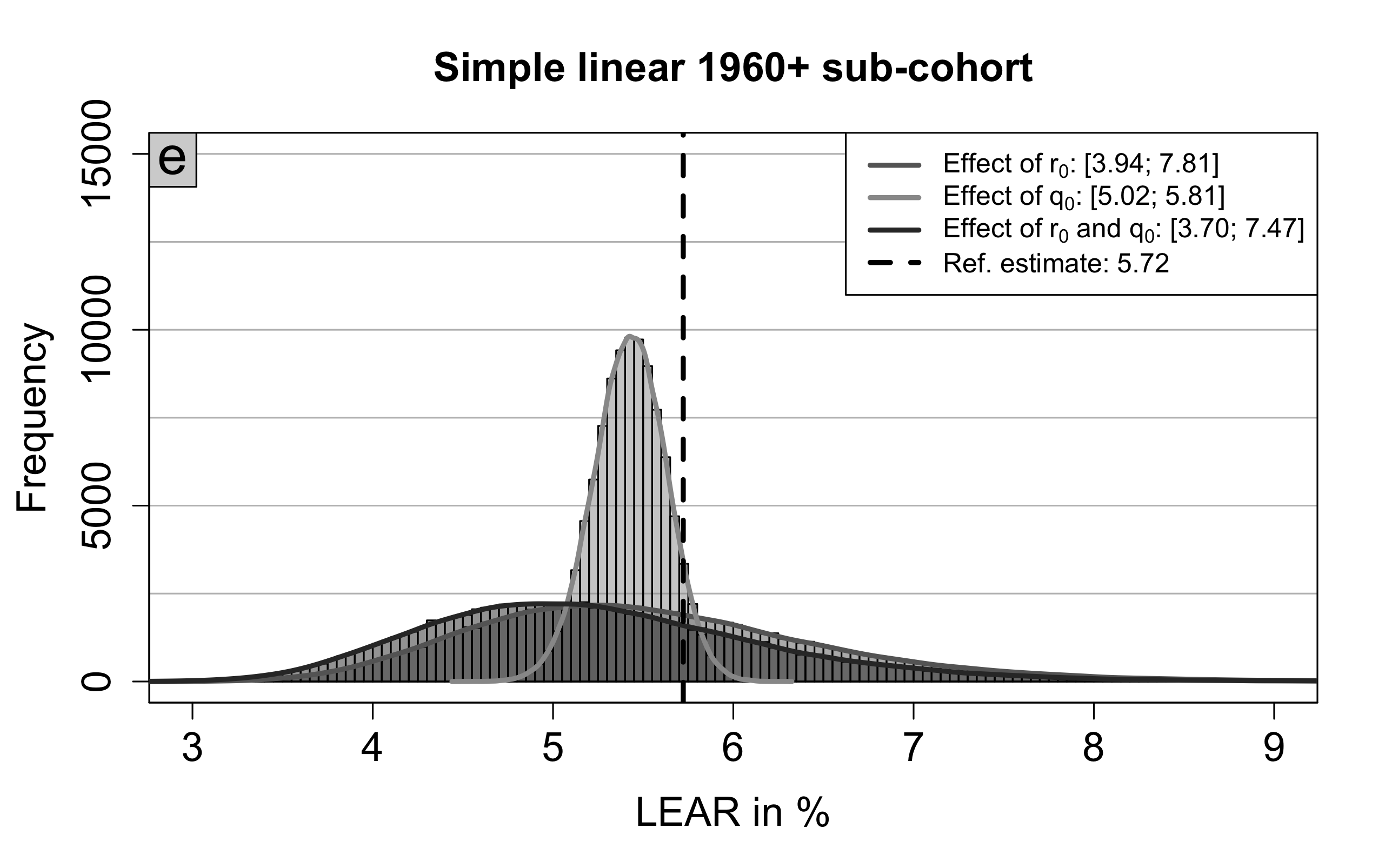}
    \caption{Histogram of the $100,000$ sampled $LEAR$ estimates with kernel density estimate (solid lines) for different risk models and varying uncertainty in mortality rates by grayscale. Lung cancer mortality rates $r_0(t)$ and all-cause mortality rates $q_0(t)$ are assumed to follow a log-normal distribution with parameter estimates derived from WHO data. The joint effect results from independent sampling from both corresponding probability distributions. The 95\% uncertainty interval is presented in the legend.}
    \label{fig:LEAR_MR_Histogramm_LN}
\end{figure}

\begin{figure}[htbp]
    \centering
    \includegraphics[scale=0.15]{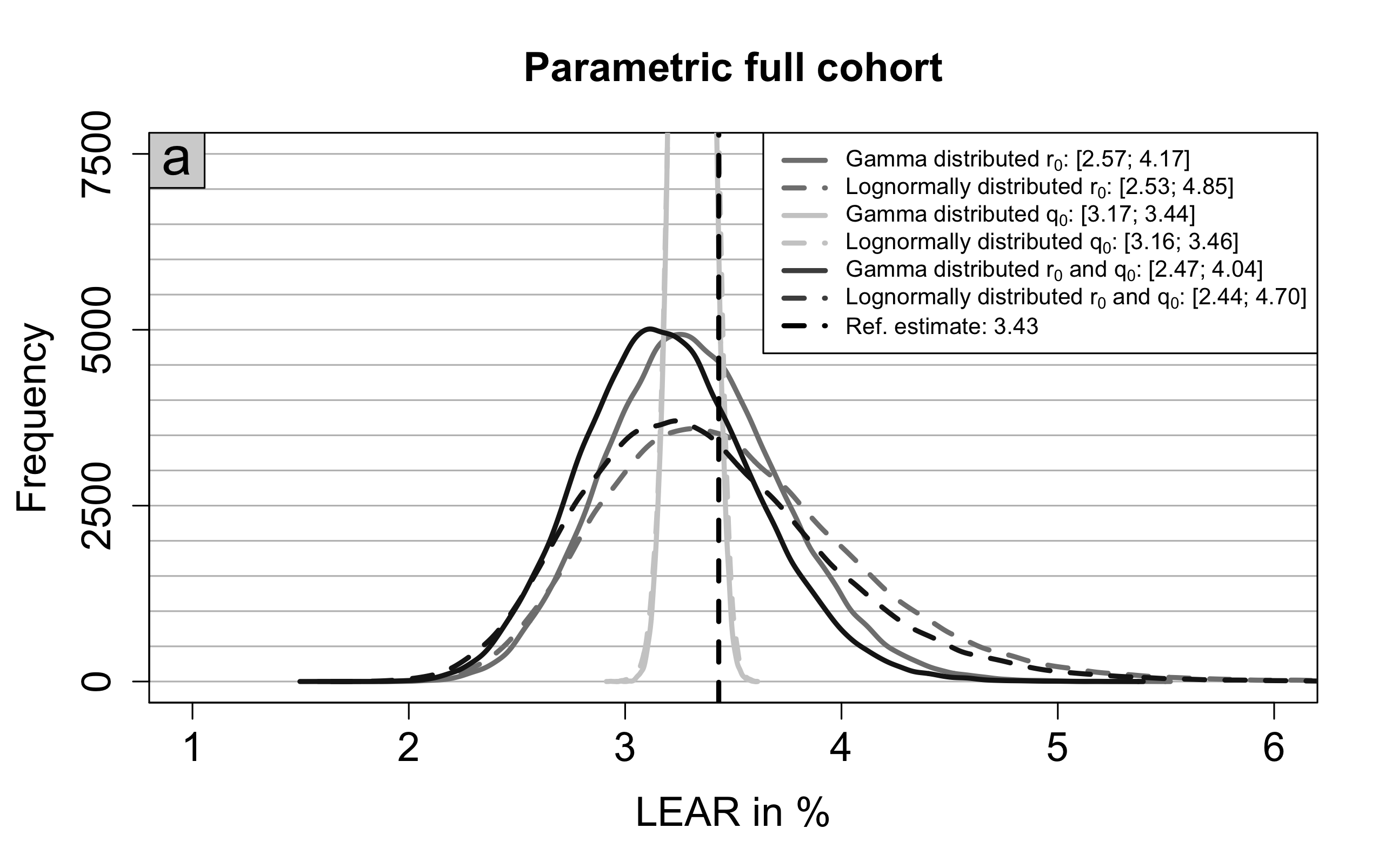}
    \caption{Comparison of densities based on the distribution of $100,000$ sampled $LEAR$ estimates for the parametric full cohort risk model and varying uncertainty in mortality rates derived from WHO data by grayscale. The solid (dashed) lines show the distribution for gamma (log-normally) distributed baseline mortality rates (lung cancer $r_0(t)$, all-cause $q_0(t)$ or both). The joint effect results from independent sampling from both corresponding probability distributions. The 95\% uncertainty intervals are presented in the legend. The dashed vertical line indicates the reference LEAR in \% estimate.}
    \label{fig:LEAR_MR_Histogramm_G_vs_LN}
\end{figure}

Both gamma and log-normal distributions fit the WHO data comparably well (Suppl. Figures \ref{fig:WHO_lungrates_5EAA_ln_vs_g} and  \ref{fig:WHO_allcauserates_5EAA_ln_vs_g}). Resulting 95\% uncertainty intervals are similar (Suppl. Table \ref{tab:LEAR_MR_intervals_LN_vs_G}), with log-normal distributions yielding slightly wider intervals (Suppl. Figure \ref{fig:LEAR_MR_Histogramm_LN}). This is because log-normal distributions inherit heavier tails  (direct comparison in Suppl. Figure \ref{fig:LEAR_MR_Histogramm_G_vs_LN}). The lower bounds are very similar but the upper bounds are larger for log-normal mortality rates. All-cause mortality rates have minimal impact on $LEAR$ uncertainty, regardless of the chosen distribution, especially compared to lung cancer rates. \\

Acknowledging the impact of the chosen distribution on uncertainty intervals highlights the need for cautious interpretation. Sampling results and the data source (here: WHO mortality database) influence the intervals. Therefore, the overall tendency of interval span provides more valuable information than precise interval bounds. The computed intervals offer a quantitative sense of how mortality rate variability can influence lifetime risk estimates.

\begin{table}[htbp]
    \centering\resizebox{\columnwidth}{!}{ 
 \begin{tabular}[h]{llllll}
\addlinespace
\hline
\addlinespace
Risk model& LEAR in \% & Distribution& Effect of $r_0$& Effect of $q_0$  & Effect of $r_0$ and $q_0$  \\
\addlinespace
\hline
\addlinespace
\multirow{ 2}{*}{Parametric full cohort} & \multirow{ 2}{*}{$3.43$}  & Gamma & $[2.57; 4.17]$ $(0.47)$& $[3.17; 3.44]$ $(0.08)$&$[2.47; 4.04]$ $(0.46)$\\
& &  Log-normal & $[2.53; 4.85]$ $(0.68)$& $[3.16; 3.46]$ $(0.09)$&$[2.44; 4.70]$ $(0.66)$ \\
\multirow{ 2}{*}{BEIR VI full cohort} & \multirow{ 2}{*}{$2.95$} & Gamma & $[2.18; 3.47]$ $(0.44)$& $[2.64; 2.97]$ $(0.11)$&$[2.06; 3.33]$ $(0.43)$\\
& &  Log-normal & $[2.14; 4.02]$ $(0.64)$& $[2.63; 2.98]$ $(0.12)$&$[2.04; 3.86]$ $(0.62)$\\
\multirow{ 2}{*}{Parametric 1960+ sub-cohort} & \multirow{ 2}{*}{$6.70$}  & Gamma & $[4.96; 8.11]$ $(0.47)$ & $[6.16; 6.72]$ $(0.08)$&$[4.76; 7.85]$ $(0.46)$\\
& &  Log-normal & $[4.88; 9.45]$ $(0.68)$& $[6.15; 6.74]$ $(0.09)$ &$[4.69; 9.15]$ $(0.67)$ \\
\multirow{ 2}{*}{BEIR VI 1960+ sub-cohort} & \multirow{ 2}{*}{$5.74$}  & Gamma & $[4.29; 7.36]$ $(0.53)$& $[5.36; 5.73]$ $(0.06)$&$[4.14; 7.14]$ $(0.52)$\\
& &  Log-normal & $[4.20; 8.58]$ $(0.76)$& $[5.35; 5.75]$ $(0.07)$&$[4.06; 8.36]$ $(0.75)$\\
\multirow{ 2}{*}{Simple linear 1960+ sub-cohort} & \multirow{ 2}{*}{$5.72$}  & Gamma & $[4.02; 6.66]$ $(0.46)$& $[5.05; 5.79]$ $(0.13)$&$[3.76; 6.36]$ $(0.45)$\\
& &  Log-normal & $[3.94; 7.81]$ $(0.68)$& $[5.02; 5.81]$ $(0.14)$&$[3.70; 7.47]$ $(0.66)$\\
\addlinespace
\hline
\addlinespace
\end{tabular}}
    \caption{Reference $LEAR$ estimates in \% with 95\% uncertainty interval (relative uncertainty span in brackets) derived from $100,000$ sampled values for different risk models and distributional assumptions on the mortality rates $r_0(t),q_0(t)$ for all ages $t$ derived from WHO data.}
    \label{tab:LEAR_MR_intervals_LN_vs_G}
\end{table}
\clearpage
\section{Sex-specific uncertainty}
\label{Sex_specific_uncertainty_SDC_E}
This section examines sex-specific $LEAR$ uncertainties by quantifying female- and male-specific mortality rate uncertainties and their impact on $LEAR$ estimates. Further, risk model parameter uncertainties are quantified with the Bayesian approach for sex-specific ICRP reference mortality rates.

\subsection{Mortality rate uncertainty}
\label{Sex_specific_uncertainty_mortality_rates_SDC_E1}
Sex-specific uncertainty in mortality rates $r_0(t),q_0(t)$  and its impact on $LEAR$ are examined by deriving gamma distributions fit to sex-specific mortality rates from WHO data. Risk model parameters are fixed and not uncertain here. \\

\begin{figure}[htbp]
    \centering
    \includegraphics[scale=0.0925]{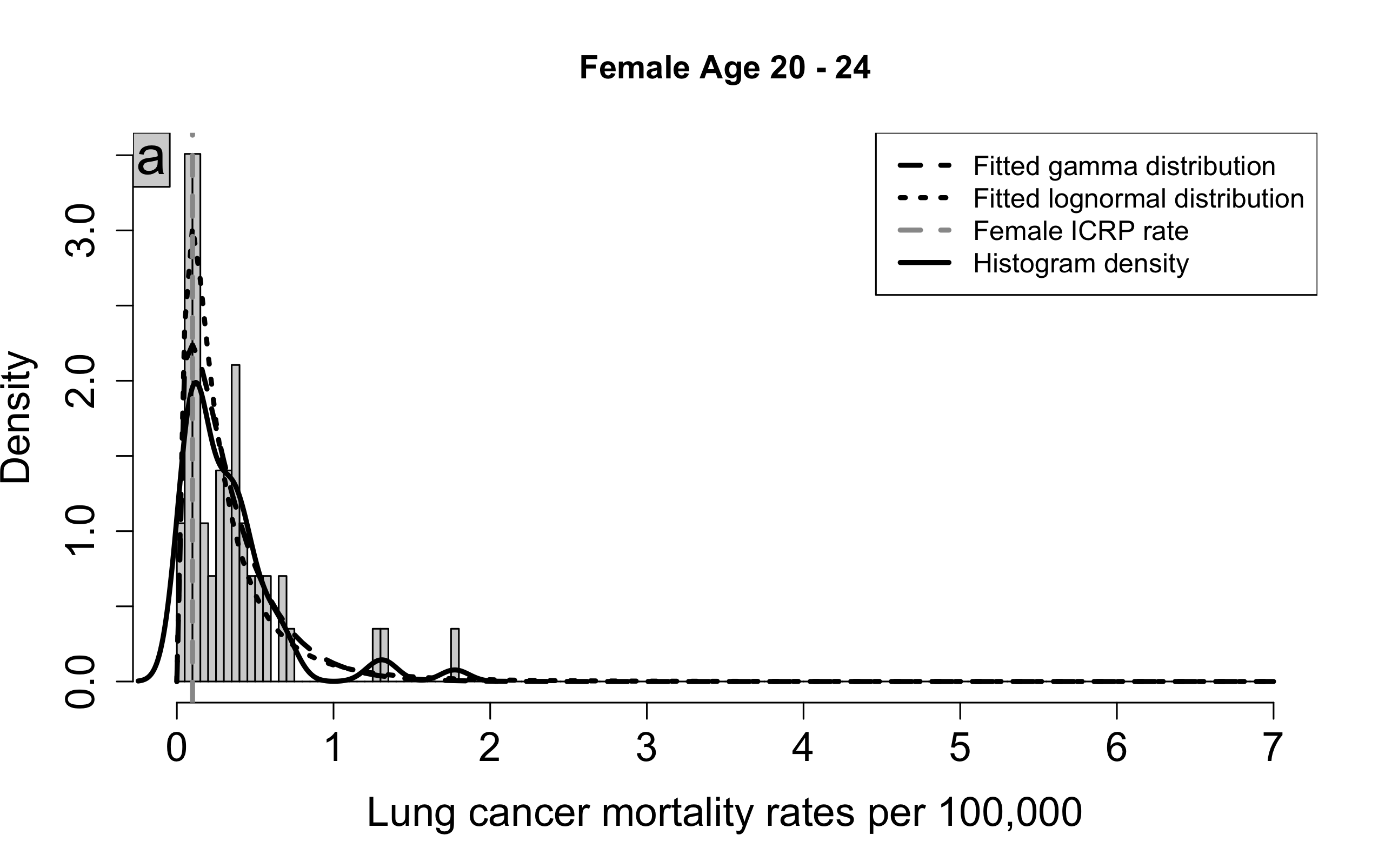}
    \includegraphics[scale=0.0925]{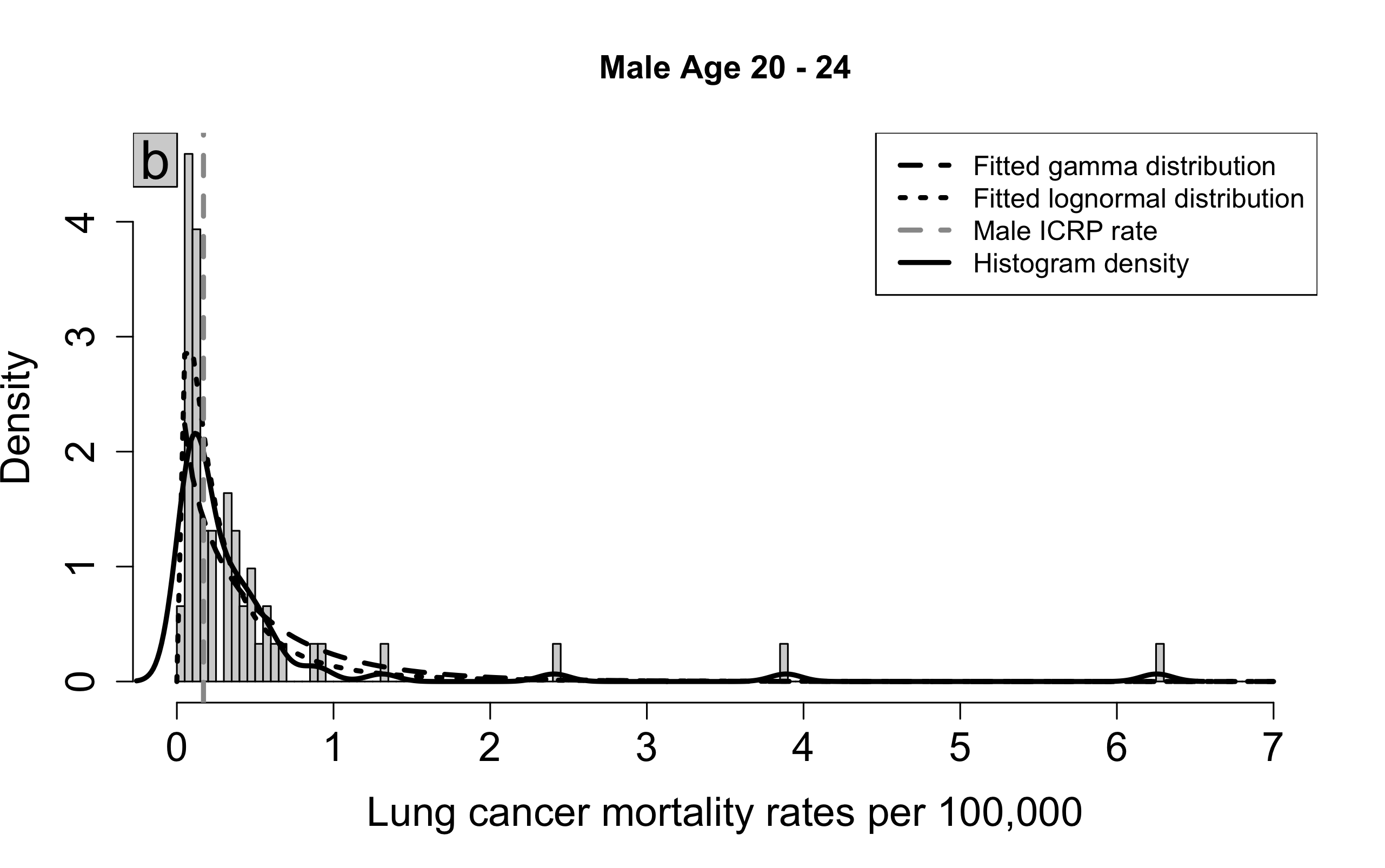}
    \includegraphics[scale=0.0925]{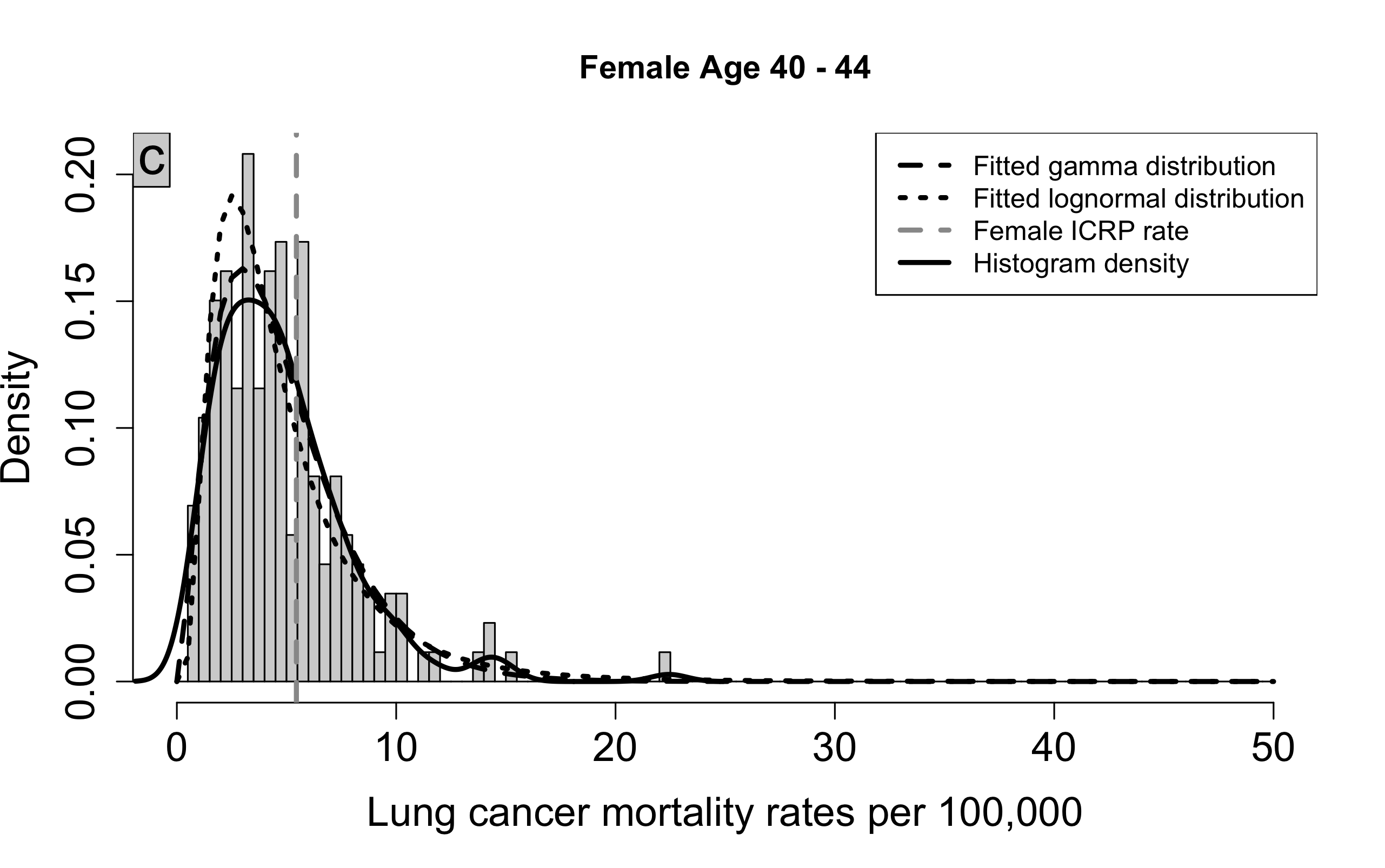}
    \includegraphics[scale=0.0925]{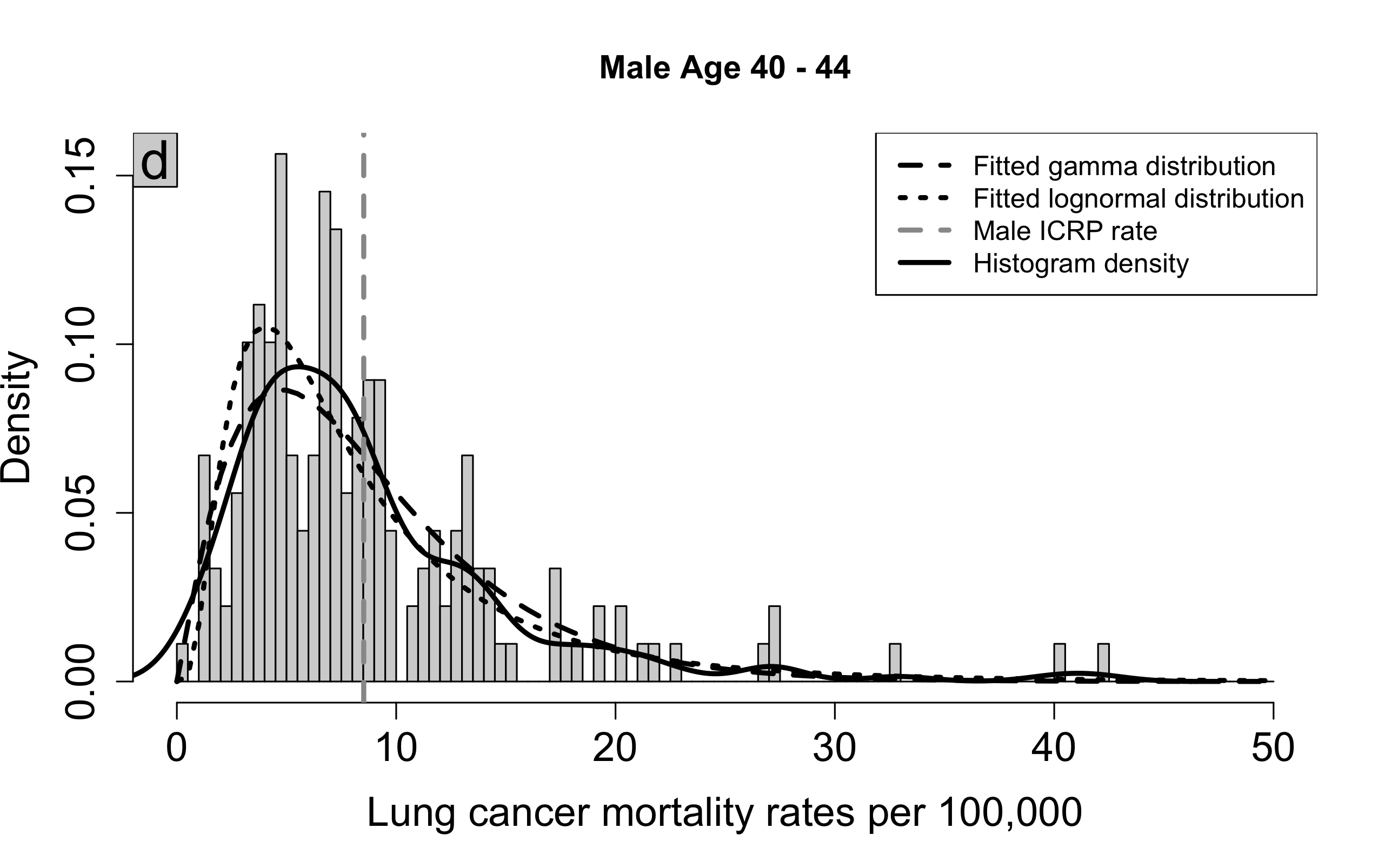}
    \includegraphics[scale=0.0925]{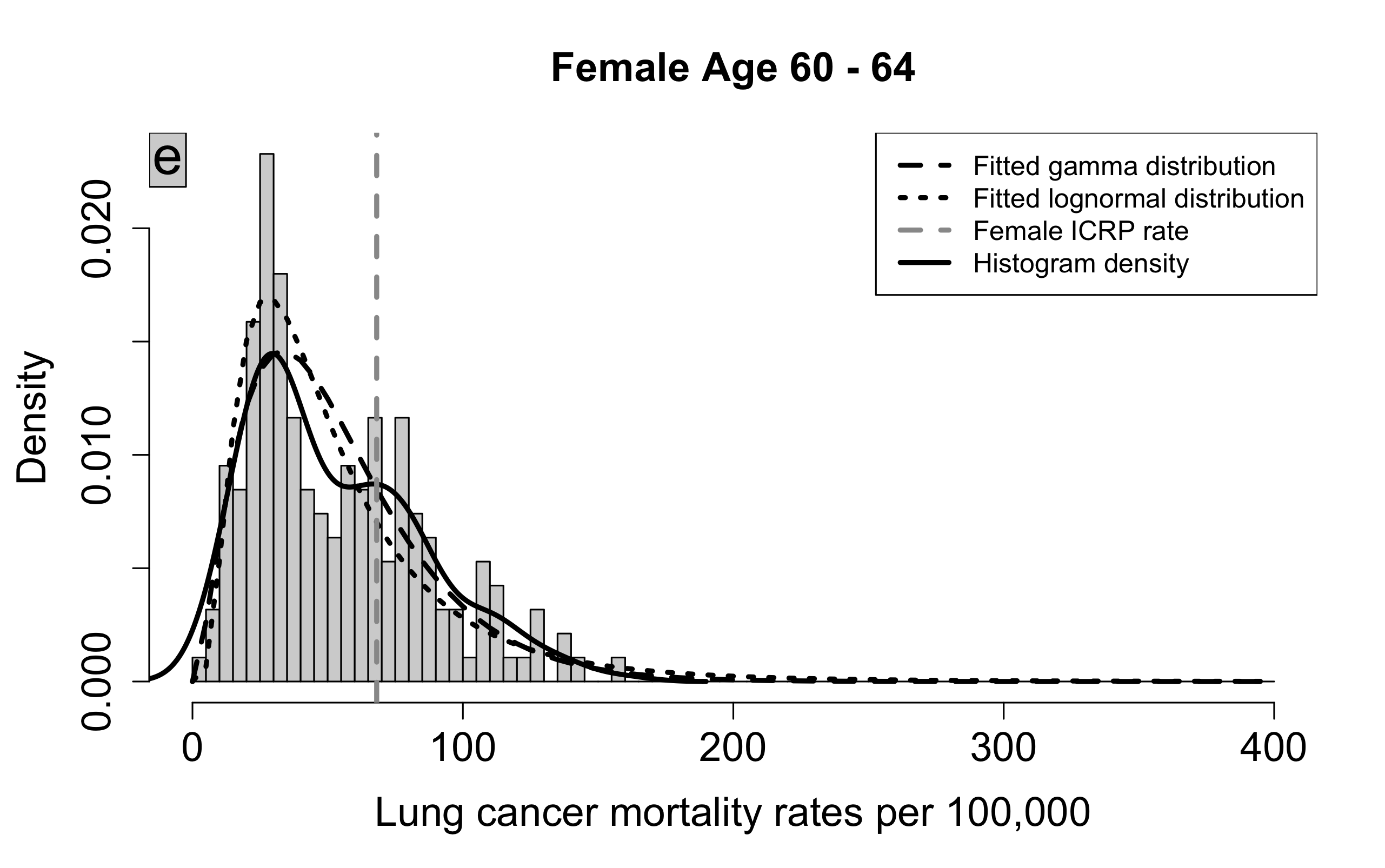}
    \includegraphics[scale=0.0925]{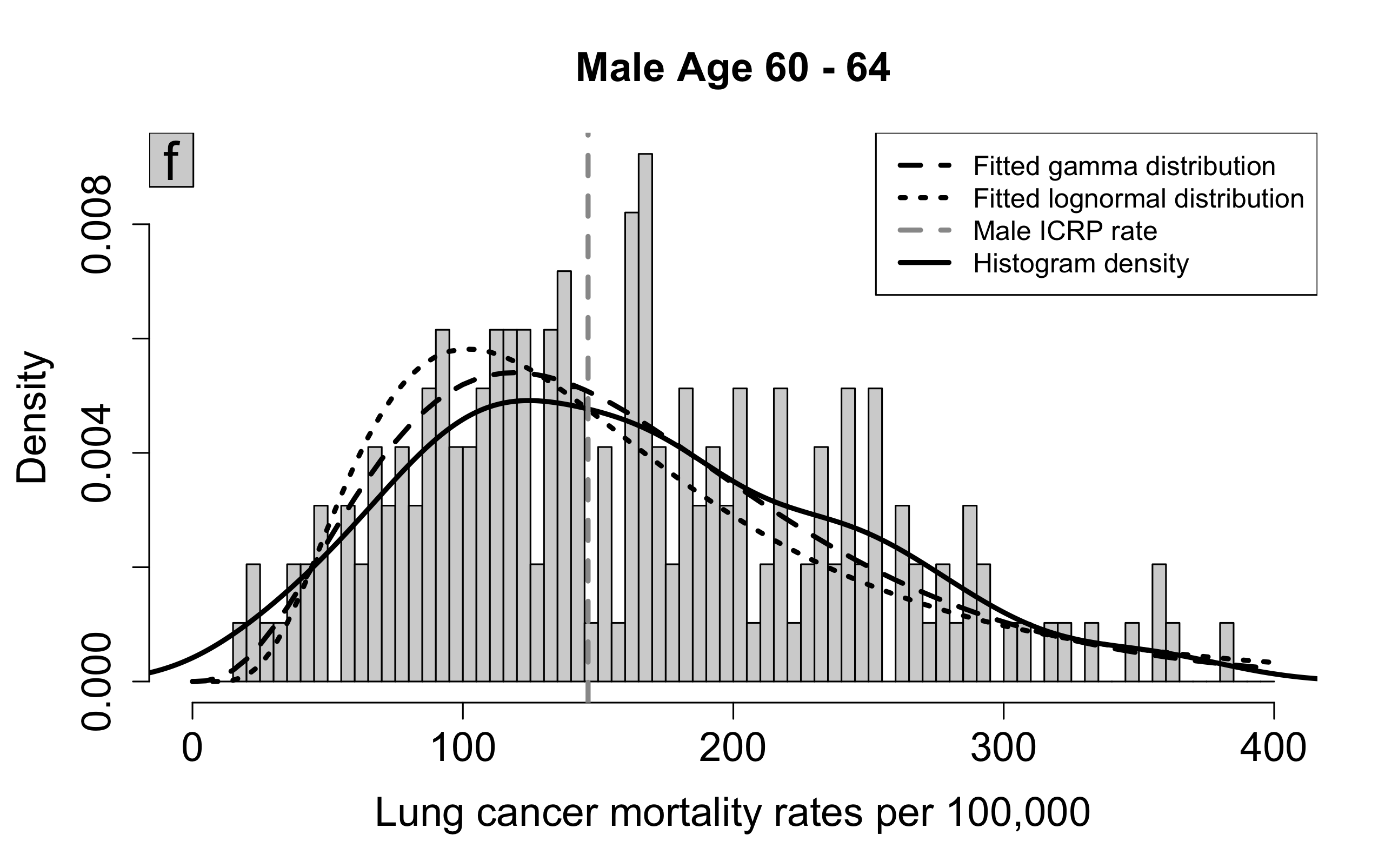}
    \includegraphics[scale=0.0925]{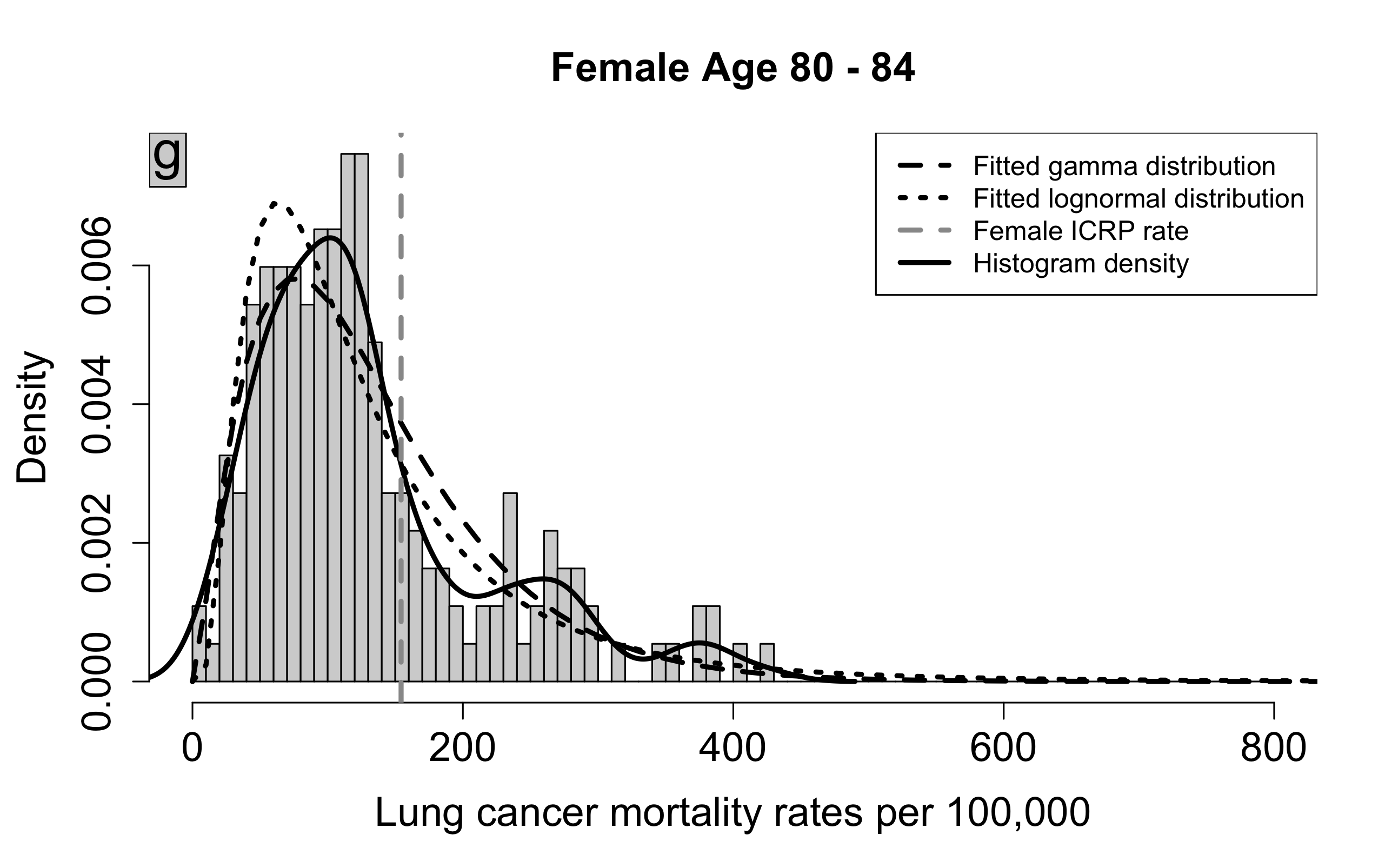}
    \includegraphics[scale=0.0925]{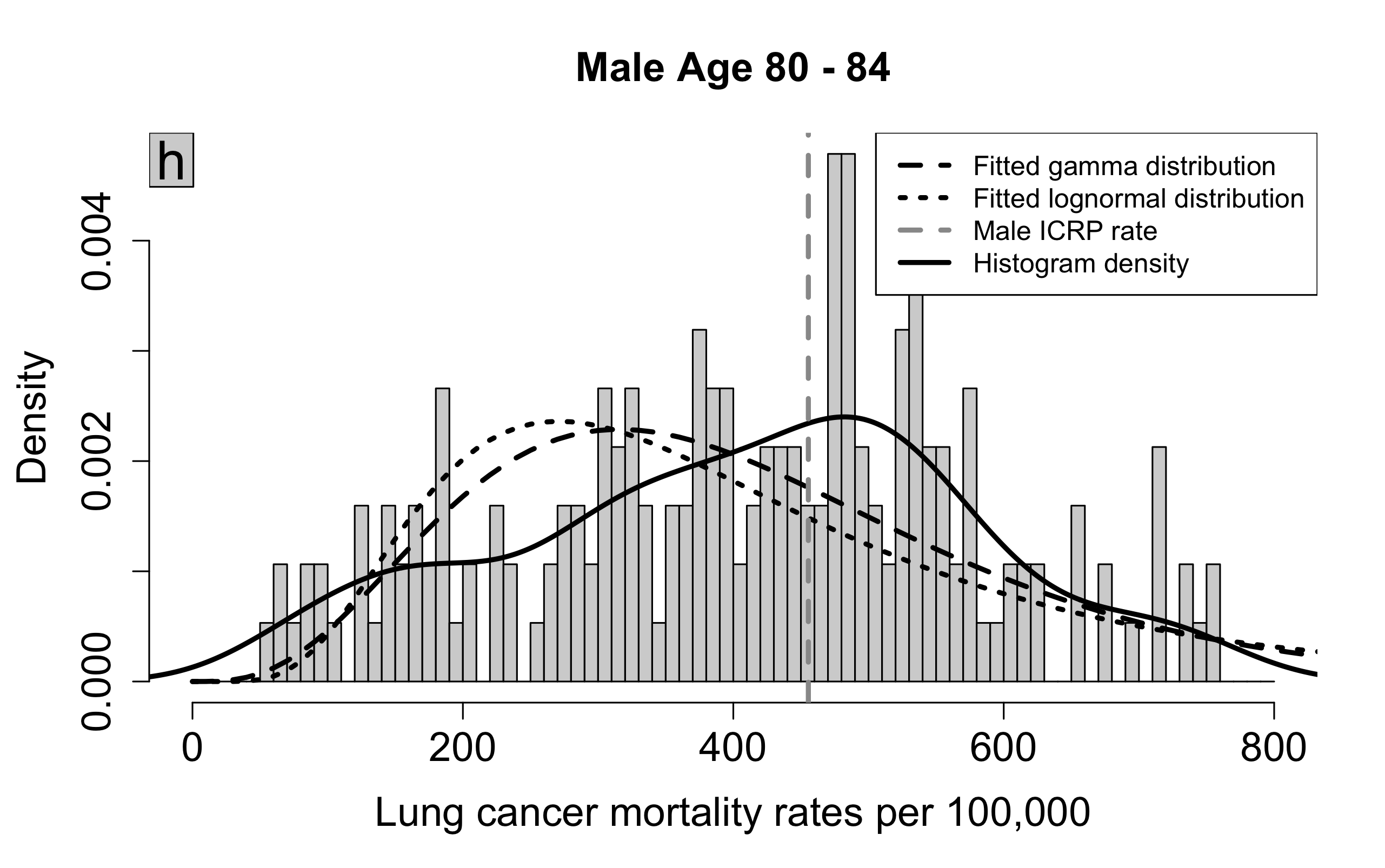}
    \caption{Histogram of male and female lung cancer mortality rates for the ages $20-24,40-44, 60-64$ and $80-84$ derived from WHO data. Every histogram is equipped with a corresponding kernel density estimate. For comparison, the male and female lung cancer mortality rates from the ICRP Euro-American-Asian mixed population are shown with a dashed vertical line. The dashed (dotted) curve shows the density for a gamma (log-normal) distribution fitted to the histogram data.}
    \label{fig:WHO_lungrates_5EAA_male_female}
\end{figure}

\begin{figure}[htbp]
    \centering
    \includegraphics[scale=0.0925]{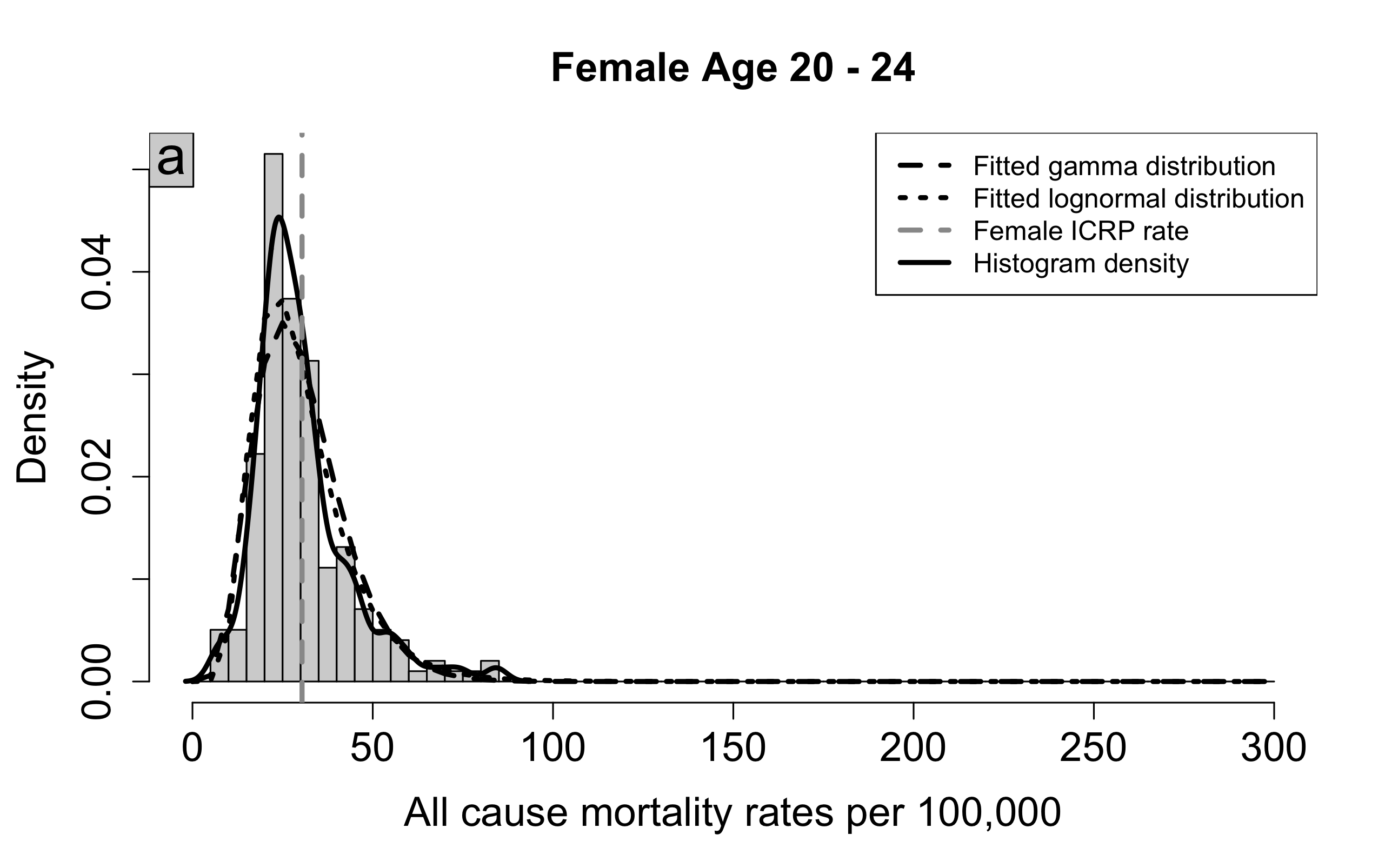}
    \includegraphics[scale=0.0925]{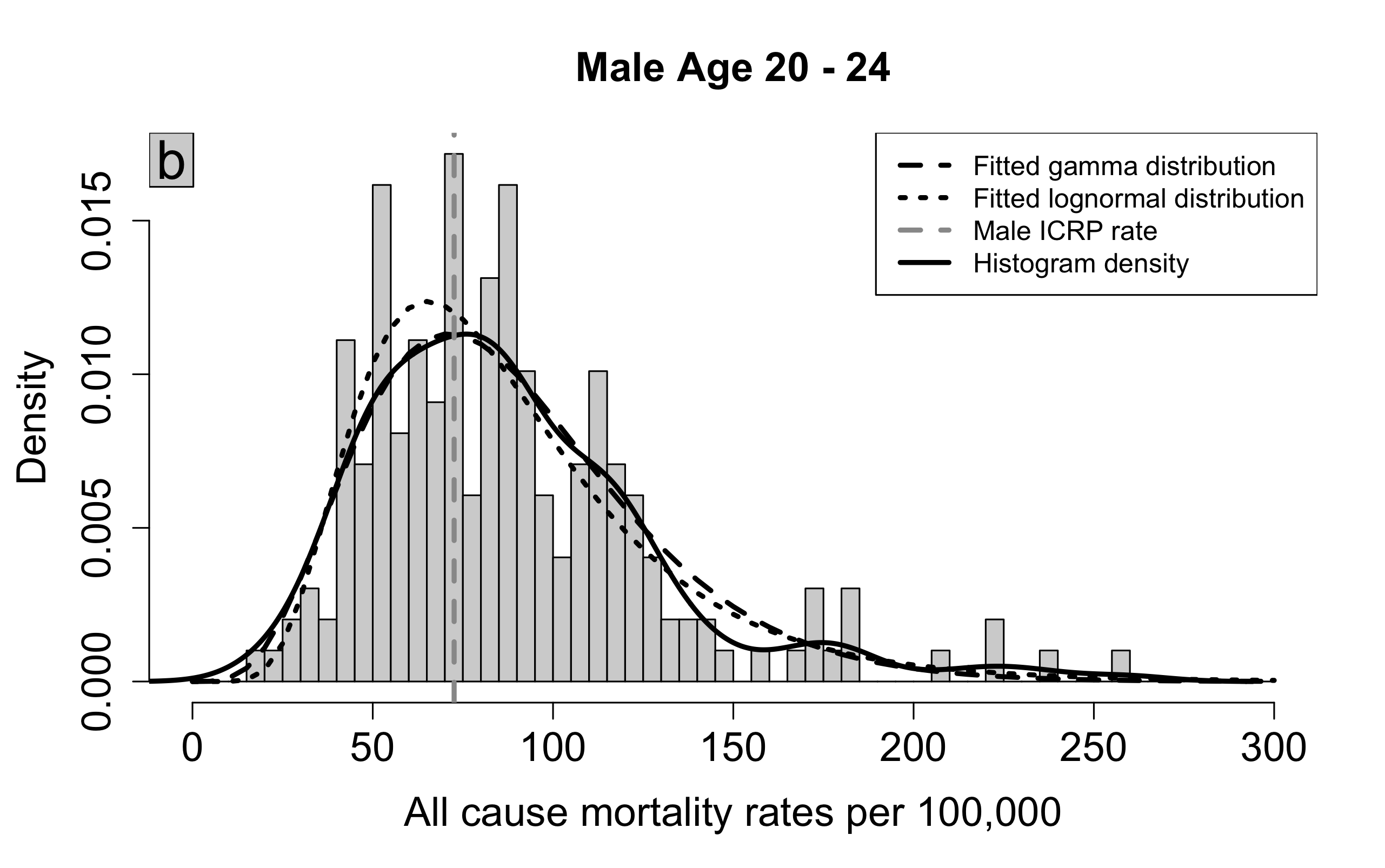}
    \includegraphics[scale=0.0925]{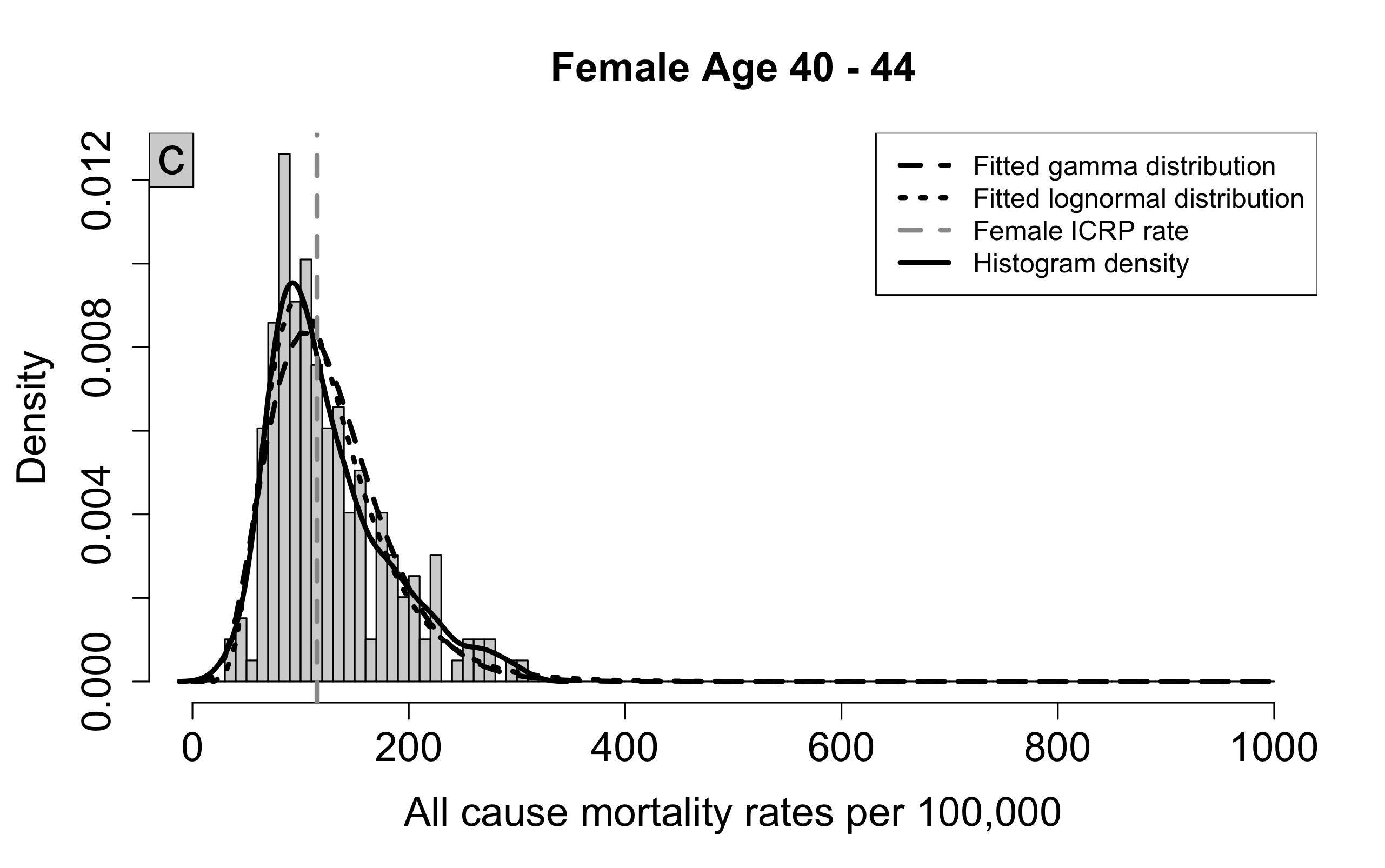}
    \includegraphics[scale=0.0925]{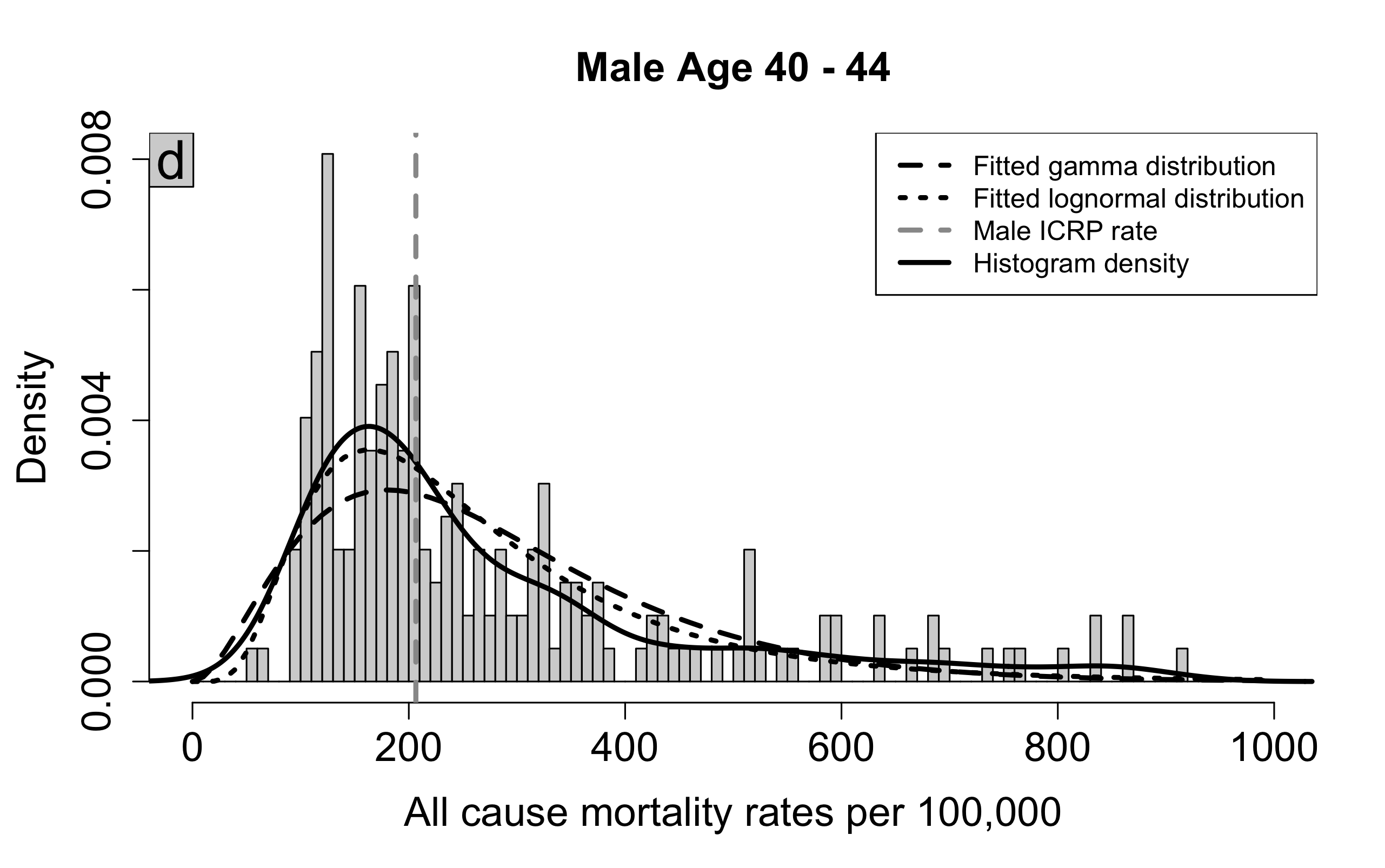}
    \includegraphics[scale=0.0925]{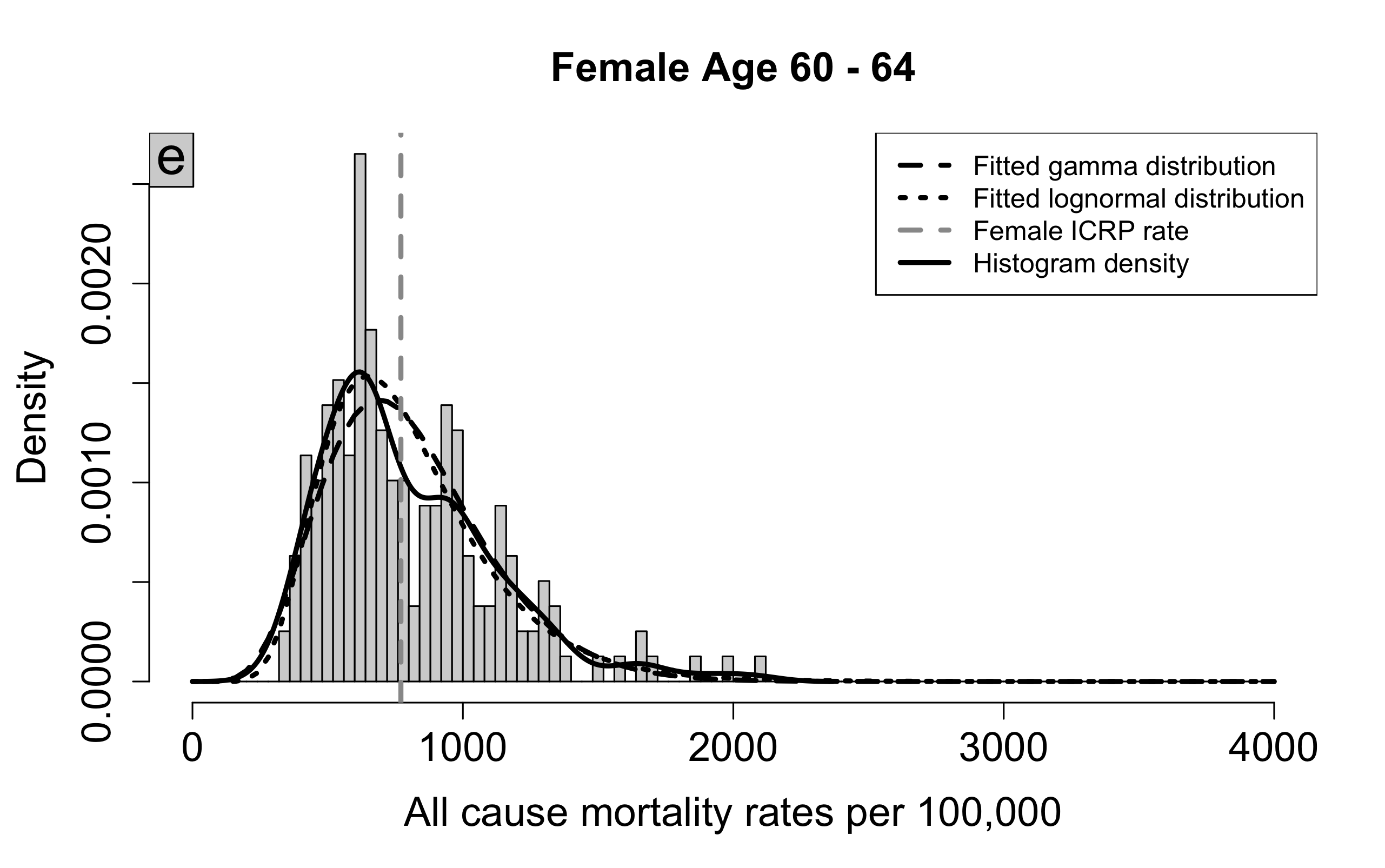}
    \includegraphics[scale=0.0925]{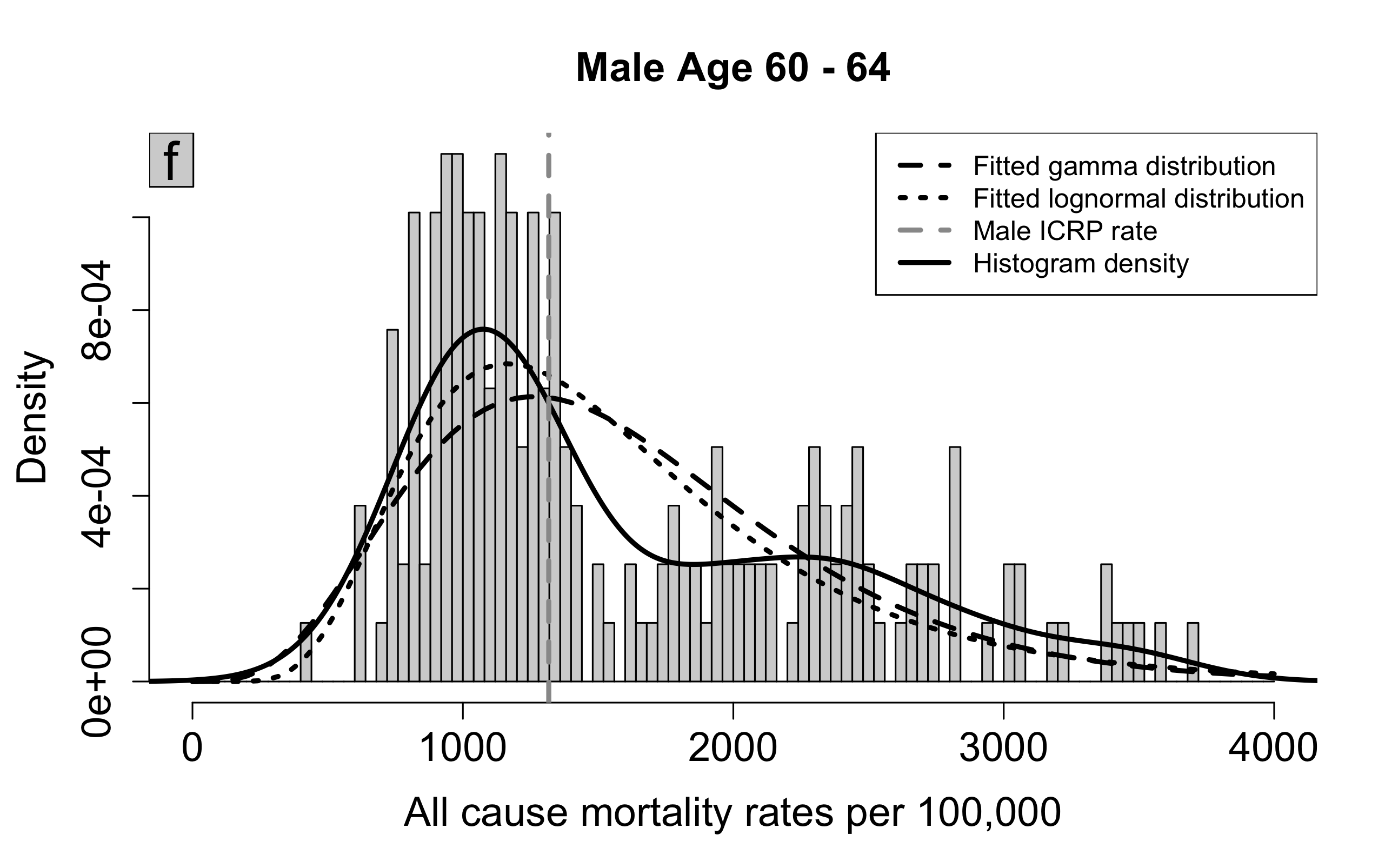}
    \includegraphics[scale=0.0925]{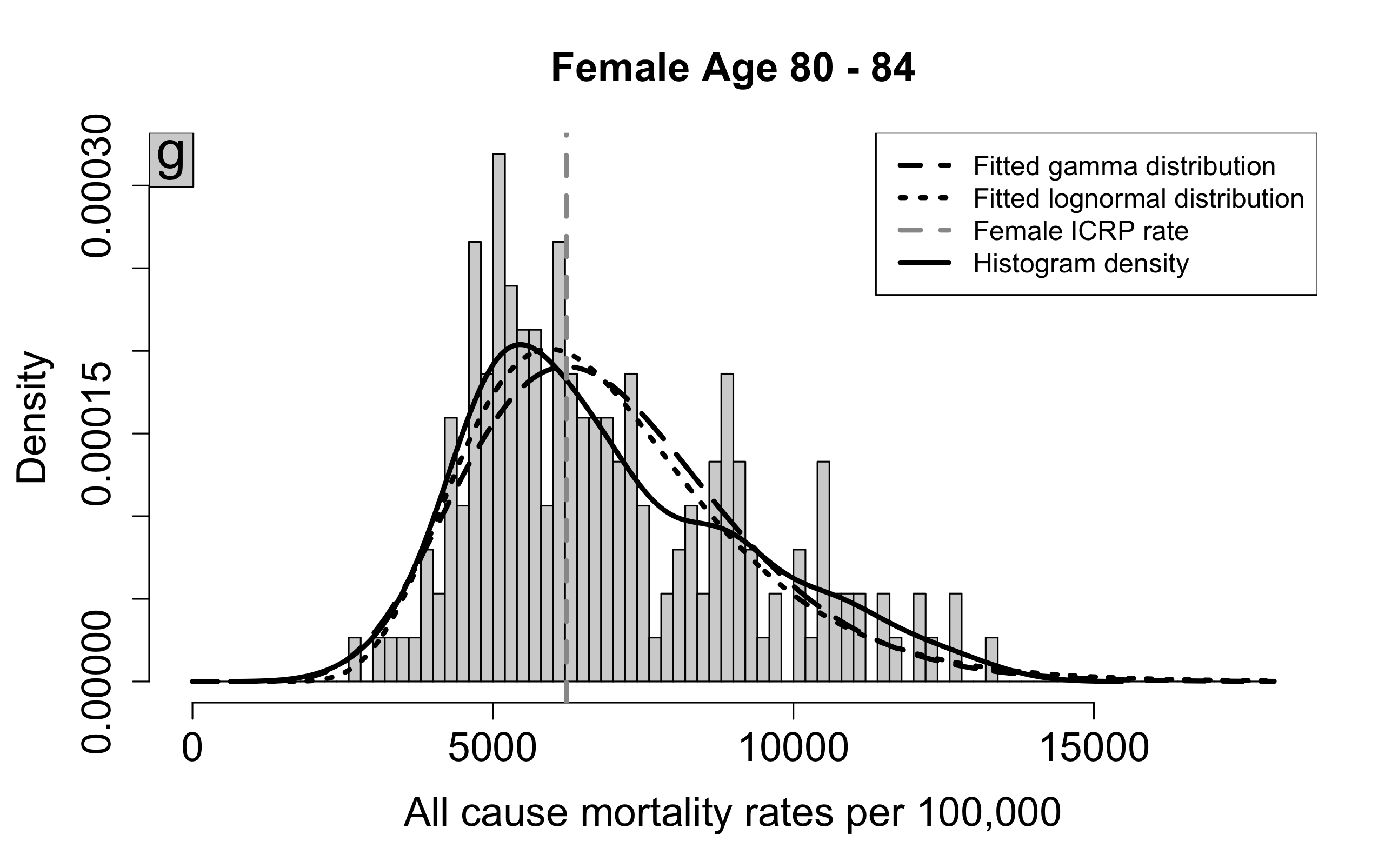}
    \includegraphics[scale=0.0925]{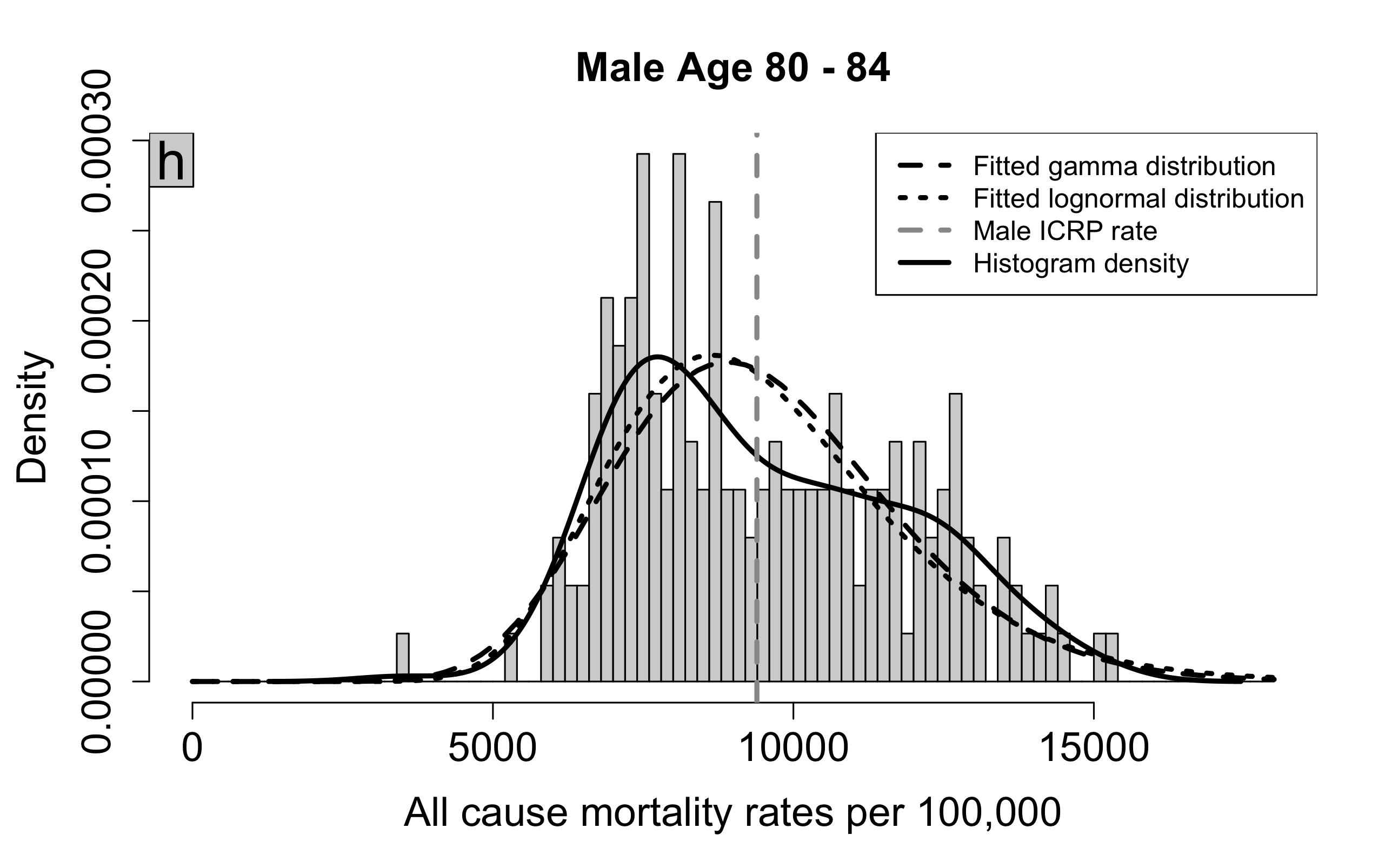}
    \caption{Histogram of male and female all-cause mortality rates for the ages $20-24,40-44, 60-64$ and $80-84$ derived from WHO data. Every histogram is equipped with a corresponding kernel density estimate. For comparison, the male and female all-cause mortality rates from the ICRP Euro-American-Asian mixed population are shown with a dashed vertical line. The dashed (dotted) curve shows the density for a gamma (log-normal) distribution fitted to the histogram data.}
    \label{fig:WHO_allcauserates_5EAA_male_female}
\end{figure}

\begin{figure}[htbp]
    \centering
    \includegraphics[scale=0.0925]{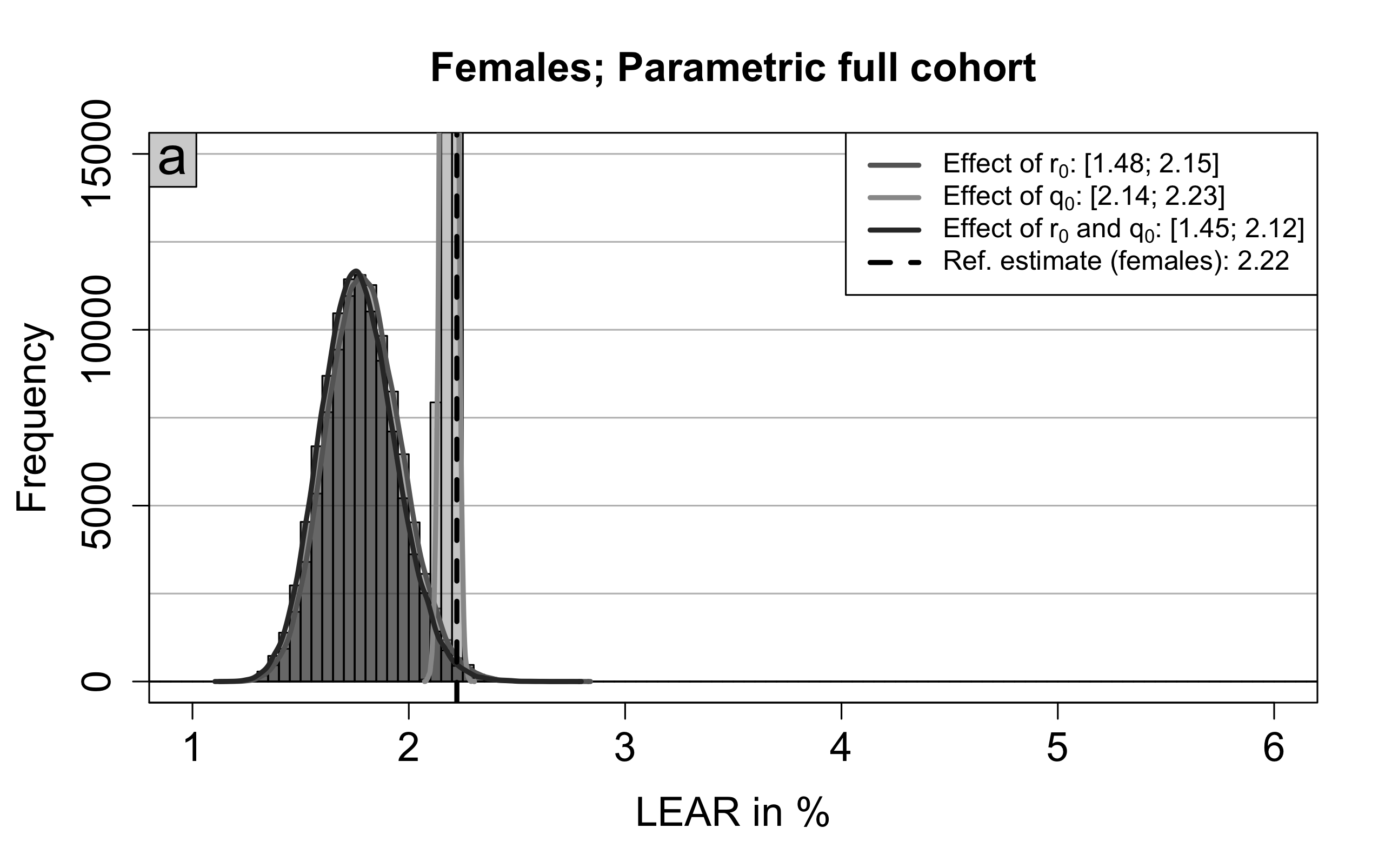}
    \includegraphics[scale=0.0925]{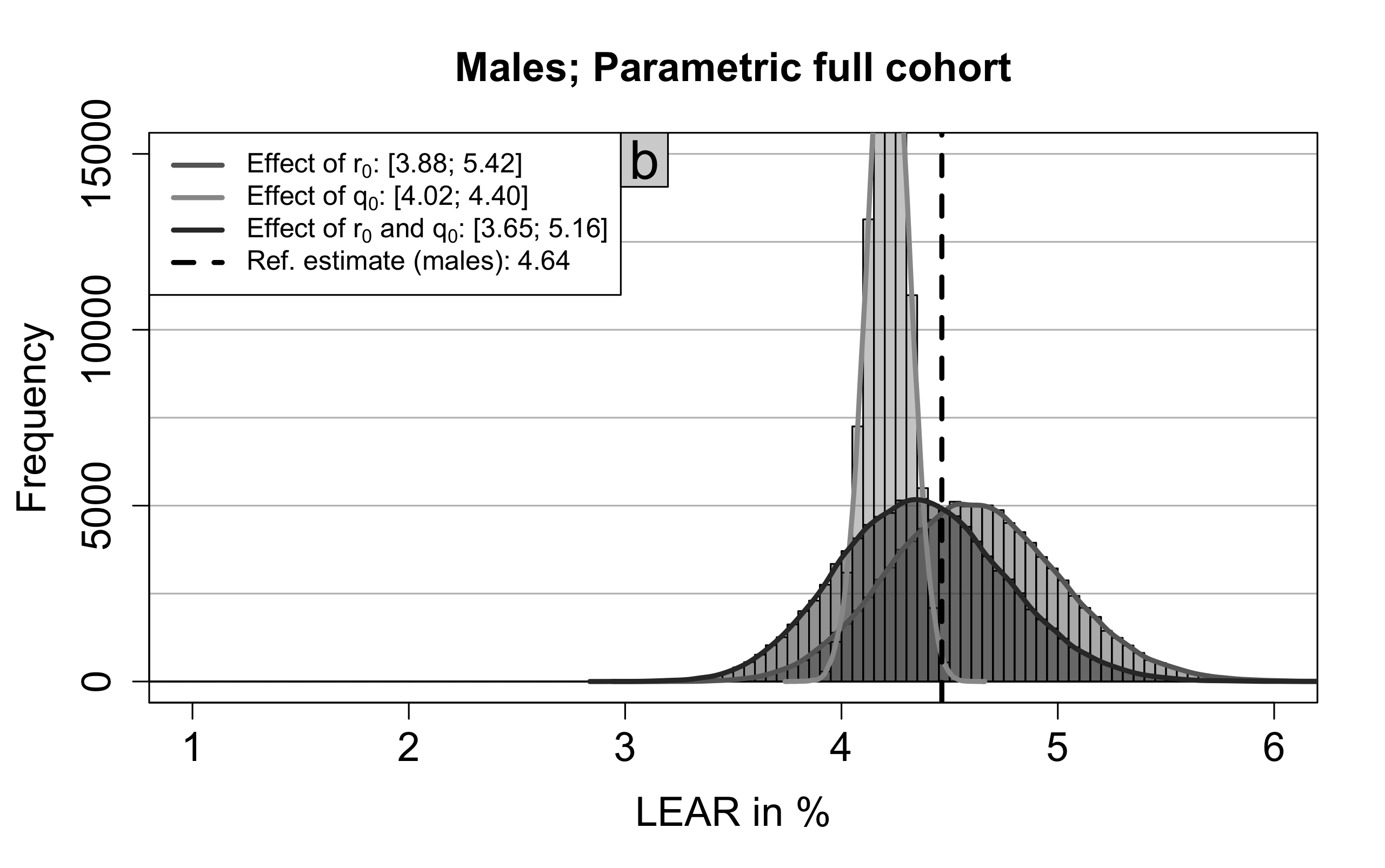}
    \includegraphics[scale=0.0925]{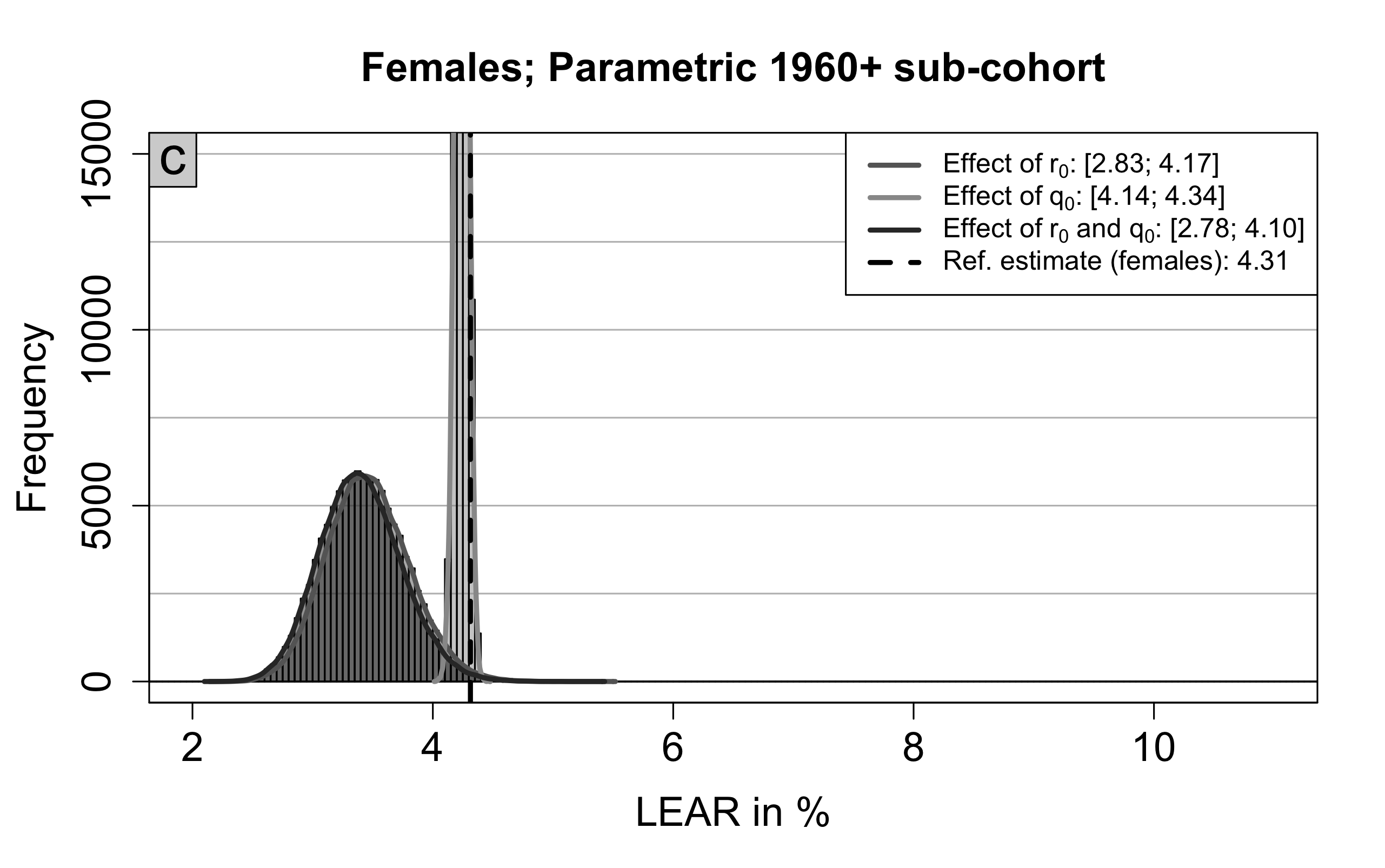}
    \includegraphics[scale=0.0925]{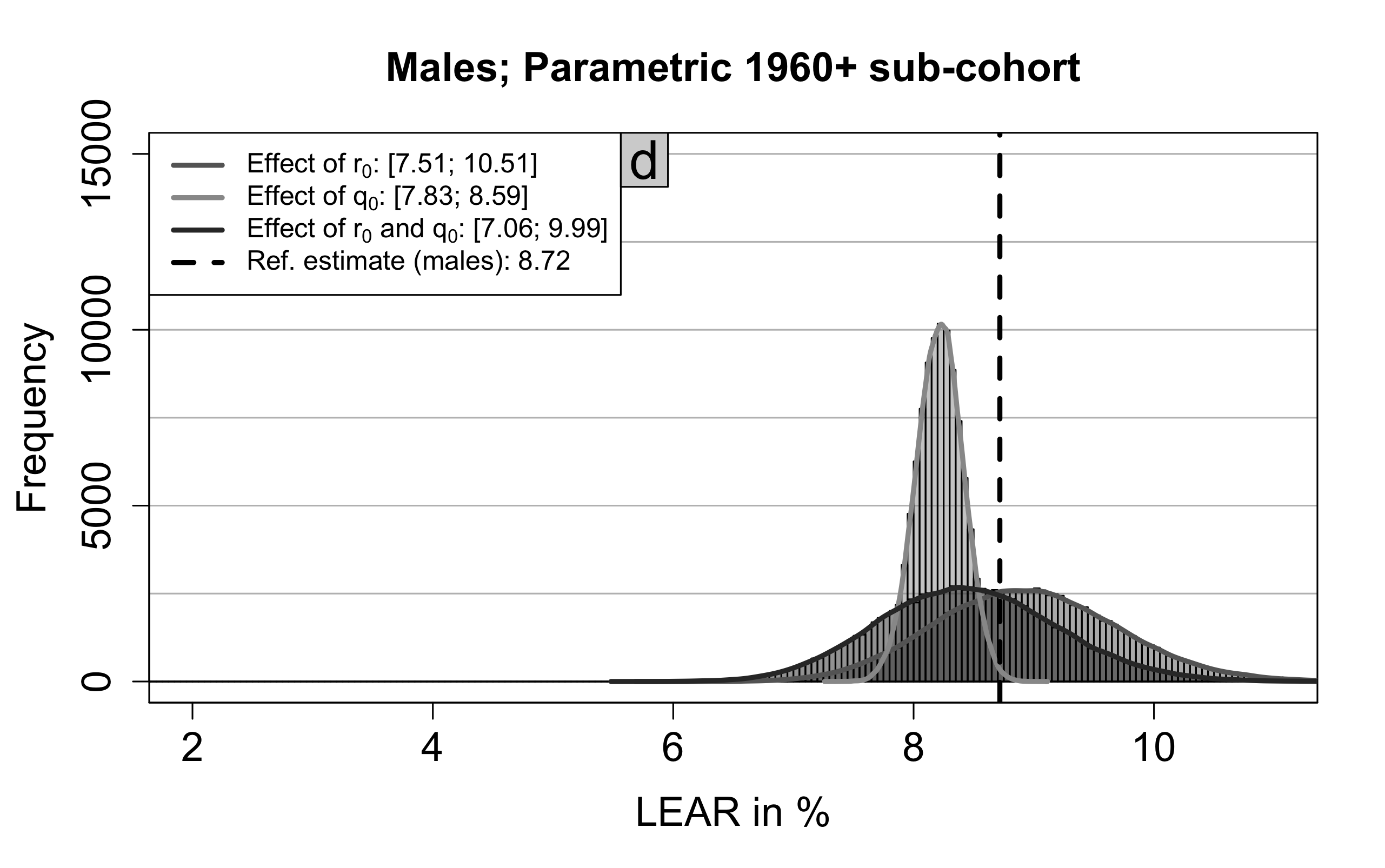}
    \includegraphics[scale=0.0925]{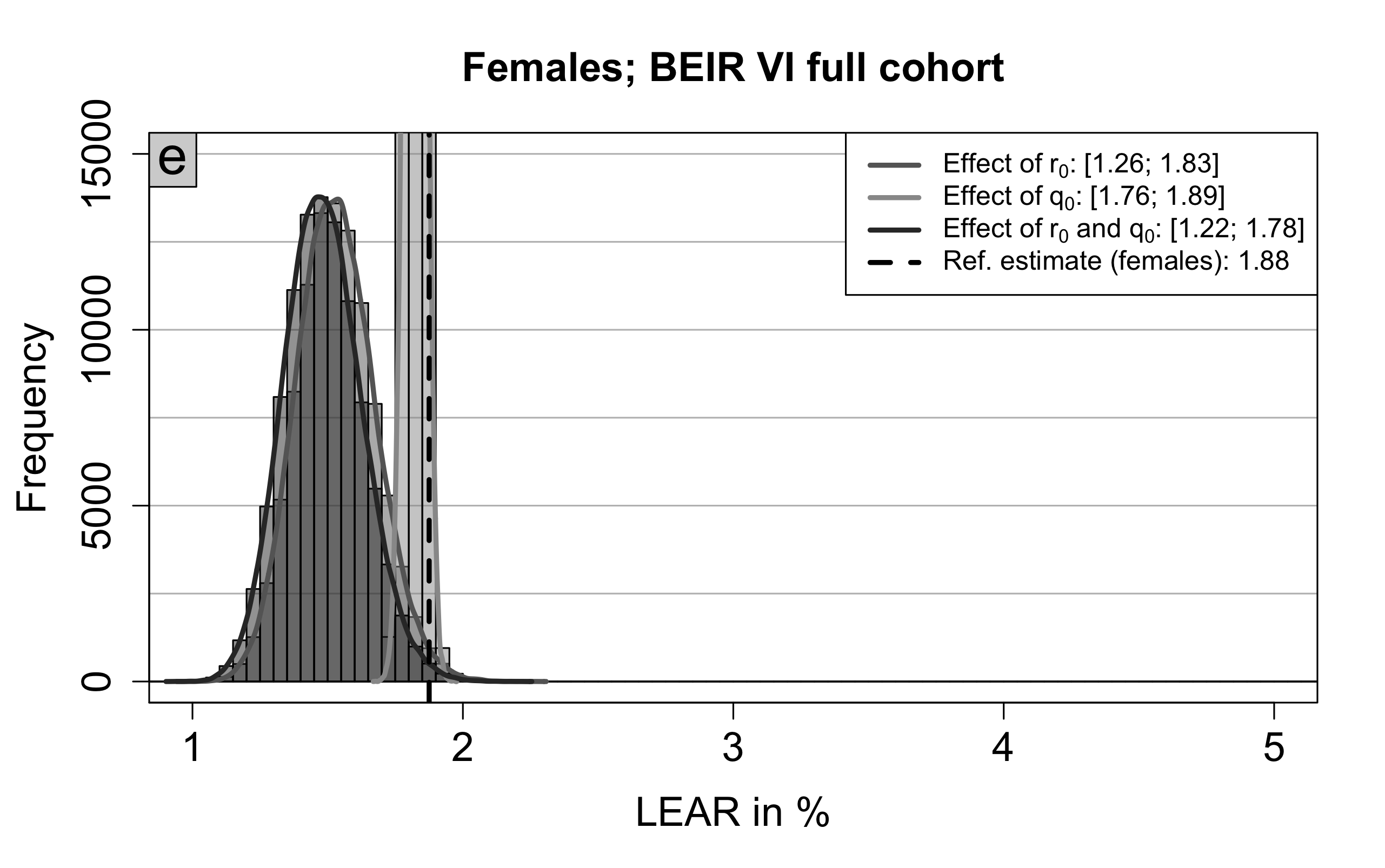}
    \includegraphics[scale=0.0925]{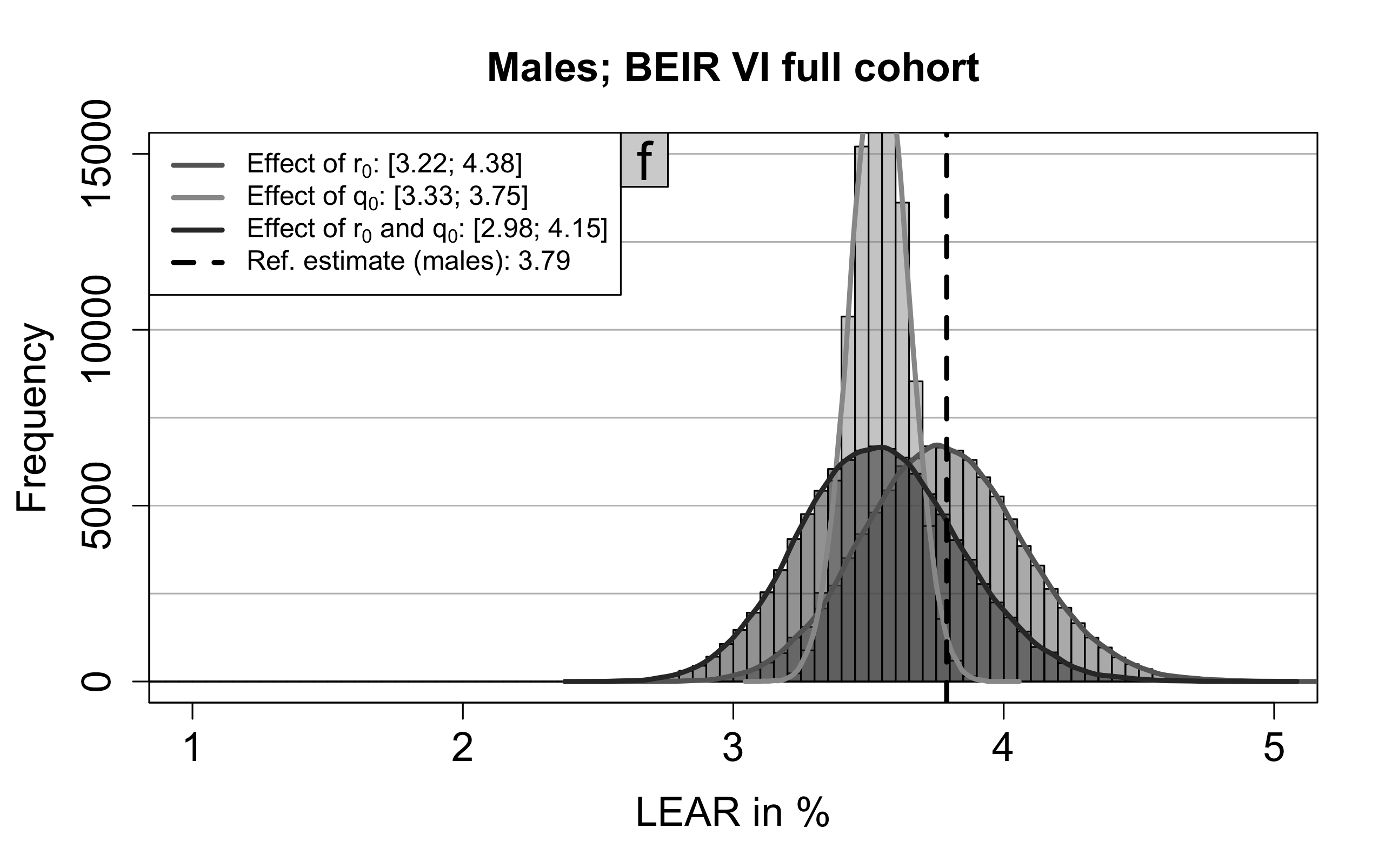}
    \includegraphics[scale=0.0925]{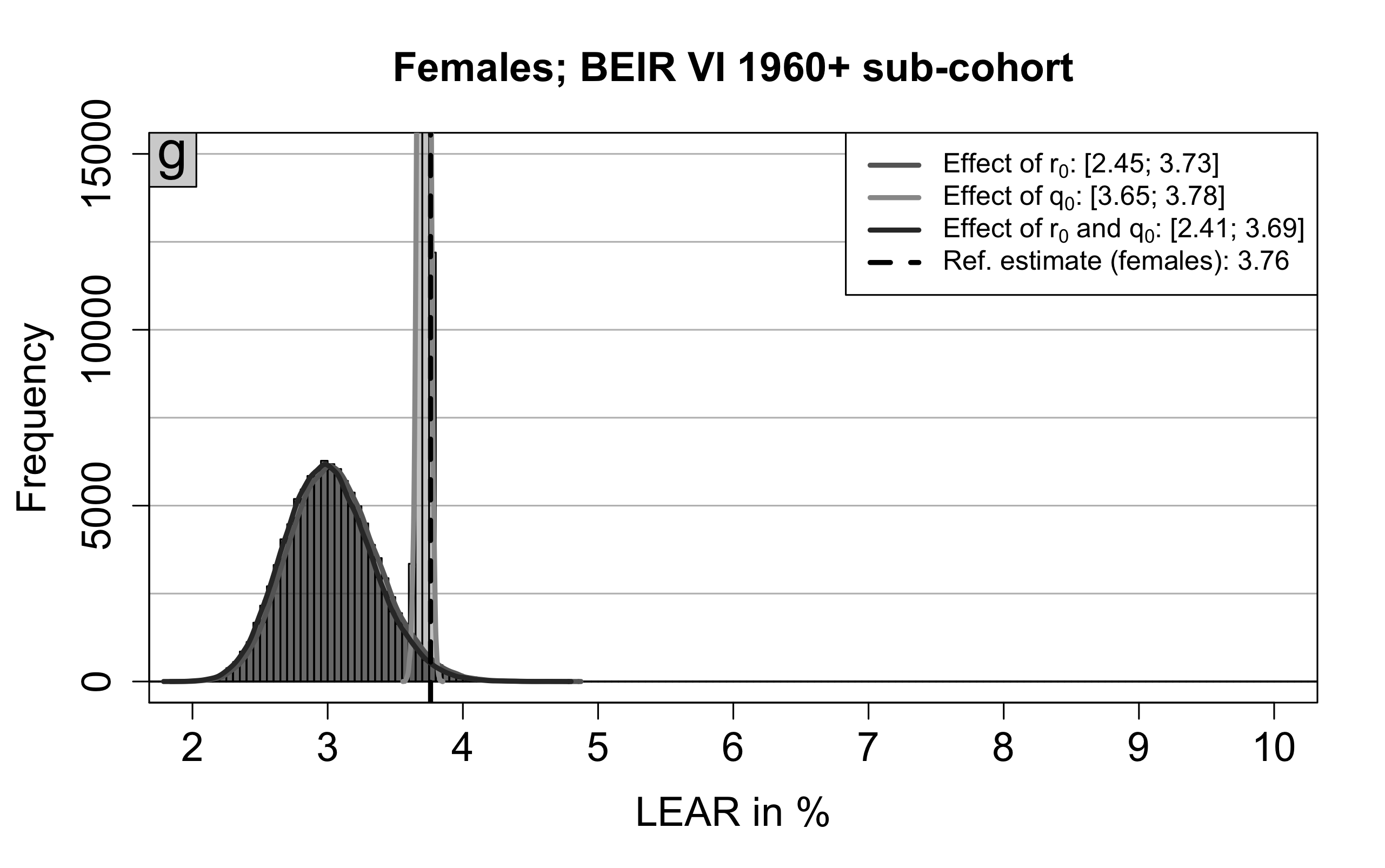}
    \includegraphics[scale=0.0925]{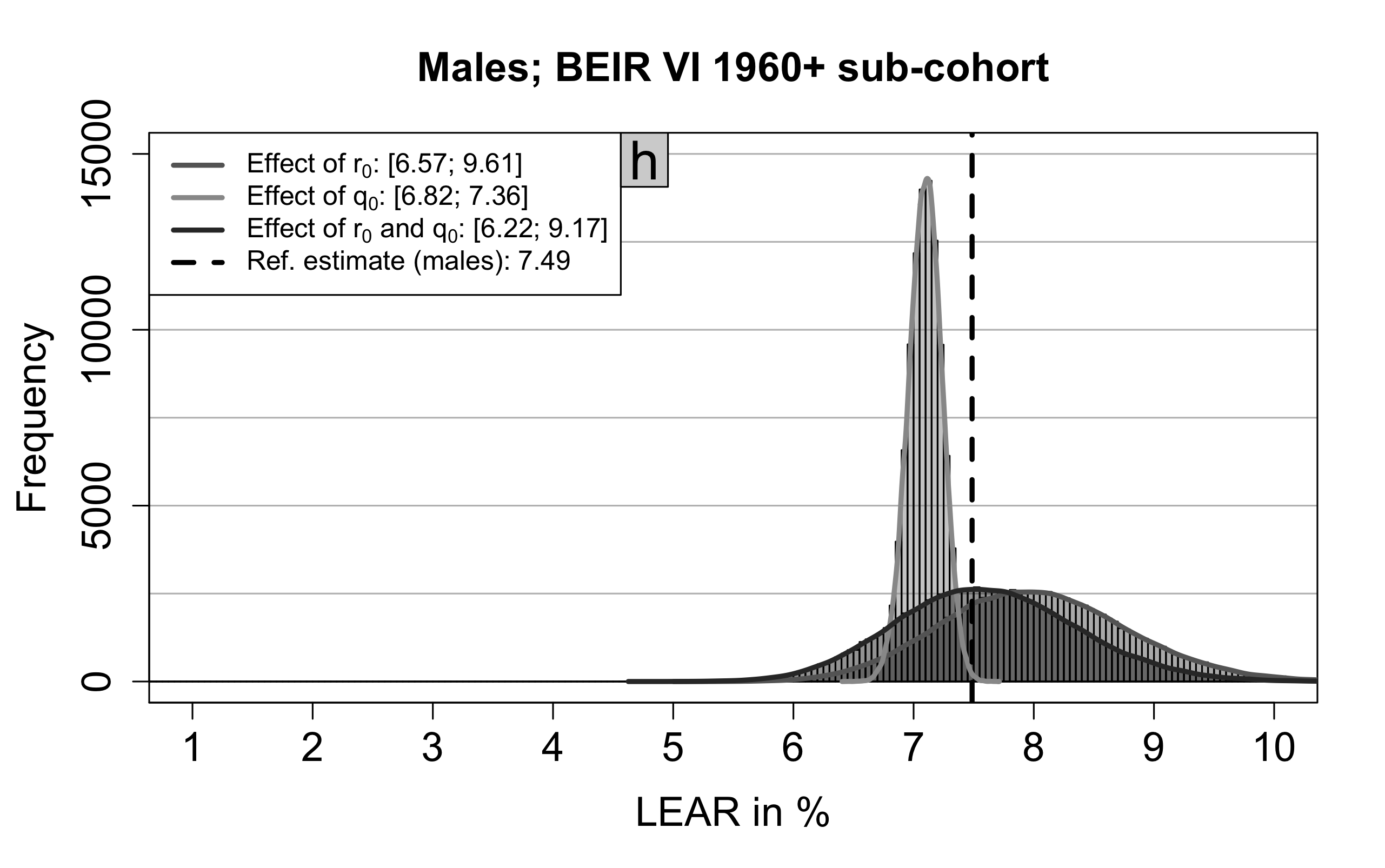}
    \caption{Histogram of the $100,000$ sampled $LEAR$ estimates with kernel density estimate (solid lines) for four risk models and varying uncertainty in sex-specific mortality rates by grayscale. Lung cancer mortality rates $r_0(t)$ and all-cause mortality rates $q_0(t)$ are assumed to follow a gamma distribution with parameter estimates derived from the histograms from Suppl. Figure \ref{fig:WHO_lungrates_5EAA_male_female} and Suppl. Figure \ref{fig:WHO_allcauserates_5EAA_male_female}, respectively. The joint effect results from independent sampling from both corresponding probability distributions. The 95\% uncertainty interval is presented in the legend.}
    \label{fig:LEAR_MR_Histogramm_male_female}
\end{figure}

Sex-specific mortality rates from the WHO data (Suppl. Figures \ref{fig:WHO_lungrates_5EAA_male_female} and \ref{fig:WHO_allcauserates_5EAA_male_female}) show higher mortality for males across all ages compared to females. Both gamma and log-normal distributions fit the data comparably well. We decided to use the gamma distribution for sex-specific mortality rate uncertainty in LEAR estimates (Suppl. Figure \ref{fig:LEAR_MR_Histogramm_male_female}) to be comparable to the main paper approach. \\

The heavy differences in mortality rates result in almost non-overlapping histograms between female- and male-specific $LEAR$ estimates. WHO data reveals lower female all-cause and lung cancer mortality rates compared to ICRP reference rates, resulting in $LEAR$ estimates skewed leftward for females (Suppl. Figure \ref{fig:LEAR_MR_Histogramm_male_female}). This left-skewing is even more pronounced for male all-cause mortality, though male lung cancer rates align with the male ICRP reference rates. \\

\begin{figure}[htbp]
    \centering
    \includegraphics[scale=0.0925]{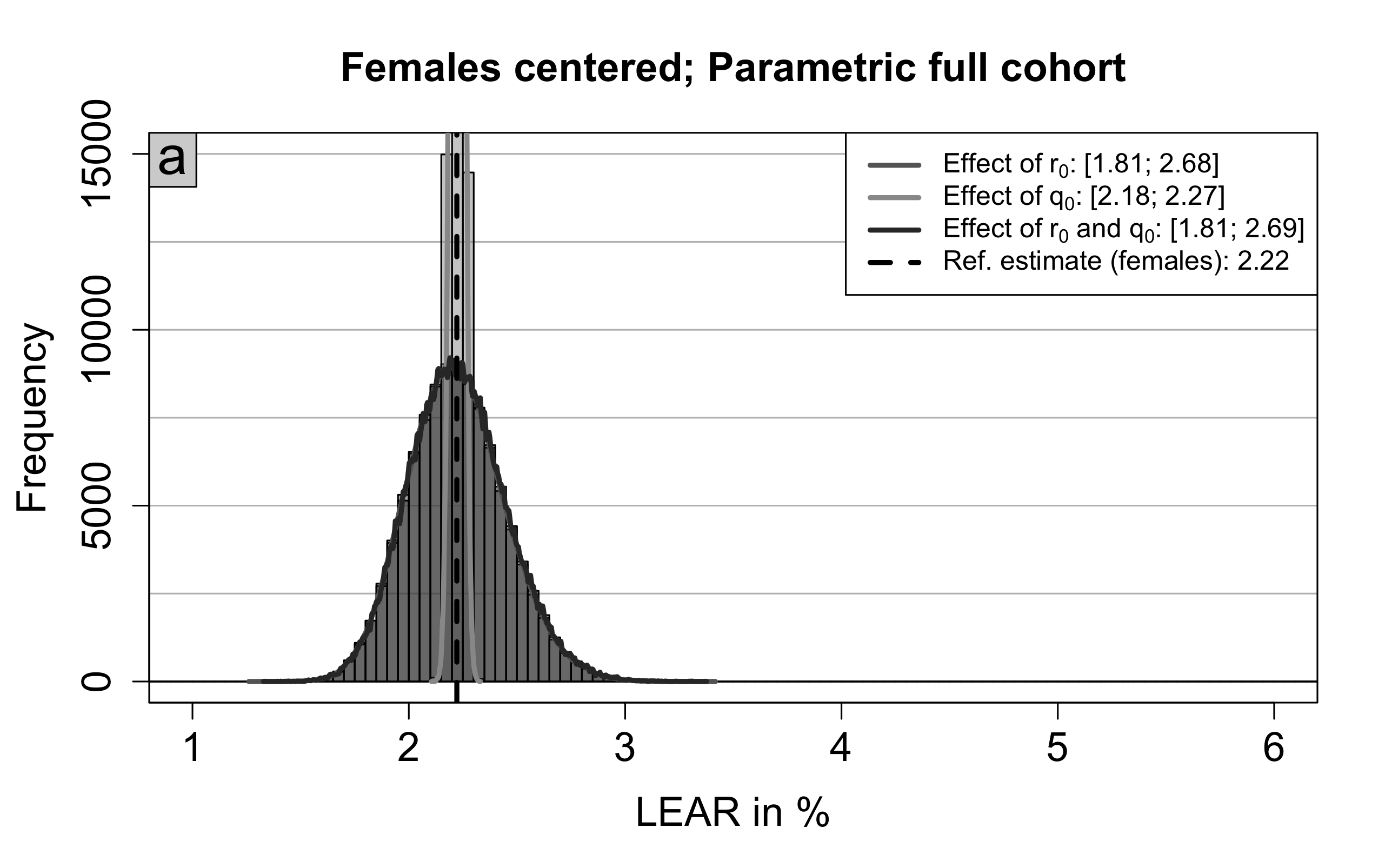}
    \includegraphics[scale=0.0925]{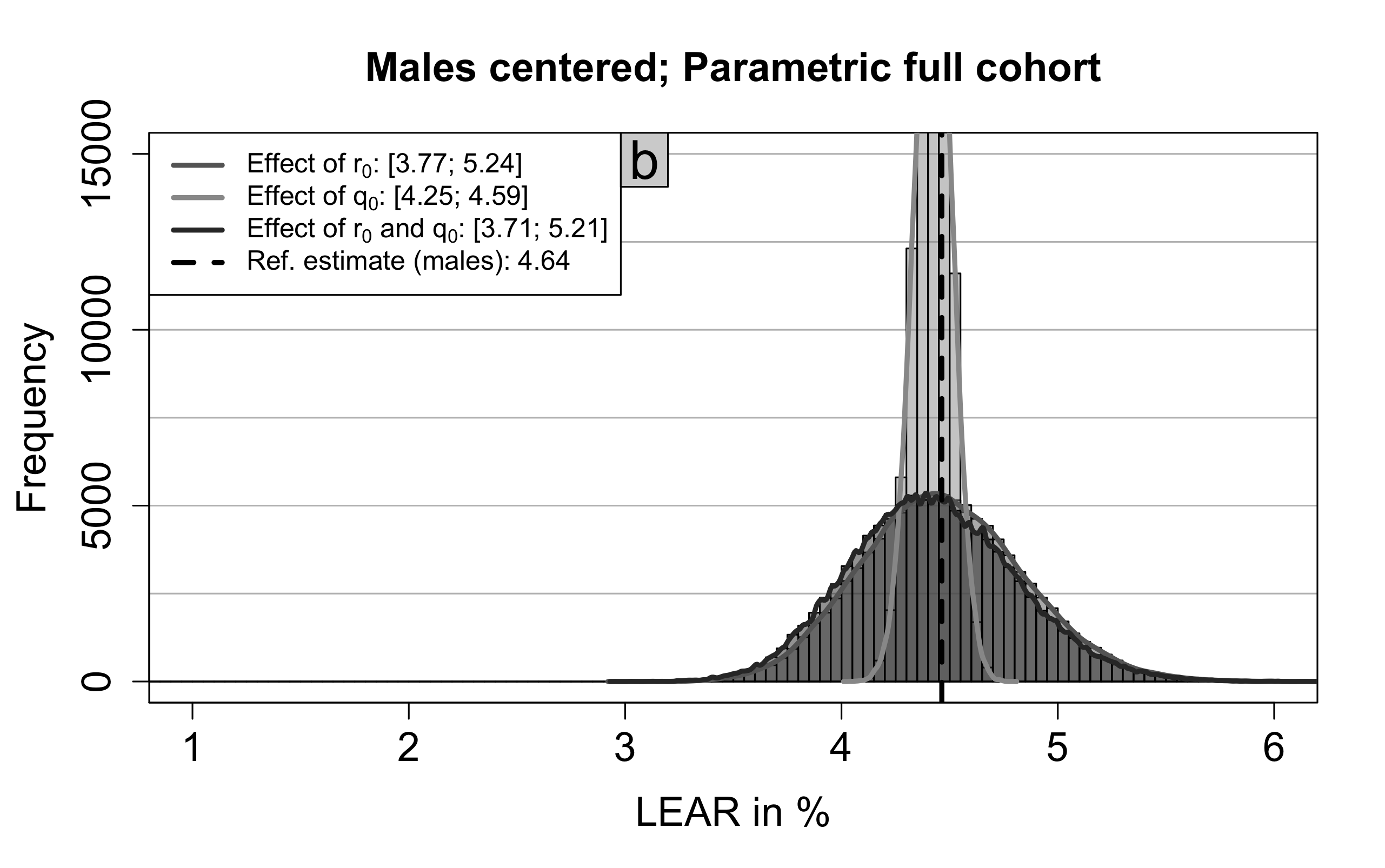}
    \includegraphics[scale=0.0925]{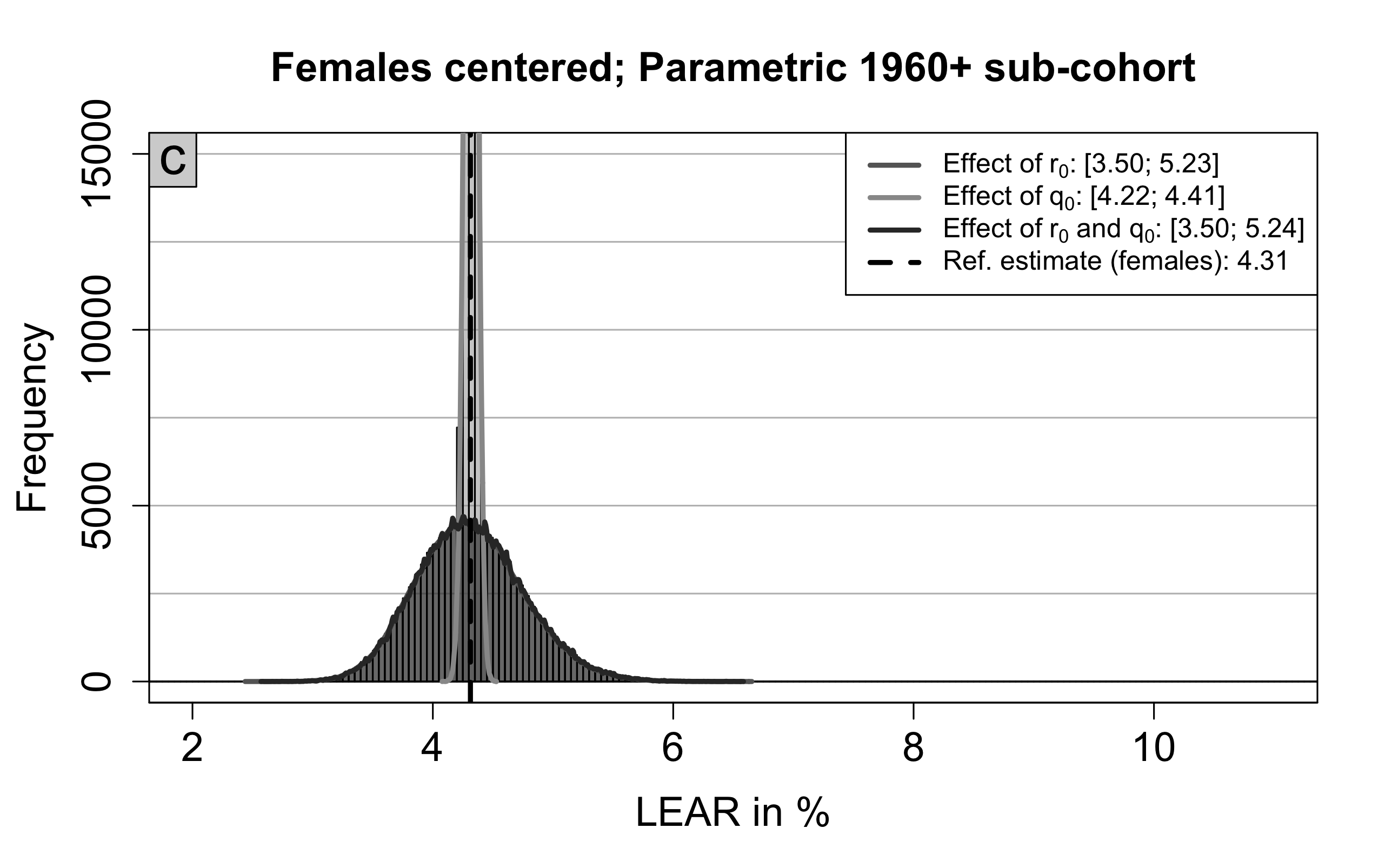}
    \includegraphics[scale=0.0925]{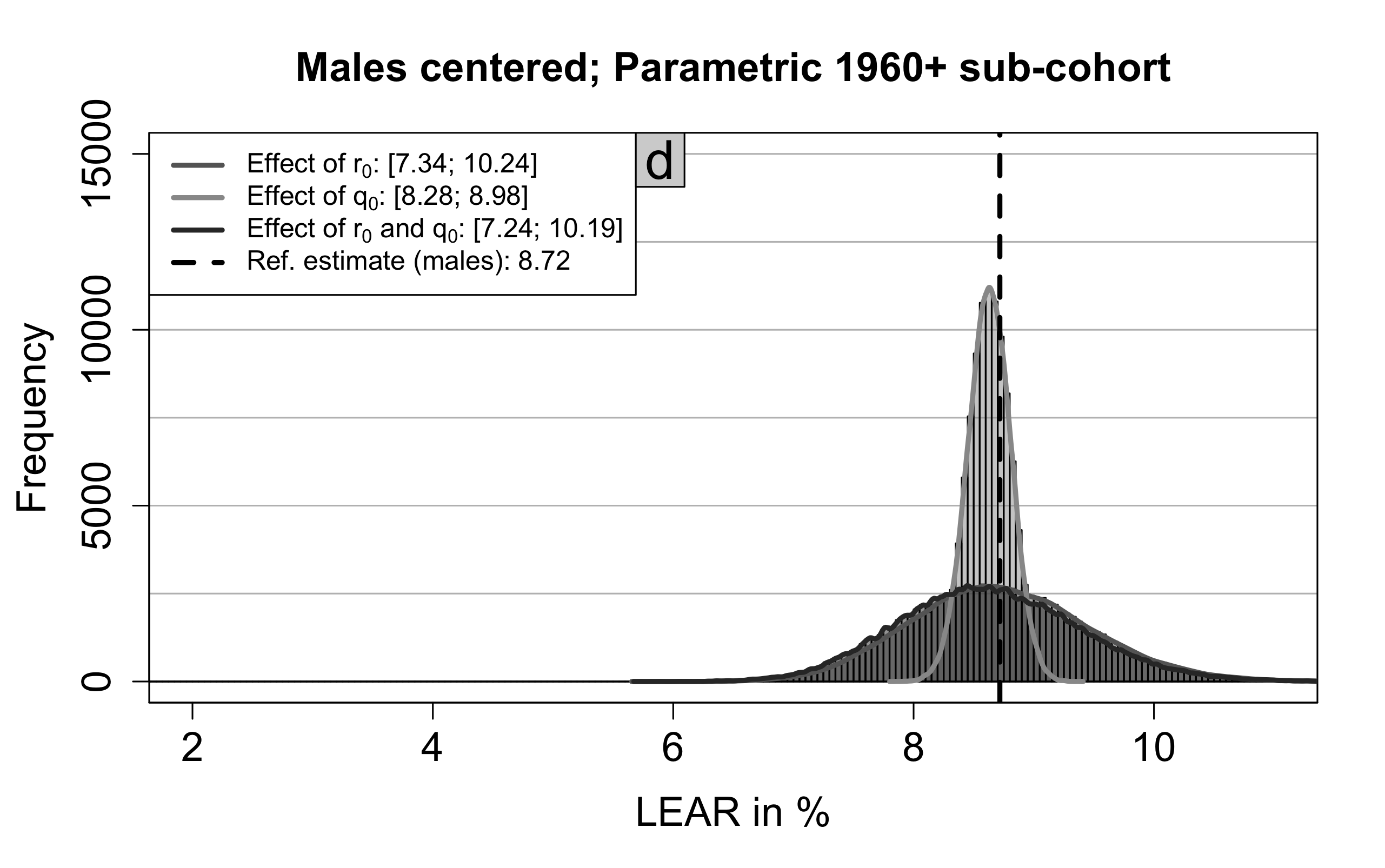}
    \includegraphics[scale=0.0925]{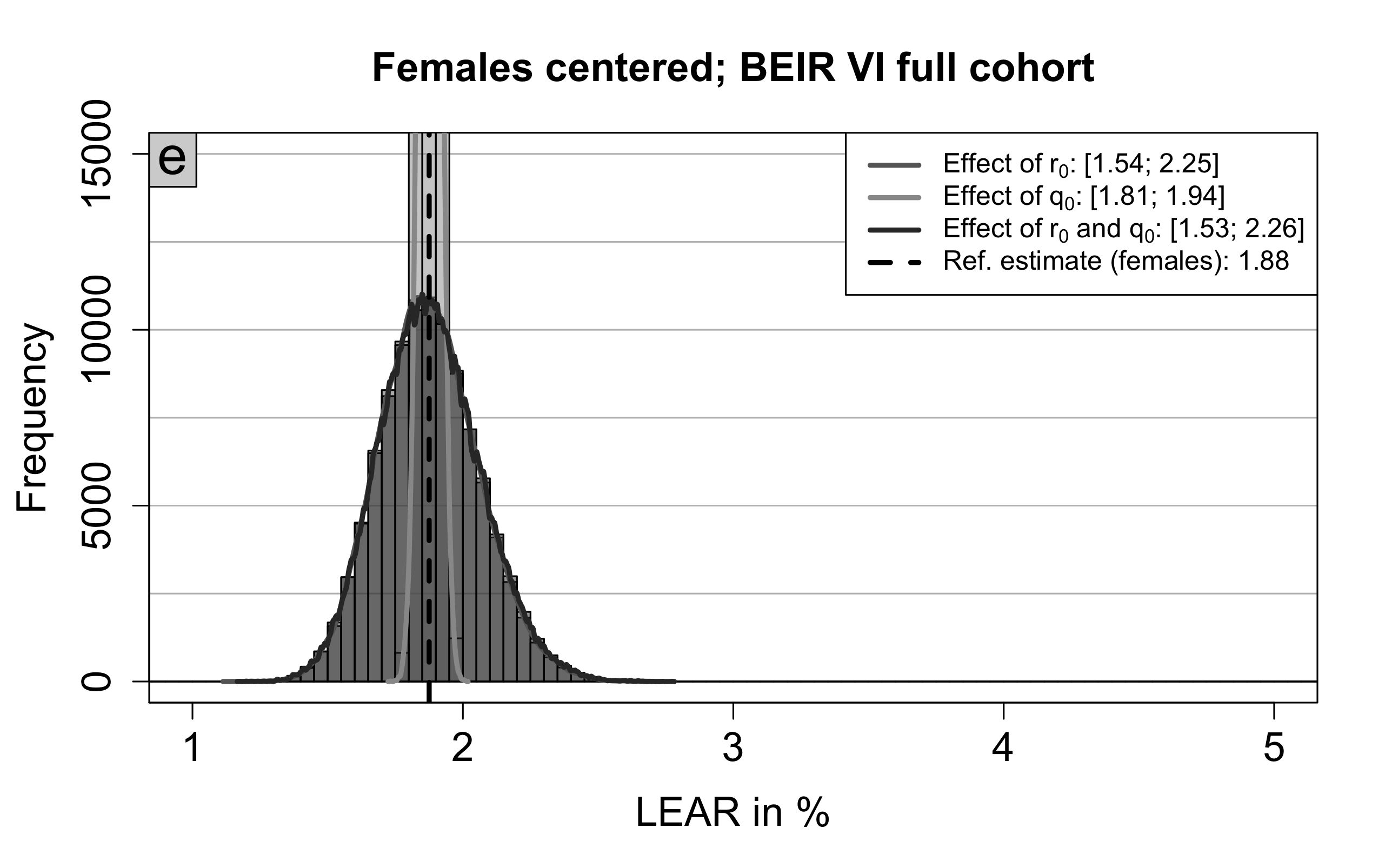}
    \includegraphics[scale=0.0925]{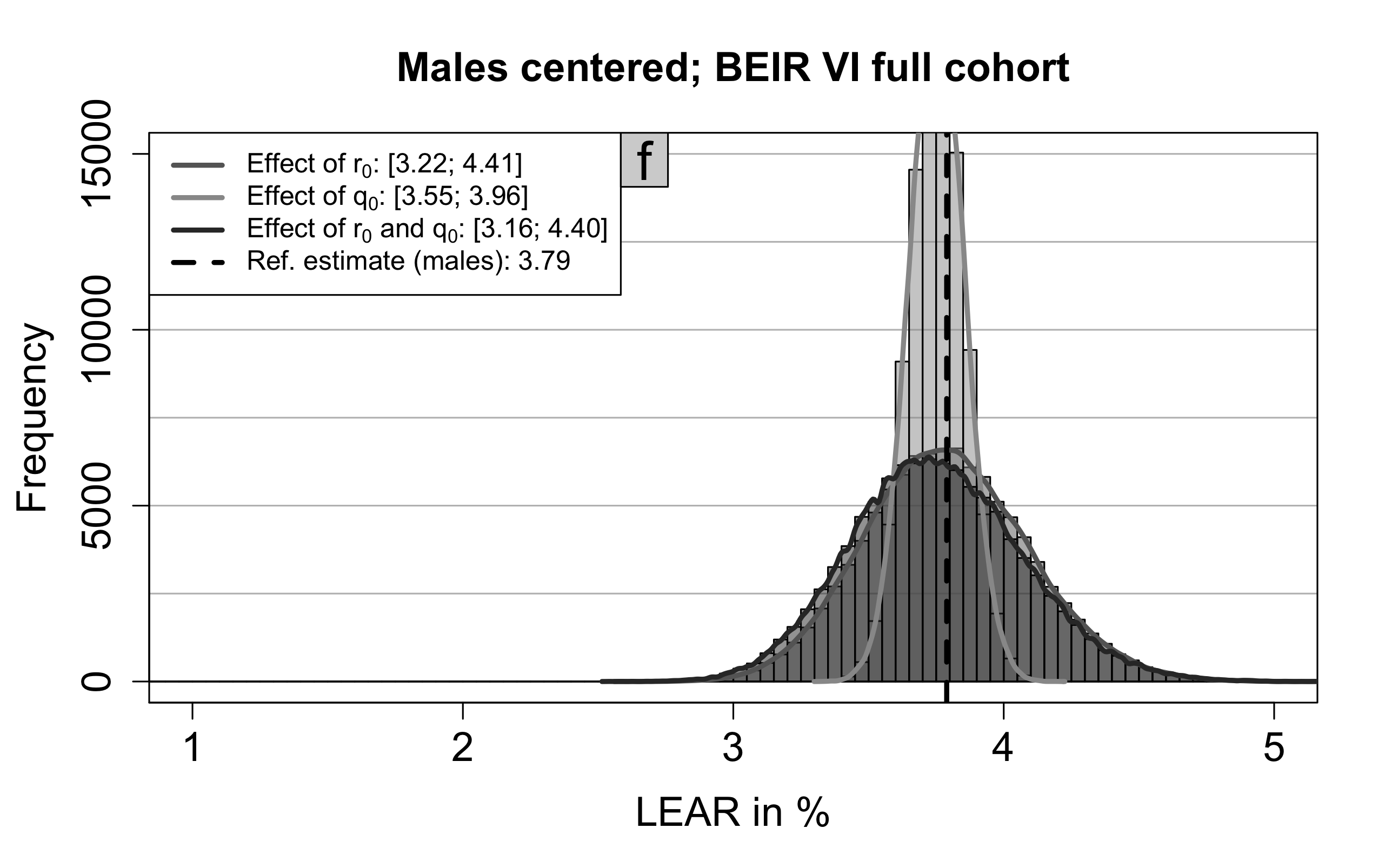}
    \includegraphics[scale=0.0925]{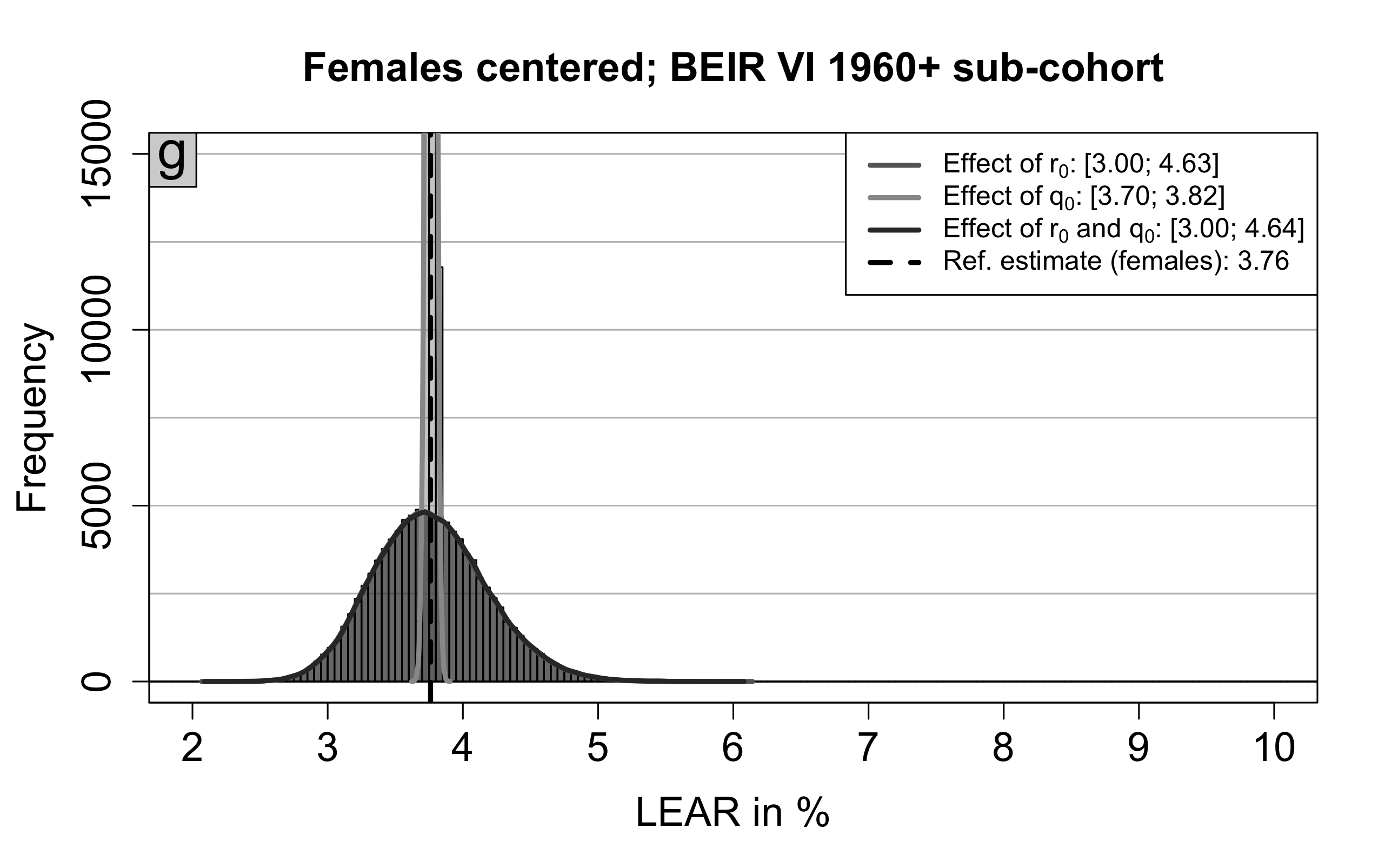}
    \includegraphics[scale=0.0925]{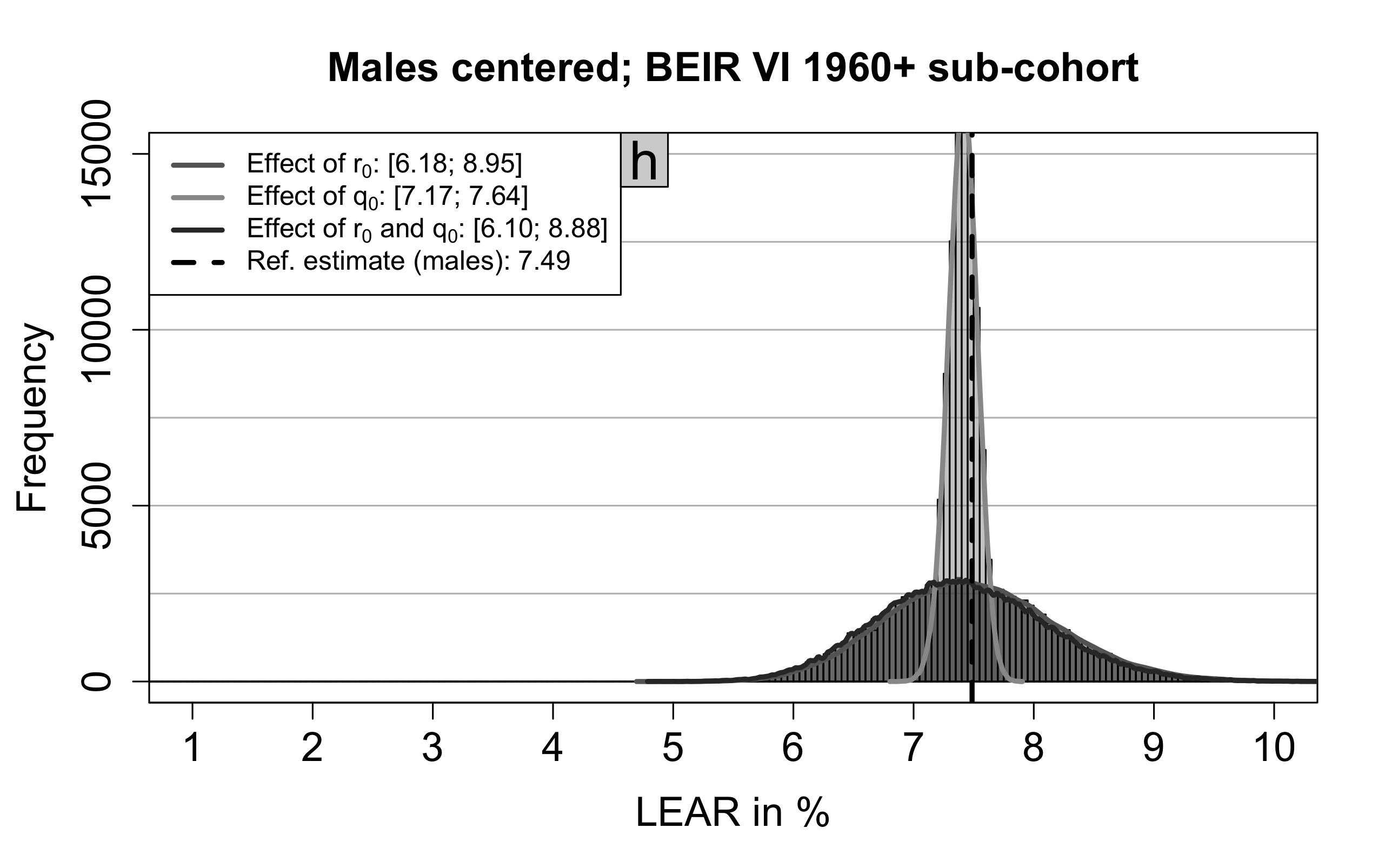}
    \caption{Histogram of the $100,000$ sampled $LEAR$ estimates with kernel density estimate (solid lines) for four risk models and varying uncertainty in sex-specific mortality rates by grayscale. Lung cancer mortality rates $r_0(t)$ and all-cause mortality rates $q_0(t)$ are assumed to follow a gamma distribution with parameter estimates derived from the histograms from Suppl. Figure \ref{fig:WHO_lungrates_5EAA_male_female} and Suppl. Figure \ref{fig:WHO_allcauserates_5EAA_male_female} centered such that the mean is equal to the corresponding ICRP reference rate, respectively. The joint effect results from independent sampling from both corresponding probability distributions. The 95\% uncertainty interval is presented in the legend.}
    \label{fig:LEAR_MR_Histogramm_male_female_centered}
\end{figure}

In contrast to the sex-averaged case, here large discrepancies between sample histogram means and ICRP rates motivate analysis of centered\footnote{The centering is carried out by setting the mean of the gamma-distributed mortality rate equal to the corresponding ICRP reference rate (compare Suppl. Section \ref{Mortality_rate_uncertainty_preliminaries_centered_SDC_D12}).} mortality rates (Suppl. Figure \ref{fig:LEAR_MR_Histogramm_male_female_centered}). However, centering minimally affects relative uncertainty spans. In particular, low all-cause rate variation results in a negligible variation of $LEAR$ estimates. Notably, male-specific uncertainty remains slightly larger than female-specific uncertainty. The relative uncertainty span is similar for lung cancer rates between sexes, but about twice as large for male all-cause rates. Overall, the combined effect of both mortality rate uncertainties remains comparable between females and males. \\

This analysis showed that male-specific ICRP reference rates align better with observed male rates from WHO data compared to corresponding female rates. The derived uncertainty intervals suggest statistically significant differences in sex-specific $LEAR$ estimates. However, the overall relative uncertainty spans of $LEAR$ (incorporating lung cancer and all-cause mortality rates) remains comparable for both sexes. Notably, female-specific estimates have limited interpretability due to risk transfer issues since our risk models from \cite{Kreuzer_2023_lc} are derived entirely from male miners cohort data.

\clearpage

\subsection{Risk model parameter uncertainty for sex-specific ICRP reference mortality rates}
\label{Sex_specific_uncertainty_risk_model_SDC_E2}
\begin{figure}[htbp]
    \centering
    \includegraphics[scale=0.0925]{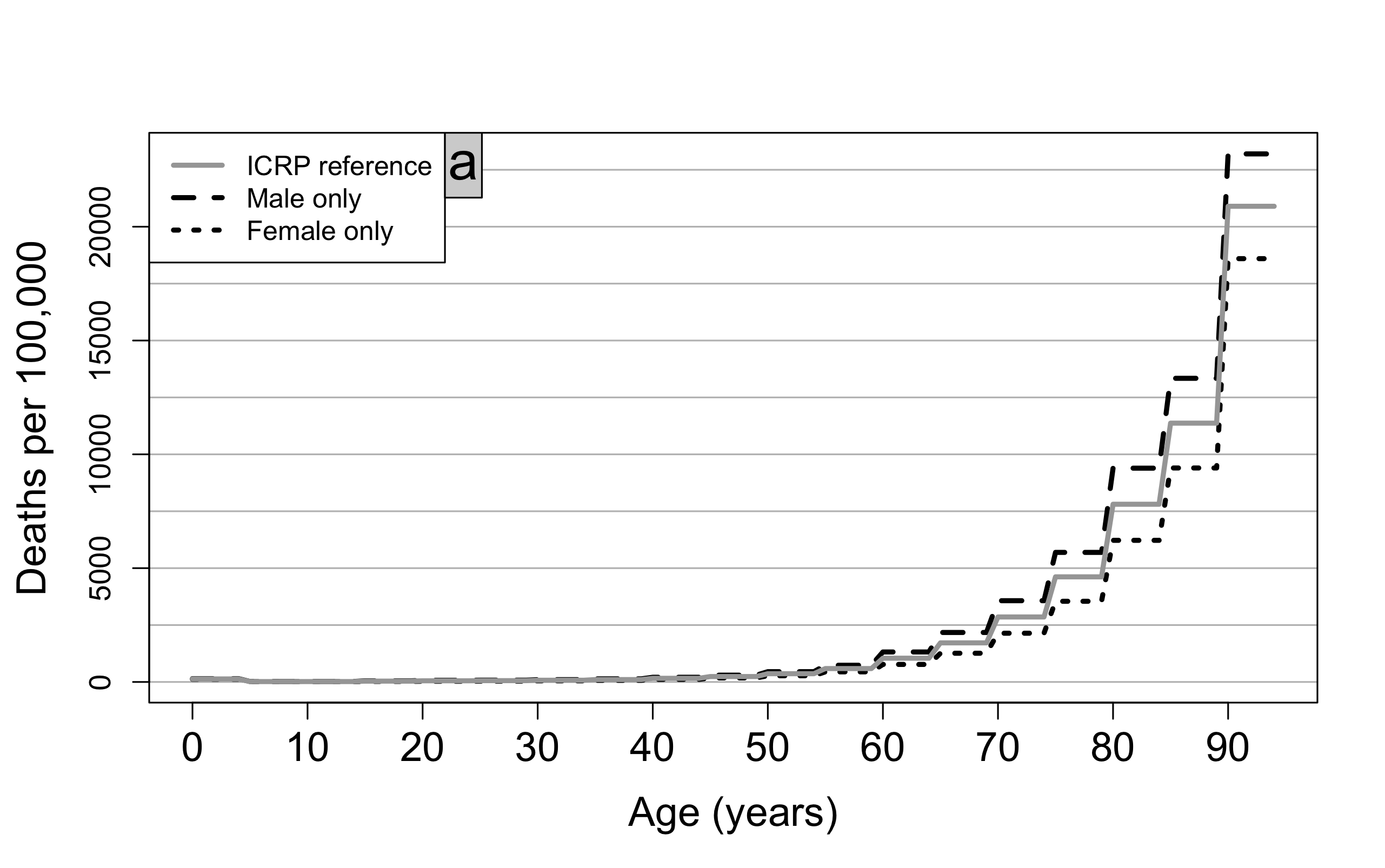}
    \includegraphics[scale=0.0925]{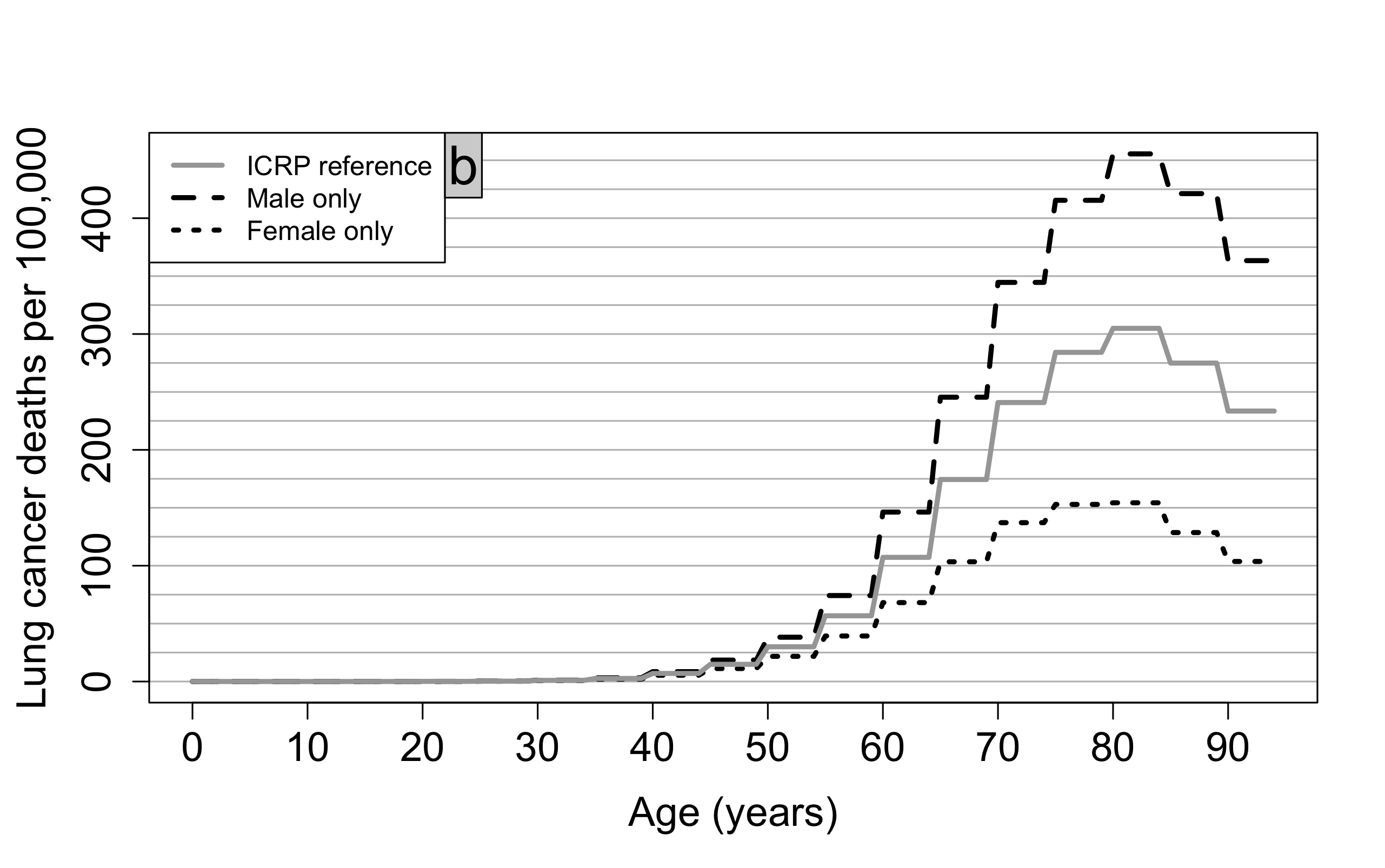}
    \caption{Deaths and lung cancer deaths per 100,000 persons by age (i.e. $q_0(t) \times 10^5, r_0(t) \times 10^5$ for all $t$) in the ICRP reference population \cite{ICRP103_2007} and specifically for males and females only.}
    \label{fig:icrp_reference_mortality_rates_sex_specific}
\end{figure}

ICRP reference mortality rates are sex-averaged (mean of male and female specific rates). We calculate male- and female-specific $LEAR$ estimates using the sex-specific mortality rates (Suppl. Figure \ref{fig:icrp_reference_mortality_rates_sex_specific}). Notably, male reference lung cancer rates are notably higher at older ages (important for $LEAR$). We explore how these differences in mortality rates impact $LEAR$ using the Bayesian approach for the simple linear 1960+ sub-cohort risk model and the parametric 1960+ sub-cohort model, but without introducing randomness on mortality rates themselves. Note that we do not apply the ANA approach here because as other results here have shown, it yields results very similar to those of the Bayesian approach with a uniform prior. \\

\subsubsection{Simple linear 1960+ sub-cohort risk model}

For the simple linear risk model structure $ERR(t; \beta)=\beta W(t)$, sampling is not necessary to assess differences in sex-specific risk model parameter uncertainties. It holds as in equation (\ref{LEAR_linearERR_ana}),
\begin{equation}
    LEAR \approx \sum_{t \geq 0} r_0(t) ERR(t; \beta ) \Tilde{S}(t) = \beta \sum_{ t \geq 0} r_0(t) W(t) e^{-\sum_{u=0}^{t-1}q_0(u)} = \beta \cdot  C,
\end{equation}
with $C=\sum_{t \geq 0} r_0(t) W(t) e^{-\sum_{u=0}^{t-1}q_0(u)}$. ICRP reference sex-mixed mortality rates and an exposure of 2 WLM from 18 to 64 years yields $C=4.27$. Sex-specific mortality rates give different values for $C$ with $C_M=5.49$ and $C_F=2.67$ for the male- and female-specific ICRP mortality rates, respectively. Results for $LEAR$ with the linear risk model are translated to a male or female ICRP reference population by multiplying the results with $C_M/C=1.29$ or $C_F/C=0.62$ for males or females, respectively. Hence, male-specific lifetime risk estimates are roughly twice as high as female-specific lifetime risk estimates due to higher male baseline mortality rates. This directly translates to sex-specific uncertainty intervals incorporating uncertain risk model parameters. However, by definition, the span of uncertainty intervals relative to the reference estimate (ICRP 103) remains identical between female- and male-specific $LEAR$ uncertainty intervals.  Note that this result does not depend on the specific parameter estimate $\hat{\beta}$.

\subsubsection{Parametric 1960+ sub-cohort risk model}

\begin{figure}[htbp]
    \centering
    \includegraphics[scale=0.0925]{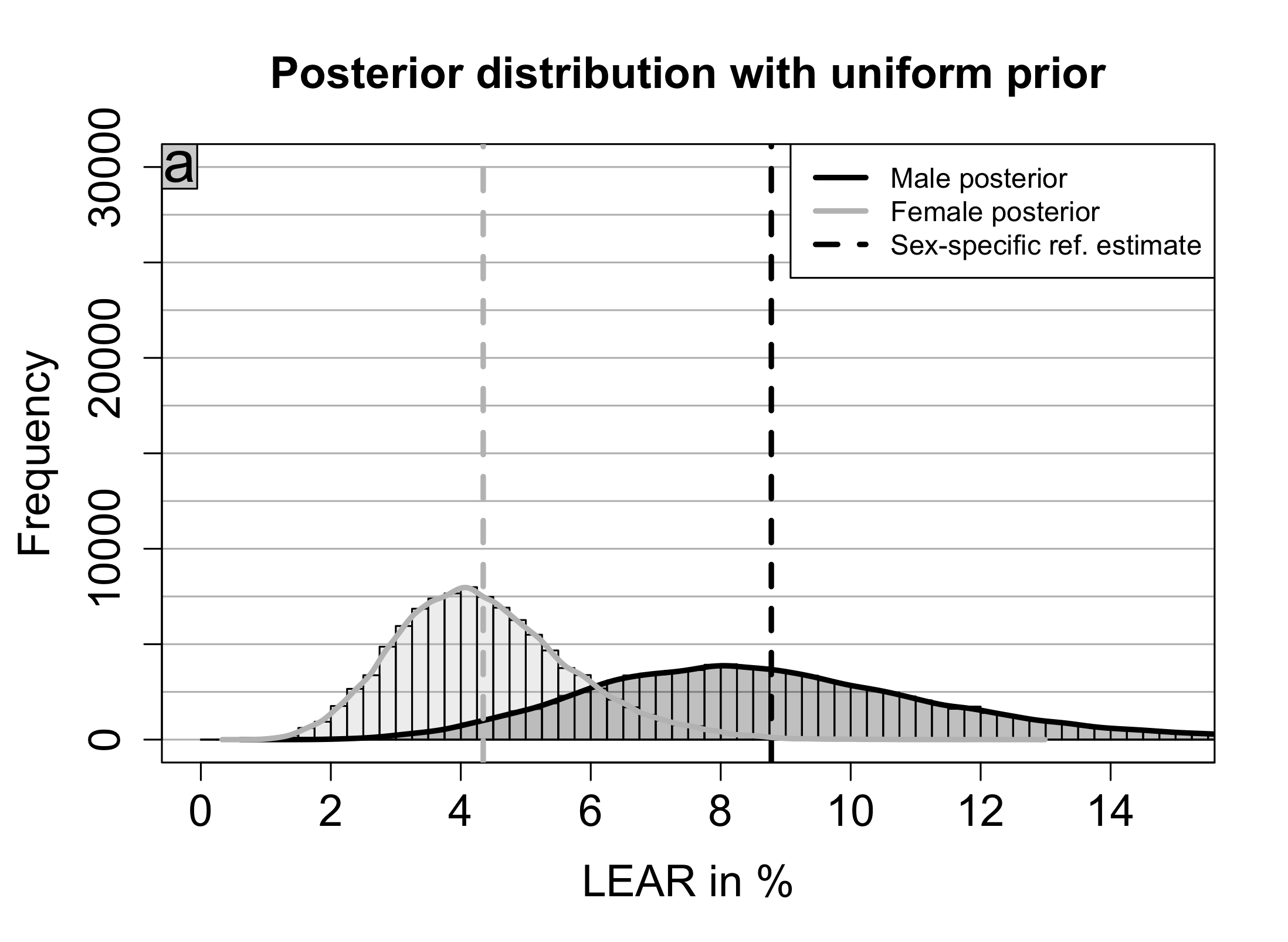}
    \includegraphics[scale=0.0925]{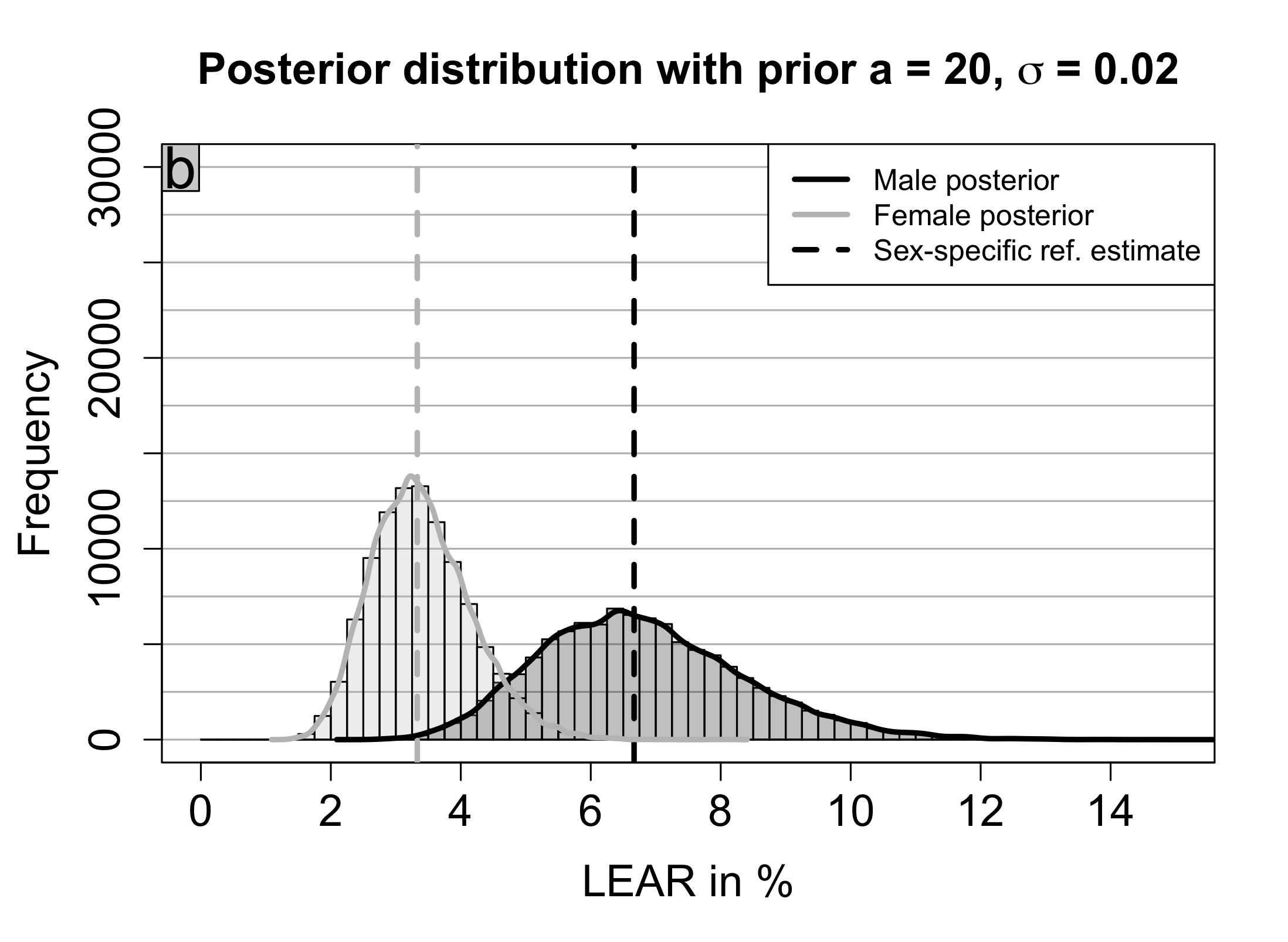}
    \includegraphics[scale=0.0925]{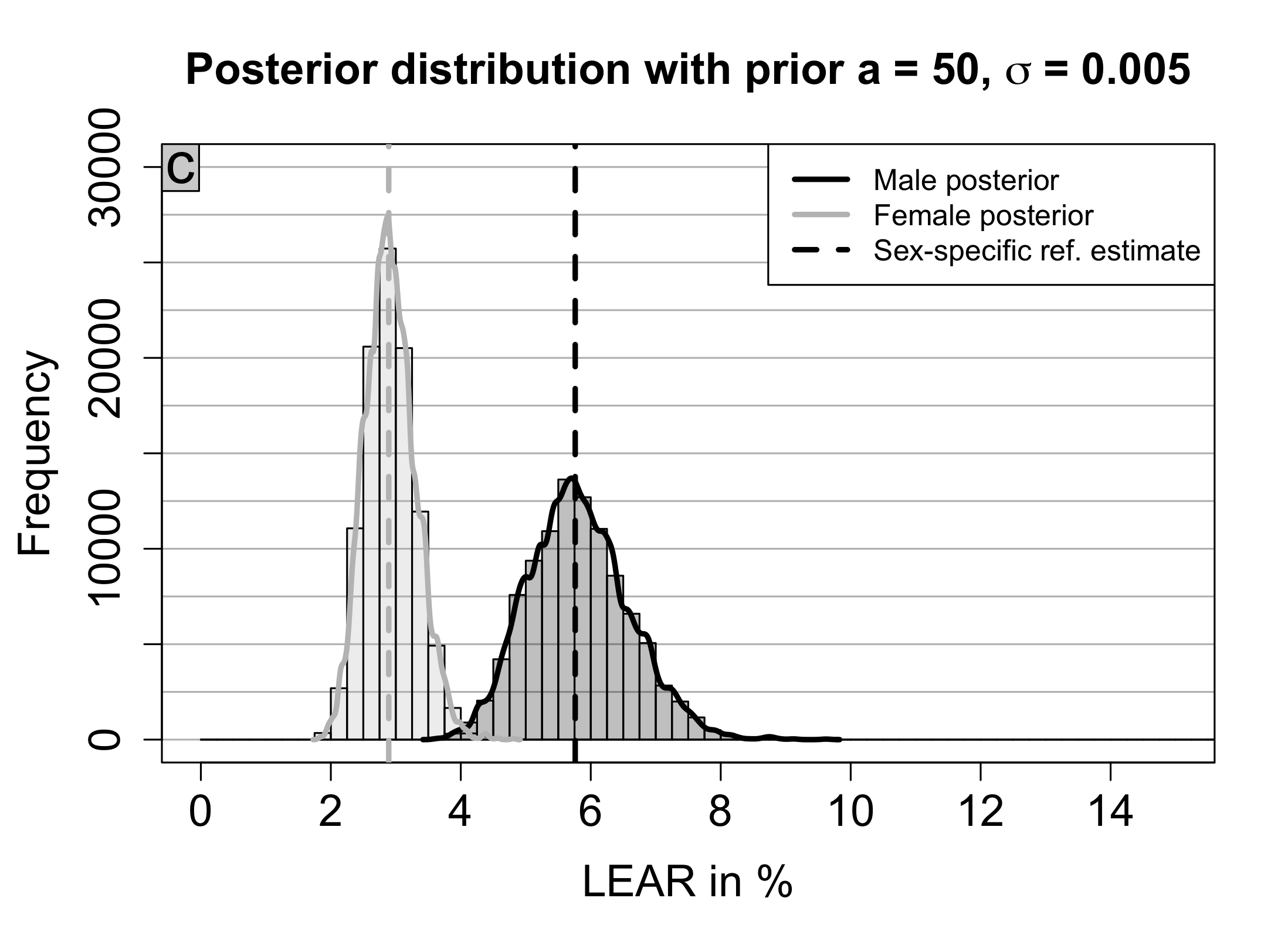}
    \caption{Distributions of $LEAR$ estimates using ICRP reference male and female mortality rates, respectively calculated from $100,000$ risk model parameter estimates drawn from the posterior distribution. The risk model parameter prior $P(\Theta)=P(\beta)P(\alpha)P(\varepsilon)$ is varied for different combinations of gamma-distributed $\beta$ and normally distributed $\alpha,\varepsilon$ for different shape parameters $a$ and standard deviations $\sigma$.}
    \label{fig:linlogERR_bayes_full_LEAR_sex_specific}
\end{figure}

For the parametric 1960+ sub-cohort risk model Markov Chain Monte Carlo (MCMC) methods are applied to obtain $N=100,000$ samples from the posterior distribution $P(\Theta \vert X)$ analogously to the main paper approach. The overall higher baseline mortality for men directly shows in the $LEAR$ estimates and corresponding 95\% HPDIs (Suppl. Table \ref{tab:linlogERR_bayes_ki_sex_specific}, Suppl. Figure \ref{fig:linlogERR_bayes_full_LEAR_sex_specific}). At very high prior certainty ($a=50,\sigma=0.005$) the 95\% HPDIs do not intersect, indicating a statistically significant difference between female- and male-specific $LEAR$ estimates. However, it depends on subjective reasoning on the prior information. Further, as seen for the simple linear risk model, the relative uncertainty span is very similar between female- and male-specific $LEAR$ uncertainty intervals.

\begin{table}[htbp]
\centering
 \begin{tabular}[h]{lll}
\addlinespace
\hline
\addlinespace
Prior information & Male $LEAR$ in \%  & Female $LEAR$ in \%  \\
\addlinespace
\hline
\addlinespace
Uniform prior & $8.78$ $[3.79; 14.40]$ $(1.21)$&$4.34$ $[1.92; 7.09]$ $(1.19)$\\
$a=20, \sigma = 0.02$ &  $6.67$ $[3.89; 9.87]$ $(0.90)$& $3.33$ $[1.99; 4.91]$ $(0.88)$\\
$a=50, \sigma = 0.005$ &  $5.76$ $[4.41; 7.42]$ $(0.52)$&$2.89$  $[2.19; 3.69]$ $(0.52)$\\
\addlinespace
\hline
\addlinespace
\end{tabular}
    \caption{Sex-specific $LEAR$ estimate with 95\% highest posterior density interval (HPDI) (relative uncertainty span in brackets) for underlying Bayesian random risk model parameters $\Theta=\left(\beta, \alpha, \varepsilon\right)$  with posterior distribution  $P\left(\Theta \vert X \right)$  for different values of prior gamma shape parameters $a$ and standard deviation $\sigma$. The male and female $LEAR$ in \% with the risk model derived from the Joint Czech+French cohort (prior information) is $5.59$ and $2.81$, respectively.}
    \label{tab:linlogERR_bayes_ki_sex_specific}
\end{table}

Overall, our analysis revealed that $LEAR$ estimates differ roughly by a factor of two between female- and male-specific mortality rates. However, the relative uncertainty span remained comparable across sexes. Acknowledging that current lifetime risk calculations in the literature employ sex-mixed reference mortality rates, shows that mortality rate uncertainty should not be underestimated. Although there was no clear difference in sex-specific (relative) uncertainty, supporting the assessment of sex-mixed mortality rate uncertainties, the inherent uncertainty associated with risk transfer uncertainty between excess relative risk terms and female-specific mortality rates remains. 

\section{Radon exposure uncertainty}
\label{Radon_exposure_uncertainty_SDC_F}

Although lifetime risks like $LEAR$ related to radon exposure are often used with a fixed exposure scenario (e.g. for dose conversion purposes $2$ WLM from age 18-64 years), we briefly explore how exposure uncertainty may affect lifetime risk estimates. This exploration aims to provide a more comprehensive assessment of how variability in calculation components impacts lifetime risk variability. A typical situation with exposure uncertainty arises for compensation claims, where an individual lifetime risk is estimated based on given exposure data. However, for compensation claims for radiation-induced cancer, the specialized software "ProZES" \cite{proZES_2020} is recommended and typically applied in Germany. \\

Inspired by results from \cite{Jablon_1971} that errors in the Japanese atomic bomb dosimetry are approximately log-normally distributed, we assume a log-normally distributed yearly exposure in WLM to explore its impact on lifetime risk estimates for different lifetime risk measures and different risk models. Note that also normal distributions are viable options \cite{2020_Walsh, Hafner_2021}. However, this is rather an explorative approach and results are interpreted with care. \\

The reference case is the exposure scenario of 2 WLM from age 18-64 years. Let $w(t)$ be the exposure in WLM at age $t$ and $W(t)=\sum_{u=0}^t w(u)$ be the corresponding cumulative exposure in WLM at age $t$. We assume for every age $t$,
\begin{equation}
    w(t) \sim \mathcal{LN}\left( \log(2)-\frac{\sigma^2}{2}, \sigma^2 \right) \label{ln_w_random}
\end{equation} to guarantee that the mean of the distribution is at $2$ WLM. The standard deviation $\sigma$ governs the uncertainty and 95\% uncertainty intervals are obtained as the span of observed estimates via simulations by calculating 10,000 estimates and discarding the 500 (2.5\%) highest and lowest values (Suppl. Table \ref{tab:expo_uncertainty_s1_s05_s01}). For each lifetime risk estimate, $w(t)$ is sampled from the log-normal distribution (\ref{ln_w_random}) for all ages $18 \leq t \leq 64$. 

\begin{table}[htbp]
    \centering
    \centering\resizebox{\columnwidth}{!}{ \begin{tabular}[h]{lllll}
        \toprule
       Risk model & $ELR$ in \%  & $REID$ in \% & $LEAR$ in \% &$RADS$ in \%  \\
       \midrule
        \multicolumn{5}{c}{\textbf{Exposure variability $\sigma=1$}} \\
        \midrule
Parametric full cohort &  $3.24$ $[2.24; 4.78]$ $(0.78) $&$3.37$  $[2.33; 4.97]$ $(0.78)$& $3.43$ $[2.36; 5.11]$ $(0.80)$&$4.65$ $[3.21; 6.88]$ $(0.79)$\\
Parametric 1960+ sub-cohort &  $6.21$ $[4.08; 9.94]$ $(0.94)$& $6.45$ $[4.24; 10.32]$ $(0.94)$& $6.70$ $[4.34; 10.98]$ $(0.99)$&$8.94$ $[5.82; 14.39]$ $(0.96)$\\
BEIR VI full cohort &  $2.80$ $[1.95; 4.06]$ $(0.75)$& $2.89$  $[2.02; 4.20]$ $(0.75)$& $2.95$ $[2.04; 4.31]$ $(0.77)$&$4.85$ $[3.39; 7.02]$ $(0.75)$\\
BEIR VI 1960+ sub-cohort &  $5.34$ $[3.70; 7.70]$ $(0.75)$&$5.56$  $[3.86; 8.03]$ $(0.75)$ & $5.74$ $[3.94; 8.40]$ $(0.78)$&$7.11$ $[4.95; 10.24]$ $(0.74)$\\
Simple linear 1960+ sub-cohort& $5.35$ $[3.75; 7.74]$ $(0.75)$ & $5.52$ $[3.87; 7.98]$ $(0.74)$& $5.72$ $[3.97; 8.43]$ $(0.78)$&$9.98$ $[7.04; 14.33]$ $(0.73)$\\
        \midrule
        \multicolumn{5}{c}{\textbf{Exposure variability $\sigma=0.5$}} \\
        \midrule
Parametric full cohort & $[2.77; 3.78]$ $(0.31)$& $[2.88; 3.93]$ $(0.31)$& $[2.92; 4.02]$ $(0.32)$& $[3.98; 5.43]$ $(0.31)$\\
Parametric 1960+ sub-cohort &  $[5.17; 7.45]$ $(0.37)$& $[5.38; 7.74]$ $(0.37)$& $[5.54; 8.10]$ $(0.38)$& $[7.42; 10.77]$ $(0.37)$\\
BEIR VI full cohort &  $[2.41; 3.24]$ $(0.30)$& $[2.49; 3.35]$ $(0.30)$& $[2.53; 3.42]$ $(0.30)$& $[4.18; 5.62]$ $(0.30)$\\
BEIR VI 1960+ sub-cohort & $[4.58; 6.20]$ $(0.30)$& $[4.77; 6.46]$ $(0.30)$& $[4.90; 6.69]$ $(0.31)$& $[6.12; 8.24]$ $(0.30)$\\
Simple linear 1960+ sub-cohort&   $[4.61; 6.18]$ $(0.29)$& $[4.75; 6.37]$ $(0.29)$& $[4.91; 6.65]$ $(0.30)$& $[8.63; 11.49]$ $(0.29)$\\
        \midrule
  \multicolumn{5}{c}{\textbf{Exposure variability $\sigma=0.1$}} \\
        \midrule
Parametric full cohort & $[3.15; 3.34]$ $(0.06)$& $[3.27; 3.47]$ $(0.06)$& $[3.33; 3.54]$ $(0.06)$& $[4.52; 4.79]$ $(0.06)$\\
Parametric 1960+ sub-cohort&  $[6.00; 6.43]$ $(0.07)$& $[6.23; 6.68]$ $(0.07)$& $[6.46; 6.94]$ $(0.07)$& $[8.63; 9.26]$ $(0.07)$\\
BEIR VI full cohort &  $[2.72; 2.88]$ $(0.06)$& $[2.81; 2.97]$ $(0.06)$& $[2.86; 3.03]$ $(0.06)$& $[4.72; 4.99]$ $(0.06)$\\
BEIR VI 1960+ sub-cohort&  $[5.19; 5.49]$ $(0.06)$& $[5.41; 5.72]$ $(0.06)$& $[5.57; 5.91]$ $(0.06)$& $[6.92; 7.32]$ $(0.06)$\\
Simple linear 1960+ sub-cohort& $[5.21; 5.50]$ $(0.05)$& $[5.37; 5.67]$ $(0.05)$& $[5.56; 5.89]$ $(0.06)$& $[9.72; 10.25]$ $(0.05)$\\
        \bottomrule
    \end{tabular}}
    \caption{Lifetime risk estimates with 95\% uncertainty intervals  (relative uncertainty span in brackets) derived from $10,000$ sampled estimates for log-normally distributed yearly exposure in WLM (assumption (\ref{ln_w_random})), with corresponding reference estimates for different risk models, lifetime measures, and exposure variability $\sigma$. Reference estimates are only stated for $\sigma=1$ as they do not change with varying $\sigma$.}
    \label{tab:expo_uncertainty_s1_s05_s01}
\end{table}

\begin{figure}[htbp]
\centering
\begin{subfigure}[t]{0.48\textwidth}
    \includegraphics[width=\textwidth]{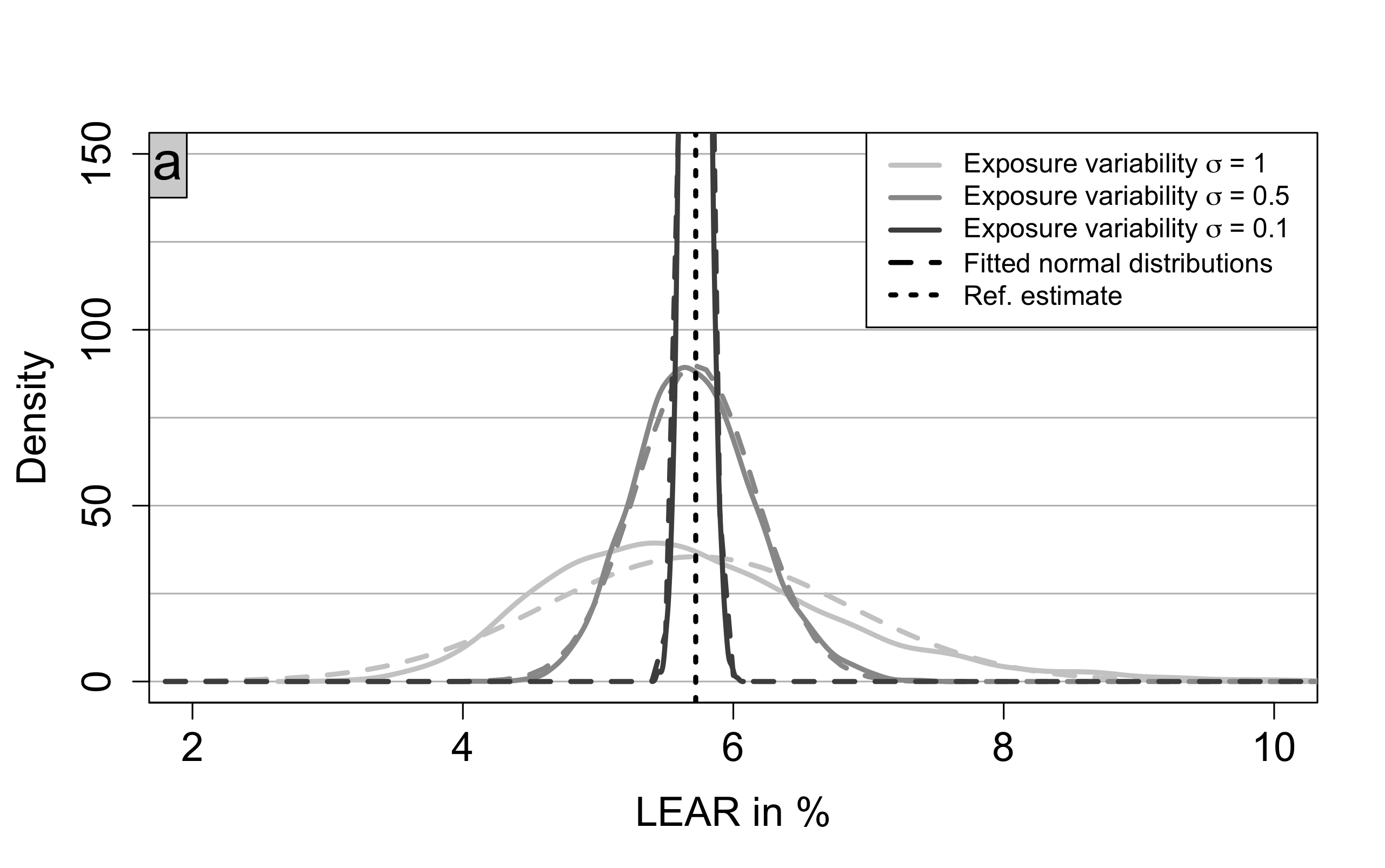}
    \caption{$LEAR$ densities for the simple linear risk model $ERR(t; \beta)=\beta W(t)$ with $\beta=0.0134$ and different $\sigma$ values. A normal distribution fitted on the $10,000$ $LEAR$ samples is shown for comparison (dashed lines).}
    \label{fig:expo_uncertainty_on_LEAR1}
\end{subfigure}
\hspace{\fill}
\begin{subfigure}[t]{0.48\textwidth}
    \includegraphics[width=\textwidth]{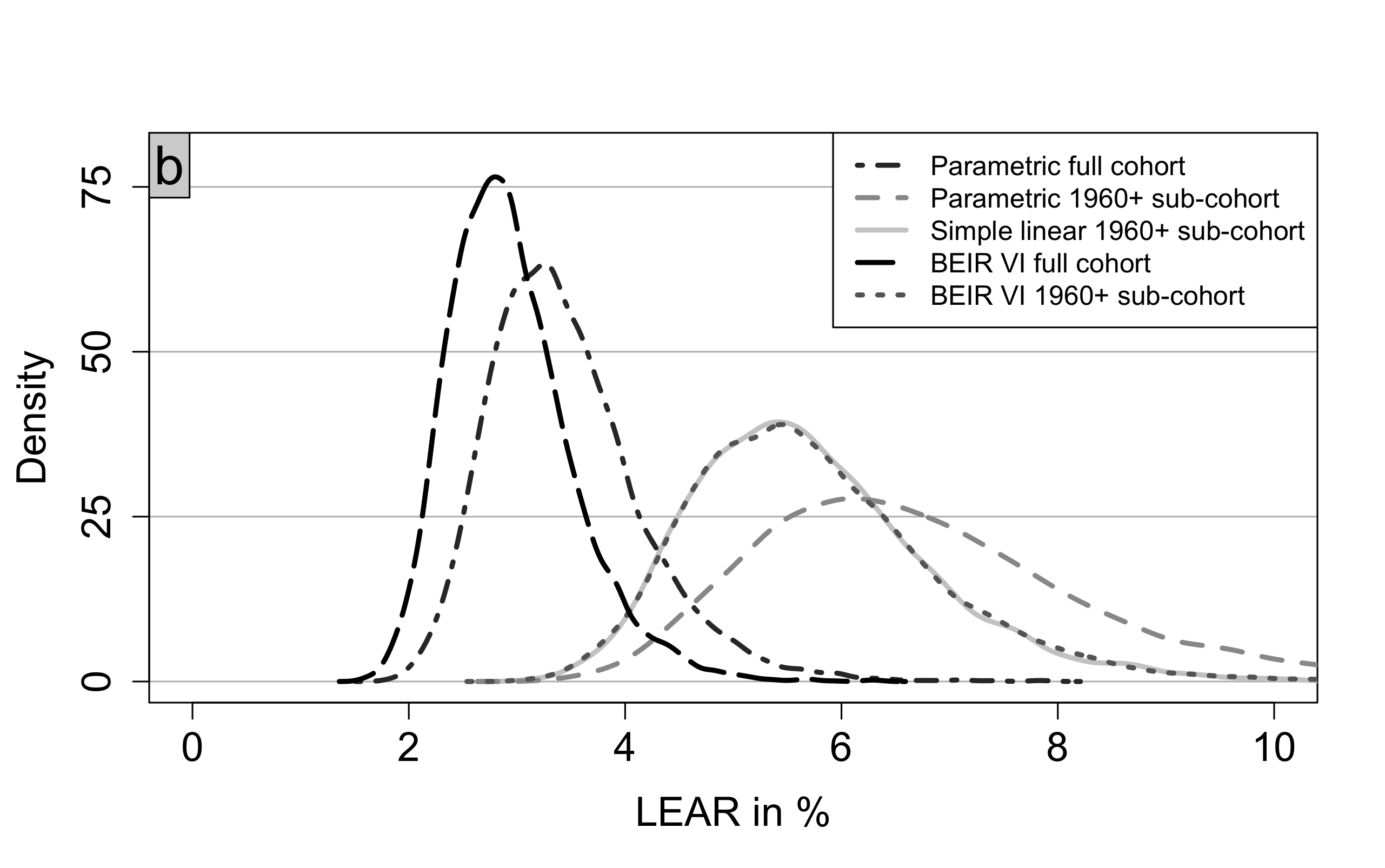}
     \caption{$LEAR$ densities for different risk models and fixed log-normal standard deviation $\sigma=1$.}
        \label{fig:expo_uncertainty_on_LEAR2}
\end{subfigure}
\caption{Density of $LEAR$ estimates based on the distribution of $10,000$ sampled values with random annual exposure in WLM from age 18-64 years (assumption (\ref{ln_w_random})) for varying $\sigma$ and different risk models.}
\label{fig:expo_uncertainty_on_LEAR}
\end{figure}

Similar to previous findings, differences between lifetime risk measures are minimal with only $RADS$ being considerably larger. Relative uncertainty interval spans are comparable across all lifetime risk measures and risk models, except for the parametric 1960+ sub-cohort model (wider intervals). Exposure-affected effect-modifying variables do not impose additional uncertainty (compare simple linear model to others, Suppl. Figure \ref{fig:expo_uncertainty_on_LEAR2}). As expected, smaller $\sigma$ leads to narrower and more symmetrical intervals due to less influence from right-skewed random exposure. This effect is stronger, the larger $\sigma$. For small $\sigma$, the probability mass of log-normal random variables is very centered with less heavy tails. The span of lifetime risk uncertainty intervals decreases proportionally with decreasing $\sigma$.\\

Lower exposure uncertainty (smaller $\sigma$) leads all lifetime risk measures (e.g., LEAR, Suppl. Figure \ref{fig:expo_uncertainty_on_LEAR1}) to resemble a normal distribution, regardless of the chosen risk model. This is because the assumed log-normal distribution approximates normality for smaller $\sigma$, which translates to the lifetime risk measures. After all, sums of independent normally distributed random variables are again normally distributed. Notably, this applies even to complex risk models with non-linear exposure-risk relations. \\

Note that the observed magnitude of uncertainty does not translate to the $LEAR$ per WLM, defined as the $LEAR$ divided by total cumulative exposure accrued over the entire exposure scenario in WLM. The quantity $LEAR$ per WLM is hardly affected by varying radon exposure \cite{Sommer_2024_Sensi}.
In conclusion, accounting for uncertainties in annual radon exposure (on a simple scale) results in an approximately normally distributed lifetime risk, regardless of the chosen lifetime risk measure or risk model. However, this approach requires knowledge or assumptions on exposure errors directly influencing uncertainty intervals. While this analysis can be extended to more complex exposure scenarios, it is beyond the scope here as exposure uncertainty assessment is not a core focus for lifetime risk calculations. Nevertheless, accurate exposure assessment and uncertainty quantification remain crucial for uranium miners cohorts and risk model derivation \cite{Wismut_Unsicherheiten_Teil1_2018, Wismut_Unsicherheiten_Teil2_2022}.

\end{document}